\documentstyle[12pt]{article}
\begin{document}
\renewcommand{\thefootnote}{\arabic{footnote}}
\newcommand{\fn}{\footnote}
\newcommand{\be}{\begin{equation}}
\newcommand{\bea}{\begin{eqnarray}}
\newcommand{\ee}{\end{equation}}
\newcommand{\eea}{\end{eqnarray}}
\newcommand{\sect}[1]{\setcounter{equation}{0}\bigskip\medskip\section{#1}\smallskip}
\newcommand{\subsect}[1]{\medskip\subsection{#1}\smallskip}
\newcommand{\subsubsect}[1]{\medskip\subsubsection{#1}\smallskip}
\renewcommand{\theequation}{\thesection.\arabic{equation}}
\newtheorem{proposition}{Proposition}[section]  
\newcommand{\bprop}{\medskip\begin{proposition} ~~\\ \it}
\newcommand{\eprop}{\end{proposition} \hfill $\Box$ }
\newcommand{\proof}{\medskip \noindent \\{\it Proof.} \rm }  
\newtheorem{naming}{Definition}[section]   
\newcommand{\bdefi}{\medskip\begin{naming} ~~\\ \it}
\newcommand{\edefi}{\end{naming} \hfill $\Diamond$ }
\newtheorem{example}{Example}[section]   
\def\bexam{\medskip\begin{example} ~~\\ \rm}
\def\eexam{\end{example} \hfill $\triangle$ }
\def\norm#1{{\vert\vert#1\vert\vert}}
\def\abs#1{{\vert#1\vert}}
\def\ha{\widehat{\cal A}}
\def\hc{\widehat{\cal C}}
\def\prim{Prim{\cal A} }
\def\comp{{\cal K}({\cal H})}
\def\inf1{{\cal L}^{(1,\infty)}}
\def\omca{\Omega {\cal A}}
\def\oca#1{\Omega^{#1}{\cal A}}
\def\omcad{\Omega_D {\cal A}}
\def\ocad#1{\Omega_D^{#1}{\cal A}}
\def\ota{\otimes_{\cal A}}
\def\otc{\otimes_{\IC}}
\def\join{{\protect{\footnotesize $\backslash\!/~$}}}  
\def\meet{{\protect{\footnotesize $/\!\backslash~$}}}
\def\bar#1{\overline{#1}}
\def\wh{\widehat}
\def\wt{\widetilde}
\def\bra#1{\left\langle #1\right|}
\def\ket#1{\left| #1\right\rangle}
\def\hs#1#2{\left\langle #1,#2\right\rangle}
\def\vev#1{\left\langle #1\right\rangle}
%
%
\def\sdp{\hbox{ \raisebox{.25ex}{\tiny $|$}\hspace{.2ex}{$\!\!\times $}} }
\def\pa{\partial}
\def\del{\nabla}
\def\a{\alpha}
\def\b{\beta}
\def\c{\raisebox{.4ex}{$\chi$}}
\def\d{\delta}
\def\e{\epsilon}
\def\ve{\varepsilon}
\def\f{\phi}
\def\vf{\varphi}
\def\g{\gamma}
\def\h{\eta}
\def\j{\psi}
\def\k{\kappa}
\def\l{\lambda}
\def\m{\mu}
\def\n{\nu}
\def\o{\omega}
\def\p{\pi}
\def\q{\theta}
\def\r{\rho}
\def\s{\sigma}
\def\t{\tau}
\def\u{\upsilon}
\def\x{\xi}
\def\z{\zeta}
\def\D{\Delta}
\def\F{\Phi}
\def\G{\Gamma}
\def\J{\Psi}
\def\L{\Lambda}
\def\O{\Omega}
\def\P{\Pi}
\def\Q{\Theta}
\def\S{\Sigma}
\def\U{\Upsilon}
\def\X{\Xi}
\def\Z{\Zeta}
\def\ca{{\cal A}}
\def\cb{{\cal B}}
\def\cc{{\cal C}}
\def\cd{{\cal D}}
\def\ce{{\cal E}}
\def\cf{{\cal F}}
\def\cg{{\cal G}}
\def\ch{{\cal H}}
\def\ci{{\cal I}}
\def\cj{{\cal J}}
\def\ck{{\cal K}}
\def\cl{{\cal L}}
\def\cm{{\cal M}}
\def\cn{{\cal N}}
\def\co{{\cal O}}
\def\cp{{\cal P}}
\def\cs{{\cal S}}
\def\ct{{\cal T}}
\def\cu{{\cal U}}
\def\cv{{\cal V}}
\def\cw{{\cal W}}
\def\cx{{\cal X}}
\def\cuai{{\cal Y}}
\def\cz{{\cal Z}}
\def\bca{\overline{{\cal A}}}
%
%
%
\def\inbar{\,\vrule height1.5ex width.4pt depth0pt}
\def\IC{\relax\,\hbox{$\inbar\kern-.3em{\rm C}$}}
\def\ID{\relax{\rm I\kern-.18em D}}
\def\IF{\relax{\rm I\kern-.18em F}}
\def\IH{\relax{\rm I\kern-.18em H}}
\def\II{\relax{\rm I\kern-.17em I}}
\def\I1{\relax{\rm 1\kern-.28em l}}
\def\IK{\relax{\rm I\kern-.18em K}}
\def\IM{\relax{\rm I\kern-.18em M}}
\def\IN{\relax{\rm I\kern-.18em N}}
\def\IP{\relax{\rm I\kern-.18em P}}
\def\IQ{\relax\,\hbox{$\inbar\kern-.3em{\rm Q}$}}
\def\IZ{\relax\,\hbox{$\inbar\kern-.3em{\rm Z}$}}
\def\IR{\relax{\rm I\kern-.18em R}}
\font\cmss=cmss10 \font\cmsss=cmss10 at 7pt
\def\Z{\relax\ifmmode\mathchoice {\hbox{\cmss Z\kern-.4em Z}}{\hbox{\cmss
Z\kern-.4em Z}} {\lower.9pt\hbox{\cmsss Z\kern-.4em Z}} {\lower1.2pt\hbox{\cmsss
Z\kern-.4em Z}}\else{\cmss Z\kern-.4emZ}\fi}
\def\otc{\otimes_{\IC}}
\def\otca{\otimes_{{\cal A}}}
\def\bc{{\bf C}}
\def\br{{\bf R}}
\def\bz{{\bf Z}}
\def\bn{{\bf N}}
\def\bm{{\bf M}}
\def\bt{{\bf T}}
\def\Up{\Uparrow}
\def\up{\uparrow}
\def\Dn{\Downarrow}
\def\dn{\downarrow}
\def\Ra{\Rightarrow}
\def\ra{\rightarrow}
\def\La{\Leftarrow}
\def\la{\leftarrow}
\def\lra{\longrightarrow}
\def\iff{\Leftrightarrow}
\thispagestyle{empty}
\pagenumbering{roman}
\setcounter{page}{0}
%
%
\vspace{.5cm}
\begin{center}{\Large \bf An Introduction to \\
~\\ 
Noncommutative Spaces and their Geometry}
\end{center}
\vspace{1cm}
\centerline{\large Giovanni Landi}
\vspace{.25cm}
\begin{center}
{\it Dipartimento di Scienze Matematiche, Universit\`a di Trieste, \\
P.le Europa 1, I-34127, Trieste, Italia.}
\end{center}
\begin{center}
{\it INFN, Sezione di Napoli, \\
Mostra d' Oltremare pad. 20, I-80125, Napoli, Italia.}
\end{center}
\vspace{1cm}
\centerline{\large Trieste, January 16, 1997}
\vspace{2cm}
\centerline{\large hep-th/9701078}
\vfill\eject
\tableofcontents

\vfill\eject
\section*{Preface}

These notes arose from a series of introductory seminars on noncommutative
geometry I gave at the University of Trieste in September 1995 during the X
Workshop on  Differential Geometric Methods in Classical Mechanics. It was Beppe
Marmo's suggestion that I wrote notes of the lectures.  

The notes are mainly an introduction to Connes' noncommutative geometry.  They could
serve as a `first aid kit' before one ventures into the beautiful but bewildering
landscape of Connes' theory \cite{Co1}.  The main difference with other available
introductions to Connes's work, notably Kastler's papers \cite{kastler} and also
Gracia-Bond\'{\i}a and Varilly paper \cite{VG}, is the emphasis on noncommutative
spaces seen as concrete spaces.

Important examples of noncommutative spaces are provided by noncommutative lattices.
The latter are the subject of intensive work I am doing in collaboration with A.P.
Balachandran, Giuseppe Bimonte, Elisa Ercolessi, Fedele Lizzi, Gianni
Sparano and Paulo Teotonio-Sobrinho. These notes are also meant to be an introduction
to these researches. There is still a lot of work in progress and by no means they
can be considered as a review of everything we have achieved so far. Rather, I hope
they will show the relevance and potentiality for physical theories of noncommutative
lattices.

\bigskip

\noindent
Acknowledgments. \\
I am indebted to several people for help and suggestions at different stages of this
project: A.P.~Balachandran, G.~Bimonte, U.~Bruzzo, M.~Carfora, R.~Catenacci,
L.~Dabrowski, G.F.~Dell'Antonio, B.~Dubrovin, E.~Ercolessi, J.M.~Gracia-Bond\'{\i}a,
P.~Hajac, F.~Lizzi, G.~Marmo, C.~Reina, C. Rovelli, G.~Sewell, P.~Siniscalco,
G.~Sparano  P.~Teotonio-Sobrinho, J.C.~V\'{a}rilly.

\vfill\eject

\renewcommand{\thefootnote}{\arabic{footnote}}
\setcounter{footnote}{0}
\pagenumbering{arabic}
\sect{Introduction}\label{se:int}

In the last fifteen years, there has been an increasing interest in
noncommutative (and/or quantum geometry) both in mathematics and physics. 

In A. Connes' functional analytic approach \cite{Co1}, noncommutative
$C^*$-algebras are the  `dual' arena for noncommutative
topology. The (commutative)
Gel'fand-Naimark theorem (see for instance \cite{FD}) states
that there is a complete equivalence between the category of (locally) compact
Hausdorff spaces and (proper and) continuous maps and the category of commutative
(non necessarily) unital
$C^*$-algebras and
$^*$-homomorphisms. Any commutative
$C^*$-algebra can be realized as the $C^*$-algebra of complex valued functions over a
(locally) compact Hausdorff space. A noncommutative $C^*$-algebra will be now thought
of as the algebra of continuous functions on some `virtual noncommutative space'.  
The attention will be switched from spaces, which in general do not even
exist `concretely', to algebras of functions. 

Connes has also developed a new calculus which replaces the usual differential
calculus. It is based on the notion of real spectral triple $(\ca, \ch, D, J)$ where
$\ca$ is a noncommutative $^*$-algebra (in fact, in general not necessarily a
$C^*$-algebra), $\ch$ is a Hilbert space on which $\ca$ is realized as an
algebra of bounded operators, and $D$ is an operator on $\ch$ with suitable
properties and which contains (almost all) the `geometric' information. The
antilinear isometry $J$ on $\ch$ will provide a real structure on the triple.
With any closed $n$-dimensional Riemannian spin manifold $M$ there is associated a
canonical spectral triple with $\ca=C^\infty(M)$, the algebra of complex valued
smooth functions on $M$; $\ch=L^2(M,S)$, the Hilbert space of square integrable
sections of the irreducible spinor bundle over $M$;
$D$ ~the Dirac operator associated with the Levi-Civita connection. For this triple,
Connes' construction gives back the usual differential calculus on $M$.
In this case $J$ is the composition of the charge conjugation operator with usual 
complex conjugation.

Yang-Mills and gravity theories stem from the notion of connection (gauge or
linear) on vector bundles. The possibility of extending these notions to the
realm of noncommutative geometry relies on another classical duality. Serre-Swan
theorem \cite{Sw} states that there is a complete equivalence between the
category of (smooth) vector bundles over a (smooth) compact space and bundle
maps and the category of projective modules of finite type over commutative
algebras and module morphisms. The space $\G(E)$ of (smooth) sections of a vector
bundle $E$ over a compact space is a  projective module of finite type over the
algebra $C(M)$ of (smooth) functions over $M$ and any finite projective
$C(M)$-module can be realized as the module of sections of some bundle over $M$.
        
With a noncommutative algebra $\ca$ as the starting ingredient, the (analogue
of) vector bundles will be projective modules of finite type over $\ca$. One
then develops a full theory of connections which culminates in the definition of
a Yang-Mills action. Needless to say, starting with the
canonical triple associated with an ordinary manifold one recovers the usual
gauge theory. But now, one has a much more general setting.  In \cite{CL}
Connes and Lott computed the Yang-Mills action for a space $M \times Y$ which is
the product of a Riemannian spin manifold $M$ by a `discrete' internal space
$Y$ consisting of two points. The result is a Lagrangian which reproduces the
Standard Model with its Higgs sector with quartic
symmetry breaking self-interaction and the parity violating Yukawa coupling with
fermions. A nice feature of the model is a geometric interpretation of the Higgs
field which appears as the component of the gauge field in the internal
direction. Geometrically, the space $M \times Y$ consists of two sheets which
are at a distance of the order of the inverse of the mass scale of the theory.
Differentiation on $M \times Y$ consists of differentiation on each copy of $M$
together with a finite difference operation in the $Y$ direction. A gauge
potential $A$ decomposes as a sum of an ordinary differential part $A^{(1,0)}$
and a finite difference part $A^{(0,1)}$ which gives the Higgs field.  
    
Quite recently Connes \cite{Co3} has proposed a purely `gravity' action which, for
a suitable noncommutative algebra $\ca$ (noncommutative geometry of the Standard
Model) yields the Standard Models Lagrangian coupled with Einstein gravity. The group
$Aut(\ca)$ plays the r\^ole of the diffeomorphism group while the normal subgroup
$Inn(\ca) \subset Aut(\ca)$ gives the gauge transformations. Internal fluctuations of
the geometry, produced by the action of inner automorphisms, gives the gauge degrees
of freedom.
 
A theory of linear connections and Riemannian geometry, culminating in the analogue of
the Hilbert-Einstein action in the context of noncommutative geometry has been
proposed in \cite{CFF}. Again, for the canonical triple one recovers the usual
Einstein gravity.  When computed for a Connes-Lott space $M \times Y$ as in
\cite{CFF}, the action produces a Kaluza-Klein model which contains the usual
integral of the scalar curvature of the metric on $M$, a minimal coupling for the
scalar field to such a metric, and a kinetic term for the scalar field.  A somewhat
different model of geometry on the space $M \times Y$ produced an action which is is
just the Kaluza-Klein action of unified gravity-electromagnetism consisting  of the
usual gravity term, a kinetic term for a minimally coupled scalar field and an
electromagnetic term \cite{LNW}.

Algebraic $K$-theory of an algebra $\ca$ as the study of equivalence classes of
projective module of finite type over $\ca$ provides analogues of topological
invariants of the `corresponding virtual spaces'. On the other hand, cyclic
cohomology provides analogues of differential geometric invariants. $K$-theory and
cohomology are connected by the Chern characters. This has found a beautiful
application by Bellissard \cite{Be} to quantum Hall effect. He has constructed a
natural cyclic $2$-cocycle on the noncommutative algebra of function on the Brillouin
zone. The Hall conductivity is just the pairing between this cyclic $2$-cocycle and
an idempotent in the algebra: the spectral projection of the Hamiltonian. A crucial
r\^ole is played by the noncommutative torus \cite{Ri}.

In this notes we present a self-contained introduction to a limited part of
Connes' noncommutative theory, without even trying to cover all aspects of the
theory and finalized to the presentation of some of the physical applications.

In Section~\ref{se:saf}, we introduce $C^*$-algebras and the (commutative) 
Gel'fand-Naimark theorem. We then pass to structure spaces of noncommutative 
$C^*$-algebras. We describe to some extent the space $\prim$ of an algebra
$\ca$ with its natural Jacobson topology. Examples of such spaces turn out to be
relevant in an approximation scheme to `continuum' topological spaces by means of
lattices with a non trivial $T_0$ topology \cite{So}. Such lattices are truly
noncommutative lattices since their algebras of continuous functions are
noncommutative $C^*$-algebras of operator valued functions. Techniques from
noncommutative geometry have been used to constructs models of gauge theory on these
noncommutative lattices \cite{ncl,bbllt}. Noncommutative lattices are described at
length in Section~\ref{se:ncl}.

Section~\ref{se:spc} is devoted to the theory of infinitesimals and the spectral
calculus. We first describe the Dixmier trace which play a fundamental r\^ole in
the theory of integration. Then the notion of spectral triple is introduced
with the associated definition of distance and integral on a `noncommutative
space'. We work out in detail the example of the canonical triple associated
with any Riemannian spin manifold. Noncommutative forms are then introduced in
Section~\ref{se:ndf}. Again, we show in detail how to recover the usual
exterior calculus of forms. 

In the first part of Section~\ref{se:vbg}, we describe abelian gauge theories in
order to get some feelings about the structures. We then develop the theory  of
projective modules and describe the Serre-Swan theorem. Also the notion of
Hermitian structure, an algebraic counterpart of a metric, is described.
We finish by presenting the connections, compatible connections, and gauge
transformations.

In Sections~\ref{se:ftm} and~\ref{se:gra} we present field theories on modules. In
particular we show how to construct Yang-Mills and fermionic models. Gravity models
are treated in Sections~\ref{se:gra}.  In Section~\ref{se:qmm} we describe a simple
quantum mechanical system on a noncommutative lattice, namely the
$\theta$-quantization of a particle on a noncommutative lattice approximating the
circle. 

We feel we should warn the interested reader that we shall not give any detailed
account of the construction of the Standard Model in noncommutative geometry nor of
the use of the latter for model building in particle physics. We shall limit ourself
to a very sketchy overview while referring to the existing and rather useful
literature on the subject.
 
The appendices contain related material to the one developed in the
text. 

As alluded to before, the territory of noncommutative or quantum geometry is so
vast and new regions are discovered at a high speed that the number of relevant
papers is overwhelming. It is impossible to even think of cover `everything. We just
finish this introduction with a very partial list of `further readings'. 
The generalization from classical (differential) geometry to noncommutative
(differential) geometry it is not unique. This is a consequence of the existence of
several type of noncommutative algebras. A different  approach to noncommutative
calculus is the so called `derivation based calculus' proposed in \cite{D-V}. 
Given a non commutative algebra $\ca$ one takes as the analogue of vector fields the
Lie algebra $Der\ca$ of derivations of $\ca$. Beside the fact that, due to
noncommutativity, $Der\ca$ is a module only over the center of
$\ca$, there are several algebras which admits only few derivations. We refer to
\cite{Mad} for details and several applications to Yang-Mills models and gravity
theories. For Hopf algebras and quantum groups and their applications to Quantum
Field Theory we refer to \cite{Dr,FK,Jo,Kas,Maj,Pi,Swe}. Twisted (or pseudo) groups
have been proposed in \cite{Wo}. For other interesting quantum spaces such as the
quantum plane we refer to \cite{Man} and \cite{WZ}. Very interesting work on the
structure of the space-time has been done in \cite{DFR}.

The reference for Connes' noncommutative geometry is `par excellence'
his book \cite{Co1}. Very helpful has been the paper \cite{VG}.       

\vfill\eject
\sect{Noncommutative Spaces and Algebras of Functions}\label{se:saf}

The starting idea of noncommutative geometry is the shift from spaces to algebras of
functions defined on them. In general, one has only the algebra and there is no
analogue of space whatsoever. In this section  we shall give some general facts
about algebras of  (continuous) functions on (topological) spaces. In particular we
shall try to make some sense  of the notion of  `noncommutative space'. 

\subsect{Algebras}\label{se:csa}
Here we present mainly the objects that we shall need later on while referring
to \cite{BR, Di1, Pe} for details. In the sequel, any algebra~\index{algebra} $\ca$
will be an algebra over the  field of complex numbers $\IC$. This means that $\ca$
is a vector space over $\IC$, so that objects like $\a a + \b b$ with $a, b \in \ca$
and
$\a,
\b \in \IC$, make sense. Also, there is a product $\ca \times \ca \ra \ca$,
$\ca \times \ca \ni(a,b) \mapsto ab \in \ca$, which is distributive over 
addition,
\be
a(b+c) = ab + ac~, ~~~(a+b)c=ac+bc~, ~~~\forall ~a,b,c \in \ca~.
\ee
In general, the product is not commutative so that 
\be
ab \not= ba~.
\ee
We shall assume that $\ca$ has a unit $\II$. Here and there we shall comment on
the  situations for which this is not the case.\\
The algebra $\ca$ is called a {\it $^*$-algebra} if it admits an
(antilinear) involution $^* : \ca \ra \ca$ with the properties,
\bea
&& a^{**} = a~, \nonumber \\
&& (ab)^* = b^* a^*~, \nonumber \\
&& (\a a + \b b)^* = \bar{\a} a^* + \bar{\b}b^*~, 
\eea
for any $a,b \in \ca$ and $\a, \b \in \IC$ and bar denoting usual
complex conjugation. \\
A {\it normed algebra}~\index{normed algebra} $\ca$ is an algebra with a norm 
$\norm{\cdot} : \ca \ra \IR$\index{norm} which has the properties, 
\bea
&& \norm{a} \geq 0~, ~~~ \norm{a} = 0 ~\iff ~ a = 0~, \nonumber \\
&& \norm{\a a} = |\a| \norm{a}, \nonumber \\
&& \norm{a+b} \leq \norm{a} + \norm{b}, \nonumber \\
&& \norm{ab} \leq \norm{a} \norm{b}, \label{normprop}
\eea
for any $a,b \in \ca$ and $\a \in \IC$. The third condition is called the
triangle inequality while the last one is called the product inequality. The
topology defined by the norm is called the {\it norm}~\index{topology!norm} or 
{\it uniform topology}~\index{topology!uniform}.  The
corresponding  neighborhoods of any $a \in \ca$ are given by
\be
U(a, \ve) = \{ b \in \ca ~|~ \norm{a-b} < \ve \}~, ~~~ \ve > 0~. 
\ee
A {\it Banach algebra}~\index{Banach algebra} is a normed algebra which is complete
in the uniform  topology.\\
A {\it Banach $^*$-algebra} is a normed $^*$-algebra which is complete
and such that 
\be\label{ss1}
\norm{a^*} = ~\norm a,~~~ \forall~ a \in \ca~.
\ee
A {\it $C^*$-algebra} $\ca$ is a Banach $^*$-algebra
whose norm satisfies the additional identity~\index{cstaral@$C^*$-algebra}
\be\label{ss2}
\norm{a^*a} = \norm{a}^2,~~~ \forall~ a \in \ca~.
\ee
In fact, this property, together with the product inequality yields
(\ref{ss1}) automatically. Indeed, $\norm{a}^2 = \norm{a^*a} \leq
\norm{a^*}\norm{a}$ from which  ${\norm a} \leq {\norm{a^*}}$. By interchanging $a$
with $a^*$ one gets $\norm{a^*}\leq \norm{a}$ and in turn (\ref{ss1}).    

\bexam\label{ex:cofu}
The commutative algebra $\cc(M)$ of continuous functions on a compact Hausdorff
topological space $M$, with $^*$ denoting complex conjugation and the norm given
by the supremum norm~\index{norm!supremum}, 
\be\label{suno}
\norm f _\infty = \sup_{x\in M}|f(x)|~.
\ee 
If $M$ is not compact but only locally compact, then one should take  the
algebra $\cc_0(M)$ of continuous functions vanishing at
infinity;  this algebra has no unit. Clearly
$\cc(M) =
\cc_0(M)$ if $M$ is compact. One can prove that $\cc_0(M)$ (and a fortiori $\cc(M)$
if $M$ is compact) is complete in the supremum norm
\fn{Recall that a function $f: M \ra \IC$ on a locally compact Hausdorff space is 
said to {\it vanish at infinity} if for every $\e > 0$ there exists a compact set 
$K \subset M$ such that $|f(x)| < \e$ for all $x \notin K $. As
mentioned in Appendix~\ref{se:bnt}, the algebra $\cc_0(M)$ is the closure in the norm
(\ref{suno}) of the algebra of functions with compact support. The function $f$ is
said to have compact support if the space $K_f =: \{x \in M ~|~ f(x) \not= 0 \}$ is
compact\cite{Ru}.}. 
\eexam
\bexam
The noncommutative algebra $\cb(\ch)$ of  bounded linear
operators on an infinite dimensional Hilbert space $\ch$ with  involution
$^*$ given by the adjoint and the norm given by  
the operator norm~\index{norm!operator}, 
\be\label{opnorm}
\norm{B} = \sup\{\norm{B \chi} : \chi \in \ch, \norm{\chi} \leq 1 \}~.
\ee 
\eexam
\bexam
As a particular case of the previous, consider the noncommutative algebra 
${\bf M}_n(\IC)$ of $n\times n$ matrices $T$ with complex entries, with $T^*$ given
by the Hermitian conjugate of $T$. The norm (\ref{opnorm}) can also be
equivalently written as 
\be\label{opnormfin}
\norm{T} = {\rm ~the ~positive ~square ~root ~of ~the ~largest  ~eigenvalue ~of
~T^* T}~.
\ee
On the algebra ${\bf M}_n(\IC)$ one could also define a different norm,
\be\label{opnormfin1}
\norm{T}' = sup\{T_{ij}\}~, ~~~ T = (T_{ij})~. 
\ee
One can easily convince oneself that this norm is not a $C^*$-norm, the property
(\ref{ss2}) being not fulfilled. It is worth noticing though, that the two norms
(\ref{opnormfin}) and (\ref{opnormfin1}) are equivalent as Banach norm in the sense
that they define the same topology on ${\bf M}_n(\IC)$: any ball in the topology of
the norm (\ref{opnormfin}) is contained in a ball in the topology of the norm
(\ref{opnormfin1}) and viceversa.
\eexam

\bigskip
A (proper, norm closed) subspace $\ci$ of the algebra $\ca$ is a {\it left ideal}
(respectively a {\it right ideal}) if $a \in \ca$ and $b \in \ci$ imply that $a b
\in \ci $ (respectively $b a \in \ci $). A {\it two-sided ideal}\index{ideal} is a
subspace which is both a left and a right ideal. The ideal $\ci$ (left, right or
two-sided) is called {\it maximal}\index{ideal!maximal} if there exists no other
ideal of the same kind in which $\ci$ is contained. Each ideal is automatically an
algebra. If the algebra
$\ca$ has an involution, any
$^*$-ideal (namely an ideal which contains the $^*$ of any of its elements) is
automatically two-sided. If $\ca$ is a Banach $^*$-algebra and $\ci$ is a two-sided
$^*$-ideal which is also closed (in the norm topology), then the quotient $\ca /
\ci$ can be made a Banach $^*$-algebra. Furthermore, if $\ca$ is a $C^*$-algebra,
then the quotient $\ca / \ci$ is also a $C^*$-algebra. The $C^*$-algebra $\ca$ is
called {\it simple}\index{algebra!simple} if it has no nontrivial two-sided ideals. A
two-sided ideal
$\ci$ in the $C^*$-algebra $\ca$ is called {\it essential}\index{ideal!essential} in
$\ca$ if any other non-zero ideal in $\ca$ has a non-zero intersection with it. 

If $\ca$ is any algebra, the {\it resolvent set}\index{resolvent set} $r(a)$ of an
element
$a
\in
\ca$ is the subset of complex numbers given by
\be
r(a) = \{ \l \in \IC ~|~ a - \l \II ~{\rm is ~invertible} \}~.   
\ee 
For any $\l \in r(a)$, the inverse $(a - \l \II)^{-1}$ is called the 
{\it resolvent}\index{resolvent}
of $a$ at $\l$. The complement of $r(a)$ in $\IC$ is called the 
{\it spectrum}\index{spectrum}
$\s(a)$ of $a$. While for a general algebra, the spectra of its elements may be
rather complicate, for $C^*$-algebras they are quite nice. If $\ca$ is a 
$C^*$-algebra, it
turns out that the spectrum of any of its element $a$ is a nonempty compact
subset of $\IC$. The {\it spectral radius}\index{spectral radius} $\r(a)$ of $a\in
\ca$ is given by
\be
\r(a) = sup\{ |\l| ~, ~\l \in r(a) \}
\ee 
and, $\ca$ being a $C^*$-algebra, it turns out that
\be\label{unique}
\r(a) = \norm{a}~, ~~~\forall ~a\in\ca~.
\ee 
A $C^*$-algebra~\index{cstaral@$C^*$-algebra} is really such for a unique
norm given by the spectral radius~\index{spectral radius} as in (\ref{unique}): the
norm is uniquely determined by the algebraic structure. 

An element $a \in \ca$ is called {\it self-adjoint} if $a=a^*$. The spectrum of any
such element is real and $\s(a) \subseteq [ -\norm{a},\norm{a}]$, 
$\s(a^2) \subseteq [0, \norm{a}^2]$. An element $a \in \ca$ is called {\it
positive} if it is self-adjoint and its spectrum is a subset of the positive
half-line. It turns out that the element $a$ is positive if and only if
$a=b^*b$ for some $b \in \ca$. If $a \not= 0$ is positive, one also writes 
$a > 0$. 

\bigskip
A {\it $^*$-morphism}~\index{cstaral@$C^*$-algebra!morphism} between two
$C^*$-algebras
$\ca$ and $\cb$ is any
$\IC$-linear map $\pi : \ca \ra \cb$ which in addition satisfies the
conditions,
\bea
&& \pi(ab) = \pi(a) \pi(b)~, \nonumber \\
&& \pi(a^*) = \pi(a)^* ~, ~~~ \forall ~a, b \in \ca~.
\eea       
These conditions automatically imply that $\pi$ is positive, namely $\pi(a) \geq
0$ if $a \geq 0$. Indeed, if $a \geq 0$, then $a=b^* b $ for some $b\in\ca$; as
a consequence,
$\pi(a) = \pi(b^* b) = \pi(b)^* \pi(b) \geq 0$. It also turns out that
$\pi$ is automatically continuous, norm decreasing,
\be
\norm{\pi(a)}_{\cb} ~\leq~ \norm{a}_{\ca}~, ~~~\forall ~a \in \ca~, 
\ee
and the image $\pi(\ca)$ is a $C^*$-subalgebra of $\cb$. A $^*$-morphism $\pi$
which is also bijective as a map, is called a {\it $^*$-isomorphism} (the inverse
map $\pi^{-1}$ is automatically a $^*$-morphism).  

\bigskip

A {\it representation}~\index{cstaral@$C^*$-algebra!representation}
of a $C^*$-algebra $\ca$~\index{representation}
 is a pair $(\ch, \pi)$ where 
$\ch$ is a Hilbert space and $\pi$ is a $^*$-morphism
\be
\pi : \ca \lra \cb(\ch)~,
\ee
with $\cb(\ch)$ the $C^*$-algebra of bounded operators on $\ch$. \\ 
The representation $(\ch, \pi)$ is called 
{\it faithful}\index{representation!faithful} if $ker(\pi) = \{0\}$, so
that $\pi$ is a $^*$-isomorphism between $\ca$ and $\pi(\ca)$. One proves that a
representation is faithful if and only if $\norm{\pi(a)} = \norm{a}$ for any $a\in
\ca$ or $\pi(a) > 0$ for all $a > 0$. \\ 
The representation $(\ch, \pi)$ is called 
{\it irreducible}\index{representation!irreducible} if the only closed
subspaces of $\ch$ which are invariant under the action of $\pi(\ca)$ are the
trivial subspaces $\{0\}$ and $\ch$. One proves that a representation is
irreducible if and only if the commutant $\pi(\ca)'$ of $\pi(\ca)$, i.e. the
set of of elements in $\cb(\ch)$ which commute with each element in
$\pi(\ca)$, consists of multiples of the identity operator.\\
Two representations $(\ch_1, \pi_1)$ and $(\ch_2, \pi_2)$ are said to be
{\it equivalent}\index{representation!equivalence of} (or more precisely, {\it
unitary equivalent}) if there exists a unitary operator $U : \ch_1 \ra \ch_2$, such 
that
\be
\pi_1(a) = U^* \pi_2(a) U ~, ~~~ \forall ~ a \in \ca~.
\ee
In the Appendix~\ref{se:gns} we describe the notion of states of a
$C^*$-algebra and the representations associated with them via the
Gel'fand-Naimark-Segal construction.  

The subspace $\ci$ of the $C^*$-algebra $\ca$ is called a {\it primitive
ideal}\index{ideal!primitive} if
$\ci = ker(\pi)$ for some irreducible representation $(\ch, \pi)$ of $\ca$. Notice
that $\ci$ is automatically a two-sided ideal which is also closed. If 
$\ca$ has a faithful irreducible representation on some Hilbert space so that the
set $\{0 \}$ is a primitive ideal, it is called a {\it primitive
$C^*$-algebra}~\index{cstaral@$C^*$-algebra!primitive}. The set $\prim$ of all
primitive ideals of the $C^*$-algebra $\ca$ will play a crucial role in the
following.

\subsect{Commutative Spaces}\label{se:gnt}

The content of the commutative Gel'fand-Naimark theorem
~\index{Gel'fand-Naimark theorem!commutative|(} is precisely  the fact that given
{\it any} commutative
$C^*$-algebra
$\cc$, one can  reconstruct a Hausdorff topological space $M$ such that $\cc$ is 
isometrically $*$-isomorphic to the algebra of continuous functions $\cc(M)$
\cite{Di1,FD}.

In this section $\cc$ denotes a fixed commutative $C^*$-algebra with unit. Given such a
$\cc$, we let $\hc$ denote the {\it structure space}  of $\cc$, namely the space
of equivalence classes of irreducible  representations  of $\cc$.
The trivial representation given by $\cc\rightarrow \{0\}$ is not included in $\hc$. 
The $C^*$-algebra $\cc$ being commutative, every irreducible representation is
one-dimensional. It is then a (non-zero) $^*$-linear  functional $\f : \cc \ra \IC$
which is multiplicative, i.e. it satisfies $\f(ab) = \f(a)\f(b)$, for any $a, b \in 
\cc$. It follows that
$\f(\II) = 1, ~\forall~ \f\in \hc$.  Any such  multiplicative functional is also
called a {\it character}~\index{character}  of $\cc$.  The space $\hc$ is then also
the space of all characters of $\cc$. 

The space $\hc$ is made a topological space, called the 
{\it Gel'fand space}~\index{Gel'fand!space} of
$\cc$, by endowing it with the {\it Gel'fand topology}~\index{Gel'fand!topology},
namely with the topology of pointwise convergence on $\cc$. A sequence
$\{\f_\l \}_{\l \in \L}$ ($\Lambda$ is any directed set) of elements of $\hc$
converges to $\f \in \hc$ if and only if for any $c \in \cc$, the sequence
$\{\f_\l(c)\}_{\l \in \L}$ converges to $\f(c)$ in the topology of $\IC$. The
algebra $\cc$ having a unit,
$\hc$ is a compact Hausdorff space
\fn{Recall that a topological space is 
called Hausdorff~\index{topology!Hausdorff} if for any two points of
the space there are two open disjoint neighborhoods each containing one of the
points \cite{Ke}.}. The space $\hc$ would be only locally compact if $\cc$ is
without unit. 

Equivalently, $\hc$ could be taken to be the space of maximal ideals
(automatically two-sided) of $\cc$ instead of the space of irreducible
representations
\fn{If there is no unit, one needs to consider ideals which are {\it
regular}\index{ideal!regular} 
(also called {\it modular}\index{ideal!modular})
as well. An ideal $\ci$ of a general algebra $\ca$ being called regular if there is a
unit in $\ca$ modulo $\ci$, namely an element $u \in
\ca$ such that $a - a u$ and $a - u a$ are in $\ci$ for all $a \in \ca$ \cite{FD}.
If $\ca$ has a unit, then any ideal is automatically regular.}. 
The $C^*$-algebra $\cc$ being commutative, these two  constructions agree because, on
one side, kernels of  (one-dimensional) irreducible representations are maximal
ideals, and, on  the other side, any maximal ideal is the kernel of an irreducible 
representation \cite{FD}. Indeed, consider $\f \in \hc$.  Then, since
$\cc = Ker(\f) \oplus \IC$, the ideal $Ker(\f)$ is of  codimension one and so is a
maximal ideal of $\cc$. Conversely, suppose that $\ci$ is a maximal ideal of
$\cc$. Then, the natural representation of $\cc$ on $\cc / \ci$ is irreducible,
hence one-dimensional. It follows that $\cc / \ci \cong \IC$, so that the
quotient homomorphism $\cc \ra \cc / \ci$ can be identified with an element $\f
\in \hc$. Clearly, $\ci = Ker(\f)$. When thought of as a space of maximal
ideals, $\hc$ is given the 
{\it Jacobson topology}~\index{Jacobson topology} (or {\it hull kernel topology})
producing a space which is homeomorphic to the one constructed by means of the
Gel'fand topology. We shall later describe in details the Jacobson topology. 

\bexam
Let us suppose that the algebra $\cc$ is generated by $N$-commuting 
self-adjoint elements $x_1, \dots, x_N$. Then the structure space 
$\hc$ can be identified with a compact subset of $\IR^N$ by the map \cite{Co2},
\be
\f \in \hc ~ \lra ~ (\f(x_1), \dots, \f(x_N) ) \in \IR^N~,
\ee and the range of this map is the joint spectrum~\index{spectrum!joint} of 
$x_1, \dots, x_N$, namely the set of all $N$-tuples of eigenvalues  corresponding
to common eigenvectors. 
\eexam

In general, if $c \in \cc$, its {\it Gel'fand transform}~\index{Gel'fand!transform}
$\hat{c}$ is the  complex-valued function on $\hc$, $\hat{c} : \hc \ra \IC$, given
by 
\be\label{gn1}
\hat{c}(\f) = \f(c)~, ~~~ \forall ~\f \in \hc~.
\ee 
It is clear that $\hat{c}$ is continuous for each $c$. We thus get the
interpretation of elements in $\cc$ as $\IC$-valued continuous functions on $\hc$. 
The Gel'fand-Naimark theorem states that all continuous functions on 
$\hc$ are of the form (\ref{gn1}) for some $c \in \cc$ \cite{Di1,FD}.
\bprop\label{th:gn} 
Let $\cc$ be a commutative $C^*$-algebra. Then, the Gel'fand transform 
$c \ra \hat{c}$ is an isometric $*$-isomorphism of $\cc$ onto $\cc(\hc)$; isometric
meaning that 
\be\label{gn2}
\norm{\hat{c}}_\infty = \norm{c}~, ~~~\forall~ c \in \cc~,
\ee 
with $\norm{\cdot}_{\infty}$ the supremum norm on $\cc(\hc)$ as in (\ref{suno}). 
\eprop

Suppose now that $M$ is a (locally) compact topological space.  As we have seen in
Example~\ref{ex:cofu} of Section~\ref{se:csa}, we have a natural $C^*$-algebra
$\cc(M)$.  It is natural to ask what is the relation between the Gel'fand space 
$\wh{\cc(M)}$ and $M$ itself. It turns out that this two spaces can be identified 
both setwise and topologically.  First of all, each $m \in M$ gives a complex
homomorphism $\f_m \in \wh{\cc(M)}$ through the evaluation map,
\be\label{gn3}
\f_m : \cc(M) \ra \IC~,~~~ \f_m(f) = f(m)~.
\ee 
Let $\ci_m$ denote the kernel of $\f_m$, namely the maximal ideal of 
$\cc(M)$ consisting of all functions vanishing at $m$. We have the  following
\cite{Di1,FD}, 
\bprop\label{th:gn1}  
The map $\f$ of (\ref{gn3}) is a homeomorphism of
$M$ onto $\wh{\cc(M)}$. Equivalently, every maximal ideal of
$\cc(M)$ is of the form $\ci_m$ for some $m \in M$. 
\eprop 

\noindent
The previous two theorems set up a one-to-one correspondence between  the
$*$-isomorphism classes of commutative $C^*$-algebras and the  homeomorphism
classes of locally compact Hausdorff spaces. Commutative $C^*$-algebras with
unit correspond to compact Hausdorff spaces.  In fact, this correspondence is a
complete duality between the category of (locally) compact Hausdorff spaces and
(proper
\fn{Recall that a continuous map between two locally compact Hausdorff spaces $f: X
\ra Y$ is called {\it proper} if $f^{-1}(K)$ is a compact subset of $X$ when $K$ is
a compact subset of $Y$. }\index{proper map}
  and) continuous maps and the category of commutative
(non necessarily) unital $C^*$-algebras and $^*$-homomorphisms. Any commutative
$C^*$-algebra can be realized as the $C^*$-algebra of complex valued functions over
a (locally) compact Hausdorff 
space.~\index{Gel'fand-Naimark theorem!commutative|)}
Finally, we mention that the space $M$ is 
metrizable~index{topological space!metrizable},
namely its topology comes from a metric, if and only if the
$C^*$-algebra is norm 
separable,~\index{cstaral@$C^*$-algebra!separable} namely it admits a dense (in norm)
countable subset. Also  it is 
connected~index{topological space!connected} if the corresponding
algebra has no projectors,~index{projector} namely self-adjoint, $p^*=p$,
idempotents,~index{idempotent} $p^2=p$, \cite{Colh}.

\subsect{Noncommutative Spaces}\label{se:nca}

The scheme described in the previous section cannot be directly generalized to a
noncommutative $C^*$-algebra.  To show some of the features of the general case, let
us consider the  simple example (taken from \cite{Co2}) of the algebra 
\be 
{\bf M}_2(\IC) = \{ 
\left[
\begin{array}{cc} a_{11} & a_{12} \\ a_{21} & a_{22} 
\end{array}
\right]~, ~a_{ij} \in \IC \}~. 
\ee The commutative subalgebra of diagonal matrices
\be
\cc = \{ 
\left[
\begin{array}{cc}
\l & 0 \\ 0 & \m 
\end{array}
\right]~, ~\l, \m \in \IC \}~,
\ee has a structure space consisting of two points given by the characters 
\be
\f_1(\left[
\begin{array}{cc}
\l & 0 \\ 0 & \m 
\end{array}
\right]) = \l~, ~~~
\f_2(\left[
\begin{array}{cc}
\l & 0 \\ 0 & \m 
\end{array}
\right]) = \m~.
\ee 
These two characters extend as {\it pure states}~\index{state!pure} (see
Appendix~\ref{se:gns}) to the full algebra ${\bf M}_2(\IC) $ as follows,
\bea 
&& \wt{\f}_i : {\bf M}_2(\IC) \lra \IC~, ~i = 1,2~, \nonumber \\ 
&& \wt{\f}_1(  
\left[
\begin{array}{cc} a_{11} & a_{12} \\ a_{21} & a_{22} 
\end{array}
\right]) = a_{11}~,~~~
\wt{\f}_2(  
\left[
\begin{array}{cc} a_{11} & a_{12} \\ a_{21} & a_{22} 
\end{array}
\right]) = a_{22}~.
\eea 
But now, noncommutativity implies the equivalence of the irreducible
representations of ${\bf M}_2(\IC)$ associated, via the
Gel'fand-Naimark-Segal~\index{GNS construction} construction, with the pure states
$\wt{\f}_1$ and $\wt{\f}_2$.  In fact, up to equivalence, the algebra ${\bf
M}_2(\IC)$ has only one irreducible representation, i.e. the defining two
dimensional one 
\fn{As we shall mention in Appendix~\ref{se:sme}, ${\bf M}_2(\IC)$ is strongly Morita
equivalent to $\IC$. Two strongly Morita equivalent $C^*$-algebras have the same
space  of classes of irreducible representations.}. We show this in
Appendix~\ref{se:gns}. 

\bigskip
 
For a noncommutative $C^*$-algebra, there is more than one candidate for the
analogue of the topological space $M$. We shall consider the following ones:

\begin{itemize}
\item[1)] The {\it structure space}~\index{structure space} of $\ca$ or space
of all unitary equivalence  classes of irreducible $^*$-representations. Such a
space is denoted by 
$\ha$.

\item[2)] The {\it primitive spectrum}~\index{prim@$\prim$} of $\ca$ or the
space of kernels of  irreducible $^*$-representations. Such a space is denoted by
$\prim$.  Any element of  $\prim$ is automatically a two-sided $^*$-ideal of $\ca$.  
\end{itemize} While for a commutative $C^*$-algebra these two spaces agree, this is not
any  more true for a general $C^*$-algebra $\ca$, not even setwise. For instance,
$\ha$ may be very complicate while $\prim$ consisting of a single point. 
One can define natural topologies on $\ha$ and $\prim$. We shall describe  them in
the next section.

\subsubsect{The Jacobson (or hull-kernel) Topology}\label{se:jcp}

The topology on $\prim$ is given by means of a closure operation. Given any subset
$W$ of $\prim$, the closure $\bar{W}$ of $W$ is by definition the set of all
elements in $\prim$ containing the intersection $\bigcap W$
of the elements of $W$, namely 
\be\label{cl} 
\bar{W} =: \{ \ci \in Prim{\ca} : \bigcap W \subseteq \ci\}~.  
\ee  
For any $C^*$-algebra $\ca$ we have the following,
\bprop
The closure operation (\ref{cl}) satisfies the Kuratowski 
axioms~\index{Kuratowski axioms}
\begin{itemize}
\item[~] $K_1$. $\bar{\emptyset} = \emptyset$~.
\item[~] $K_2$. $W \subseteq \bar{W}~, ~~~ \forall ~W \in \prim$~;  
\item[~] $K_3$. $\bar{\bar{W}} = \bar{W}~, ~~~ \forall ~W \in \prim$~;
\item[~] $K_4$. $\bar{W_1 \cup W_2} = \bar{W}_1 \cup \bar{W}_2~, 
~~~\forall ~W_1, W_2 \in \prim$~. 
\end{itemize}
\proof 
Property $K_1$ is immediate since $\bigcap \emptyset$ `does not exists'. By
construction, also $K_2$ is immediate. Furthermore, $\bigcap\bar{W} = \bigcap W$ from
which $\bar{\bar{W}} = \bar{W}$, namely $K_3$.  To prove $K_4$, observe first that
$V \subseteq W ~\Longrightarrow~ (\bigcap V) \supseteq (\bigcap W) ~\Longrightarrow~
\bar{V} \subseteq \bar{W}$. From this it follows that 
$\bar{W}_i \subseteq \bar{W_1 \bigcup W_2}, ~i=1,2$ and in turn
\be\label{kappa4}
\bar{W}_1 \cup \bar{W}_2 \subseteq \bar{W_1 \cup W_2}
\ee  
To obtain the opposite inclusion, consider a primitive ideal $\ci$ not belonging to 
$\bar{W}_1 \bigcup \bar{W}_2$. This means that $\bigcap W_1 \not\subset \ci$ and 
$\bigcap W_2 \not\subset \ci$. Thus, if $\pi$ is a representation of $\ca$ with
$\ci=Ker(\pi)$, there are elements $a\in\bigcap W_1$ and $b\in\bigcap W_2$ such
that $\pi(a)\not=0$ and $\pi(b)\not=0$. If $\xi$ is any vector in the
representation space $\ch_\pi$ such that $\pi(a)\xi\not=0$ then, $\pi$ being
irreducible, $\pi(a)\xi$ is a cyclic vector for $\pi$ (see Appendix~\ref{se:gns}).
This, together with the fact that $\pi(b)\not=0$, ensures that there is an element
$c\in\ca$ such that $\pi(b)(\pi(c)\pi(a))\xi\not=0$ which implies that $bca \not=
Ker(\pi)=\ci$. But $bca \in (\bigcap W_1) \cap (\bigcap W_2) = \bigcap(W_1
\cup W_2)$. Therefore $\bigcap(W_1 \cup W_2) \not\subset\ci$; whence $\ci
\not\in \bar{W_1 \cup W_2}$. What we have proved is that $\ci \not\in \bar{W}_1
\bigcup \bar{W}_2 ~\Rightarrow \ci \not\in \bar{W}_1 \bigcup \bar{W}_2$, which
gives the inclusion opposite to (\ref{kappa4}). So $K_4$ follows.
\eprop 

\noindent
It follows that the closure operation (\ref{cl}) defines a topology on $\prim$,
(see Appendix~\ref{se:bnt}) which is called {\it Jacobson
topology}~\index{Jacobson topology} or {\it hull-kernel topology}. The reason for the
name is that $\bigcap W$ is also called the {\it kernel} of $W$ and then $\bar{W}$
is the {\it hull} of $\bigcap W$ \cite{FD,Di1}. 

To illustrate this topology, we shall give a simple example.  Consider the
algebra 
$\cc(I)$ of complex-valued continuous functions on an interval $I$. As we have
seen, its structure space $\wh{\cc(I)}$ can be  identified with the interval
$I$.  For  any $a, b \in I$, let $W$ be the subset of $\wh{\cc(I)}$ given by
\be 
W = \{\ci_x, ~x \in ~]a, b[~~ \}~,
\ee 
where $\ci_x$ is the maximal ideal of $\cc(I)$ consisting of all functions
vanishing at $x$,
\be
\ci_x = \{f \in \cc(I) ~|~ f(x) = 0 \}~.
\ee 
The ideal $\ci_x$ is the kernel of the evaluation homomorphism as in
(\ref{gn3}). Then
\be
\bigcap W = \bigcap_{x \in ]a,b[} \ci_x = \{f \in \cc(I)~; f(x) = 0~,  ~\forall~
x \in ~]a, b[~~ \}~,
\ee 
and, the functions being continuous,
\bea
\bar{W} &=& \{ \ci \in \hc ~|~ \bigcap W \subset \ci\} \nonumber \\
   &=& W ~\bigcup~ \{\ci_a, \ci_b \} \nonumber \\
   &=& \{\ci_x, ~x \in [a, b]~ \}~,
\eea 
which can be identified with the closure of the interval $]a, b[$.

\bigskip 
In general, the space $\prim$ has few properties which are easy to prove and that we
state as propositions \cite{Di1}. 

\bprop\label{pr1} 
Let $W$ be a subset of $\prim$. Then $W$ is closed if and only
if $W$ is exactly the set of primitive ideals containing some subset of $\ca$. 

\proof If $W$ is closed then $W=\bar{W}$ and by the very definition 
(\ref{cl}), $W$ is the set of primitive ideals containing $\bigcap W$. Conversely,
let $V \subseteq \ca$. If $W$ is the set of primitive ideals of $\ca$ containing $V$,
then $V \subseteq \bigcap W$ from which $\bar{W} \subset W$, and, in turn, $\bar{W}=
W$.  
\eprop
\bprop\label{pr1a} 
There is a bijective correspondence between closed subset $W$ of $\prim$ and
(norm-closed two sided) ideals $\cj_W$ of $\ca$. The correspondence is given by 
\be
W = \{\ci \in \prim ~:~ \cj_W \subseteq \ci \}~. \label{r1a}
\ee

\proof If $W$ is closed then $W=\bar{W}$ and by the very definition 
(\ref{cl}), $\cj_W$ is just the ideal $\bigcap W$. Conversely, from the
previous proposition, $W$ defined as in (\ref{r1a}) is closed.
\eprop
\bprop\label{pr1b} 
Let $W$ be a subset of $\prim$. Then $W$ is closed if and only
if $\ci \in W$ and $\ci \subseteq \cj~ \Rightarrow J\in W$. 

\proof 
If $W$ is closed then $W=\bar{W}$ and by the very definition 
(\ref{cl}), $\ci \in W$ and $\ci \subseteq \cj$ implies that $J\in W$.
The converse implication is also evident by the previous Proposition. 
\eprop
\bprop\label{pr2} 
The space $\prim$ is a $T_0$-space 
\fn{Recall that a topological space is called 
$T_0$\index{topology!t0@$T_0$} if for any two distinct
points of the space there is an open neighborhood of one of the points which
does not contain the other \cite{Ke}.}. 

\proof Suppose $\ci_1$ and $\ci_2$ are two distinct points of $\prim$ so that,
say, $\ci_1 \not\subset \ci_2$. Then the set $W$ of those $\ci \in \prim$ which
contain $\ci_1$ is a closed subset (by \ref{pr1}), such that
$\ci_1 \in W$ and $\ci_2 \not\in W$. The complement $W^c$ of $W$ is an open set
containing $\ci_2$ and not $\ci_1$. 
\eprop
\bprop\label{pr3} Let $\ci \in \prim$. Then the point $\{\ci\}$ is closed in
$\prim$ if and  only if $\ci$ is maximal among primitive ideals.

\proof Indeed, the closure of $\{\ci\}$ is just the set of primitive ideals of 
$\ca$ containing $\ci$.  
\eprop

In general, $\prim$ is not a $T_1$-space
\fn{Recall that a topological space is called
$T_1$$T_0$\index{topology!t1@$T_1$}
 if any point of the space is closed 
\cite{Ke}.}   and will be so if and
only if all primitive ideals in $\ca$ are also maximal. This is for instance the
case if $\ca$ is commutative. The notion of primitive ideal is more general that
the one of maximal  ideal. For a commutative $C^*$-algebra an ideal is primitive if and
only if  is maximal. In general it is not even true that a maximal ideal is also
primitive. One can prove that this is the case if $\ca$ has a unit \cite{Di1}. 


\bigskip

Let us now consider the structure space $\ha$. Now, there is a canonical  surjection 
\be
\ha \lra ~Prim{\ca}~, ~~\pi \mapsto ker(\pi)~.
\ee  
The inverse image under  this map, of the Jacobson topology on $\prim$ is a topology
for $\ha$. In this topology, a subset $S \subset \ha$ is open if and only if is of
the form $\{ \pi \in \ha ~|~ ker(\pi) \in W \}$  for some subset $W \subset \prim$
which is open in the (Jacobson) topology of $\prim$.  The resulting topological
space is still called the structure space. There is another natural topology on the
space $\ha$ called the {\it regional topology}. For a $C^*$-algebra $\ca$, the
regional~\index{regional topology} and the pullback  of the Jacobson topology on
$\ha$ coincide, \cite[page 563]{FD}.

\bprop\label{pr5} 
Let $\ca$ be a $C^*$-algebra. The following conditions are equivalent
\begin{itemize}
\item[(i)] $\ha$ is a $T_0$ space.
\item[(ii)] Two irreducible representations of $\ha$ with the same  kernel are
equivalent.
\item[(iii)] The canonical map $\ha \ra \prim$ is a homeomorphism.
\end{itemize}

\proof By construction, a subset $\cs \in \ha$ will be closed if and only if it is of
the form $\{\pi \in \ha : ker(\pi) \in W \}$ for some $W$ closed in $\prim$. As a
consequence, given any two (classes of) representations $\pi_1, \pi_2
\in \ha$,  the representation $\pi_1$ will be in the closure of $\pi_2$ if and
only if $ker(\pi_1)$ is in the closure of $ker(\pi_2)$, or, by Prop.\ref{pr1} if
and only if $ker(\pi_2) \subset ker(\pi_1)$. In turn, $\pi_1$ and $\pi_2$ are
one in the closure of the other if and only if $ker(\pi_2) = ker(\pi_1)$.
Therefore, $\pi_1$ and $\pi_2$ will not be distinguished by the topology of
$\ha$ if and only if they have the same kernel. On the other side, if
$\ha$ is $T_0$ one is able to distinguish points. It follows that $(i)$ implies
that two representations with the same kernel must be equivalent so as to
correspond to the same point of $\ha$, namely $(ii)$. 
The other implications are obvious. 
\eprop

Recall that a (non necessarily Hausdorff) topological space
$S$ is called locally compact if any point of $S$ has at least one compact
neighborhood. A compact space is automatically locally compact. If $S$ is a
locally compact space which is also Hausdorff, than the family of closed compact
neighborhoods of any point is a base for its neighborhood system.
With respect to compactness, the structure space of a 
noncommutative $C^*$-algebra algebra behaves as in the commutative situation 
\cite[page 576]{FD},

\bprop If $\ca$ is a $C^*$-algebra, then $\ha$ is locally compact. Likewise, $\prim$  is
locally compact. If $\ca$ has a unit, then both $\ha$ and $\prim$ are  compact. 
\eprop

\noindent 
Notice that in general, $\ha$ compact does not imply that $\ca$ has 
a unit. For instance, the algebra $\comp$ of compact operators  on an infinite
dimensional Hilbert space $\ch$ has no unit but its  structure space has only
one point (see next section). 

\subsect{Compact Operators}\label{se:cop}

\index{compact operators|(}
We recall \cite{RS} that an operator on the Hilbert space $\ch$ is said to
be of finite rank if the orthogonal complement of its  null space is finite
dimensional. Essentially, we may think of such an  operator as a finite
dimensional matrix even if the Hilbert space is infinite dimensional. 
\bdefi
An operator $T$ on $\ch$ is said to be {\rm compact} if it can be approximated in
norm by finite rank operators.  
\edefi

\noindent
An equivalent way to characterize a compact operator $T$
is by stating that 
\be
\forall ~\ve >0~, ~~\exists ~~{\rm ~a ~finite ~dimensional ~subspace}~~  E
\subset \ch ~:~ \norm {T |_{E^\perp}} < \ve~. \label{compact}
\ee 
Here the orthogonal subspace $E^\perp$ is of finite codimension in $\ch$. The set
$\comp$ of all compact operators $T$ on the  Hilbert space $\ch$ is the largest
two-sided ideal in the $C^*$-algebra $\cb(\ch)$ of all bounded operators.
In fact, it is the only norm closed and two-sided when $\ch$ is
separable; and it is essential \cite{FD}. 
It is also a $C^*$-algebra with no unit, since the operator
$\II$ on an infinite dimensional Hilbert space is not  compact. The defining
representation of $\comp $ by itself is irreducible \cite{FD} and it is the
{\it only} irreducible representation of $\comp$ up to equivalence
\footnote{If $\ch$ is finite dimensional, $\ch = \IC^n$ say, then $\cb(\IC^n) =
\ck(\IC^n) = \IM_n(\IC)$, the algebra of $n\times n$ matrices with complex
entries. Such algebra has only one irreducible representation (as an algebra),
namely the defining one.}. 

There is a special class of $C^*$-algebras which have been used in a scheme of
approximation by means of topological lattices \cite{ncl, bbllt, pangs}; they are
{\it postliminal} algebras. For these  algebras, a relevant r\^ole is again 
played by the compact operators. Before we give the appropriate definitions, we
state another results which shows the relevance of compact operators in the
analysis of irreducibility of representations of a general $C^*$-algebra and
which is a consequence of the fact that $\comp$ is the largest two-sided
ideal in $\cb(\ch)$ \cite{Mu}, 
\bprop\label{comp} 
Let $\ca$ be a $C^*$-algebra acting irreducibly on a Hilbert space $\ch$ and 
having non-zero intersection with $\comp$. Then $\comp \subseteq \ca$.
\eprop
\bdefi\label{limi} 
A $C^*$-algebra $\ca$ is said to be {\it
liminal}~\index{liminal algebra@liminal $C^*$-algebra} if for every irreducible
representation
$(\ch, \pi)$ of $\ca$ one has that $\pi(\ca) =
\comp$ (or equivalently, from Prop.~\ref{comp},  $\pi(\ca) \subset  \comp )$.
\edefi 

\noindent
So, the algebra $\ca$ is liminal it is mapped to the algebra of compact
operators under any irreducible representation. Furthermore, if
$\ca$ is a liminal algebra, then one can prove that each primitive ideal of $\ca$ is
automatically a maximal closed two-sided ideal of
$\ca$. As a consequence, all points of $\prim$ are closed and $\prim$ is a
$T_1$ space. In particular, every commutative $C^*$-algebra is liminal
\cite{Mu,Di1}. 
\bdefi\label{postlimi} 
A $C^*$-algebra $\ca$ is said to be 
{\it postliminal}~\index{postliminal algebra@postliminal $C^*$-algebra} if for every
irreducible representation $(\ch,
\pi)$ of $\ca$ one has that $\comp \subseteq \pi(\ca)$ (or equivalently, from
Prop.~\ref{comp}, $\pi(\ca)
\cap \comp  \not= 0$). 
\edefi 

\noindent
Every liminal $C^*$-algebra is postliminal but the converse is not true.
Postliminal algebras have the remarkable property that their irreducible
representations are completely characterized by the kernels: if $(\ch_1, \pi_1)$
and $(\ch_2, \pi_2)$ are two irreducible representations  with the same kernel,
then $\pi_1$ and $\pi_2$ are equivalent \cite{Mu,Di1}. From Prop. (\ref{pr5}), the
spaces $\ha$ and $\prim$ are homeomorphic. 

\index{compact operators|)}

\vfill\eject
\sect{Noncommutative Lattices}\label{se:ncl} 

The idea of a `discrete substratum' underpinning  the `continuum' is
somewhat spread among physicists. With particular emphasis this idea has been pushed
by R. Sorkin who in \cite{So} assumes that the substratum be a {\it finitary}
(see later) topological space which maintains some of the topological information of
the continuum. It turns out that the finitary topology can be equivalently described in
terms of a partial order. This partial order has been alternatively interpreted as
determining the causal structure in the approach to quantum gravity of \cite{BLMS}.  
Recently, finitary topological spaces have been interpreted as noncommutative lattices
and noncommutative geometry has been used to construct quantum mechanical and field
theoretical models, notably lattice fields models, on them
\cite{ncl, bbllt}.  

Given a suitable covering of a topological space $M$, by identifying any  two points of
$M$ which cannot be `distinguished' by the sets in the covering, one constructs a
lattice with a finite (or in general  a countable) number  of points. Such a lattice,
with the quotient topology becomes a $T_0$-space which turns out to be the structure
space (or equivalently, the space of primitive ideal) of a postliminar approximately
finite dimensional (AF) algebra. Therefore the lattice is truly a noncommutative
space. In the rest of this Section we shall describe noncommutative lattices in some
detail while in Section~\ref{se:qmm} we shall illustrate some of their applications in
physics.

\subsect{The Topological Approximation}\label{se:toap}

The approximation scheme that we are going to describe has really a deep physical
flavor. To get a taste of the general situation, let us consider the following
simple example. Let us suppose we are about to measure the position of a particle
which moves on a circle, of radius one say, $S^1 = \{ 0\leq\vf\leq 2\p, ~{\rm
mod}~ 2\pi\}$. Our {\it `detectors'} will be taken to be (possibly overlapping) open
subsets of $S^1$ with some mechanism which switch on the detector when the particle is
in the corresponding open set 
The number of detectors must be clearly limited and we take them to consist of the
following three open subsets whose union covers $S^1$,
\be
\begin{array}{l}
U_1 = \{ - {1 \over 3} \p < \vf < {2 \over 3} \p \}~,  \\
~ \\
U_2 = \{{1 \over 3} \p < \vf < {4 \over 3} \p \}~,  \\
~ \\
U_3 = \{ \p < \vf < 2\p \}~. 
\end{array} \label{detectors}
\ee
Now, if two detectors, $U_1$ and $U_2$ say, are on, we will know that the particles
is in the intersection $U_1 \cap U_2$  although we will be unable to distinguish any
two points in this intersection. The same will be true for the other two intersections.
Furthermore, if only one detector, $U_1$ say, is on, we can infer the presence of the
particle in the {\it closed } subset of $S^1$ given by $U_1 \setminus \{U_1 \cap U_2
\bigcup U_1 \cap U_3 \}$ but again we will be unable to distinguish any
two points in this closed set. The same will be true for the other two closed sets of
similar type. Summing up, if we have only the three detectors (\ref{detectors}), we
are forced to identify the points which cannot be distinguished and $S^1$ will be
represented by a collection of six points $P = \{\a, \b, \g, a, b, c \}$ which
correspond to the following identifications
\be\label{discre}
\begin{array}{lcl}
U_1 \cap U_3 = \{{5 \over 3} \p < \vf < 2\p \} & \ra &  \a ~,  \\
~ & ~ & ~ \\
U_1 \cap U_2 = \{ {1 \over 3} \p < \vf < {2 \over 3} \p \} & \ra & \b ~, \\
~ & ~ & ~ \\
U_2 \cap U_3 = \{ \p < \vf < {4 \over 3} \p \} & \ra & \g ~,  \\
~ & ~ & ~ \\
U_1 \setminus \{U_1 \cap U_2 \bigcup U_1 \cap U_3 \} = \{ 0 \leq \vf \leq 
{1 \over 3} \p \} & \ra & a ~, \\
~ & ~ & ~ \\
U_2 \setminus \{U_2 \cap U_1 \bigcup U_2 \cap U_3 \} = 
\{ {2 \over 3} \p \leq \vf \leq \p \} & \ra & b ~, \\
~ & ~ & ~ \\
U_3 \setminus \{U_3 \cap U_2 \bigcup U_3 \cap U_1 \} = 
\{ {4 \over 3} \p \leq \vf \leq {5 \over 3} \p \} & \ra & c ~. \\
\end{array}
\ee  
\\
We can push things a bit further and keep track of the kind of set from which a
point comes by declaring a point to be open (respectively closed) if the subset of
$S^1$ from  which it comes is open (respectively closed). This is equivalently
achieved by endowing the space $P$ with a topology  a basis of which is given by
the following open (by definition) sets,
\bea\label{top6}
&& \{\a \}, ~~\{\b \}, ~~\{\g \}~,  \nonumber \\
&& \{\a, a, \b \}, ~~\{\b, b, \g\}, ~~\{\a, c, \g \}~. 
\eea  
The corresponding topology on the quotient space $P$ is noting but the quotient
topology of the one on $S^1$ generated by the three open sets $U_1, U_2, U_3$, by the
quotient map (\ref{discre}).
 
\bigskip
In general, let us suppose we have a topological space $M$ together with 
an open covering $\cu =\{U_\l\}$ which is also a topology for $M$, namely $\cu$ is
closed under arbitrary unions and finite intersections (see Appendix~\ref{se:bnt}). 
One defines  an equivalence relation among points of $M$ by declaring that any two
points $x, y \in M$ are equivalent if every open set $U_\l$ containing either $x$ or
$y$ contains the other too,
\be
   x\sim y ~~~~{\rm if~ and~ only~ if}~~~~ x\in U_\l \iff y\in U_\l~, 
~~~\forall~~ U_\l \in \cu~ . \label{2.2}
\ee
Thus, two points of $M$ are identified if they cannot be distinguished by any
`detector' in the collection $\cu$.

The space $P_{\cu}(M) =: M /\!\!\sim$ of equivalence classes is then given the quotient
topology. If $\pi : M \ra P_{\cu}(M)$ is the natural projection, a set 
$U \subset P_{\cu}(M)$ is declared to be open if and only if $\pi^{-1}(U)$ is  open in
the topology of $M$ given by $\cu$. The quotient topology is the  finest one making
$\pi$ continuous. When $M$ is compact, the covering $\cu$ can be taken to be finite so
that $P_{\cu}(M)$ will consist of a finite number of points. If $M$ is only locally
compact the covering can be taken to be locally finite and each point has a
neighborhood intersected by only finitely many $U_\l$' s. Then the space $P_{\cu}(M)$
will consists of a countable number of points; in the terminology of \cite{So}
$P_{\cu}(M)$ would be a {\it finitary}
\index{finitary approximation} approximation of $M$. If $P_{\cu}(M)$ has
$N$ points we shall also denote it by $P_N(M)$
\fn{In fact, this notation is incomplete since it does not keep track of the
finite topology given on the set of $N$ points. However, at least for the examples
considered in these notes, the topology will be always given explicitly.}. 
For example, the finite space given by (\ref{discre}) is $P_6(S^1)$. 

In general, $P_{\cu}(M)$ is not Hausdorff: from (\ref{top6}) it is evident
that in $P_6(S^1)$, for instance, we cannot isolate the point $a$ from $\a$ by
using open sets. It is not even a $T_1$-space; again, in $P_6(S^1)$ only the points
$a$, $b$ and $c$ are closed while the points $\a$, $\b$ and $\g$ are open. In general
there will be points which are neither closed nor open. It can be shown, however, that
$P_{\cu}(M)$ is always a $T_0$-space, being, indeed, the $T_0$-quotient
of $M$ with respect to the topology $\cu$ \cite{So}. 

\subsect{Order and Topology}\label{se:orto}
The next thing we shall show is how the topology of any finitary
\index{finitary approximation}
$T_0$ topological space $P$ can be given equivalently by means of a partial order 
which makes $P$ a 
{\it partially ordered set} (or 
{\it poset}~\index{poset} for short) \cite{So}.
Consider first the case when $P$ is finite. Then, the collection $\t$ of open sets (the
topology on $P$) will be closed under arbitrary unions and arbitrary intersections.  
As a consequence, for any point $x\in P$, the intersection of all open sets
containing it,
\be\label{basis}
\L(x) =: \bigcap\{U \in \t ~:~ x \in U \}
\ee
will be the smallest open neighborhood containing the point.
A relation $\preceq$ is then defined on $P$ by
\be\label{order1} 
x \preceq y~ \Leftrightarrow ~\L(x) \subseteq \L(y)~, ~~\forall ~x,y \in P~. 
\ee
Now, $x\in \L(x)$ always, so that the previous definition is equivalent to  
\be\label{order2} 
x \preceq y~ \Leftrightarrow ~x \in \L(y)~, ~~\forall ~x,y \in P~, 
\ee
which can also be stated as saying that 
\be\label{order3}
x \preceq y  ~ \Leftrightarrow ~ {\rm every ~open ~set
~containing} ~y~ {\rm contains ~also} ~x~, ~~\forall ~x,y \in P~, 
\ee 
or, in turn that 
\be\label{order4} 
x \preceq y  ~ \Leftrightarrow ~ y \in \bar{\{ x\}}~,
\ee 
with $\bar{\{ x\}}$ the closure of the one point set $\{ x\}$
\fn{Still another equivalent definition consists in saying that $x \preceq y$ if
and only if the constant sequence $(x, x, x, \cdots)$ converges to $y$. It is
worth noticing that in a $T_0$-space the limit of a sequence needs not be unique so
that the constant sequence $(x, x, x, \cdots)$ may converge to more than one point.
}.\\ From (\ref{order1}) it is clear that the relation
$\preceq$ is reflexive and transitive,
\bea
&& x \preceq x, \nonumber \\
&& x \preceq y~, ~y \preceq z ~\Rightarrow~ x \preceq z~.
\eea  
Furthermore, being $P$ a $T_0$-space, for any two distinct points $x,y\in P$, there
is at least one open set containing $x$, say, and not $y$. This, together with
(\ref{order3}), implies that the relation $\preceq$ is antisymmetric as well,
\be
x \preceq y~, ~y \preceq x ~\Rightarrow~ x = y~.
\ee  
Summing up, we get that a $T_0$ topology on a finite space $P$ determines a reflexive,
antisymmetric and transitive relation, namely a 
{\it partial order}~\index{partial order} on $P$ which makes
the latter a {\it partially ordered set} 
({\it poset})~\index{poset}.  
~\index{partial order!and topology}
~\index{topology!and partial order}
Conversely, given a partial order $\preceq$ on the set $P$, one produces a
topology on $P$ by taking as a basis for it the finite collection of `open'
sets defined as~\index{partial order!and topology}
~\index{topology!and partial order}
\be\label{bo}
\L(x) =: \{y \in P ~:~ y \preceq x \}~, ~~ \forall ~x \in P. 
\ee  
Thus, a subset $W\subset P$ will be open if and only if is the union of sets of the
form (\ref{bo}), namely, if and only if $x\in W$ and $y\preceq x ~\Rightarrow~ y
\in W$. 
Indeed, the smallest open set containing $W$ is given by 
\be\label{smop}
\L(W) = \bigcup_{x\in W} \L(x)~, 
\ee
and $W$ is open if and only if $W = \L(W)$.\\
The resulting topological space is clearly $T_0$ by the antisymmetry of the order
relation.

It is easy to express the closure operation in terms of the partial order. From
(\ref{order4}), the closure $V(x) = \bar{\{ x\}}$, of the one point set $\{ x\}$ is
given by
\be\label{smcl}
V(x) =: \{y \in P ~:~ x \preceq y \}~, ~~ \forall ~x \in P~. 
\ee
A subset $W \subset P$ will be closed if and only if  $x\in W$ and $x\preceq y
~\Rightarrow~ y \in W$. 
Indeed, the closure of $W$ is given by 
\be
V(W) = \bigcup_{x \in W} V(x)~,
\ee
and $W$ is closed if and only if $W = V(W)$.

If one relaxes the condition of finiteness of the space $P$, there is still an
equivalence between topology and partial order
~\index{topology!and partial order}
~\index{partial order!and topology}
for any $T_0$ topological space 
which has the additional property that every intersection of open sets is an open set
(or equivalently, that every union of closet sets is a closed set), so that the sets
(\ref{basis}) are all open and provide a basis for the topology \cite{Al, Bo1}. This
would be the case if $P$ is a finitary approximation of a (locally compact) topological
space $M$, obtained then from a locally finite covering of $M$
\fn{In fact, Sorkin \cite{So} regards as 
finitary\index{finitary approximation} 
only those posets $P$ for which the
sets $\L(x)$ and $V(x)$ defined in (\ref{smop}) and (\ref{smcl}) respectively, are all
finite. This would be the case if the poset is derived from a locally compact
topological space with a locally finite covering consisting of bounded open sets.}.

Given two posets $P,Q$, it is clear that a map $f:P \ra Q$ will be continuous if and
only if it is {\it order preserving}, namely, if and only if $x \preceq_P y
~\Rightarrow ~f(x) \preceq_Q f(y)$; indeed, $f$ is continuous if and only if
preserves convergence of sequences. 
\\ In the sequel, $x\prec y$ will indicates that $x$ precedes $y$ while $x\not=y$.  

\bigskip

A pictorial representation of the topology of a poset is obtained by constructing the
associated 
{\it Hasse diagram}:~\index{Hasse diagram} one  arranges the points of the
poset at different levels and connects them by using the following rules :
\begin{enumerate}
\item[1)] if $x\prec y$, then $x$ is at a lower level than $y$;
\item[2)] if $x\prec y$ and there is no $z$ such that $x\prec z\prec y$,
then $x$ is
at the level immediately below $y$ and these two points
are connected by a link.
\end{enumerate}
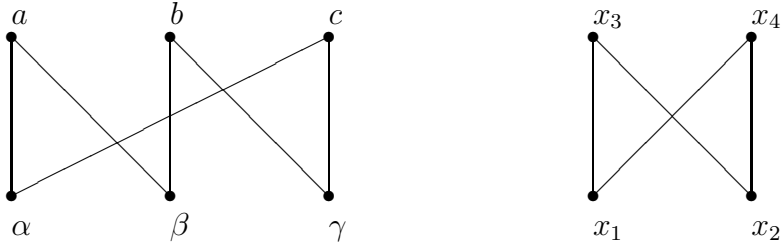
\begin{figure}[htb]
\begin{center}
\begin{picture}(220,130)(-200,-50)
\put(-30,30){\circle*{4}}
\put(30,30){\circle*{4}}
\put(-30,-30){\circle*{4}}
\put(30,-30){\circle*{4}}
\put(-30,30){\line(0,-1){60}}
\put(30,30){\line(0,-1){60}}
\put(-30,30){\line(1,-1){60}}
\put(30,30){\line(-1,-1){60}}
\put(-30,35){$x_3$}
\put(30,35){$x_4$}
\put(-30,-45){$x_1$}
\put(30,-45){$x_2$}
\put(-250,30){\circle*{4}}
\put(-190,30){\circle*{4}}
\put(-250,-30){\circle*{4}}
\put(-190,-30){\circle*{4}}
\put(-130,30){\circle*{4}}
\put(-130,-30){\circle*{4}}
\put(-250,30){\line(0,-1){60}}
\put(-190,30){\line(0,-1){60}}
\put(-130,30){\line(0,-1){60}}
\put(-250,30){\line(1,-1){60}}
\put(-190,30){\line(1,-1){60}}
\put(-250,-30){\line(2,1){120}}
\put(-250,35){$a$}
\put(-250,-45){$\a$}
\put(-190,35){$b$}
\put(-190,-45){$\b$}
\put(-130,35){$c$}
\put(-130,-45){$\g$}
\end{picture}
\caption{\label{fi:cirhas}
\protect{\footnotesize The Hasse diagrams for $P_6(S^1)$  and for $P_4(S^1)$.  }}
\end{center}
\vskip.5cm
\end{figure}

The Fig.~\ref{fi:cirhas} shows the Hasse diagram for $P_6(S^1)$ whose basis of open
sets is in (\ref{top6}) and for $P_4(S^1)$. For the former, the partial order reads 
$\a \prec a~, ~\a \prec c~, ~\b \prec a~, ~\b \prec b~, ~\g \prec b~, ~\g \prec c$.
The latter is a four points  approximation of $S^1$ obtained from a covering
consisting of two intersecting open sets. The partial order reads $x_1\prec x_3~,
~x_1\prec x_4~, ~x_2\prec x_3~, ~x_2\prec x_4~$.\\   
In that Figure, (and in general, in any 
Hasse diagram\~index{Hasse diagram}) the smallest open set
containing any point $x$ consists of all points which are below the given one $x$, and
can be connected to it by a series of links. For instance, for $P_4(S^1)$ we get for
the minimal open sets the following collection,
\bea\label{top4}
&& \L(x_1) = \{x_1\}~, \nonumber \\
&& \L(x_2) = \{x_2\}~, \nonumber \\
&& \L(x_3) = \{x_1,x_2,x_3\}~, \nonumber \\
&& \L(x_4) = \{x_1,x_2,x_4\}~,  
\eea
which are a basis for the topology of $P_4(S^1)$.

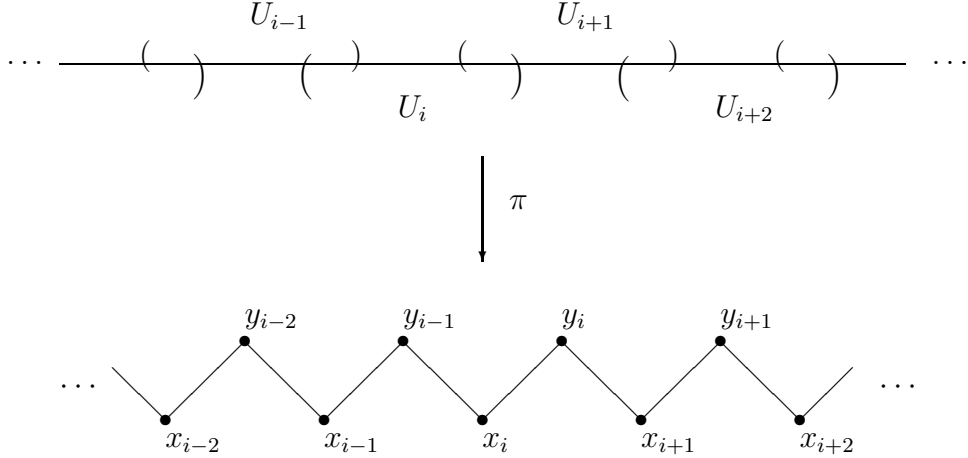
\begin{figure}[htb]
\begin{center}
\begin{picture}(400,190)(-180,-10)
\put(-88,150){$U_{i-1}$}
\put(-32,115){$U_{i}$}
\put(28,150){$U_{i+1}$}
\put(88,115){$U_{i+2}$}
\put(-180,135){$\dots$}
\put(-160,135){\line(1,0){320}}
\put(170,135){$\dots$}
\put(-130,135){$($}
\put(-110,126){{\Large$)$}}
\put(-70,126){{\Large$($}}
\put(-50,135){$)$}
\put(-10,135){$($}
\put(10,126){{\Large$)$}}
\put(50,125){{\Large$($}}
\put(70,135){$)$}
\put(110,135){$($}
\put(130,126){{\Large$)$}}
\put(10,80){$\pi$}
\put(0,100){\vector(0,-1){40}}
\put(-120,0){\circle*{4}}
\put(-60,0){\circle*{4}}
\put(0,0){\circle*{4}}
\put(60,0){\circle*{4}}
\put(120,0){\circle*{4}}
\put(-90,30){\circle*{4}}
\put(-30,30){\circle*{4}}
\put(30,30){\circle*{4}}
\put(90,30){\circle*{4}}
\put(-160,10){$\cdots$}
\put(150,10){$\cdots$}
\put(-90,30){\line(1,-1){30}}
\put(-90,30){\line(-1,-1){30}}
\put(-30,30){\line(1,-1){30}}
\put(-30,30){\line(-1,-1){30}}
\put(30,30){\line(1,-1){30}}
\put(30,30){\line(-1,-1){30}}
\put(90,30){\line(1,-1){30}}
\put(90,30){\line(-1,-1){30}}
\put(-120,0){\line(-1,1){20}}
\put(120,0){\line(1,1){20}}
\put(-120,-10){$x_{i-2}$}
\put(-60,-10){$x_{i-1}$}
\put(0,-10){$x_i$}
\put(60,-10){$x_{i+1}$}
\put(120,-10){$x_{i+2}$}
\put(-90,37){$y_{i-2}$}
\put(-30,37){$y_{i-1}$}
\put(30,37){$y_i$}
\put(90,37){$y_{i+1}$}
\end{picture}
\caption{\label{fi:linhas}
\protect{\footnotesize  The finitary poset of $\IR$. }}
\end{center}
\vskip.5cm
\end{figure}

\noindent
The generic finitary poset $P(\IR)$ associated with the real line $\IR$ is shown in
Fig.~\ref{fi:linhas}. The corresponding projection $\pi : \IR \ra P(\IR)$ is given by
\bea
U_i \cap U_{i+1} &\lra&  x_i ~,~~i \in \IZ~, \nonumber \\
U_{i+1}\setminus\{ U_i \cap U_{i+1}  \bigcup U_{i+1} \cap U_{i+2} \} 
&\lra&  y_i ~,~~i \in \IZ~.
\eea
A basis for the quotient topology is provided by the collection of all open sets of the
form
\be
\L(x_i) = \{x_i \}~, ~~\L(y_i) = \{x_{i}, y_i, x_{i+1} \} ~,~~i \in \IZ~.
\ee

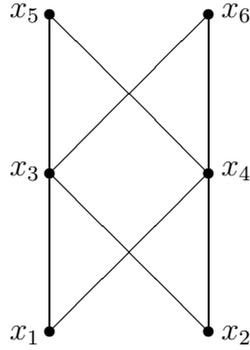
\begin{figure}[htb]
\begin{center}
\begin{picture}(220,150)(-100,-30)
\put(-30,90){\circle*{4}}
\put(30,90){\circle*{4}}
\put(-30,30){\circle*{4}}
\put(30,30){\circle*{4}}
\put(-30,-30){\circle*{4}}
\put(30,-30){\circle*{4}}
\put(-30,90){\line(0,-1){60}}
\put(30,90){\line(0,-1){60}}
\put(-30,90){\line(1,-1){60}}
\put(30,90){\line(-1,-1){60}}
\put(-30,30){\line(0,-1){60}}
\put(30,30){\line(0,-1){60}}
\put(-30,30){\line(1,-1){60}}
\put(30,30){\line(-1,-1){60}}
\put(-45,89){$x_5$}
\put(35,89){$x_6$}
\put(-45,29){$x_3$}
\put(35,29){$x_4$}
\put(-45,-33){$x_1$}
\put(35,-33){$x_2$}
\end{picture}
\caption{\label{fi:sphpos}
\protect{\footnotesize The Hasse diagram for the poset $P_6(S^2)$.   }}
\end{center}
\vskip.5cm
\end{figure}

\noindent
Fig.~\ref{fi:sphpos} shows  the Hasse diagram for the six-point poset
$P_6(S^2)$ of the two dimensional sphere, coming from a covering with four open
sets, which has been derived in \cite{So}. A basis for its topology is given by
\be
\begin{array}{ll}
\L(x_1) = \{x_1\}~, & \L(x_2) = \{x_2\}~, \\
~&~\\
\L(x_3) = \{x_1,x_2,x_3\}~, & \L(x_4) = \{x_1,x_2,x_4\}~, \\
~&~\\
\L(x_5) = \{x_1,x_2,x_3,x_4,x_5\}~, & \L(x_6) = \{x_1,x_2,x_3,x_4,x_6\}~.
\label{2.6}
\end{array}
\ee
Now, the top two points are closed, the bottom two points are open and the
intermediate ones are neither closed nor open.

\begin{figure}[htb]
\begin{center}
\begin{picture}(200,100)(0,-45)
\put(10,0){\circle{40}}
\put(5,25){{\small $0,1$}}
\put(10,18){\line(0,1){4}}
\put(80,10){$\sigma$}
\put(60,0){\vector(1,0){40}}
\put(130,30){\circle*{4}}
\put(190,30){\circle*{4}}
\put(130,-30){\circle*{4}}
\put(190,-30){\circle*{4}}
\put(130,30){\line(0,-1){60}}
\put(190,30){\line(0,-1){60}}
\put(130,30){\line(1,-1){60}}
\put(190,30){\line(-1,-1){60}}
\put(130,-30){\vector(0,1){30}}
\put(190,-30){\vector(0,1){30}}
\put(130,30){\vector(1,-1){20}}
\put(190,30){\vector(-1,-1){20}}
\put(120,40){{\small $0,1$}}
\put(190,40){{\small${1 \over 2}$}}
\put(120,-45){{\small$]{1\over 2}, 1[ $}}
\put(180,-45){{\small$]0, {1\over 2}[ $}}
\end{picture}
\caption{\label{fi:cirhom}
\protect{\footnotesize A representative of the generator of the homotopy group
$\pi_1(P_4(S^1))$.}} 
\end{center}
\vskip.5cm
\end{figure}
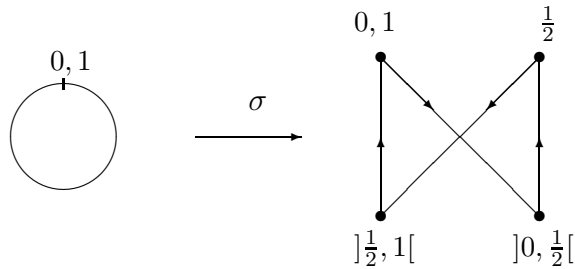

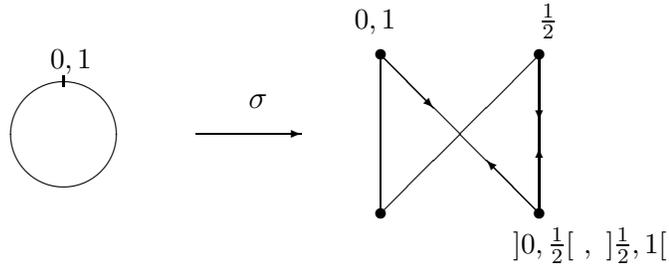
\begin{figure}[htb]
\begin{center}
\begin{picture}(200,100)(0,-45)
\put(10,0){\circle{40}}
\put(5,25){{\small $0,1$}}
\put(10,18){\line(0,1){4}}
\put(80,10){$\sigma$}
\put(60,0){\vector(1,0){40}}
\put(130,30){\circle*{4}}
\put(190,30){\circle*{4}}
\put(130,-30){\circle*{4}}
\put(190,-30){\circle*{4}}
\put(130,30){\line(0,-1){60}}
\put(190,30){\line(0,-1){60}}
\put(130,30){\line(1,-1){60}}
\put(190,30){\line(-1,-1){60}}
\put(190,30){\vector(0,-1){25}}
\put(190,-30){\vector(0,1){25}}
\put(130,30){\vector(1,-1){20}}
\put(190,-30){\vector(-1,1){20}}
\put(120,40){{\small $0,1$}}
\put(190,40){{\small ${1 \over 2}$}}
\put(180,-45){{\small $]0, {1\over 2}[~,~]{1\over 2}, 1[ $}}
\end{picture}
\caption{\label{fi:triv}
\protect{\footnotesize A representative of the trivial class in the homotopy group
$\pi_1(P_4(S^1))$.}} 
\end{center}
\vskip.5cm
\end{figure}

As alluded to before, posets retain some of the topological information of the space
they approximate. For example, one can prove that for the first homotopy group
$\p_1(P_N(S^1)) = \IZ = \pi(S^1)$ whenever $N \geq 4$ \cite{So}. Consider the case
$N=4$.  Elements of $\p_1(P_4(S^1))$ are homotopy classes of continuous maps 
$\s : [0,1] \ra P_4(S^1)$, such that $\s(0)=\s(1)$. With $a$ any real number in the
open interval $]0,1[$, consider the map
\be\label{winds}
\s(t) = 
\left\{
\begin{array}{lll}
x_3 & if & t = 0  \\
x_2 & if & 0 < t < a \\
x_4 & if & t = a  \\
x_1 & if & a < t < 1  \\
x_3 & if & t = 1  \\
\end{array}
\right. ~~.
\ee
Figure~\ref{fi:cirhom} shows this map for $a = 1/2$; the map can be seen to `winds once
around' $P_4(S^1)$. 
Furthermore, the map $\s$ in (\ref{winds}) is manifestly continuous, being
constructed in such a manner that closed (respectively open) points of $P_4(S^1)$
are the image of closed (respectively open) sets of the interval $[0,1]$ so that,
automatically, the inverse image of an open set in $P_4(S^1)$ is open in $[0,1]$.
A bit of extra analysis shows that $\s$ is not contractible to the constant map, any
such contractible map being one that skips at least one of the points of $P_4(S^1)$
like the following one,
\be\label{winds0}
\s_0(t) = 
\left\{
\begin{array}{lll}
x_3 & if & t = 0  \\
x_2 & if & 0 < t < a \\
x_4 & if & t = a  \\
x_2 & if & a < t < 1  \\
x_3 & if & t = 1  \\
\end{array}
\right. ~~,
\ee
which is shown in Fig.~\ref{fi:triv} for the values $a=1/2$.
Indeed, the not contractible map in (\ref{winds})  is a generator of the group
$\pi_1(P_4(S^1))$ which therefore can be identified with the group of integer
numbers $\IZ$. 

\begin{figure}[htb]
\begin{center}
\begin{picture}(450,150)(-225,-30)
\put(-210,90){\circle*{4}}
\put(210,90){\circle*{4}}
\put(-90,90){\circle*{4}}
\put(90,90){\circle*{4}}
\put(-210,30){\circle*{4}}
\put(210,30){\circle*{4}}
\put(-150,30){\circle*{4}}
\put(150,30){\circle*{4}}
\put(-90,30){\circle*{4}}
\put(90,30){\circle*{4}}
\put(-30,30){\circle*{4}}
\put(30,30){\circle*{4}}
\put(-210,-30){\circle*{4}}
\put(210,-30){\circle*{4}}
\put(-90,-30){\circle*{4}}
\put(90,-30){\circle*{4}}
\put(-210,90){\line(0,-1){60}}
\put(210,90){\line(0,-1){60}}
\put(-210,90){\line(1,-1){60}}
\put(210,90){\line(-1,-1){60}}
\put(-210,90){\line(3,-1){180}}
\put(210,90){\line(-3,-1){180}}
\put(-210,90){\line(5,-1){300}}
\put(210,90){\line(-5,-1){300}}
\put(-90,90){\line(-2,-1){120}}
\put(90,90){\line(2,-1){120}}
\put(-90,90){\line(2,-1){120}}
\put(90,90){\line(-2,-1){120}}
\put(-90,90){\line(3,-1){180}}
\put(90,90){\line(-3,-1){180}}
\put(-90,90){\line(4,-1){240}}
\put(90,90){\line(-4,-1){240}}
\put(-210,-30){\line(0,1){60}}
\put(210,-30){\line(0,1){60}}
\put(-210,-30){\line(1,1){60}}
\put(210,-30){\line(-1,1){60}}
\put(-210,-30){\line(2,1){120}}
\put(210,-30){\line(-2,1){120}}
\put(-210,-30){\line(4,1){240}}
\put(210,-30){\line(-4,1){240}}
\put(-90,-30){\line(-2,1){120}}
\put(90,-30){\line(2,1){120}}
\put(-90,-30){\line(0,1){60}}
\put(90,-30){\line(0,1){60}}
\put(-90,-30){\line(1,1){60}}
\put(90,-30){\line(-1,1){60}}
\put(-90,-30){\line(4,1){240}}
\put(90,-30){\line(-4,1){240}}
\end{picture}
\caption{\label{fi:torhas}
\protect{\footnotesize The Hasse diagram for the poset $P_{16}(S^1 \times S^1) =
P_{4}(S^1) \times P_{4}(S^1)$.   }}
\end{center}
\vskip.5cm
\end{figure}
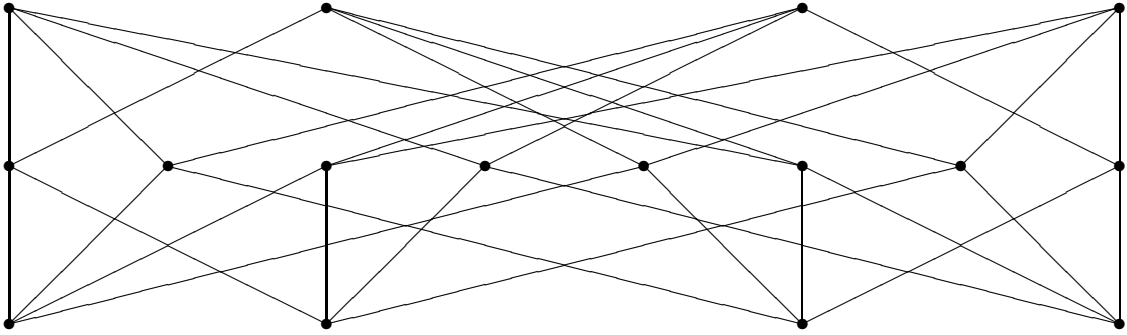

Finally, we mention the notion of Cartesian product of posets. If $P$ and $Q$ are
posets, their {\it Cartesian product} is the poset $P \times Q$ on the set $\{(x, y)
~:~ x \in P, ~y \in Q \}$ such that $(x,y) \preceq (x',y')$ in $P \times Q$ if $x
\preceq x'$ in $P$ and $y \preceq y'$ in $Q$. To draw the Hasse diagram of $P \times
Q$, one draws the diagram of $P$, replace each element $x$ of $P$ by a copy $Q_x$ of
$Q$ and connects corresponding elements of $Q_x$ and $Q_y$ 
(by identifying $Q_x \simeq Q_y$) if $x$ and $y$ are connected in the diagram of $P$.
Fig.~\ref{fi:torhas} shows the Hasse diagram of a poset $P_{16}(S^1 \times S^1)$
obtained as $P_{4}(S^1) \times P_{4}(S^1)$.

\subsect{How to Recover the Space Being Approximated}\label{se:recospa}
We shall now briefly describe how the topological space being approximated
can be recovered `in the limit' by considering a sequence of finer and finer
coverings,  the appropriated framework being that of inverse (or projective)  
systems of topological spaces \cite{So}. 

Well, let us suppose we have a topological space $M$ together with 
a sequence $\{ \cu_n \}_{n \in \IN}$ of finer and finer coverings, namely 
of coverings such that  
\be
\cu_i \subseteq \t(\cu_{i+1})~,
\ee
where $\t(\cu)$ is the topology generated by the covering $\cu$
\fn{For more general situations, such as the system of all finite open covers of $M$,
this is not enough and one needs to consider a {\it directed} collection 
$\{ \cu_i \}_{i \in \L}$ of open covers of $M$, where directed just means that 
for any two coverings $\cu_1$ and $\cu_2$, there exists a third cover $\cu_3$ such
that  $\cu_1, \cu_2 \subseteq \t(\cu_3)$. The construction of the remaining part of
the section applies to this more general situation if one defines a
partial order on the `set of indices' $\L$ by declaring that 
$ i \leq k ~\Leftrightarrow~ \cu_i \subseteq \t(\cu_j)~$.}. 
Here we are relaxing the harmless assumption made in Section~\ref{se:toap} that each
$\cu$ was already a subtopology, namely that $\cu = \t(\cu)$.

\begin{figure}[htb]
\begin{center}
\begin{picture}(220,300)(-80,-100)
\put(-120,180){$M$}
\put(-60,180){\vector(1,0){180}}
\put(-100,165){\vector(1,-1){120}}
\put(-100,165){\vector(1,-2){120}}
\put(25,30){$P_j$}
\put(25,-90){$P_i$}
\put(30,130){$\vdots$}
\put(30,110){\vector(0,-1){60}}
\put(30,10){\vector(0,-1){60}}
\put(25,190){$\p_{\infty}$}
\put(35,-20){$\p_{ij}$}
\put(-60,30){$\p_{i}$}
\put(-30,70){$\p_{j}$}
\put(105,30){$\p_{i\infty}$}
\put(80,70){$\p_{j\infty}$}
\put(170,180){$P_\infty$}
\put(160,165){\vector(-1,-1){120}}
\put(160,165){\vector(-1,-2){120}}
\end{picture}
\caption{\label{fi:invlim}
\protect{\footnotesize The inverse system.   }}
\end{center}
\vskip.5cm
\end{figure}
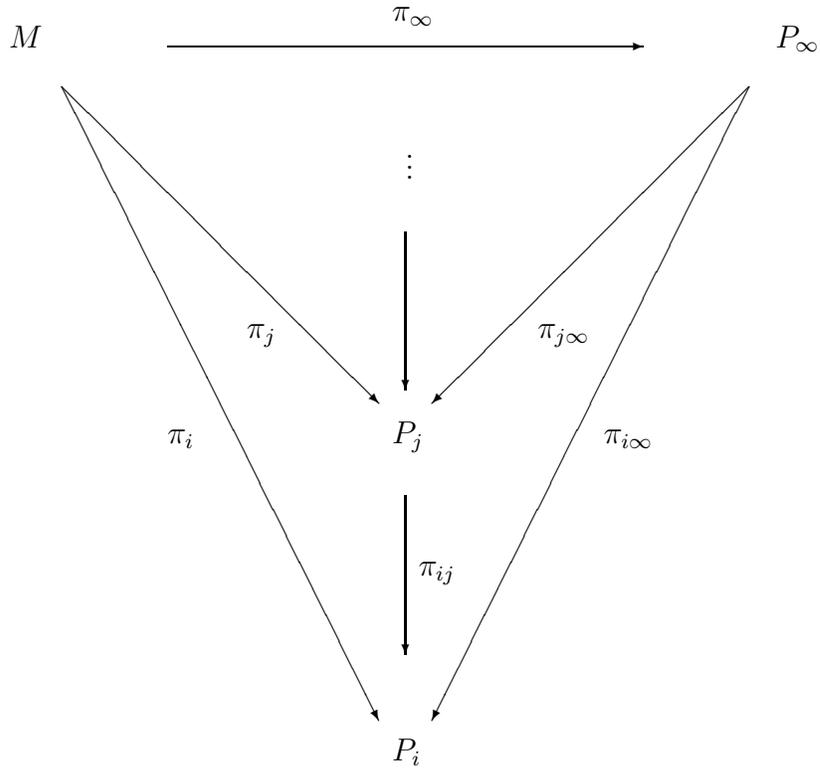

In Section~\ref{se:toap} we have associated with each covering $\cu_i$ a
$T_0$-topological space $P_i$ and a continuous surjection 
\be
\pi_i : M \ra P_i~.
\ee
We now construct an {\it inverse system of spaces $P_i$ together with continuous
maps}~\index{inverse system}
\be
\pi_{ij} : P_i \ra P_j~,
\ee defined whenever $i \leq j$ and such that 
\be
\p_i = \p_{ij} \circ \p_j~. \label{prmap}
\ee
These maps are uniquely defined by the fact that the spaces $P_i$ are $T_0$ and the
map $\p_i$ is continuous with respect to $\t(\cu_j)$ whenever $i\leq j$.
Indeed, if $U$ is open in $P_i$, then $\p_i^{(-1)}(U)$ is open in the 
$\cu_i$-topology by definition, thus it is also open in the finer $\cu_j$-topology. 
Furthermore, uniqueness also implies the compatibility conditions
\be
\p_{ij} \circ \p_{jk} = \p_{ik}~, 
\ee
whenever $i \leq j \leq k$ 
\fn{In fact, the map $\p_{ij}$ is the solution (by definition then unique) of an
universal problem of maps relating $T_0$-spaces \cite{So}.}.
Notice that from the surjectivity of the maps $\p_i$ and relation (\ref{prmap}), it
follows that all maps $\p_{ij}$ are surjective. \\
The inverse system of topological spaces and continuous maps 
$\{P_i, \p_{ij} \}_{i,j \in \IN}$ has a unique 
{\it inverse limit}~\index{inverse limit}, namely a
topological space $P_\infty$, together with continuous maps 
\be
\p_{i\infty} : P_\infty \ra P_i~,
\ee
such that 
\be
\p_{ij} \circ \p_{j\infty} = \p_{i\infty}~,
\ee
whenever $i \leq j$. The space $P_\infty$ and the maps $\p_{ij}$ can be explicitly
construct. An element $x \in P_\infty$ is an arbitrary coherent sequence of elements
$x_i  \in P_i$,
\be
x = (x_i)_{i\in\IN} ~,~ x_i \in P_i ~:~ \exists ~N_0 ~~~{\rm s.t.}~~~ x_i =
\p_{i,i+1}(x_{i+1})~, ~~\forall ~ i \geq N_0~.
\ee 
As for the map $\p_{i\infty}$, it  is just defined by
\be
\p_{i\infty}(x) = x_i~.
\ee
The space $P_{i\infty}$ is made a $T_0$ topological space by endowing it with the
weakest topology making all maps $\p_{i\infty}$ continuous: a basis for it is given
by the sets $\p^{(-1)}_{i\infty}(U)$, for all open sets $U \subset P_i$. 
The inverse system and its limit are depicted in Fig.~\ref{fi:invlim}
\\
It turns out that the limit space $P_{\infty}$ is {\it bigger} than the starting $M$
and the latter is contained as a dense subspace. Furthermore, $M$ can be
characterized as the set of all {\it closed} points of $P_{i\infty}$. Let us first
observe that we also get a unique (by universality) continuous map
\be
\p_\infty : M \ra P_\infty~,
\ee 
which satisfies
\be
\p_i = \p_{i\infty} \circ \p_\infty~, ~~~ \forall ~ i \in \IN~.
\ee
The map $\p_\infty$ is the `limit' of the maps $\p_i$. However, while the latter are
surjective, under mild hypothesis the former turns out to be {\it injective}. We have
indeed the following two propositions
\cite{So}.
\bprop
The image $\p_\infty(M)$ is dense in $P_\infty$.

\proof
If $U \subset P_\infty$ is any nonempty open set, by the definition of the topology
of $P_\infty$, $U$ is the union of sets of the form $\p^{(-1)}_{i\infty}(U_i)$, with
$U_i$ open in $P_i$. Choose $x_i \in U_i$. Since $\p_i$ is surjective, there is at
least a point $m \in M$, for which $\p_i(m) = x_i$ and let $\p_\infty(m) = x$. Then
$\p_{i\infty}(m) = \p_{i\infty}(\p_{\infty})(m) = x_j$, from which $x \in
\p^{(-1)}_{i\infty}(x_i) \subset \p^{(-1)}_{i\infty}(U_i) \subset U$. This proves
that  $\p_\infty(M) \cap W \not= \emptyset$, namely that  $\p_\infty(M)$ is dense.
\eprop
\bprop\label{lemma2}
Let $M$ be $T_0$ and the collection $\{\cu_i \}$ of coverings be such that for every $m
\in M$ and every neighborhood $N \ni m$, there exists an index $i$ and an element $U
\in \t(\cu_i)$ such that $m \in U \subset N$. Then, the map $\p_\infty$ is injective.

\proof
If $m_1, m_2$ are two distinct points of $M$, since the latter is $T_0$, there is an
open set $V$ containing $m_1$ (say) and not $m_2$. By hypothesis, there exists an
index $i$ and an open $U \in \t(\cu_i)$ such that $m_1 \in U \subset V$. Therefore
$\t(\cu_i)$ distinguishes $m_1$ from $m_2$. Since $P_i$ is the corresponding $T_0$
quotient, $\p_i(m_1) \not= \p_i(m_2)$. Then $\p_{i\infty}(\p_\infty(m_1)) \not=
\p_{i\infty}(\p_\infty(m_2))$, and in turn $\p_{\infty}(m_1) \not=
\p_{\infty}(m_2)$. 
\eprop

\noindent
We remark that in a sense, the second condition in the previous proposition just say
that the covering $\cu_i$ contains `enough small open sets', a condition one would
expect in the process of recovering $M$ by a refinement of the coverings.

As alluded to before, there is a nice characterization of the points of $M$ (or better
of $\p_\infty(M)$) as the set all all closed points of $P_\infty$. We have indeed a
further Proposition, whose easy but long proof is given in \cite{So},
\bprop
Let $M$ be $T_1$ and let the collection $\{\cu_i \}$ of coverings fulfill the
`fineness' condition of Proposition~\ref{lemma2}. Let each covering $\cu_i$ consists
only of sets which are bounded (have compact closure). Then $\p_\infty : M \ra
P_\infty$ embeds $M$ in $P_\infty$ as the subspace of closed points.
\eprop

\noindent 
We remark that the additional requirement on the element of each covering is
automatically fulfilled if $M$ is compact.

As for the extra points of $P_\infty$, one can prove that for any extra
$y \in P_\infty$, there exists an $x\in\p_\infty(M)$ to which $y$ is `infinitely
close'. Indeed, $P_\infty$ can be made a poset by defining a partial order relation as
follows
\be
x \preceq_\infty y ~~~\Leftrightarrow~~~ x_i \preceq y_i~, ~~~\forall ~i~,
\label{poinf}
\ee
where the coherent sequences $x=(x_i)$ and $y=(y_i)$ are any two elements of
$P_\infty$
\fn{In fact, one could directly construct $P_\infty$ as the 
inverse limit\index{inverse limit} of an
inverse system of posets by defining a partial order on the coherent sequences as in
(\ref{poinf}).}.
Then one can characterize $\p_\infty(M)$ as the {\it set of maximal elements of
$P_\infty$}, with respect to the order $\preceq_\infty$. Given any such maximal
element $x$, the points of $P_\infty$ which are infinitely closed to $x$ are all (non
maximal) points which converge to $x$, namely all (non maximal) $y\in P_\infty$ such
that $y \preceq_\infty x$. In $P_\infty$, these points $y$ cannot be separated from
the corresponding $x$. By identifying points in $P_\infty$ which cannot be separated
one recovers $M$.  The interpretation that emerges is that the top points of a poset
$P(M)$ (which are always closed) approximate the points of $M$ and give all of $M$ in
the limit. The role of the remaining points is to `glue' the top points together so as
to produce a topologically nontrivial approximation to $M$. They also give the extra
points in the limit.
  
\begin{figure}[htb]
\begin{center}
\begin{picture}(340,105)(-50,-40)
\put(-30,30){\circle*{4}}
\put(30,30){\circle*{4}}
\put(90,30){\circle*{4}}
\put(270,30){\circle*{4}}
\put(-30,-30){\circle*{4}}
\put(30,-30){\circle*{4}}
\put(90,-30){\circle*{4}}
\put(270,-30){\circle*{4}}
\put(160,-30){$\dots$}
\put(180,-30){$\dots$}
\put(200,-30){$\dots$}
\put(160,30){$\dots$}
\put(180,30){$\dots$}
\put(200,30){$\dots$}
\put(-30,-30){\line(5,1){300}}
\put(-30,30){\line(0,-1){60}}
\put(30,30){\line(0,-1){60}}
\put(90,30){\line(0,-1){60}}
\put(270,30){\line(0,-1){60}}
\put(-30,30){\line(1,-1){60}}
\put(30,30){\line(1,-1){60}}
\put(90,30){\line(1,-1){40}}
\put(270,-30){\line(-1,1){20}}
\put(-30,35){$x_{N+1}$}
\put(30,35){$x_{N+2}$}
\put(90,35){$x_{N+3}$}
\put(270,35){$x_{2N}$}
\put(-30,-40){$x_1$}
\put(30,-40){$x_2$}
\put(90,-40){$x_3$}
\put(270,-40){$x_{N}$}
\end{picture}
\caption{\label{fi:cirhas2n}
\protect{\footnotesize The Hasse diagram for $P_{2N}(S^1)$.  }}
\end{center}
\vskip.5cm
\end{figure}
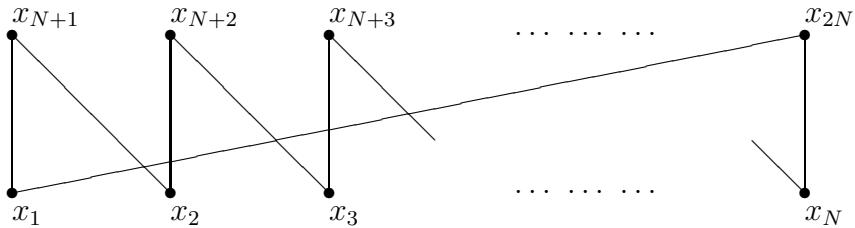
\begin{figure}
\begin{center}
\begin{picture}(220,430)(-100,-90)
\put(-30,-10){\circle*{4}}
\put(30,-10){\circle*{4}}
\put(-30,-70){\circle*{4}}
\put(30,-70){\circle*{4}}
\put(-30,-10){\line(0,-1){60}}
\put(30,-10){\line(0,-1){60}}
\put(-30,-10){\line(1,-1){60}}
\put(30,-10){\line(-1,-1){60}}
\put(-32,-5){$3$}
\put(28,-5){$4$}
\put(-32,-85){$1$}
\put(28,-85){$2$}
\put(-60,150){\circle*{4}}
\put( 0,150){\circle*{4}}
\put(-60,90){\circle*{4}}
\put( 0,90){\circle*{4}}
\put(60,150){\circle*{4}}
\put(60,90){\circle*{4}}
\put(-60,150){\line(0,-1){60}}
\put( 0,150){\line(0,-1){60}}
\put( 60,150){\line(0,-1){60}}
\put(-60,150){\line(1,-1){60}}
\put( 0,150){\line(1,-1){60}}
\put(-60,90){\line(2,1){120}}
\put(-62,155){$3$}
\put( -5,155){$2_0$}
\put( 58,155){$4$}
\put(-62,75){$1$}
\put( -2,75){$2_1$}
\put( 58,75){$2_2$}
\put(-30,310){\circle*{4}}
\put(30,310){\circle*{4}}
\put(-30,250){\circle*{4}}
\put(30,250){\circle*{4}}
\put(-90,310){\circle*{4}}
\put(90,310){\circle*{4}}
\put(-90,250){\circle*{4}}
\put(90,250){\circle*{4}}
\put(-30,310){\line(0,-1){60}}
\put(30,310){\line(0,-1){60}}
\put(-30,310){\line(1,-1){60}}
\put(30,310){\line(1,-1){60}}
\put(-90,310){\line(0,-1){60}}
\put(90,310){\line(0,-1){60}}
\put(-90,310){\line(1,-1){60}}
\put(-90,250){\line(3,1){180}}
\put(-95,315){$1_0$}
\put(-33,315){$3$}
\put(27,315){$2_0$}
\put(90,315){$4$}
\put(-95,235){$1_2$}
\put(-35,235){$1_1$}
\put(27,235){$2_1$}
\put(87,235){$2_2$}
\put(0,340){$\vdots$}
\put(0,220){$\vector(0,-1){40}$}
\put(0,60){$\vector(0,-1){40}$}
\put(5,200){$\p_{23}$}
\put(5,40){$\p_{12}$}

\end{picture}
\caption{\label{fi:cirinvlim}
\protect{\footnotesize The inverse system for $S^1$.  }}
\end{center}
\vskip.5cm
\end{figure}
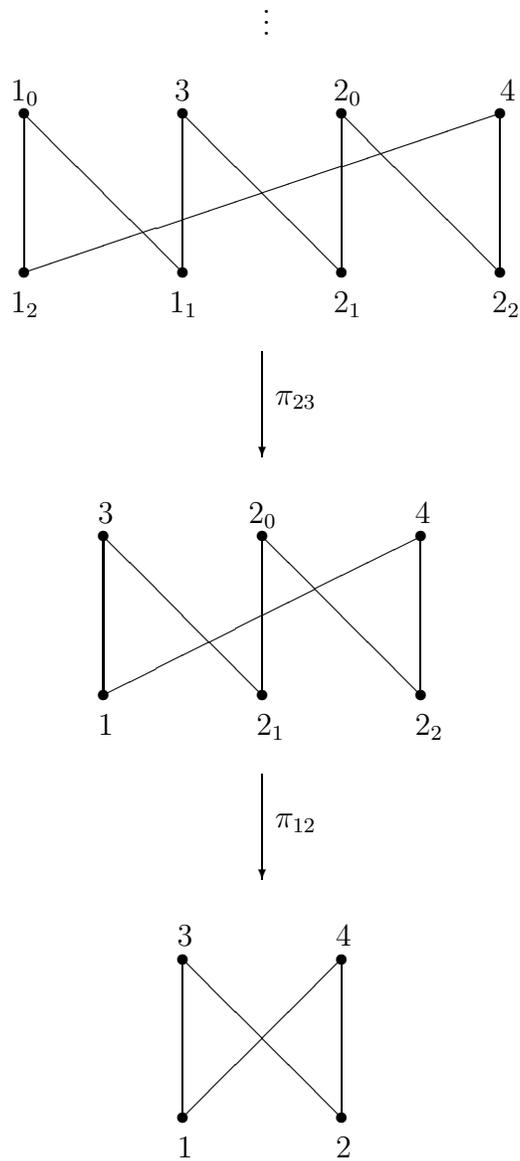

Fig.~\ref{fi:cirhas2n}
shows the $2N$-poset approximation to $S^1$ obtained with a covering consisting of $N$
open sets. In Fig.~\ref{fi:cirinvlim} we have the associated inverse system of posets.
As seen in that figure, by going from one level to the next one, only one of the bottom
points $x$ is `split' in three $\{x_0, x_1, x_1\}$ while the other are not changed.
The projection from one level to the previous one is the map which sends the triple
$\{x_0, x_1, x_1\}$ to the parent $x$ while acting as the identity on the remaining
points. The projection is easily seen to be order preserving (and then continuous). As
in the general case, the limit space $P_\infty$ consists of $S^1$ together with extra
points. These extra points come in couples anyone of which is `glued' (in the sense of
being infinitely closed) to a point in a numerable collection of points. This
collection is dense in $S^1$ and could be taken as the collection of all points of the
form  $\{ m/2^n~, ~m,n \in \IN\}$ of the interval $[0, 1]$ with endpoints identified. 

In \cite{pangs} a somewhat different interpretation of the approximation and of the
limiting procedure in terms of simplicial decompositions has been proposed. 

\subsect{Noncommutative Lattices}\label{se:ncl1}

It turns out that any (finite) poset $P$ is the structure space $\ha$ (space of
irreducible representations, see Section~\ref{se:nca}) of a noncommutative
$C^*$-algebra $\ca$ of operator valued functions which then plays the role of the
algebra of continuous functions on $P$
\fn{It is worth noticing that, a poset $P$ being non Hausdorff, there cannot be `enough'
$\IC$-valued continuous functions on $P$ since the latter separate points. For
instance, on the poset of Fig.~\ref{fi:cirhas} or Fig.~\ref{fi:sphpos} the only
$\IC$-valued continuous functions are the constant ones. In fact, the previous
statement is true for each connected component of any poset.}. Indeed, there is a
complete classification of all separable 
\fn{Recall that a $C^*$-algebra $\ca$ is called {\it separable} if it admits a
countable subset which is dense in the norm topology of $\ca$.}
$C^*$-algebras with a finite dual \cite{BL}. Given
any finite $T_0$-space $P$, it is possible to construct a $C^*$-algebra $\ca(P, d)$ of
operators on a separable 
\fn{Much as in the previous footnote, a Hilbert space $\ch$ is called {\it separable} if
it admits a countable basis.}  
Hilbert space $\ch(P, d)$ which satisfies
$\widehat{\ca(P, d)} = P$. Here $d$ is a function on $P$ with values in $\IN \cup \infty$
which is called {\it defector}. Thus there is more than one algebra with the same
structure space. We refer to \cite{BL} (see also \cite{ELTfunctions}) for the actual
construction of the algebras together with extensions to countable posets. We shall
instead describe a more general class of algebras, namely approximately finite
dimensional ones, a subclass of which is associated with posets. As the name suggests,
these algebras can be approximated by finite dimensional algebras, a fact which has
been used in the construction of physical models on posets as we shall describe in
Section~\ref{se:qmm}. They are also useful in the analysis of the $K$-theory of posets
as we shall see in Section~\ref{se:kt}. 

Before we proceed, we mention that if a separable $C^*$-algebra has a finite dual than
it is postliminar \cite{BL}. From Section~\ref{se:cop} we know that for any such algebra
$\ca$, irreducible representations are completely characterized by their kernels so that
the structure space $\ha$ is homeomorphic with the space $\prim$ of primitive ideals. As
we shall see momentarily, the Jacobson topology on $\prim$ is equivalent to the partial
order defined by inclusion of ideals. This fact in a sense `closes a circle' making
any poset, when thought of as the $\prim$ space of a noncommutative algebra, a truly
noncommutative space or, rather, a {\it  noncommutative lattice}.

\subsubsect{The space $\prim$ as a Poset}\label{jacorder}
Recall that in Section~\ref{se:jcp} we introduced the natural $T_0$-topology (the
Jacobson topology) on the space  $\prim$ of primitive ideals of a noncommutative
$C^*$-algebra $\ca$. In particular, from Prop.~\ref{pr1b}, we have that given any 
subset $W$ of $\prim$, 
\be\label{cassio}
W ~{\rm ~is ~closed}~~ \Leftrightarrow~~ \ci \in W ~~{\rm and}~~ \ci \subseteq \cj~
\Rightarrow J\in W~. 
\ee
Now, a partial order $\preceq$ is naturally introduced on $\prim$ by inclusion,
\be
\ci_1 \preceq \ci_2 ~~\Leftrightarrow~~ \ci_1 \subseteq \ci_2~, 
~~\forall ~\ci_1, \ci_2 \in \prim~.
\ee 
From what we said after (\ref{smcl}), given any subset $W$ of the topological
space $(\prim, \preceq)$, 
\be
W ~{\rm ~is ~closed}~~ \Leftrightarrow~~ \ci \in W ~~{\rm and}~~ \ci \preceq \cj~
\Rightarrow \cj \in W~, 
\ee
which is just the partial order reading of (\ref{cassio}). We infer that on $\prim$ the
Jacobson topology and the partial order 
topology~\index{Jacobson topology!and partial order} can be identified.
   
\subsubsect{AF-Algebras}\label{se:afa}
In this section we shall describe approximately finite dimensional algebras using mainly
\cite{Br1}. A general algebra of this sort may have a rather complicated ideal structure
and a complicated primitive ideal structure. As alluded to before, for applications
to posets only a special subclass is selected.
\bdefi~\index{AF algebra}~~\index{algebra!AF} 
A $C^*$-algebra $\ca$ is said to be {\it approximately finite dimensional}
(AF) if there exists an increasing sequence
\be
\ca_0 ~{\buildrel I_0 \over \hookrightarrow}~ \ca_1
      ~{\buildrel I_1 \over \hookrightarrow}~ \ca_2
      ~{\buildrel I_2 \over \hookrightarrow}~ \cdots
      ~{\buildrel I_{n-1} \over \hookrightarrow}~ \ca_n
      ~{\buildrel I_n \over \hookrightarrow} \cdots
\label{af}
\ee
of finite dimensional $C^*$-subalgebras of $\ca$, such that $\ca$ is the norm closure 
of $\bigcup_n \ca_n~, ~ \ca = \bar{\bigcup_n \ca_n}$. The maps $I_n$ are injective
$^*$-morphisms. 
\edefi

\noindent
The algebra $\ca$ is the {\it inductive} (or {\it direct}) {\it limit} of the sequence
$\{\ca_n, I_n \}_{n\in \IN}$ \cite{W-O}. As a set, $\bigcup_n \ca_n$ is made of coherent
sequences,
\be
\bigcup_n \ca_n = \{ a=(a_n)_{n \in \IN}~, a_n \in \ca_n ~|~ \exists  N_0 ~:
~a_{n+1} =  I_n(a_n)~, \forall ~n>N_0 \}.
\ee
Now the sequence $(||a_n||_{\ca_n})_{n \in \IN}$ is
eventually decreasing since $||a_{n+1}|| \leq ||a_n||$ (the maps $I_n$ are
norm decreasing) and therefore convergent. One writes for the norm on $\ca$,
\be
||(a_n)_{n \in \IN}|| = \lim_{n \ra \infty} ||a_n||_{\ca_n}~. \label{norm}
\ee
Since the maps $I_n$ are injective, the expression (\ref{norm}) gives a true
norm directly and not simply a seminorm and there is no need to quotient out the zero
norm elements. So, the algebra $\ca$ is the inductive (or direct) limit 
$\bar{\bigcup_n \ca_n}$ of the sequence  $\{\ca_n, I_n \}_{n \in \IN}$ \cite{Mu,W-O}.

\noindent
We shall assume that the algebra $\ca$ has a unit $\II$. If $\ca$ and $\ca_n$ are as
before, then $\ca_n + \IC \II$ is clearly a finite dimensional $C^*$-subalgebras
of $\ca$ and $\ca_n \subset \ca_n + \IC \II \subset \ca_{n+1} + \IC \II$. We may thus
assume that each $\ca_n$  contains the unit $\II$ of $\ca$ and that the maps $I_n$ are
unital.
\bexam
Let $\ch$ be an infinite dimensional (separable) Hilbert space. The algebra 
\be\label{alge}
\ca = \ck(\ch) + \IC\II_\ch~,
\ee
with $\ck(\ch)$ the algebra of compact operators, is an AF-algebra \cite{Br1}. The
approximating algebras are given by
\be
\ca_n = \IM_{n}(\IC) \oplus \IC~, ~~n > 0~, 
\ee
with embedding
\be\label{dentro}
\IM_{n}(\IC) \oplus \IC \ni (\L, \l) \mapsto
\left( \left\{
\begin{array}{ll}
\L & 0 \\
0 & \l
\end{array}
\right\}, \l 
\right)
\in \IM_{n+1}(\IC) \oplus \IC~. 
\ee
Indeed, let $\{\xi_n\}_{n\in\IN}$ be an orthonormal basis in $\ch$ and let $\ch_n$ be
the subspace generated by the first $n$ basis elements, $\xi_1, \cdots, \xi_n$. 
With $\cp_n$ the orthogonal projection onto $\ch_n$, define
\bea
\ca_n &=& \{T \in \cb(\ch) ~:~ T(\II - \cp_n) = (\II - \cp_n)T \in \IC (\II - \cp_n) \}
\nonumber \\
&\simeq&  \cb(\ch_n) \oplus \IC \simeq \IM_{n}(\IC) \oplus \IC~.  
\eea
Then $\ca_n$ embeds in $\ca_{n+1}$ as in (\ref{dentro}). Since each $T\in\ca_n$ is a
sum of a finite rank operator and a multiple of the identity, one has that  
$\ca_n \subseteq \ca = \ck(\ch) + \IC\II_\ch$ and, in turn,  
$\bar{\bigcup_n \ca_n} \subseteq \ca = \ck(\ch) + \IC\II_\ch$. Conversely, since finite
rank operators are norm dense in $\ck(\ch)$, and finite linear combinations of strings  
$\xi_1, \cdots, \xi_n$ are dense in $\ch$, one gets that 
$\ck(\ch) + \IC\II_\ch \subset \bar{\bigcup_n \ca_n}$.

\begin{figure}[htb]
\begin{center}
\begin{picture}(220,90)(-100,-30)
\put(30,30){\circle*{4}}
\put(0,-30){\circle*{4}}
\put(30,30){\line(-1,-2){30}}
\put(35,29){$\ci_2$}
\put(5,-33){$\ci_1$}
\end{picture}
\caption{\label{fi:point}
\protect{\footnotesize The two point poset of the interval.  }}
\end{center}
\vskip.5cm
\end{figure}
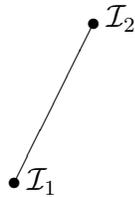

The algebra (\ref{alge}) has only two irreducible representations
\cite{BL},
\be\label{repalge}
\begin{array}{ll}
\pi_1 : \ca \lra \cb(\ch)  ~, & a = (k + \l \II_\ch) 
\mapsto \pi_1(a) = a~, \\
\pi_2 : \ca \lra \cb(\IC) \simeq \IC ~, & a = (k + \l \II_\ch)
\mapsto \pi_2(a) = \l~, 
\end{array}
\ee
with $\l_1,\l_2 \in \IC$ and $k \in \ck(\ch)$. The corresponding kernels are
\bea\label{kerint}
&& \ci_1 =: ker(\pi_1) = \{0\} ~, \nonumber \\
&& \ci_2 =: ker(\pi_2) = \ck(\ch) ~. 
\eea
The partial order given by the inclusions $\ci_1 \subset \ci_2$ produces the two points
poset shown in Fig.~\ref{fi:point}. 
As we shall see, this space is really the fundamental building
block for all posets. A comparison with the poset of the line in
Fig.~\ref{fi:linhas}, shows that it can be thought of as a two points approximation of
an interval. 
\eexam 

In general, each subalgebra $\ca_n$, being a finite dimensional $C^*$-algebra, is a
direct sum of  matrix algebras, 
\be
\ca_n = \bigoplus_{k=1}^{k_n} \IM_{d_k^{(n)}}(\IC)~,
\ee
where $\IM_d(\IC)$ is the algebra of $d \times d $ matrices with complex coefficients.
In order to study the embedding $\ca_1 \hookrightarrow \ca_2$ of any two such algebras 
$\ca_1 = \bigoplus_{j=1}^{n_1} \IM_{d_j^{(1)}}(\IC)$ and
$\ca_2 = \bigoplus_{k=1}^{n_2} \IM_{d_k^{(2)}}(\IC)$, it is useful the following
proposition \cite{Ef, W-O}.
\bprop\label{pr:afemb}
Let $\ca$ and $\cb$ be the direct sum of two matrix algebras, 
\be
\ca = \IM_{p_1}(\IC) \oplus \IM_{p_2}(\IC)~, 
~~~\cb = \IM_{q_1}(\IC) \oplus \IM_{q_2}(\IC)~. 
\ee
Then, any (unital) morphism $\a : \ca \ra \cb$ can be written as the direct sum of
the representations $\a_j : \ca \ra \IM_{q_j}(\IC) \simeq \cb(\IC^{q_j}),
j=1,2$. If $\p_{ji}$ is the unique irreducible representation of 
$\IM_{p_i}(\IC)$ in $\cb(\IC^{q_j})$, then $\a_j$ breaks into a direct sum of
the $\p_{ji}$ with multiplicity $N_{ji}$, the latter being non-negative integers.

\proof
This proposition just says that, by suppressing the symbols $\p_{ji}$, and modulo
a change of basis, the morphism $\a : \ca \ra \cb$ is of the form
\be
A\bigoplus B \mapsto \underbrace{A \oplus \cdots \oplus A}_{N_{11}} \oplus 
\underbrace{B \oplus \cdots \oplus B}_{N_{12}} \bigoplus
\underbrace{A \oplus \cdots \oplus A}_{N_{21}} \oplus 
\underbrace{B \oplus \cdots \oplus B}_{N_{22}} ~,  
\ee
with $A\bigoplus B \in \ca$. Moreover, the dimensions $(p_1,p_2)$ and $(q_1,q_2)$
are related by
\bea
N_{11} p_1 + N_{12} p_2 = q_1~, \nonumber \\
N_{21} p_1 + N_{22} p_2 = q_2~.
\eea
\eprop

Given a unital embedding  $\ca_1 \hookrightarrow \ca_2$ of the algebras 
$\ca_1 = \bigoplus_{j=1}^{n_1} \IM_{d_j^{(1)}}(\IC)$ and
$\ca_2 = \bigoplus_{k=1}^{n_2} \IM_{d_k^{(2)}}(\IC)$, by using
Proposition~\ref{pr:afemb} one can always choose suitable bases in $\ca_1$ and $\ca_2$
in such a manner to identify $\ca_1$ with a subalgebra of
$\ca_2$ of the following form
\be
\ca_1 \simeq \bigoplus_{k=1}^{n_2} \left( \bigoplus_{j=1}^{n_1} N_{kj}
\IM_{d_j^{(1)}}(\IC) \right)  \; .
\ee
Here, with any two nonnegative integers $p,q$, the symbol $p\IM_{q}(\IC)$ stands for
\be\label{azzo}
p\IM_{q}(\IC) \simeq \IM_{q}(\IC) \otc \II_p~,
\ee 
and one identifies $\bigoplus_{j=1}^{n_1} N_{kj} \IM_{d_j^{(1)}}(\IC)$ with a
subalgebra of $\IM_{d_k^{(2)}}(\IC)$. The nonnegative integers $N_{kj}$ satisfies the
condition
\be
\sum_{j=1}^{n_1} N_{kj} d^{(1)}_j = d^{(2)}_k  \; . \label{dim}
\ee
One says that the algebra $\IM_{d_j^{(1)}}(\IC)$ is 
{\it partially embedded}~index{partial embedding} in
$\IM_{d_k^{(2)}}(\IC)$ with {\it multiplicity} $N_{kj}$. A useful way to represent the
algebras $\ca_1$, $\ca_2$ and the embedding $\ca_1 \hookrightarrow \ca_2$ is by means
of a diagram, the 
{\it Bratteli diagram}~\index{Bratteli diagram} \cite{Br1}, which can be constructed out
of the dimensions $d_j^{(1)}~, ~j=1,\ldots,n_1$ and $d_k^{(2)}~, ~k=1,\ldots,n_2$, of the
diagonal blocks of the two algebras and of the numbers $N_{kj}$ that describe the
partial embeddings. One draws two horizontal rows of vertices, the top (bottom) ones
representing $\ca_1$ ($\ca_2$) and consisting of $n_1$ ($n_2$) vertices, one for each
block, labeled by the corresponding dimensions
$d_1^{(1)}, \ldots, d_{n_1}^{(1)}$ ($d_1^{(2)},\ldots,d_{n_2}^{(2)}$). Then, for each
$j=1,\ldots,n_1$ and $k=1,\ldots,n_2$, one has a relation 
$d^{(1)}_j \searrow^{N_{kj}} d^{(2)}_k$ to denote the fact that
$\IM_{d^{(1)}_j}(\IC)$ is embedded in $\IM_{d^{(2)}_k}(\IC)$ with
multiplicity $N_{kj}$.

For any AF-algebra $\ca$ one repeats the procedure for each
level so obtaining a semi-infinite diagram denoted by $\cd(\ca)$ which completely
defines $\ca$ up to isomorphism. The diagram 
$\cd(\ca)$~\index{Bratteli diagram!of an AF algebra|(} depends not only on $\ca$
but also on the particular sequence $\{\ca_n\}_{n \in \IN}$ which generate
$\ca$. However, one can obtain an algorithm which allows one to construct from a given
diagram all diagrams which define AF-algebras which are isomorphic with the
original one \cite{Br1}. The problem of identifying the limit algebra or of
determining whether or not two such limits are isomorphic can be very subtle. Elliot
\cite{El} has devised a complete invariant for AF-algebras in terms of the
corresponding $K$ theory which distinguishes among them (see also \cite{Ef}). We shall
elaborate a bit on this in Section~\ref{se:kt}. It is worth remarking that the
isomorphism class on an AF-algebra $\bar{\bigcup_n \ca_n}$ depends not only on the
$\ca_n$ but also on the way they are embedded into each other. 

Any AF-algebra is clearly separable but the converse is not true. Indeed, one can
prove that a separable $C^*$-algebra $\ca$ is an  AF-algebra if and only if and it has
the following approximation property: for each finite set $\{a_1, \dots , a_n\}$ of
elements of $\ca$ and $\ve > 0$, there exists a finite dimensional $C^*$-algebra $ \cb
\subseteq \ca$ and elements $b_1, \dots, b_n \in \cb$ such that $\norm{a_k - b_k} < \ve
~, k = 1, \dots, n $~.

Given a set $\cd$ of ordered pairs $(n, k), k = 1, \cdots, k_n~, ~n = 0, 1, \cdots$,
with $k_0=1$, and a sequence $\{ \searrow^p \}_{p = 0, 1, \cdots }$ of relations on
$\cd$, the latter is the diagram $\cd(\ca)$ of an AF-algebras when the following
conditions are satisfied, \label{`brdiagram'}
\begin{enumerate}
\item[(i)] If $(n, k), (m, q) \in \cd$ and $m = n+1$, there exists one and only one
nonnegative (or equivalently, at most a positive) integer $p$ such that 
$(n, k) \searrow^p (n+1, q)$.
\item[(ii)] If $m \not= n+1$ not such integer exists.
\item[(iii)] If $(n, k) \in \cd$ there exists $q \in \{1, \cdots, n_{n+1} \}$ and a
nonnegative integer $p$ such that $(n, k) \searrow^p (n+1, q)$.
\item[(iv)] If $(n, k) \in \cd$ and $n > 0$, there exists $q \in \{1, \cdots, n_{n-1}
\}$ and a nonnegative integer $p$ such that $(n-1, q) \searrow^p (n, k) $.
\end{enumerate}

It is easy to see that the diagram of a given AF-algebra satisfies the previous
conditions. Conversely, if the set $\cd$ of ordered pairs satisfies these properties,
one constructs by induction a sequence of finite dimensional
$C^*$-algebras $\{ \ca_n \}_{n\in\IN}$  and of injective morphisms $I_n : \ca_n \ra
\ca_{n+1}$ in such a manner that the inductive limit $\{ \ca_n, I_n \}_{n\in\IN}$ will
have diagram $\cd$. Explicitly, one defines
\be
\ca_n = \bigoplus_{k; (n,k) \in \cd} \IM_{d_k^{(n)}}(\IC) = \bigoplus_{k=1}^{k_n}
\IM_{d_k^{(n)}}(\IC)~, 
\ee
and morphisms  
\bea
&& I_n : \bigoplus_{j=1}^{j_n} \IM_{d_j^{(n)}}(\IC) \lra \bigoplus_{k=1}^{k_{n+1}}
\IM_{d_k^{(n+1)}}(\IC) ~, \nonumber \\
&& A_1 \oplus \cdots \oplus A_{j_n} ~\mapsto~ (\oplus_{j=1}^{j_n} N_{1j} A_j) \bigoplus
\cdots \bigoplus  (\oplus_{j=1}^{j_n} N_{k_{n+1} j} A_j)~, 
\eea
where the integers $N_{kj}$ are such that  $(n, j) \searrow^{N_{kj}} (n+1, k)$ and we
have used the notation (\ref{azzo}). Notice that the dimension $d_k^{(n+1)}$ of the
factor $\IM_{d_k^{(n+1)}}(\IC)$ is not arbitrary but it is determined by a relation like
(\ref{dim}), namely $d_k^{(n+1)} = \sum_{j=1}^{j_n} N_{kj} d^{(n)}_j$.
~\index{Bratteli diagram!of an AF algebra|)}
\bexam
An AF-algebra $\ca$ is abelian if and only if all factors $\IM_{d_k^{(n)}}(\IC)$ are one
dimensional, $\IM_{d_k^{(n)}}(\IC) \simeq \IC$. Thus the corresponding diagram $\cd$ has
the property that for each $(n,k) \in \cd, n > 0$, there is exactly one $(n-1,j) \in \cd$
such that $(n-1, j) \searrow^{1} (n, k)$. 
\eexam

\noindent
There is a very nice characterization of
commutative AF-algebras and of their primitive spectra \cite{Br2}.
\bprop
Let $\ca$ be a commutative $C^*$-algebra with unit $\ci$. Then the following
statements are equivalent.
\begin{itemize}
\item[(i)] The algebra $\ca$ is AF.
\item[(ii)] The algebra $\ca$ is generated in the norm topology by a sequence of
projectors $\{\cp_i \}$, with $\ci_0 = \ci$.
\item[(iii)] The space $\prim$ is a second-countable, totally disconnected, compact
Hausdorff space
\fn{\rm We recall that a topological space is called totally disconnected if the
connected component of each point consists only of the point itself. Also, a
topological space is called second-countable is it admits a countable basis of open
sets.}.
\end{itemize}

\begin{figure}[t]
\begin{center}
\begin{picture}(120,180)(0,30)
\put(30,150){\circle*{4}}
\put(30,120){\circle*{4}}
\put(30,90){\circle*{4}}
\put(30,60){\circle*{4}}
\put(60,180){\circle*{4}}
\put(60,120){\circle*{4}}
\put(60,90){\circle*{4}}
\put(60,60){\circle*{4}}
\put(90,150){\circle*{4}}
\put(90,120){\circle*{4}}
\put(90,90){\circle*{4}}
\put(90,60){\circle*{4}}
\put(30,150){\line(0,-1){30}}
\put(30,120){\line(0,-1){30}}
\put(30,90){\line(0,-1){30}}
\put(60,120){\line(0,-1){30}}
\put(60,90){\line(0,-1){30}}
\put(90,150){\line(0,-1){30}}
\put(90,120){\line(0,-1){30}}
\put(90,90){\line(0,-1){30}}
\put(30,60){\line(0,-1){10}}
\put(60,60){\line(0,-1){10}}
\put(90,60){\line(0,-1){10}}
\put(60,180){\line(1,-1){30}}
\put(30,150){\line(1,-1){30}}
\put(30,120){\line(1,-1){30}}
\put(30,90){\line(1,-1){30}}
\put(60,180){\line(-1,-1){30}}
\put(90,150){\line(-1,-1){30}}
\put(90,120){\line(-1,-1){30}}
\put(90,90){\line(-1,-1){30}}
\put(30,60){\line(1,-1){10}}
\put(90,60){\line(-1,-1){10}}
\put(54,185){{\small$1$}} 
\put(16,148){{\small$1$}} 
\put(16,118){{\small$1$}}
\put(16,88){{\small$1$}} 
\put(16,58){{\small$1$}} 
\put(97,148){{\small$1$}}
\put(97,118){{\small$1$}} 
\put(97,88){{\small$1$}} 
\put(97,58){{\small$1$}}
\put(63,113){$2$}
\put(63,83){$4$}
\put(63,53){$6$}
\put(45,30){$\vdots$}
\put(75,30){$\vdots$}
\end{picture}
\caption{\label{fi:veealg}
\protect{\footnotesize The Bratteli diagram of the algebra $\ca_\vee$. The labels
indicate the dimension of the corresponding matrix algebras.}}
\end{center}
\vskip.5cm
\end{figure}
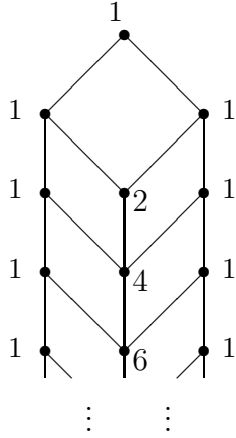

\proof
The equivalence of $(i)$ and $(ii)$ is clear. To prove that $(iii)$ implies $(ii)$, let
$X$ be a second-countable, totally disconnected, compact Hausdorff space. Then $X$ has a
countable basis $\{X_n\}$ of open-closed sets. Let $\cp_n$ be the characteristic
function of $X_n$. The $^*$-algebra generated by the projector $\{\cp_n\}$ is dense in
$C(X)$: since $PrimC(X) = X$, $(iii)$ implies $(ii)$. The converse, that $(ii)$ implies
$(iii)$, follows from the fact that projectors in a commutative $C^*$-algebra
correspond to open-closed subset in its primitive spaces. 
\eprop
\bexam
Let us consider the subalgebra $\ca$ of the algebra
$\cb(\ch)$ of bounded operators on an infinite dimensional (separable) Hilbert space
$\ch = \ch_1
\oplus \ch_2$, given in the following manner. Let $\cp_j$ be the projection operators
on $\ch_j~, j = 1, 2$ and $\ck(\ch)$ the algebra of compact operators on $\ch$. Then,
the algebra $\ca$ is 
\be
\ca_\vee = \IC\cp_1 + \ck({\ch}) + \IC\cp_2~. \label{alvee}
\ee
The use of the symbol $\ca_\vee$ is due to the fact that, as we shall see below, this
algebra is associated with any part of the poset of the line in Fig.~\ref{fi:linhas},
of the form 
\be
\bigvee = \{y_{i-1}, x_i, y_i\}~, \label{pvee}
\ee
in the sense that this poset is identified with the space of primitive ideals of
$\ca_\vee$. The $C^*$-algebra (\ref{alvee}) can be obtained as the direct limit of
the following sequence of finite dimensional algebras:
\bea
& & \ca_0 = \IM_{1}(\IC)   \nonumber \\
& & \ca_1 = \IM_{1}(\IC) \oplus \IM_{1}(\IC)  \nonumber \\
& & \ca_2 = \IM_{1}(\IC) \oplus \IM_{2}(\IC) \oplus \IM_{1}(\IC)  \nonumber \\
& & \ca_3 = \IM_{1}(\IC) \oplus \IM_{4}(\IC) \oplus \IM_{1}(\IC) \nonumber \\
& & ~~~ \vdots \nonumber \\
& & \ca_n = \IM_{1}(\IC) \oplus \IM_{2n-2}(\IC) \oplus \IM_{1}(\IC) \nonumber \\
& & ~~~ \vdots
\label{vee}
\eea
where, for $n \geq 1$, $\ca_n$ is embedded in $\ca_{n+1}$ as follows
\bea
&& \IM_{1}(\IC) \oplus \IM_{2n-2}(\IC) \oplus \IM_{1}(\IC) ~\hookrightarrow \nonumber
\\ && ~~~~~~~~~~~~~~~ \hookrightarrow ~ \IM_{1}(\IC) \oplus (\IM_{1}(\IC)
\oplus \IM_{2n-2}(\IC) \oplus \IM_{1}(\IC)) \oplus \IM_{1}(\IC) \nonumber \\ 
&& ~ \nonumber \\
&& \left[
\begin{array}{ccc}
\l_1 & 0                  & 0    \\
0    & B_{(2n-2)\times(2n-2)} & 0    \\
0    & 0                  & \l_2
\end{array}
\right]~~ \mapsto ~
\left[
\begin{array}{ccccc}
\l_1 & 0    & 0                  & 0       & 0      \\
0    & \l_1 & 0                  & 0       & 0      \\
0    & 0    & B_{(2n-2)\times(2n-2)} & 0       & 0      \\
0    & 0    & 0                  & \l_2    & 0       \\
0    & 0    & 0                  & 0       & \l_2
\end{array}
\right]~.
\label{vee1}
\eea
The corresponding Bratteli diagram is in Fig.~\ref{fi:veealg}.

The algebra (\ref{alvee}) has three irreducible representations,
\be\label{repvee}
\begin{array}{ll}
\pi_1 : \ca_\vee \lra \cb(\ch)  ~, & a = (\l_1 \cp_1 + k + \l_2 \cp_2) 
\mapsto \pi_1(a) = a~, \\
\pi_2 : \ca_\vee \lra \cb(\IC) \simeq \IC ~, & a = (\l_1 \cp_1 + k + \l_2 \cp_2)
\mapsto \pi_2(a) = \l_1~, \\
\pi_3 : \ca_\vee \lra \cb(\IC) \simeq \IC ~, & a = (\l_1 \cp_1 + k + \l_2 \cp_2)
\mapsto \pi_3(a) = \l_2~, 
\end{array}
\ee
with $\l_1,\l_2 \in \IC$ and $k \in \ck(\ch)$. The corresponding kernels are
\bea\label{kervee}
&& \ci_1 = \{0\} ~, \nonumber \\
&& \ci_2 = \ck(\ch) + \IC\cp_2 ~, \nonumber \\
&& \ci_3 = \IC\cp_1 + \ck(\ch) ~. 
\eea
The partial order given by the inclusions $\ci_1 \subset \ci_2$ and $\ci_1 \subset
\ci_3$ (which, as shown in Section~\ref{jacorder} is an equivalent way to provide the
Jacobson topology) produces a topological space $Prim{\cal A}_{\vee}$ which is just the
$\bigvee$ poset in (\ref{pvee}).
\eexam 

\subsubsect{From Bratteli Diagrams to Noncommutative Lattices}\label{se:blb}

{}From the Bratteli diagram of an AF-algebra $\ca$ one can also obtain the (norm closed
two-sided) ideals of the latter and determine which ones are primitive. On the set of
such ideals the topology is then given by constructing a poset whose partial order is
provided by inclusion of ideals. Therefore, both $Prim(\ca)$ and its topology can be
determined from the Bratteli diagram of $\ca$. This is possible thanks to the following
results by Bratteli \cite{Br1}.
\bprop\label{bratteli}
Let $\ca = \bar{\bigcup_n \cu_n}$ be any AF-algebra with associated Bratteli diagram
$\cd(\ca)$. Let $\ci$ be an ideal of $\ca$. Then $\ci$ has the form 
\be\label{idla}
\ci = \bar{\bigcup_{n=1}^{\infty} \oplus_{k; (n,k) \in \Lambda_{\ci}}
~\IM_{d_k^{(n)}}(\IC)}~  
\ee
with the subset $\Lambda_{\ci} \subset \cd(\ca)$ satisfying the following two properties:
\begin{itemize}
\item[i)]
if $(n, k) \in \Lambda_{\ci}$ and $(n, k) \searrow^p (n+1, j)~, p > 0$, then $(n+1, j)$
belongs to $\Lambda_{\ci}$;
\item[ii)]
if all factors $(n+1, j)~, j = 1, \dots, n_{n+1}, $ in which $(n, k)$ is partially
embedded belong to $\Lambda_{\ci}$, then $(n, k)$ belongs to $\Lambda_{\ci}$.
\end{itemize}
Conversely, if $\Lambda \subset \cd(\ca)$ satisfies properties $(i)$ and $(ii)$ above,
then the subset $\ci_\Lambda$ of $\ca$ defined by (\ref{idla}) (with $\Lambda$
substituted for $\Lambda_\ci$) is an ideal in $\ca$ such that $ \ci \bigcap
\ca_n = \oplus_{k; (n,k) \in \Lambda_{\ci}} ~\IM_{d_k^{(n)}}(\IC)$.
\eprop
\bprop\label{primitive}
Let $\ca = \bar{\bigcup_n \cu_n}$, let $\ci$ be an ideal of $\ca$ and let
$\Lambda_{\ci} \subset \cd(\ca)$ be the associated subdiagram.
Then the following conditions are equivalent
\fn{In fact, the equivalence of $(i)$ and $(ii)$ is true for any separable
$C^*$-algebra \cite{Di1}}. 
\begin{itemize}
\item[(i)] The ideal $\ci$ is primitive.~\index{ideal!primitive} 
\item[(ii)] There does not exist two ideals $\ci_1, \ci_2 \in \ca$ such that $\ci_1
\not= \ci \not= \ci_2$ and $\ci = \ci_1 \cap \ci_2$.
\item[(iii)] If $(n, k), (m, q) \notin \Lambda_{\ci}$, there exists an integer 
$p \geq n, m$ and a couple $(p,r) \notin \Lambda_{\ci}$ such that
$\IM_{d_{k}^{(n)}}(\IC)$ and $\IM_{d_{q}^{(m)}}(\IC)$ are both partially embedded in
$\IM_{d_{r}^{(p)}}(\IC)$  (equivalently, there are two sequences along the diagram
$\cd(\ca)$ starting at the points $(n, k)$ and $(m, q)$  both ending at the point
$(p,r)$). 
\end{itemize}
\eprop

\noindent
We recall that the whole $\ca$ is an ideal which, by definition, is not primitive
since the trivial representation $\ca \rightarrow 0$ is not irreducible. 
Furthermore, the ideal $\{0\} \subset \ca$ is primitive if and only if $\ca$ is
primitive, namely it has an irreducible faithful representation. This fact can also be
inferred from the Bratteli diagram. Now, the ideal $\{0\}$, being represented by 
the element $0 \in \ca_n$ at each level 
\fn{In fact one could think of $\Lambda_{\{0\}}$ as being the empty set.}, is not
associated with any  subdiagram of $\cd(\ca)$. Therefore, to check if
$\{0\}$ is primitive we have the following corollary of Proposition~\ref{primitive}.
\bprop \label{zeroprim}
Let $\ca = \bar{\bigcup_n \cu_n}$.
Then the following conditions are equivalent.
\begin{itemize}
\item[(i)] The algebra $\ca$ is 
primitive~\index{cstaral@$C^*$-algebra!primitive} (namely the ideal $\{0\}$ is
primitive). 
\item[(ii)] There does not exist two ideals in $\ca$ different from $\{0\}$ whose
intersection is $\{0\}$. 
\item[(iii)] If $(n, k), (m, q) \in \cd(\ca)$, there exists an integer 
$p \geq n, m$ and a couple $(p,r) \in \cd(\ca)$ such that
$\IM_{d_{k}^{(n)}}(\IC)$ and $\IM_{d_{q}^{(m)}}(\IC)$ are both partially embedded in
$\IM_{d_{r}^{(p)}}(\IC)$  (equivalently, {\rm any} two points of the diagram
$\cd(\ca)$ can be connected to a {\rm single} point at a later level of the diagram). 
\end{itemize}
\eprop

\noindent
For instance, from the diagram of Fig.~\ref{fi:veealg} we infer that the
corresponding algebra is primitive, namely the ideal $\{0\}$ is primitive. 

\begin{figure}[t]
\begin{center}
\begin{picture}(340,200)(0,10)
\put(57,117){$\star$}
\put(57,87){$\star$}
\put(57,57){$\star$}
\put(87,147){$\star$}
\put(87,117){$\star$}
\put(87,87){$\star$}
\put(87,57){$\star$}
\put(30,150){\line(0,-1){30}}
\put(30,120){\line(0,-1){30}}
\put(30,90){\line(0,-1){30}}
\put(60,120){\line(0,-1){30}}
\put(60,90){\line(0,-1){30}}
\put(90,150){\line(0,-1){30}}
\put(90,120){\line(0,-1){30}}
\put(90,90){\line(0,-1){30}}
\put(30,60){\line(0,-1){10}}
\put(60,60){\line(0,-1){10}}
\put(90,60){\line(0,-1){10}}
\put(60,180){\line(1,-1){30}}
\put(30,150){\line(1,-1){30}}
\put(30,120){\line(1,-1){30}}
\put(30,90){\line(1,-1){30}}
\put(60,180){\line(-1,-1){30}}
\put(90,150){\line(-1,-1){30}}
\put(90,120){\line(-1,-1){30}}
\put(90,90){\line(-1,-1){30}}
\put(30,60){\line(1,-1){10}}
\put(90,60){\line(-1,-1){10}}
\put(147,147){$\star$}
\put(147,117){$\star$}
\put(147,87){$\star$}
\put(147,57){$\star$}
\put(177,117){$\star$}
\put(177,87){$\star$}
\put(177,57){$\star$}
\put(150,150){\line(0,-1){30}}
\put(150,120){\line(0,-1){30}}
\put(150,90){\line(0,-1){30}}
\put(180,120){\line(0,-1){30}}
\put(180,90){\line(0,-1){30}}
\put(210,150){\line(0,-1){30}}
\put(210,120){\line(0,-1){30}}
\put(210,90){\line(0,-1){30}}
\put(150,60){\line(0,-1){10}}
\put(180,60){\line(0,-1){10}}
\put(210,60){\line(0,-1){10}}
\put(180,180){\line(1,-1){30}}
\put(150,150){\line(1,-1){30}}
\put(150,120){\line(1,-1){30}}
\put(150,90){\line(1,-1){30}}
\put(180,180){\line(-1,-1){30}}
\put(210,150){\line(-1,-1){30}}
\put(210,120){\line(-1,-1){30}}
\put(210,90){\line(-1,-1){30}}
\put(150,60){\line(1,-1){10}}
\put(210,60){\line(-1,-1){10}}
%
\put(297,117){$\star$}
\put(297,87){$\star$}
\put(297,57){$\star$}
\put(270,150){\line(0,-1){30}}
\put(270,120){\line(0,-1){30}}
\put(270,90){\line(0,-1){30}}
\put(300,120){\line(0,-1){30}}
\put(300,90){\line(0,-1){30}}
\put(330,150){\line(0,-1){30}}
\put(330,120){\line(0,-1){30}}
\put(330,90){\line(0,-1){30}}
\put(270,60){\line(0,-1){10}}
\put(300,60){\line(0,-1){10}}
\put(330,60){\line(0,-1){10}}
\put(300,180){\line(1,-1){30}}
\put(270,150){\line(1,-1){30}}
\put(270,120){\line(1,-1){30}}
\put(270,90){\line(1,-1){30}}
\put(300,180){\line(-1,-1){30}}
\put(330,150){\line(-1,-1){30}}
\put(330,120){\line(-1,-1){30}}
\put(330,90){\line(-1,-1){30}}
\put(270,60){\line(1,-1){10}}
\put(330,60){\line(-1,-1){10}}
\put(52,10){$(a)$}
\put(172,10){$(b)$}
\put(292,10){$(c)$}
\put(75,170){$\ci_2$}
\put(195,170){$\ci_3$}
\put(315,170){$\ci_\ck$}
\put(45,30){$\vdots$}
\put(75,30){$\vdots$}
\put(165,30){$\vdots$}
\put(195,30){$\vdots$}
\put(285,30){$\vdots$}
\put(315,30){$\vdots$}
\end{picture}
\caption{\label{fi:veealgid}
\protect{\footnotesize The three ideals of the algebra
$\ca_\vee$. }}
\end{center}
\vskip.5cm
\end{figure}
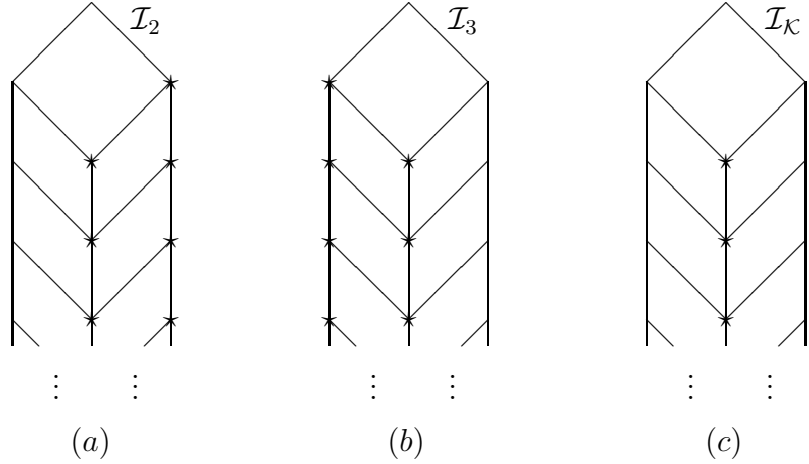

\bexam
As a simple example, consider the diagram of Fig.~\ref{fi:veealg}. The
corresponding AF-algebra $\ca_\vee$ in (\ref{alvee}) contains only three nontrivial
ideals, whose diagrammatic representation is in Fig.~\ref{fi:veealgid}. In this
pictures the points belonging to the same ideal are marked with a $``\star"$. It is
not difficult to check that only $\ci_2$ and $\ci_3$ are primitive ideals, since
$\ci_\ck$ does not satisfy property $(iii)$ above. Now $\ci_1 = \{0\}$ is an
ideal which clearly belongs to both $\ci_2$ and $\ci_3$ so that $Prim(\ca)$ is any
$\bigvee$ part of Fig.~\ref{fi:linhas} of the form  ~$\bigvee = \{y_{i-1}, x_i, y_i\}$. 
From the diagram of Figure~\ref{fi:veealgid} one immediately obtains
\bea
&& \ci_2 = \IC \II_\ch + \ck(\ch)~, \nonumber \\
&& \ci_1 = \IC \II_\ch + \ck(\ch)~, 
\eea 
$\ch$ being an infinite dimensional Hilbert space. Thus, $\ci_2$ and $\ci_3$ can be
identified with the corresponding ideals of $\ca_\vee$ given in (\ref{kervee}). As
for $\ci_\ck$, from Figure~\ref{fi:veealgid} one gets $\ci_\ck = \ck(\ch)$ which is
not  a primitive ideal of $\ca_\vee$.    
\eexam

\subsubsect{From Noncommutative Lattices to Bratteli Diagrams}\label{se:nlb}

Bratteli \cite{Br2} has also a reverse procedure which allows one to construct an 
AF-algebra (or rather its Bratteli diagram $\cd(\ca)$) whose primitive ideal space is a
given (finitary, noncommutative) lattice $P$. We shall briefly describe this procedure
while referring to \cite{ELTkappa,ELTfunctions} for more details and several examples.
\bprop\label{most}
Let $P$ be a topological space with the following properties,
\begin{itemize}
\item[(i)] The space $P$ is $T_0$.
\item[(ii)] If $F \subset P$ is a closed set which is not the union of two proper
closed subset, then $F$ is the closure of a one-point set.
\item[(iii)] The space $P$ contains at most a countable number of closed sets.
\item[(iv)] If $\{F_n\}_n$ is a decreasing ($F_{n+1} \subset F_n)$ sequence of closed
subsets of $P$, then $\bigcap_n F_n$ is an element in $\{F_n\}_n$.
\end{itemize} 
Then, there exists an AF algebra $\ca$ whose primitive space $\prim$ is
homeomorphic to $P$.

\proof
The proof consists in constructing explicitly the Bratteli diagram $\cd(\ca)$ of the
algebra $\ca$. We shall sketch the main passages while referring to \cite{Br2} for more
details.
\\ \\
Let $\{K_0, K_1, K_2, \ldots \}$ be the collection of all closed sets in the lattice
$P$, with $K_0 = P$.
\\ \\
Consider the subcollection $\ck_n = \{K_0,K_1,\ldots, K_n\}$ and
let $\ck_n'$ be the smallest collection of (closed) sets in $P$ containing 
$\ck_n$ which is closed under union and intersection. 
\\ \\
Consider the algebra of sets
\fn{We recall that a non empty collection $R$ of subsets of a set $X$ is called an
algebra if $R$ is closed under the operation of union, namely $E,F \in R \Rightarrow
E \cup F \in R$ and the operation of complement, namely $E \in R \Rightarrow E^c =: X
\setminus E \in R$.}  generated by the collection
$\ck_n$. Then, the minimal sets $\cuai_n = \{Y_n(1), Y_n(2), \ldots, Y_n(k_n) \}$ of the
algebra form a partition of $P$. 
\\ \\
Let $F_n(j)$ be the smallest set in the subcollection $\ck_n'$ which contains
$Y_n(j)$. Define $\cf_n = \{F_n(1), F_n(2), \ldots, F_n(k_n)\}$. 
\\ \\
As a consequence of the assumptions in the propositions one has that
\bea
&& Y_n(k) \subseteq F_n(k)~, \label{a1} \\
&& \bigcup_k Y_n(k) = P~,    \label{a2} \\
&& \bigcup_k F_n(k) = P~,    \label{a3} \\
&& Y_n(k) = F_n(k) \setminus \bigcup_{p\not=k} 
\{ F_n(p) ~|~ F_n(p) \subset F_n(k)\}~, \label{a4} \\
&& F_n(k) = \bigcup_{p } 
\{ F_{n+1}(p) ~|~ F_{n+1}(p) \subseteq F_n(k)\}~, \label{a5}\\
&& {\rm If}~~ F\subset P ~~{\rm is ~closed}~~, ~\exists ~n \geq 0~, ~~{\rm s.t.}~
F_n(k) = \bigcup_{p } \{ F_{n}(p) ~|~ F_{n}(p) \subseteq F  \}~. \label{a6}
\eea 

\noindent
The diagram $\cd(\ca)$ is constructed as follows.
\begin{itemize}
\item[1.]
{\it The $n$-th level of $\cd(\ca)$ has $k_n$ points, one for each set 
$Y_n(k), k=1, \cdots, k_n$}.\\ 
Thus $\cd(\ca)$ is the set of all ordered pairs $(n,k)~,
k =1, \ldots, k_n~, n = 0, 1, \ldots $. 
\item[2.]
{\it The point corresponding to $Y_n(k)$ at the level $n$ of the diagram is linked to
the point corresponding to $Y_{n+1}(j)$ at the level $n+1$, if and only if
$Y_n(k) \cap~F_{n+1}(j) \neq \emptyset$. The multiplicity of the embedding is always
$1$}. \\
Thus, the partial embeddings of the diagram are given by 
\bea
(n,k) &\searrow^p& (n+1,j)~, ~~~{\rm with} \nonumber \\
&& p = 1 ~~{\rm if}~~ Y_n(k) \cap~F_{n+1}(j) \neq \emptyset ~, \nonumber \\
&& p = 0 ~~{\rm otherwise}~.
\eea
\end{itemize}
That the diagram $\cd(\ca)$ is really the diagram of an AF algebra $\ca$, namely that
conditions $(i)-(iv)$ of page~\pageref{`brdiagram'} are satisfied, follows from
conditions (\ref{a1})-(\ref{a6}) above.

\noindent
Before we proceed, recall from Proposition~(\ref{pr1a}) that there is a bijective   
correspondence between ideals in a $C^*$-algebra and closed sets in $\prim$, the
correspondence being given by (\ref{r1a}). We shall then construct a similar
correspondence between closed subsets $F \subseteq P$ and  the ideals $\ci_F$ in the
AF-algebra $\ca$ with subdiagram  $\L_F \subseteq \cd(\ca)$. Given then, a closed subset
$F \subseteq P$, from (\ref{a6}), there exists an $m$ such that $F \subseteq \ck'_m$.
Define
\be
(\L_F)_n = \{(n,k) ~|~ n \geq m~, ~Y_n(k) \cap F = \emptyset \}~. 
\ee 
By using (\ref{a4}) one proves that conditions $(i)$ and $(ii)$ of
Proposition~\ref{bratteli} are satisfied. As a consequence, if $\L_F$ is the smallest
subdiagram corresponding to an ideal $\ci_F$, namely satisfying conditions $(i)$ and
$(ii)$ of Proposition~\ref{bratteli}, which also contains $(\L_F)_n$, one has that
\be\label{mah}
(\L_F)_n = \L_F \bigcap \{(n, k) ~|~ (n, k) \in \cd(\ca), ~n \geq m \}~,
\ee
which, in turn, implies that the mapping $F \mapsto \L_F \leftrightarrow \ci_F$ is
injective. \\ 
To show surjectivity, let $\ci$ be an ideal in $\ca$ with associated
subdiagram $\L_\ci$. Define
\be
F_n = P ~\setminus~  \bigcup_k \{Y_n(k) ~|~ \exists (n-1, p) \in \L_\ci~, ~(n-1, p)
\searrow^1 \in \L_\ci \}~, ~~~ n = 0, 1, \ldots ~.
\ee
Then $\{F_n \}_n$ is a decreasing sequence of closet sets in $P$. By assumption $(iv)$,
there exists an $m$ such that $F_m = \bigcap_n F_n$. By defining $F = F_n$, one has $F_n
= F$ for $n\geq m$ and 
\be
\L_\ci \bigcap \{(n, k) ~|~  n \geq m \} =: (\L_F)_m~.
\ee
Thus, $\L_\ci = \L_F$ and the mapping $F \mapsto \ci_F$ is surjective. \\
Finally, from definition it follows that 
\be\label{cor}
F_1 \subseteq F_2 ~\Longleftrightarrow~ \ci_{F_1} \supseteq \ci_{F_2}~.
\ee
For any point $x \in P$, the closure $\bar{\{x \}}$ is not the union of two proper
closed subset. From (\ref{cor}), the corresponding ideal $\ci_{\bar{\{x \}}}$ is not the
intersection of two ideals different from itself, thus it is primitive (see
Proposition~\ref{primitive}). Conversely, if $\ci_F$ is primitive, it is not the
intersection of two ideals different from itself, thus from (\ref{cor}) $F$ is not the
union of two proper closed subsets, and from assumption $(ii)$, it is the closure of a
one-point set. We have then proved that the ideal $\ci_F$ is primitive if and only if
$F$ is the closure of a one-point set. \\
By taking into account the bijection between closed sets of the space $P$ and ideals of
the algebra $\ca$ and the corresponding bijection between of closed sets of the space
$\prim$ and ideals of the algebra $\ca$, we see that the bijection between $P$ and
$\prim$ which associates to any point of $P$ the corresponding primitive ideal, is a
homeomorphism.
\eprop

\noindent
We know that different algebras could give the same space of primitive ideals (see the
notion of strong Morita equivalence in Section~\ref{se:sme}). It may happen that by
changing the order in which the closed sets of $P$ are taken in the construction of the
previous proposition, one produces different algebras, all of which having the same
space of primitive ideals though, so all producing spaces which are homeomorphic to
the starting $P$ (any two of these spaces being, a fortiori, homeomorphic). 
\bexam
As a simple example, consider again the lattice, 
$\bigvee = \{y_{i-1}, x_i, y_i\} \equiv \{x_2, x_1, x_3 \}$. 
This topological space contains four closed sets: 
\be
K_0=\{x_2, x_1, x_3 \}~, K_1 =\{x_2\}~, K_2 =\{x_3\}~, K_3 =\{x_2, x_3\} =
K_1
\cup K_2~.
\ee
Thus, it is not difficult to check that:
\be
\begin{array}{llll}
\ck_0=\{K_0\} & \ck_0'=\{K_0\} & Y_0(1)=\{x_1, x_2, x_3\} & F_0(1)=K_0 \\ 
\ck_1=\{K_0,K_1\} & \ck_1'=\{K_0,K_1\} & Y_1(1)=\{x_2\}  & F_1(1)=K_1 \\
              &            & Y_1(2)=\{x_1, x_3\} & F_1(2)=K_0 \\
\ck_2=\{K_0,K_1,K_2\} & \ck_2'=\{K_0,K_1,K_2,K_3\}         
                                         & Y_2(1)=\{x_2\}  & F_2(1)=K_1 \\
                 &                       & Y_2(2)=\{x_1\}    & F_2(2)=K_0 \\
                 &                       & Y_2(3)=\{x_3\}   & F_2(3)=K_2 \\
\ck_3=\{K_0,K_1,K_2,K_3\} & \ck_3'=\{K_0,K_1,K_2,K_3\} 
                                     & Y_3(1)=\{x_2\}   & F_3(1)=K_1 \\
                  &                    & Y_3(2)=\{x_1\}    & F_3(2)=K_0 \\
                  &                    & Y_3(3)=\{x_3\}  & F_3(2)=K_2 \\
  &  \vdots & &
\end{array}
\ee
Since $\bigvee$ has only a finite number of points (three) and hence a finite number
of closed sets (four), the partition of $\bigvee$ repeats itself after the third 
level.
\begin{figure}[t]
\begin{center}
\begin{picture}(120,180)(0,30)
\put(30,150){\circle*{4}}
\put(30,120){\circle*{4}}
\put(30,90){\circle*{4}}
\put(30,60){\circle*{4}}
\put(60,180){\circle*{4}}
\put(60,120){\circle*{4}}
\put(60,90){\circle*{4}}
\put(60,60){\circle*{4}}
\put(90,150){\circle*{4}}
\put(90,120){\circle*{4}}
\put(90,90){\circle*{4}}
\put(90,60){\circle*{4}}
\put(30,150){\line(0,-1){30}}
\put(30,120){\line(0,-1){30}}
\put(30,90){\line(0,-1){30}}
\put(60,120){\line(0,-1){30}}
\put(60,90){\line(0,-1){30}}
\put(90,150){\line(0,-1){30}}
\put(90,120){\line(0,-1){30}}
\put(90,90){\line(0,-1){30}}
\put(30,60){\line(0,-1){10}}
\put(60,60){\line(0,-1){10}}
\put(90,60){\line(0,-1){10}}
\put(60,180){\line(1,-1){30}}
\put(30,150){\line(1,-1){30}}
\put(30,120){\line(1,-1){30}}
\put(30,90){\line(1,-1){30}}
\put(60,180){\line(-1,-1){30}}
\put(90,150){\line(-1,-1){30}}
\put(90,120){\line(-1,-1){30}}
\put(90,90){\line(-1,-1){30}}
\put(30,60){\line(1,-1){10}}
\put(90,60){\line(-1,-1){10}}
\put(54,185){{\small$01$}} 
\put(13,148){{\small$11$}} 
\put(13,118){{\small$21$}}
\put(13,88){{\small$31$}} 
\put(13,58){{\small$41$}} 
\put(93,148){{\small$13$}}
\put(93,118){{\small$23$}} 
\put(93,88){{\small$33$}} 
\put(93,58){{\small$43$}}
\put(63,113){$22$}
\put(63,83){$32$}
\put(63,53){$41$}
\put(45,30){$\vdots$}
\put(75,30){$\vdots$}
\end{picture}
\caption{\label{fi:bigbra}
\protect{\footnotesize The Bratteli diagram associated with the poset $\bigvee$; the
label $nk$ stands for $Y_n(k)$. }}
\end{center}
\vskip.5cm
\end{figure}
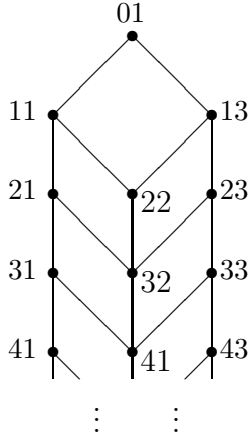
Fig.~\ref{fi:bigbra} shows the corresponding diagram, obtained through rules (1.) and
(2.) in Proposition~\ref{most} above. By using the fact that the first matrix algebra
$\ca_0$ is $\IC$ and the fact that all the embeddings have multiplicity one, the diagram
of Fig.~\ref{fi:bigbra} is seen to coincide with the diagram of Fig.~\ref{fi:veealg}. 
As we have said previously, the latter corresponds to the AF-algebra  $\ca_\vee =
\IC\cp_1 + \ck({\ch}) + \IC\cp_2~, ~ \ch = \ch_1 \oplus \ch_2.$
\eexam
\bexam 
Another interesting example is provided by the lattice $P_4(S^1)$ for the
one-dimensional sphere in Fig.~\ref{fi:cirhas}. 
This topological space contains six closed sets: 
\bea
&&K_0 = \{x_1, x_2, x_3, x_4 \}~, K_1 = \{x_1, x_3, x_4 \}~, K_2 =\{x_3\}~, 
K_3 =\{x_4\}~, \nonumber \\
&&K_5 = \{x_2, x_3, x_4 \}~, K_6 = \{x_3, x_4 \} = K_2 \cup K_3~.
\eea
Thus, one finds,
\be
\begin{array}{lll}
\ck_0=\{K_0\}                      & ~ &  \ck_0'=\{K_0\}\\  
\ck_1=\{K_0,K_1\}                  & ~ &  \ck_1'=\{K_0,K_1\} \\ 
\ck_2=\{K_0,K_1,K_2\}              & ~ &  \ck_2'=\{K_0,K_1,K_2\}  \\       
\ck_3=\{K_0,K_1,K_2,K_3\}          & ~ &  \ck_3'=\{K_0,K_1,K_2,K_3,K_5\}  \\
\ck_4=\{K_0,K_1,K_2,K_3,K_4\}      & ~ &  \ck_4'=\{K_0,K_1,K_2,K_3,K_4,K_5\} \\
\ck_5=\{K_0,K_1,K_2,K_3,K_4,K_5\}  & ~ &  \ck_5'=\{K_0,K_1,K_2,K_3,K_4,K_5\} \\
\vdots & & \\
~ & ~ & ~   \\
~ & ~ & ~   \\
Y_0(1)=\{x_1, x_2, x_3, x_4\}  & ~~~~~ & F_0(1)=K_0  ~ \\
~ & ~ & ~  \\
Y_1(1)=\{x_1, x_3, x_4\}~~~~~ Y_1(2)=\{x_2\}  
                                     & ~~~~~ & F_1(1)=K_1~~~~~ F_1(2)=K_0  \\
~ & ~ & ~   \\
Y_2(1)=\{x_3\}~~~~~ Y_2(2)=\{x_2\}  
                                     & ~~~~~ & F_2(1)=K_2~~~~~ F_2(2)=K_0  \\
Y_2(3)=\{x_1, x_4\}  ~                
                                     & ~~~~~ & F_2(3)=K_1  ~            \\
~ & ~ & ~   \\
 Y_3(1)=\{x_3\}~~~~~ Y_3(2)=\{x_2\}
                                     & ~~~~~ & F_3(1)=K_2~~~~~ F_3(2)=K_0 \\ 
Y_3(3)=\{x_1\}~~~~~ Y_3(4)=\{x_4\}
                                     & ~~~~~  & F_3(3)=K_1~~~~~ F_3(4)=K_3 \\ 
~ & ~ & ~   \\
Y_4(1)=\{x_3\}~~~~~ Y_4(2)=\{x_2\}                  
                                     & ~~~~~ & F_4(1)=K_2~~~~~ F_4(2)=K_4 \\ 
Y_4(3)=\{x_1\}~~~~~ Y_4(4)=\{x_4\} 
                                     & ~~~~~ & F_4(3)=K_1~~~~~ F_4(4)=K_3 \\ 
~ & ~ & ~   \\
Y_5(1)=\{x_3\}~~~~~ Y_5(2)=\{x_2\}   
                                     & ~~~~~ & F_5(1)=K_2~~~~~ F_5(2)=K_4 \\ 
Y_5(3)=\{x_1\}~~~~~ Y_5(4)=\{x_4\}  
                                     & ~~~~~ & F_5(3)=K_1~~~~~ F_5(4)=K_3 \\ 
\vdots &  &    
\end{array}
\ee
Since there is a finite number of points (four) and hence a finite number
of closed sets (six), the partition of $P_4(S^1)$ repeats itself after the fourth 
level. The corresponding Bratteli diagram is in Fig.~\ref{fi:cirbra}.  
The ideal $\{0\}$ is not primitive.
\begin{figure}[t]
\begin{center}
\begin{picture}(120,210)(20,0)
\put(30,180){\circle*{4}}
\put(30,150){\circle*{4}}
\put(30,120){\circle*{4}}
\put(30,90){\circle*{4}}
\put(30,60){\circle*{4}}
\put(30,30){\circle*{4}}
\put(60,150){\circle*{4}}
\put(60,120){\circle*{4}}
\put(60,90){\circle*{4}}
\put(60,60){\circle*{4}}
\put(60,30){\circle*{4}}
\put(90,90){\circle*{4}}
\put(90,60){\circle*{4}}
\put(90,30){\circle*{4}}
\put(120,30){\circle*{4}}
\put(120,60){\circle*{4}}
\put(120,90){\circle*{4}}
\put(120,120){\circle*{4}}
\put(30,180){\line(0,-1){30}}
\put(30,180){\line(1,-1){30}}
\put(30,150){\line(0,-1){30}}
\put(30,150){\line(1,-1){30}}
\put(30,150){\line(3,-1){90}}
\put(60,150){\line(0,-1){30}}
\put(30,60){\line(0,-1){30}}
\put(60,60){\line(0,-1){30}}
\put(90,60){\line(0,-1){30}}
\put(120,60){\line(0,-1){30}}
\put(30,120){\line(0,-1){30}}
\put(60,120){\line(0,-1){30}}
\put(30,90){\line(0,-1){30}}
\put(60,90){\line(0,-1){30}}
\put(90,90){\line(0,-1){30}}
\put(120,120){\line(0,-1){30}}
\put(120,90){\line(0,-1){30}}
\put(30,120){\line(1,-1){30}}
\put(30,90){\line(1,-1){30}}
\put(30,60){\line(1,-1){30}}
\put(120,120){\line(-1,-1){30}}
\put(120,90){\line(-1,-1){30}}
\put(120,60){\line(-1,-1){30}}
\put(120,120){\line(-2,-1){60}}
\put(120,90){\line(-2,-1){60}}
\put(120,60){\line(-2,-1){60}}
\put(30,120){\line(2,-1){60}}
\put(30,90){\line(2,-1){60}}
\put(30,60){\line(2,-1){60}}
\put(30,30){\line(0,-1){10}}
\put(30,30){\line(1,-1){10}}
\put(30,30){\line(2,-1){20}}
\put(60,30){\line(0,-1){10}}
\put(90,30){\line(0,-1){10}}
\put(120,30){\line(-1,-1){10}}
\put(120,30){\line(0,-1){10}}
\put(120,30){\line(-2,-1){20}}
\put(75,0){$\vdots$}
\end{picture}
\caption{\label{fi:cirbra}
\protect{\footnotesize The Bratteli diagram for the circle poset
$P_4(S^1)$  . }}
\end{center}
\vskip.5cm
\end{figure}
The algebra is given by
\bea
& & \ca_0 = \IM_{1}(\IC)   \nonumber \\
& & \ca_1 = \IM_{1}(\IC) \oplus \IM_{1}(\IC)  \nonumber \\
& & \ca_2 = \IM_{1}(\IC) \oplus \IM_{2}(\IC) \oplus \IM_{1}(\IC)
\nonumber \\ 
& & \ca_3 = \IM_{1}(\IC) \oplus \IM_{4}(\IC) \oplus \IM_{2}(\IC) \oplus
\IM_{1}(\IC) \nonumber \\ 
& & \ca_4 = \IM_{1}(\IC) \oplus \IM_{6}(\IC) \oplus \IM_{4}(\IC) \oplus
\IM_{1}(\IC) \nonumber \\ 
& & ~~~ \vdots \nonumber \\
& & \ca_n = \IM_{1}(\IC) \oplus \IM_{2n-2}(\IC) \oplus \IM_{2n-4}(\IC) \oplus
\IM_{1}(\IC) \nonumber \\ 
& & ~~~ \vdots
\label{cirafl}
\eea
where, for $n > 2$, $\ca_n$ is embedded in $\ca_{n+1}$ as follows 
\be\left[
\begin{array}{cccc}
\l_1 &   &   &   \\
     & B &   &  \\
     &   & C &   \\
     &   &   & \l_2  
\end{array}
\right]~~ \mapsto ~
\left[
\begin{array}{cccccccc}
\l_1 &      &     &      &       &    &      &   \\
     & \l_1 & 0   & 0    &       &    &      &   \\
     & 0    & B   & 0    &       &    &      &   \\
     & 0    & 0   & \l_2 &       &    &      &   \\
     &      &     &      & \l_1  & 0  & 0    &   \\
     &      &     &      & 0     & C  & 0    &   \\
     &      &     &      & 0     & 0  & \l_2 &   \\
     &      &     &      &       &    &      & \l_2
\end{array}
\right]~,
\label{cirafl1}
\ee
with $B \in \IM_{2n-2}(\IC)$ and $C \in \IM_{2n-4}(\IC)$; elements which are not shown
are equal to zero. The algebra limit $\ca_{P_4(S^1)}$ can be realized explicitly as  a
subalgebra of bounded operators on an infinite dimensional Hilbert space $\ch$ naturally
associated with the poset $P_4(S^1)$. Firstly, to any {\it link} $(x_i, x_j), x_i \succ
x_j,$  of the latter one associates an Hilbert space $\ch_{ij}$; for the case at hand,
one has four Hilbert spaces, $\ch_{31}, \ch_{32}, \ch_{41}, \ch_{42}$. Then, since all
links are at the same level, $\ch$ is just given by the direct sum
\be
\ch = \ch_{31} \oplus \ch_{32} \oplus \ch_{41} \oplus \ch_{42}~.
\ee
The algebra $\ca_{P_4(S^1)}$ is given by \cite{ELTfunctions},
\be
\ca_{P_4(S^1)} = \IC\cp_{\ch_{31} \oplus \ch_{32}} + \ck_{\ch_{31} \oplus \ch_{41}} 
+ \ck_{\ch_{32} \oplus \ch_{42}} + \IC\cp_{\ch_{41} \oplus \ch_{42}}~.
\label{ciralg}
\ee
Here $\ck$ denotes compact operators and $\cp$ orthogonal projection.
The algebra (\ref{ciralg}) has four irreducible representations. Any element
$a\in\ca_{P_4(S^1)}$ is of the form 
\be
a = \l \cp_{3,12} + k_{34,1} + k_{34,2} + \m\cp_{4,12}~,
\ee 
with $\l,\m \in \IC$, $k_{34,1}\in\ck_{\ch_{31} \oplus \ch_{41}}$ and $k_{34,2} 
\in \ck_{\ch_{32} \oplus \ch_{42}}$. The representations are the following ones,
\be\label{repcir}
\begin{array}{ll}
\pi_1 : \ca_{P_4(S^1)} \lra \cb(\ch)  ~, & a \mapsto \pi_1(a) = 
\l \cp_{3,12} + k_{34,1} + \m\cp_{4,12}~, \\
\pi_2 : \ca_{P_4(S^1)} \lra \cb(\ch)  ~, & a \mapsto \pi_2(a) = 
\l \cp_{3,12} + k_{34,2} + \m\cp_{4,12}~, \\
\pi_3 : \ca_{P_4(S^1)} \lra \cb(\IC) \simeq \IC ~, & a \mapsto \pi_3(a) = \l~, \\
\pi_4 : \ca_{P_4(S^1)} \lra \cb(\IC) \simeq \IC ~, & a \mapsto \pi_4(a) = \m~. 
\end{array}
\ee
The corresponding kernels are
\bea\label{kercir}
&& \ci_1 = \ck_{\ch_{32} \oplus \ch_{42}}~, \nonumber \\
&& \ci_2 = \ck_{\ch_{31} \oplus \ch_{41}}~, \nonumber \\
&& \ci_3 = \ck_{\ch_{31} \oplus \ch_{41}} + \ck_{\ch_{32} \oplus \ch_{42}} +
                         \IC\cp_{\ch_{41} \oplus \ch_{42}}~, \nonumber \\ 
&& \ci_4 = \IC\cp_{\ch_{31} \oplus \ch_{32}} + \ck_{\ch_{31} \oplus \ch_{41}} 
                          + \ck_{\ch_{32} \oplus \ch_{42}}~. 
\eea
The partial order given by the inclusions $\ci_1 \subset \ci_3$, $\ci_1 \subset \ci_4$
and $\ci_2 \subset \ci_3$, $\ci_2 \subset \ci_4$ produces a topological space
$Prim{\cal A}_{P_4(S^1)}$ which is just the circle poset in Fig.~\ref{fi:cirhas}.
\eexam
\bexam 
We shall now give an example of a three-level poset. It would correspond to an
approximation of a two dimensional topological space (or a portion thereof). 
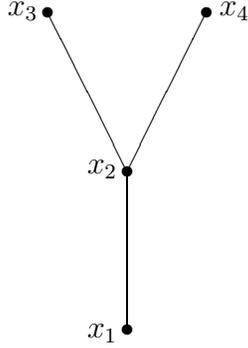
\begin{figure}[htb]
\begin{center}
\begin{picture}(220,150)(-100,-30)
\put(-30,90){\circle*{4}}
\put(30,90){\circle*{4}}
\put(0,30){\circle*{4}}
\put(0,-30){\circle*{4}}
\put(-45,89){$x_3$}
\put(35,89){$x_4$}
\put(-15,29){$x_2$}
\put(-15,-33){$x_1$}
\put(-30,90){\line(1,-2){30}}
\put(30,90){\line(-1,-2){30}}
\put(0,30){\line(0,-1){60}}
\end{picture}
\caption{\label{fi:ipshas}
\protect{\footnotesize A poset approximating a two dimensional space.}}
\end{center}
\vskip.5cm
\end{figure}

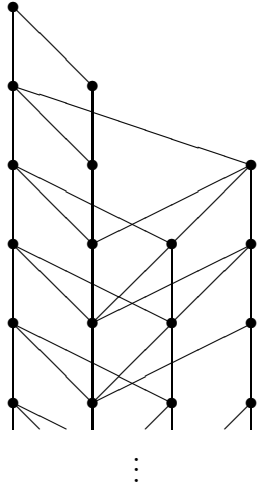
\begin{figure}[t]
\begin{center}
\begin{picture}(120,210)(20,0)
\put(30,180){\circle*{4}}
\put(30,150){\circle*{4}}
\put(30,120){\circle*{4}}
\put(30,90){\circle*{4}}
\put(30,60){\circle*{4}}
\put(30,30){\circle*{4}}
\put(60,150){\circle*{4}}
\put(60,120){\circle*{4}}
\put(60,90){\circle*{4}}
\put(60,60){\circle*{4}}
\put(60,30){\circle*{4}}
\put(90,90){\circle*{4}}
\put(90,60){\circle*{4}}
\put(90,30){\circle*{4}}
\put(120,30){\circle*{4}}
\put(120,60){\circle*{4}}
\put(120,90){\circle*{4}}
\put(120,120){\circle*{4}}
\put(30,180){\line(0,-1){30}}
\put(30,180){\line(1,-1){30}}
\put(30,150){\line(0,-1){30}}
\put(30,150){\line(1,-1){30}}
\put(30,150){\line(3,-1){90}}
\put(60,150){\line(0,-1){30}}
\put(30,120){\line(0,-1){30}}
\put(60,120){\line(0,-1){30}}
\put(120,120){\line(0,-1){30}}
\put(30,120){\line(1,-1){30}}
\put(120,120){\line(-1,-1){30}}
\put(120,120){\line(-2,-1){60}}
\put(30,120){\line(2,-1){60}}
\put(30,90){\line(0,-1){30}}
\put(60,90){\line(0,-1){30}}
\put(90,90){\line(0,-1){30}}
\put(120,90){\line(0,-1){30}}
\put(30,90){\line(1,-1){30}}
\put(120,90){\line(-2,-1){60}}
\put(30,90){\line(2,-1){60}}
\put(120,90){\line(-1,-1){30}}
\put(90,90){\line(-1,-1){30}}
\put(30,60){\line(1,-1){30}}
\put(30,60){\line(0,-1){30}}
\put(60,60){\line(0,-1){30}}
\put(90,60){\line(0,-1){30}}
\put(120,60){\line(0,-1){30}}
\put(120,60){\line(0,-1){30}}
\put(90,60){\line(-1,-1){30}}
\put(120,60){\line(-2,-1){60}}
\put(30,60){\line(2,-1){60}}
\put(30,30){\line(0,-1){10}}
\put(30,30){\line(1,-1){10}}
\put(30,30){\line(2,-1){20}}
\put(60,30){\line(0,-1){10}}
\put(90,30){\line(0,-1){10}}
\put(120,30){\line(-1,-1){10}}
\put(120,30){\line(0,-1){10}}
\put(120,30){\line(0,-1){10}}
\put(90,30){\line(-1,-1){10}}
\put(75,0){$\vdots$}
\end{picture}
\caption{\label{fi:ipsbra}
\protect{\footnotesize The Bratteli diagram for the poset $Y$ of
previous Figure.}}
\end{center}
\vskip.5cm
\end{figure}

\noindent
This topological space, shown in Fig.~\ref{fi:ipshas}, contains five closed sets: 
\bea
&& K_0 =: \bar{\{x_1\}} = \{x_1, x_2, x_3, x_4 \}~, ~~~K_1 =: \bar{\{x_2\}} = 
\{x_2, x_3, x_4 \}~, \nonumber \\
&& K_2 =: \bar{\{x_3\}} =\{x_3\}~, ~~~K_3 =: \bar{\{x_4\}} = \{x_4\}~, \nonumber \\ 
&& K_4 = \{x_3, x_4 \} = K_2 \cup K_3~.
\eea
Thus, one finds,
\be
\begin{array}{lll}
\ck_0=\{K_0\}                      & ~ &  \ck_0'=\{K_0\}\\  
\ck_1=\{K_0,K_1\}                  & ~ &  \ck_1'=\{K_0,K_1\} \\ 
\ck_2=\{K_0,K_1,K_2\}              & ~ &  \ck_2'=\{K_0,K_1,K_2\}  \\       
\ck_3=\{K_0,K_1,K_2,K_3\}          & ~ &  \ck_3'=\{K_0,K_1,K_2,K_3,K_4\}  \\
\ck_4=\{K_0,K_1,K_2,K_3,K_4\}      & ~ &  \ck_4'=\{K_0,K_1,K_2,K_3,K_4\} \\
\vdots & & \\
~ & ~ & ~   \\
~ & ~ & ~   \\
Y_0(1)=\{x_1, x_2, x_3, x_4\}        & ~~~~~ & F_0(1)=K_0  ~ \\
~ & ~ & ~  \\
Y_1(1)=\{x_2, x_3, x_4\}~~~~~ Y_1(2)=\{x_1\}  
                                     & ~~~~~ & F_1(1)=K_1~~~~~ F_1(2)=K_0  \\
~ & ~ & ~   \\
Y_2(1)=\{x_3\}~~~~~ Y_2(2)=\{x_1\}  
                                     & ~~~~~ & F_2(1)=K_2~~~~~ F_2(2)=K_0  \\
Y_2(3)=\{x_2, x_4\}  ~                
                                     & ~~~~~ & F_2(3)=K_1  ~            \\
~ & ~ & ~   \\
 Y_3(1)=\{x_3\}~~~~~ Y_3(2)=\{x_1\}
                                     & ~~~~~ & F_3(1)=K_2~~~~~ F_3(2)=K_0 \\ 
Y_3(3)=\{x_2\}~~~~~ Y_3(4)=\{x_4\}
                                     & ~~~~~  & F_3(3)=K_1~~~~~ F_3(4)=K_3 \\ 
~ & ~ & ~   \\
Y_4(1)=\{x_3\}~~~~~ Y_4(2)=\{x_1\}                  
                                     & ~~~~~ & F_4(1)=K_2~~~~~ F_4(2)=K_0 \\ 
Y_4(3)=\{x_2\}~~~~~ Y_4(4)=\{x_4\} 
                                     & ~~~~~ & F_4(3)=K_1~~~~~ F_4(4)=K_3 \\ 
~ & ~ & ~   \\
\vdots &  &    
\end{array}
\ee
Since there is a finite number of points (four) and hence a finite number
of closed sets (five), the partition of the poset is the same after the
fourth  level. The corresponding Bratteli diagram is in Fig.~\ref{fi:ipsbra}.  
The ideal $\{0\}$ is primitive.
The corresponding algebra is given by
\bea
& & \ca_0 = \IM_{1}(\IC)   \nonumber \\
& & \ca_1 = \IM_{1}(\IC) \oplus \IM_{1}(\IC)  \nonumber \\
& & \ca_2 = \IM_{1}(\IC) \oplus \IM_{2}(\IC) \oplus \IM_{1}(\IC)
\nonumber \\ 
& & \ca_3 = \IM_{1}(\IC) \oplus \IM_{4}(\IC) \oplus \IM_{2}(\IC) \oplus
\IM_{1}(\IC) \nonumber \\ 
& & \ca_4 = \IM_{1}(\IC) \oplus \IM_{8}(\IC) \oplus \IM_{4}(\IC) \oplus
\IM_{1}(\IC) \nonumber \\ 
& & ~~~ \vdots \nonumber \\
& & \ca_n = \IM_{1}(\IC) \oplus \IM_{n^2-3n+4}(\IC) \oplus \IM_{2n-4}(\IC) \oplus
\IM_{1}(\IC) \nonumber \\ 
& & ~~~ \vdots
\label{ipsafl}
\eea
where, for $n > 2$, $\ca_n$ is embedded in $\ca_{n+1}$ as follows
\be\left[
\begin{array}{cccc}
\l_1 &   &   &   \\
     & B &   &   \\
     &   & C &   \\
     &   &   & \l_2  
\end{array}
\right]~~ \mapsto ~
\left[
\begin{array}{ccccccccc}
\l_1 &      &     &      &       &       &      &       &   \\
     & \l_1 & 0   & 0    & 0     &       &      &       &   \\
     & 0    & B   & 0    & 0     &       &      &       &   \\
     & 0    & 0   & C    & 0     &       &      &       &   \\
     & 0    & 0   & 0    & \l_2  &       &      &       &   \\
     &      &     &      &       & \l_1  & 0    & 0     &   \\
     &      &     &      &       & 0     & C    & 0     &   \\
     &      &     &      &       & 0     & 0    & \l_2  &   \\
     &      &     &      &       &       &      &       & \l_2
\end{array}
\right]~,
\label{ipsafl1}
\ee
with $B \in \IM_{n^2-3n+4}(\IC)$ and $C \in \IM_{2n-4}(\IC)$; elements which are not
shown are equal to zero. Again, the algebra limit
$\ca_{Y}$ can be given as a subalgebra of bounded operators on a
Hilbert space $\ch$. The Hilbert spaces associated with the links of the poset will be 
$\ch_{32}, \ch_{42}, \ch_{21}$.  The difference with the previous example is that now
there are links at different levels. On passing from a level to the next (or previous
one) one introduces tensor products. The Hilbert space $\ch$ is given by
\be
\ch = \ch_{32} \otimes \ch_{21} \oplus \ch_{42} \otimes \ch_{21}
\simeq
(\ch_{32} \oplus \ch_{42}) \otimes \ch_{21}
\ee 
The algebra $\ca_{Y}$ is then given by \cite{ELTfunctions},
\be
\ca_{Y} = \IC\cp_{\ch_{32} \otimes \ch_{21}} + \ck_{\ch_{32} \oplus \ch_{42}} 
\otimes \cp_{\ch_{21}} + \ck_{(\ch_{32} \oplus \ch_{42}) \otimes \ch_{21}} + 
\IC\cp_{\ch_{42} \otimes \ch_{21}}~.
\label{ipsalg}
\ee
Here $\ck$ denotes compact operators and $\cp$ orthogonal projection.
This algebra has four irreducible representations. Any element of it 
is of the form 
\be
a = \l \cp_{321} + k_{34,2} \otimes \cp_{21} + k_{34,21} + \m \cp_{421}~,
\ee 
with $\l,\m \in \IC$, $k_{34,2}\in\ck_{\ch_{32} \oplus \ch_{42}}$ and 
$k_{34,21} \in \ck_{(\ch_{32} \oplus \ch_{42}) \otimes \ch_{21}}$. 
The representations are the following ones,
\be\label{ipscir}
\begin{array}{ll}
\pi_1 : \ca_{Y} \lra \cb(\ch)  ~, & a \mapsto \pi_1(a) = 
\l \cp_{321} + k_{34,2} \otimes \cp_{21} + k_{34,21} + \m \cp_{421}~, \\
\pi_2 : \ca_{Y} \lra \cb(\ch)  ~, & a \mapsto \pi_2(a) = 
\l \cp_{321} + k_{34,2} \otimes \cp_{21} + \m \cp_{421}~, \\
\pi_3 : \ca_{Y} \lra \cb(\IC) \simeq \IC ~, & a \mapsto \pi_3(a) = \l~, \\
\pi_4 : \ca_{Y} \lra \cb(\IC) \simeq \IC ~, & a \mapsto \pi_4(a) = \m~. 
\end{array}
\ee
The corresponding kernels are
\bea\label{keripscir}
&& \ci_1 = \{0 \}~, \nonumber \\
&& \ci_2 = \ck_{(\ch_{32} \oplus \ch_{42}) \otimes \ch_{21}}~, \nonumber \\
&& \ci_3 = \ck_{\ch_{32} \oplus \ch_{42}} \otimes \cp_{\ch_{21}} + 
\ck_{(\ch_{32} \oplus \ch_{42}) \otimes \ch_{21}} + \IC\cp_{\ch_{42} \otimes \ch_{21}}~,
        \nonumber \\ 
&& \ci_4 = \IC\cp_{\ch_{32} \otimes \ch_{21}} + \ck_{\ch_{32} \oplus \ch_{42}} 
\otimes \cp_{\ch_{21}} + \ck_{(\ch_{32} \oplus \ch_{42}) \otimes \ch_{21}}~. 
\eea
The partial order given by the inclusions $\ci_1 \subset \ci_2 \subset \ci_3$
and $\ci_1 \subset \ci_2 \subset \ci_4$ produces a topological space
$Prim{\cal A}_{Y}$ which is just the poset of Fig.~(\ref{fi:ipshas}).
\eexam

In fact, by looking at the previous examples a bit more carefully one can infer the
algorithm by which one goes from a (finite) poset $P$ to the corresponding Bratteli
diagram $\cd(\ca_P)$. Let $(x_1, \cdots , x_N)$ be the points of $P$ and for $k=1,
\cdots, N$, let $S_k =: \bar{\{x_k \}}$ be the smallest closet subset of $P$ containing
the point $ x_j$. Then, the Bratteli diagram repeats itself after the level $N$ and the
partition $Y_n(k)$ of Proposition~\ref{most} is just given by 
\be
Y_n(k) = Y_{n+1}(k) = \{x_k \}~, ~k = 1, \dots, N, ~~~\forall ~n \geq N~. 
\ee
As for the associated $F_n(k)$, from the level $N+1$ on, they are given by the $S_k$,
\be
F_n(k) = F_{n+1}(k) = S_k~, ~k = 1, \dots, N, ~~~\forall ~n \geq N+1~. 
\ee
In the diagram $\cd(\ca_P)$, for any $n \geq N$, $(n,k) \searrow (n+1,j)$ if and only
if $\{x_k \} \bigcap S_j \not= \emptyset$, namely if and only if $x_k \in S_j$. \\
We also sketch the algorithm to construct the algebra limit $\ca_P$ determined by the
Bratteli diagram $\cd(\ca_P)$
\fn{This algebra is really defined only modulo Morita equivalence.}
\cite{BL,ELTfunctions}. The idea is to associate to the poset $P$ an infinite
dimensional separable Hilbert space
$\ch(P)$ out of tensor products and direct sums of infinite dimensional
(separable) Hilbert spaces $\ch_{ij}$ associated with each link $(x_i, x_j), x_i \succ
x_j$, in the poset
\fn{The Hilbert spaces could be taken to be all the same. The label is there just to
distinguish among them.}. Then for each point
$x \in P$ there is a subspace $\ch(x) \subset \ch(P)$ and an algebra $\cb(x)$ of bounded
operators acting on $\ch(x)$. The algebra $\ca_P$ is the one generated by all of the
$\cb(x)$ as
$x$ varies in $P$. In fact, the algebra $\cb(x)$ can be made to act on the whole of
$\ch(P)$ by defining its action on the complement of $\ch(x)$ to be zero. 
Consider any {\it maximal chain} $C_\a$ in $P$: $C_\a = \{x_a, \dots, x_2, x_1 ~|~ x_j
\succ x_{j-1}\}$ for any maximal point $x_\a$. To this chain one associates the Hilbert 
space
\be
\ch(C_\a) = \ch_{\a,\a-1} \otimes \cdots \otimes \ch_{3,2} \otimes \ch_{2,1}~.
\ee 
By taking the direct sum over all maximal chains, one gets the Hilbert space $\ch(P)$,
\be\label{htot}
\ch(P) = \bigoplus_\a \ch(C_\a)~.
\ee
The subspace $\ch(x) \subset \ch(P)$ associated with any point $x \in P$ is
constructed in a similar manner by restricting the sum to all maximal chains
containing the point $x$. It can be split in two parts,
\be\label{fibhil}
\ch(x) = \ch(x)^u \otimes \ch(x)^d~, 
\ee 
with,
\bea
&& \ch(x)^u = \ch(P^u_x)~, ~~~ P^u_x = \{y \in P ~|~ y \succeq x\}~,  \nonumber \\
&& \ch(x)^d = \ch(P^d_x)~, ~~~ P^d_x = \{y \in P ~|~ y \preceq x\}~.
\eea 
Here $\ch(P^u_x)$ and $\ch(P^d_x)$ are constructed as in (\ref{htot}); also, 
$\ch(x)^u =\IC$ if $x$ is a maximal point and $\ch(x)^d =\IC$ if $x$ is a minimal point.
Consider now the algebra
$\cb(x)$ of bounded operators on $\ch(x)$ given by
\be\label{fibalg}
\cb(x) = \ck(\ch(x)^u) \otimes \IC \cp(\ch(x)^d) \simeq \ck(\ch(x)^u) \otimes 
\cp(\ch(x)^d)~. 
\ee
As before, $\ck$ denotes compact operators and $\cp$ orthogonal projection. We see
that  $\cb(x)$ acts by compact operators on the Hilbert space $\ch(x)^u$ determined by
the points which follow $x$ and by multiplies of the identity on the Hilbert space
$\ch(x)^d$ determined by the points which precede $x$. These algebras satisfy the
rules: $\cb(x) \cb(y) \subset \cb(x)$ if $x\preceq y$ and $\cb(x) \cb(y) = 0$ if $x$ and
$y$ are not comparable. As already mentioned, the algebra $\ca(P)$ of the poset $P$ is the algebra
of bounded operators on $\ch(P)$ generated by all $\cb(x)$ as $x$ varies over $P$. It
can be shown that $\ca(P)$ has a space of primitive ideals which is homeomorphic to
the poset $P$ \cite{BL,ELTfunctions}. We refer to \cite{ELTkappa,ELTfunctions} for
additional details and examples.

\subsect{How to Recover the Algebra Being Approximated}\label{se:raba}

In Section~\ref{se:recospa} we have described how to recover a topological space $M$ in
the limit, by considering a sequence of finer and finer coverings of $M$. We constructed
an inverse system of finitary topological spaces and continuous maps $\{P_i, \p_{ij}
\}_{i,j \in \IN}$ associated with the coverings; the maps $\p_{ij} : P_j \ra P_i~, ~j
\geq i$, being continuous surjections. The limit of the system is a topological space
$P_\infty$, in which  $M$ is embedded as the subspace of closed points. On each point
$m$ of (the image of) $M$ there is a fiber of `extra points' ; the latter are all
points of $P_\infty$ which `cannot be separated' by $m$. 

Well, dually we get a direct system of algebras and homomorphisms $\{\ca_i,
\f_{ij} \}_{i,j \in \IN}$; the maps $\f_{ij} : \ca_i \ra \ca_j~, ~j \geq i$, being
injective homeomorphisms. The system has a unique {\it inductive limit} $\ca^\infty$.
Each algebra $\ca_i$ is such that $\ha_i = P_i$ and is associated with $P_i$ as 
described
in previous Section, $\ca_i = \ca(P_i)$. The map $\f_{ij}$ is a `suitable pullback' of
the corresponding surjection $\p_{ij}$. The limit space $P_\infty$ is the structure
space of the limit algebra $\ca^\infty$, $P_\infty = \ha^\infty$. And, finally the
algebra $C(M)$ of continuous functions on $M$ can be identified with the {\it center} of
$\ca^\infty$.

We get also a direct system of Hilbert spaces and isometries $\{\ch_i,
\t_{ij} \}_{i,j \in \IN}$; the maps $\t_{ij} : \ch_i \ra \ch_j~, ~j \geq i$, being
injective isometries onto the image. The system has a unique {\it inductive limit}
$\ch^\infty$. Each Hilbert space $\ch_i$ is associated with the space $P_i$ as in
(\ref{htot}), $\ch_i = \ch(P_i)$, the algebra $\ca_i$ being the corresponding subalgebra
of bounded operators. The map $\t_{ij}$ are constructed out of the corresponding
$\f_{ij}$. The limit Hilbert space $\ch^\infty$ is associated with the space
$P_\infty$ as in (\ref{htot}), $\ch^\infty = \ch(P_\infty)$, the algebra $\ca^\infty$
being again the corresponding subalgebra of bounded operators. And, finally the
Hilbert space $L^2(M)$ of square integrable functions algebra is `contained' in
$\ch^\infty$ : $\ch^\infty = L^2(M) \oplus_\a \ch_\a$, the sum being on the `extra
points' in $P_\infty$.

All of previous is described in details in \cite{pangs}. Here we only add few
additional remarks. By improving the approximation (by increasing the number of
`detectors') one gets a noncommutative lattice whose Hasse diagram has a bigger 
number of points and links. The associated Hilbert space gets `more refined' : one may
thing of a {\it unique and the same} Hilbert space which is being refined while being 
split by means of tensor products and direct sums. In the limit the information is
enough to recover completely the Hilbert space (in fact, to recover more than it).
Further considerations along these lines and possible applications to quantum mechanics
have to await another time.

\subsect{Operator Valued Functions on Noncommutative Lattices}\label{se:ovf} 

Much in the
same way as it happens for commutative algebras described in Section~\ref{se:gnt},
elements of a noncommutative $C^*$-algebra whose primitive spectrum $\prim$ is a
noncommutative lattice can be realized as operator-valued functions on $\prim$. The
values of $a\in\ca$ at the `point' $\ci\in \prim$ is just the image of $a$ under the
representation $\pi_\ci$ associated with $\ci$,
$\ker(\pi_\ci) = \ci$,
\be
a(\ci) = \pi_\ci(a) \simeq a/\ci~, ~~~ \forall ~a \in \ca, ~\ci \in \prim~. 
\ee
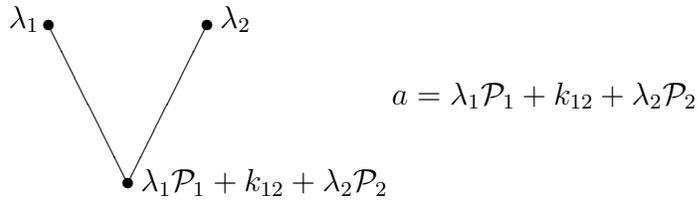
\begin{figure}[htb]
\begin{center}
\begin{picture}(220,90)(0,-30)
\put(-30,30){\circle*{4}}
\put(30,30){\circle*{4}}
\put(0,-30){\circle*{4}}
\put(-30,30){\line(1,-2){30}}
\put(30,30){\line(-1,-2){30}}
\put(-45,29){$\l_1$}
\put(35,29){$\l_2$}
\put(5,-33){$\l_1\cp_1 + k_{12} + \l_2\cp_2$}
\put(100,0){$a = \l_1\cp_1 + k_{12} + \l_2\cp_2$}
\end{picture}
\caption{\label{fi:veefun}
\protect{\footnotesize A function over the lattice $\bigvee$.  }}
\end{center}
\vskip.5cm
\end{figure}
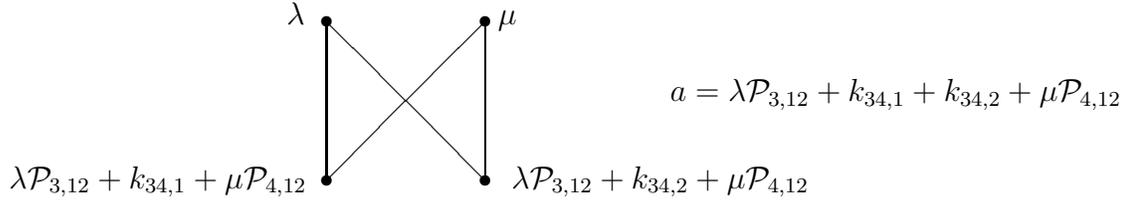
\begin{figure}[htb]
\begin{center}
\begin{picture}(220,90)(-50,-30)
\put(-30,30){\circle*{4}}
\put(30,30){\circle*{4}}
\put(-30,-30){\circle*{4}}
\put(30,-30){\circle*{4}}
\put(-30,30){\line(0,-1){60}}
\put(30,30){\line(0,-1){60}}
\put(-30,30){\line(1,-1){60}}
\put(30,30){\line(-1,-1){60}}
\put(-45,29){$\l$}
\put(35,29){$\m$}
\put(-150,-33){$\l\cp_{3,12} + k_{34,1} + \m\cp_{4,12}$}
\put(40,-33){$\l\cp_{3,12} + k_{34,2} + \m\cp_{4,12}$}
\put(100,0){$a = \l \cp_{3,12} + k_{34,1} + k_{34,2} + \m\cp_{4,12}$}
\end{picture}
\caption{\label{fi:cirfun}
\protect{\footnotesize A function over the lattice $P_4(S^1)$.  }}
\end{center}
\vskip.5cm
\end{figure}
All this is shown pictorially in Figures \ref{fi:veefun} and \ref{fi:cirfun} for the
$\bigvee$ and a circle lattices respectively. As it is evident in those Figures, the
values of a function at points which cannot be separated by the topology differ by a
compact operator. This is an illustration of the fact that compact operators play the
role of `infinitesimals' as we shall discussed at length in Section~\ref{se:mcp}.

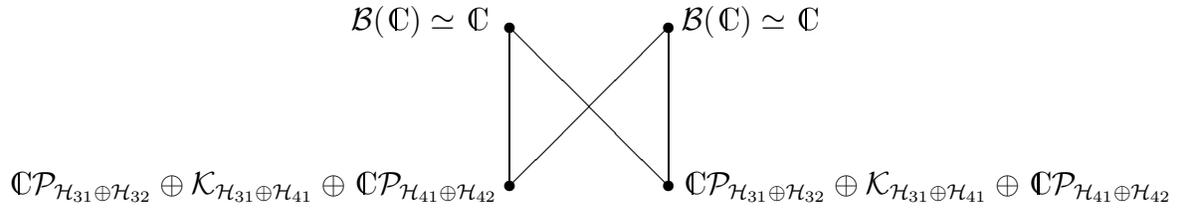
\begin{figure}[t]
\begin{center}
\begin{picture}(220,90)(-100,-30)
\put(-30,30){\circle*{4}}
\put(30,30){\circle*{4}}
\put(-30,-30){\circle*{4}}
\put(30,-30){\circle*{4}}
\put(-30,30){\line(0,-1){60}}
\put(30,30){\line(0,-1){60}}
\put(-30,30){\line(1,-1){60}}
\put(30,30){\line(-1,-1){60}}
\put(-90,29){$\cb(\IC) \simeq \IC$}
\put(35,29){$\cb(\IC) \simeq \IC$}
\put(-220,-33){$\IC\cp_{\ch_{31} \oplus \ch_{32}} \oplus \ck_{\ch_{31} \oplus \ch_{41}} 
\oplus \IC\cp_{\ch_{41} \oplus \ch_{42}}$}
\put(35,-33){$\IC\cp_{\ch_{31} \oplus \ch_{32}} \oplus \ck_{\ch_{31} \oplus \ch_{41}} 
\oplus \IC\cp_{\ch_{41} \oplus \ch_{42}}$}
\end{picture}
\caption{\label{fi:cirbun}
\protect{\footnotesize The fibers of the trivial bundle over the lattice $P_4(S^1)$. }}
\end{center}
\vskip.5cm
\end{figure}

In fact, as we shall see in Section~\ref{se:mod}, the correct way of thinking of any
noncommutative $C^*$-algebra $\ca$ is as the module of section of the `rank
one trivial vector bundle' over the associated noncommutative space. For the kind of
noncommutative lattices we are interested in, it is possible to explicitly construct
the bundle over the lattice. Such bundles are examples of {\it bundles of
$C^*$-algebras} \cite{bca}, the fiber over any point $\ci \in \prim$ being just the
algebra of bounded operators $\p_\ci(\ca) \subset \cb(\ch_\ci)$, with $\ch_\ci$ the
representation space. The Hilbert space and the algebra are given
explicitly by the Hilbert space in (\ref{fibhil}) and the algebra in (\ref{fibalg})  
respectively, by taking for $x$ just the point
$\ci$. It is also possible to endow the total space with a topology in such a manner that
elements of $\ca$ are realized as continuous sections. We refer to \cite{ELTbundle} for
more details. Fig.~\ref{fi:cirbun} shows the trivial bundle over the lattice $P_4(S^1)$.

\vfill\eject
\sect{Modules as Bundles}\label{se:vbg}

The algebraic analogue of vector bundles has its origin in the fact that a
vector bundle $E \ra M$ over a manifold  $M$ is completely characterized by the space
$\ce = \G(E,M)$ of its smooth sections thought of as a (right) module over the
algebra $C^\infty(M)$ of smooth functions over $M$.
Indeed, by the Serre-Swan theorem \cite{Sw}, locally 
trivial, finite-dimensional complex vector bundles over a compact 
Hausdorff space $M$ correspond canonically to finite projective 
modules over the algebra $\ca = C^\infty(M)$
\fn{In fact, in \cite{Sw} the correspondence is stated in the continuous category,
namely for functions and sections which are continuous. However, it can be
extended to the smooth case, see \cite{Colh}.}. 
To the vector bundle
$E$ one  associates the $C^\infty(M)$-module $\ce=\G(M,E)$ of smooth sections of
$E$. Conversely, if $\ce$ is a finite projective modules over
$C^\infty(M)$, the fiber $E_m$ of the associated bundle $E$ over the point 
$m \in M$ is 
\be
E_m = \ce / \ce \ci_m~,
\ee
where the ideal $\ci_m \subset \cc(M)$, corresponding to the point $m \in M$,
is given by \cite{Co1} 
\be
\ci_m = ker\{ \c_m : C^\infty(M)
\ra \IC~; ~ \c_m(f) =  f(m) \} = \{ f \in C^\infty(M) ~|~ f(m) = 0 \}~.
\ee 
Given an algebra $\ca$ playing the r\^ole of the algebra of smooth 
functions on some noncommutative space, the analogue of a vector bundle 
is provided by a {\it projective module of finite type} 
(or {\it finite projective module}) over $\ca$. 
Hermitian vector bundles, namely bundles with an Hermitian structure, correspond
to projective modules of finite type $\ce$ endowed with an $\ca$-valued sesquilinear
form. For $\ca$ a $C^*$-algebra, the appropriate notion is that of Hilbert module
that we describe at length in Appendix~\ref{se:hm}. 

We start with some machinery from the theory of modules which we take mainly from
\cite{Bo}.

\subsect{Modules }\label{se:0mod}
\bdefi
Suppose we are given an algebra $\ca$ over (say) the complex numbers $\IC$. A
vector space $\ce$ over $\IC$ is also a {\rm right module} over $\ca$ if it
carries a right representation of $\ca$,
\bea
\ce \times \ca \ni (\h, a) \mapsto \h a \in \ce~, && 
\h (ab) = (\h a)b ~, ~~ a, b \in \ca~, \nonumber \\ 
&& \h (a + b) = \h a + \h b~, \nonumber \\ 
&& (\h + \x)a = \h a + \x a~, 
\eea
for any $\h, \x \in \ce$ and $a, b \in \ca~$.
\edefi
\bdefi
Given two right $\ca$-modules $\ce$ and $\cf$, a {\rm morphism} of $\ce$ into
$\ce$ is any linear map $\r : \ce \ra \cf$ which in addition is $\ca$-linear, namely
\be
\r(\h a)= \r(\h) a~, ~~~\forall ~\h \in \ce, ~a \in \ca~.
\ee 
\edefi 

\noindent
A {\it left module} and a {\it morphism of left modules} are defined in a
similar manner. Since, in general, $\ca$ is not commutative, a right module
structure and a left module one should be taken to be distinct. 
A {\it bimodule} over the algebra $\ca$ is a vector space $\ce$ which carries both a
left and a right module structure. For each algebra $\ca$, the {\it opposite
algebra} $\ca^o$ has elements $a^o$ in bijective correspondence with the elements
$a \in \ca$ while the multiplication is given by $a^o b^o = (b a)^o$. Any right
(respectively left) $\ca$-module $\ce$ can be regarded as a left (respectively
right) $\ca^o$-module by setting $a^o \h = \h a$ (respectively $a \h = \h a^o$), for
any $\h \in \ce, a \in \ca$. The algebra $\ca \otc \ca^o$ is called the {\it
enveloping algebra} of $\ca$ and is denoted by $\ca^e$. Any $\ca$-bimodule
$\ce$ can be regarded as a  right $\ca^e$-module by setting $\h (a \otimes b^o) =
b \h a$, for any $\h \in \ce, a \in \ca, b^o \in \ca^o$. One can also regard 
$\ce$ as a  left $\ca^e$-module by setting $(a \otimes b^o) \h =
a \h b$, for any $\h \in \ce, a \in \ca, b^o \in \ca^o$. 

A family $(e_t)_{t\in T}$, with $T$ any (finite or infinite) directed set, is
called a {\it generating family} for the right module $\ce$ if any element of
$\ce$ can be written (possibly in more that one manner) as a combination
$\sum_{t\in T}e_t a_t$, with $a_t \in \ca$ and only a finite number of terms in the
sum being different from zero. The family $(e_t)_{t\in T}$ is called {\it free} if
it is made of linearly independent elements (over $\ca$), and it is a {\it
basis} for the module $\ce$ if it is a free generating family, so that
any $\h \in \ce$ can be written uniquely
as a combination $\sum_{t\in T} e_t a_t$, with $a_t \in \ca$. The module is
called free if it admits a basis.  \\
The module $\ce$ is said to be of {\it finite type} if it is finitely generated,
namely if it admits a generating family of finite cardinality. 

Consider now the module $\IC^N \otc \ca =:\ca^N$. Any of its elements $\h$ can
be thought of as an $N$-dimensional vector with entries in $\ca$ and can be
written uniquely as a linear combination $\h = \sum_{j=1}^N e_j a_j$, with $a_j \in
\ca$  and the basis $\{e_j, ~j=1, \dots, N\}$ being identified with the canonical
basis of $\IC^N$. This module is clearly both free and of finite type. \\
A general free module (of finite type) {\it might} admits basis of different
cardinality and it does not make sense to talk of dimension. If the free module is
such that any two basis have the same cardinality, the latter is called the
dimension of the module
\fn{A sufficient condition for this to happen is the existence of a (ring)
homomorphism $\r : \ca \ra \ID$, with $\ID$ any field. This is for instance the
case if $\ca$ is commutative (since then $\ca$ admits at least a maximal ideal
$\IM$ and $\ca / \IM $ is a field) or if $\ca$ may be written as a (ring) direct sum
$\ca = \IC \oplus \bar{\ca}$ \cite{Bo1}.}. \\
However, if the module  $\ce$ is of finite type there is always
an integer $N$ and a (module) surjection  $\r : \ca^N \ra \ce$. In this case one
finds a basis $\{\e_j~, j=1, \dots, N \}$ which is the image of the free basis,
$\e_j = \r(e_j)~,~j=1,
\dots, N$. Notice that in general it is not possible to solve the constraints among
the basis element so as to get a free basis. For example, consider the algebra
$C^\infty(S^2)$ of smooth functions on the two-dimensional sphere $S^2$ and
the Lie algebra $\X(S^2)$ of smooth vector fields on $S^2$. Then,
$\X(S^2)$ is a module of finite type over $C^\infty(S^2)$, a basis of three
elements being given by $\{Y_i = \sum_{j,k=1}^3\varepsilon_{ijk} x_k {\pa \over
\pa x_k}~, i=1,2,3\}$ with $x_1, x_2, x_3$, such that $\sum_j^3 (x_j)^2 = 1$,
just the natural coordinates of $S^2$. 
The basis is not free, since $\sum_{j=1}^3 x_j Y_j = 0$ but there are not two
globally defined vector fields on $S^2$ which could serve as a basis of
$\X(S^2)$. Of course this is nothing but the statement that the tangent bundle
$TS^2$ over $S^2$ is not trivial.

\subsect{Projective Modules of Finite Type}\label{se:mod}

\bdefi
A right $\ca$-module $\ce$ is said to be {\rm projective} if it 
satisfy one of the following equivalent properties:
\begin{enumerate}
\item[1.](Lifting property.)
Given a surjective homomorphism $\r : \cm \ra \cn$ of right $\ca$-modules, any
homomorphism $\l : \ce \ra \cn$ can be lifted to a homomorphism $\wt{\l} :
\ce \ra \cm$ such that $\r \circ \wt{\l} = \l$,
\be
\begin{array}{rccc}
id : & \cm & \longleftrightarrow & \cm \\
~ & ~ & ~ & ~ \\
~ & \wt{\l} \up & ~ & \dn \r \\
~ & ~ & ~ & ~ \\
\l : & \ce & \lra & \cn \\
~ & ~ & ~ & ~ \\
~ & ~ & ~ & \dn \\
~ & ~ & ~ & ~ \\
~ & ~ & ~ & 0 \\
\end{array}~~, ~~~~~ \r \circ \wt{\l} = \l~.
\ee   
\item[2.] Every surjective module morphism $\r : \cm \ra \ce$ splits, namely
there exists a module morphism $s : \ce \ra \cm$ such that $\r \circ s =
id_{\ce}$. 
\item[3.] The module $\ce$ is a direct summand of a free module, namely there exists
a free module $\cf$ and a module (a fortiori projective) $\ce '$, such that 
\be
\cf = \ce \oplus \ce '~.
\ee 
\end{enumerate}
\edefi           

\noindent
To prove that 1. implies 2. it is enough to apply 1. to the case $\cn =
\ce$, 
$\l = id_\ce$, and get for $\wt{\l}$ the splitting map $s$. To prove that 2. 
implies 3. one first observe that 2. implies that $\ce$ is a direct summand of $\cm$
through $s$, namely $\cm = s(\ce) \oplus ker \r$. Also, as mentioned before, for any
module $\ce$ it is possible to construct a surjection from a free module $\cf$, 
$\r : \cf \ra \ce$ (in fact $\cf = \ca^N$ for some $N$). One then applies 2. to this
surjection. To prove that 3. implies 1. one observe that a free module is projective
and that a direct sum of modules is projective if and only if any summand is.

\bigskip

Suppose now that $\ce$ is both projective and of finite type with
surjection
$\r : \ca^N \ra \ce$. Then, the projective properties allow one to find a lift
$\wt{\l} :
\ce
\ra
\ca^N$ such that
$\r \circ \wt{\l} = id_{\ce}$,
\be
\begin{array}{rccc}
id : & \ca^N & \longleftrightarrow & \ca^N \\
~ & ~ & ~ & ~ \\
~ & \wt{\l} \up & ~ & \dn \r \\
~ & ~ & ~ & ~ \\
id : & \ce & \lra & \ce
\end{array}~~, ~~~~~ \r \circ \wt{\l} = id_{\ce}~.
\label{profin}
\ee  
We can then construct an idempotent $p \in End_{\ca} \ca^N \simeq \IM_N(\ca)$,
$\IM_N(\ca)$ being the algebra of $N \times N$ matrices with entry in $\ca$,
given by
\be
p = \wt{\l} \circ \r~. \label{projector}
\ee
Indeed, from (\ref{profin}), $p^2 = \wt{\l} \circ \r \circ \wt{\l} \circ
\r = \wt{\l} \circ \r = p$. The idempotent $p$ allows one to decompose the
free module $\ca^N$ as a direct sum of submodules,
\be
\ca^N = p \ca^N + (1-p) \ca^N 
\ee 
and $\r$ and $\wt{\l}$ are isomorphisms (one the inverse of the other) between
$\ce$ and $p\ca^N$. 
The module $\ce$ is then projective of finite type over $\ca$ if and only if
there exits an idempotent $p \in \IM_N(\ca),~ p^2 = p$~, such that $\ce= p\ca^N$.
We may think of elements of $\ce$ as $N$-dimensional column vectors whose
elements are in $\ca$, the collection of which being invariant under the action
of $p$,
\be 
\ce = \{ \x = (\x_1, \dots, \x_N)~; ~\x_i \in \ca~,~p\x =\x\}~.
\ee
In the following, we shall use the name {\it finite projective} to
mean projective of finite type. 
         
The crucial link between finite projective modules and vector bundles is
provided by the following central  result which is named after Serre and Swan
\cite{Sw} (see also \cite{VG}). As mentioned before, Serre-Swan theorem was
established for functions and sections which are continuous; but it can be 
extended to the smooth case \cite{Colh}. 
\bprop  
Let $M$ be a compact finite dimensional manifold.  A $C^\infty(M)$-module $\ce$
is isomorphic to a module $\G(E,M)$ of smooth sections of a bundle $E \ra M$, if and
only if it is finite projective.

\proof 
We first prove that a module $\G(E,M)$ of sections is finite projective. 
If $E \simeq M \times \IC^k$ is the rank $k$ trivial vector bundle, then $\G(E,M)$ is
just the free module $\ca^k$, $\ca$ being the algebra  $C^\infty(M)$. In general,
from what said before, one has to construct two maps 
$\l : \G(E,M) \ra \ca^N$ (this was called $\wt{\l}$ before), and  
$\r : \ca^N \ra \G(E,M), N$ being a suitable integer, such that $\r \circ \l =
id_{\G(E,M)}$. Then
$\G(E,M) = p\ca^N$, with the idempotent $p$ given by $p = \l \circ \r$.
Let $\{ U_i, i = 1, \cdots, q \}$ be an open covering of $M$. Any element $s
\in \G(E,M)$ can be represented by $q$ smooth maps $ s_i = s_{| U_i} : U \ra
\IC^k$, which satisfies the compatibility conditions
\be
s_j(m) = \sum_j g_{ji}(m) s_i(m)~, ~~ m \in U_i \cap U_j~,
\ee
with $g_{ji} : U_i \cap U_j \ra GL(k, \IC)$ the transition functions of the
bundle. Consider now a partition of unity $\{ h_i, , i = 1, \cdots, q \}$
subordinate to the covering $\{ U_i \}$. By a suitable rescaling we can alway
assume that $h_1^2 + \cdots + h_q^2 = 1$ so that $h_j^2$ as well is a partition 
of unity subordinate to $\{ U_i \}$. Set now $N=kq$, write
$\IC^N = \IC^k \oplus \cdots \oplus \IC^k$ ($q$ summands), and define
\bea
&&\l : \G(E,M) \ra \ca^N~,~~ \l(s_1, \cdots, s_q) =: (h_1 s_1, \cdots, h_q s_q)~, 
\nonumber \\ 
&&\r : \ca^N \ra \G(E,M)~,~~ \r(t_1, \cdots, t_q) =: (\wt{s}_1,
\cdots,
\wt{s}_q)~,~~ \wt{s}_i = \sum_{j} g_{ij} h_j t_j~.
\eea     
Then 
\be
\r \circ \l(s_1, \cdots, s_q) = (\wt{s}_1, \cdots, \wt{s}_q)~,
~~ \wt{s}_i = \sum_{j} g_{ij} h_j h_j s_j~,
\ee
which, $\{ h^2_j\}$ being a partition of unity, amounts to $\r \circ \l =
id_{\G(E,M)}$.  

Conversely, suppose that $\ce$ is a finite projective $C^\infty(M)$-module. Then,
with $\ca = C^\infty(M)$, one can find an integer $N$ and an idempotent $p \in
\IM_N(\ca)$, such that $\ce = p \ca^N$. Now, $\ca^N$ can be identified with
the module of section of the trivial bundle $M \times \IC^N$, $\ca^N \simeq 
\G(M \times \IC^N)$. Since $p$ is a module map, one has that $p(s f) = p(s)f~, 
~f \in C^\infty(M)$. If $m \in M$ and $\ci_m$ is the ideal $\ci_m = \{ f \in
C^\infty(M) ~|~ f(m) = 0 \}$, then $p$ preserves the submodule $\ca^N \ci_m$. 
Since $s \mapsto s(m)$ induces a linear isomorphism of $\ca^N / \ca^N \ci_m$
onto the fiber $(M \times \IC^N)_m$, we have that $p(s)(m) \in (M \times
\IC^N)_m$ for all $s \in \ca^N$. Then the map $\p : M \times \IC^N \ra M \times
\IC^N, ~s(m) \mapsto p(s)(m)$, defines a bundle homomorphism satisfying
$p(s) = \pi \circ s$. Since $p^2=p$, one has that $\p^2=\p$. Suppose now
that $dim~\p((M \times \IC^N)_m) = k$. Then one can find $k$ linearly independent
smooth local sections $s_j \in \ca^N, j=1, ~\cdots, k$, near $m \in M$, such that
$\p\circ s_j(m)=s_j(m)$. Then, $\p\circ s_j, ~j=1, \cdots, k$ are
linearly independent in a neighborhood $U$ of $m$, so that $dim~\p((M \times
\IC^N)_{m'}) \geq k$, for any $m' \in U$. Similarly, by considering the
idempotent $(1-\p ): M \times \IC^N \ra M \times \IC^N$, one gets
that $dim~(1-\p)((M \times \IC^N)_{m'}) \geq N-k$, for any $m' \in U$. The integer
$N$ being constant, one infers that $dim~\p((M \times \IC^N)_{m'})$ is (locally)
constant, so that $\p(M \times\IC^N)$ is the total space of a vector bundle $E
\ra M$ for which $M \times\IC^N = E \oplus \ker \pi$. From its definition, one
gets that $\G(E,M)=\{\pi \circ s ~|~ s \in \G(M \times\IC^N) \} = Im\{p : \ca^N \ra
\ca^N \} = \ce$.
\eprop

If $E$ is a (complex) vector bundle over a compact manifold $M$ of
dimension $n$, there exists a finite cover $\{U_i~, ~i = 1, \cdots, n\}$ of $M$ such
that $E_{|_{U_i}}$ is trivial \cite{Kar}. Thus, the integer
$N$ which determines the rank of the free bundle from which to project onto the
sections of the bundle $E \ra M$ is determined by the equality $N = k n$ where $k$
is the rank of the bundle $E \ra M$ and $n$ is the dimension of $M$.

\subsect{Hermitian Structures over Projective Modules}\label{se:hsp}
Suppose the vector bundle $E \ra M$ is also endowed with an Hermitian structure.
Then, the Hermitian inner product $\hs{\cdot}{\cdot}_m$ on each fiber $E_m$ of the
bundle gives a $C^\infty(M)$-valued sesquilinear map on the module of smooth
sections $\G(E, M)$,
\bea
&& \hs{\cdot}{\cdot} ~: \ce \times \ce \ra C^\infty(M)~, \nonumber \\
&& \hs{\h_1}{\h_2}(m) =: \hs{\h_1(m)}{\h_2(m)}_m~, 
~~~\forall ~ \h_1, \h_2 \in \G(E, M)~.
\label{prod}
\eea   
For any $\h_1, \h_2 \in \G(E, M)$ and $a, b \in C^\infty(M)$, the map
(\ref{prod}) is easily seen to satisfy the following properties
\bea
&&  \hs{\h_1 a}{\h_2 b} = a^* \hs{\h_1}{\h_2} b~, \\ \label{her1} 
&&  \hs{\h_1}{\h_2}^*  = \hs{\h_2}{\h_1}~,  \\ \label{her2}
&&  \hs{\h}{\h} \geq 0~, ~~\hs{\h}{\h} = 0~ \Leftrightarrow ~\h = 0~.
\label{her3} 
\eea

Suppose now that we have a (finite projective right) module $\ce$ over an algebra
$\ca$ with involution $^*$. Then, equations (\ref{her1})-(\ref{her3}) are just
the definition of an {\it Hermitian structure} over $\ce$, a module being
called {\it Hermitian} is it admits an Hermitian structure. We recall that an
element $a \in \ca$ is said to be positive if can be written in the form
$a=b^*b$ for some $b\in\ca$.  \\
A condition of non degeneracy of an Hermitian structure is expressed
in term of the dual module
\be\label{dual}
\ce ' = \{ \f : \ce \ra \ca ~~|~~ \f(\h a ) = \f(\h) a~, ~~\h \in \ce, a \in
\ca \}~.
\ee
which has a natural right $\ca$-module structure given by 
\be\label{dualmodu}
\ce ' \times \ca \ni (\f, a) \mapsto \f \cdot a =: a^* \f \in \ce ' ~.
\ee
We have the following definition.
\bdefi\label{nondeg}
The Hermitian structure $\hs{\cdot}{\cdot}$ on the (right, finite projective)
$\ca$-module $\ce$ is called non degenerate if the map 
\be
\ce \ra \ce ' ~, ~~~ \h \mapsto \hs{\h }{\cdot}~,
\ee
is an isomorphism.
\edefi

\noindent
On the free module $\ca^N$ there is a canonical Hermitian structure given by
\be\label{fhs}
\hs{\h}{\x} = \sum_{j=1}^N \h_j^* \x_j~,
\ee
where $\h = (\h_1, \cdots, \h_N)$ and $\x = (\x_1, \cdots, \x_N)$ are any two
elements $\ca^N$. \\
Under suitable regularity conditions on the algebra $\ca$
all Hermitian structures on a given finite projective module $\ce$ over $\ca$ 
are isomorphic to each other and are obtained from the canonical structure 
(\ref{fhs}) on $\ca^N$ by restriction. We refer to \cite{Co1} for additional
considerations and details on this point. Moreover, if $\ce = p\ca^N$, then $p$
is self-adjoint
\fn{\rm Self-adjoint idempotents are also called projectors.}. We have indeed the
following proposition. 

\bprop
Hermitian finite projective modules are of the form $p \ca^N$ with $p$ a
self-adjoint idempotent, namely $p^* = p$, the operation $^*$ being the 
composition of the $^*$-operation in the algebra $\ca$ with usual matrix
transposition.

\proof With respect to the canonical structure (\ref{fhs}), one easily finds that
$\hs{p^* \x}{\h } = \hs{\x}{p \h }$ for any matrix $p \in \IM_N(\ca)$.
Suppose now that $p$ is an idempotent and consider the module $\ce = p\ca^N$. The
orthogonal space $\ce^{\perp} =: \{ u \in \ca^N ~|~ \hs{u}{\h } = 0~, ~\forall~ \h
\in \ce\}$ is again a right $\ca$-module since $\hs{ua}{\h} = a^* \hs{u}{\h } $.
If $u \in \ca^N$ and $\h \in \ce$, then $\hs{(1-p^*)u}{\h} = \hs{u}{(1-p)\h} = 0$
which states that $\ce^{\perp} = (1-p^*) \ca^N$. On the other side, since $\ca^N = p
\ca^N \oplus (1-p) \ca^N$, the pairing $\hs{\cdot}{\cdot}$ on $\ca^N$ gives an
Hermitian structures on $\ce = p\ca^N$ if and only if this is an orthogonal direct
sum, namely, if and only if $(1-p^*)=(1-p)$ or $p=p^*$.
\eprop

\subsect{Few Elements of $K$-theory}\label{se:kt} 
We have seen in the previous Sections that the algebraic substitutes for bundles are
projective modules of finite type over an algebra $\ca$. The (algebraic) $K$-theory
of $\ca$ is the natural framework for the analogue of bundle invariants.    
Indeed, both the notions of isomorphism and of stable isomorphism have a meaning in 
the context of finite projective (right) modules and the  group $K_0(\ca)$ will be
the group of (stable) isomorphism classes of such modules. In this Section we shall
give few fundamentals of the $K$-theory of $C^*$-algebras while  referring to
\cite{W-O} for more details. In particular, we shall have in mind AF algebras.

\subsubsect{The Group $K_0$}\label{se:kzero}
Given a unital $C^*$-algebra $\ca$ we shall indicate by
$\IM_N(\ca) \simeq \ca \otc \IM_N(\IC)$ the $C^*$-algebra of $N\times N$ matrices
with entries in $\ca$. Two projectors $p, q \in \IM_N(\ca)$ are said to be equivalent
(in the sense of Murray - von Neumann) if there exists a matrix (a partial isometry
\fn{An element $u$ in a $^*$-algebra $\cb$ is called a {\it partial isometry} if
$u^* u$ is a projector (called the support projector). Then automatically $u u^*$
is a projector \cite{W-O} (called the range projector). If $\cb$ is unital and $u^*
u = \II$, then $u$ is called an {\it isometry}. })
$u
\in
\IM_N(\ca)$ such that
$p=u^* u$ and
$q=u u^*$. In order to be able to `add' equivalence classes of projectors, one
considers all finite matrix algebras over $\ca$ at the same time by considering
$\IM_\infty(\ca)$ which is the non complete $^*$-algebra obtained as the inductive
limit of finite matrices 
\fn{The completion of $\IM_\infty(\ca)$ is $\ca \otimes \ck$, with $\ck$ the
algebra of compact operators on the Hilbert space $l_2$. The algebra $\ca \otimes
\ck$ is also called the stabilization of $\ca$.}, 
\bea
&& \IM_\infty(\ca) = \bigcup_{n=1}^{\infty} \IM_n(\ca)~, \nonumber \\
&& \f : \IM_n(\ca) \ra \IM_{n+1}(\ca)~, ~a \mapsto 
\f(a) =
\left\{
\begin{array}{cc}
a & 0 \\
0 & 0 
\end{array}
\right\}~.
\eea
Now, two projectors $p, q \in \IM_\infty(\ca)$ are said to be equivalent, $p
\sim q$, when there exists a $u \in \IM_\infty(\ca)$ such that $p=u^* u$ and
$q=u u^*$. The set $V(\ca)$ of equivalence classes $[ ~\cdot~]$ is made an
abelian semigroup by defining an {\it addition} by
\be
[p] + [q] =: [
\left\{
\begin{array}{cc}
p & 0 \\
0 & q 
\end{array}
\right\}
] ~, ~~~\forall ~[p], [q] \in V(\ca).
\ee
The additive identity is just $0=:[0]$. \\
The groups $K_0(\ca)$ is the universal canonical group (also called enveloping or
Grothendieck group) associated with the abelian semigroup $V(\ca)$. It may be defined
as a collection of equivalence classes, 
\bea
&&K_0(\ca) =: V(\ca) \times V(\ca) / \sim ~, \nonumber \\ 
&&([p], [q]) \sim ([p'], [q'])~ \iff~~ \exists ~[r] \in V(\ca)
~~{\rm s.t.}~~ [p] + [q'] + [r] = [p'] + [q] + [r]~.~~~~~~~ \label{keya}
\eea
It is straightforward to check reflexivity, symmetry and transitivity, the extra
$[r]$ in (\ref{keya}) being inserted just to get the latter property, so that
$\sim$ is an equivalence relation. The presence of the extra $[r]$ is the reason
why one is classifying only stable classes. \\
The addition in $K_0(\ca)$ is defined by
\be
[([p], [q])] + [([p'], [q'])] =: [([p] + [p'], [q] + [q'])]~,
\ee
for any $[([p], [q])], [([p'], [q'])] \in K_0(\ca)$, and does not depends on the
representatives. As for the neutral element, it is given by the class
\be
0 = [([p], [p])]
\ee
for any $[p]\in V(\ca)$. Indeed, all such elements are equivalents. Finally, the
inverse $-[([p], [q])]$ of the class $[([p], [q])]$ is given by the the class
\be
-[([p], [q])] =: [([q], [p])]~,
\ee
since,
\be
[([p], [q])] + (- [([p], [q])]) = [([p], [q])] + ([([q], [p])]) = 
[([p] + [q], [p] + [q])] = 0~. 
\ee
From all said previously, it is useful to think of the class
$[([p], [q])] \in K_0(\ca)$ as a formal difference $[p] -
[q]$.  \\   There is a natural homomorphism 
\be
\k_{\ca} : V(\ca) \ra K_0(\ca)~,~~\k_{\ca}([p]) =: ([p], [0]) = [p] - [0]~. 
\label{homkey}
\ee 
However, this map is injective if and only if the addition in $V(\ca)$ has
cancellations, namely if and only if $[p] + [r] = [q] + [r]~ \Rightarrow 
[p] = [q]$. Independently of the fact that $V(\ca)$ has cancellations, any   
$\k_{\ca}([p]), [p] \in V(\ca)$, has an inverse in $K_0(\ca)$ and any element of the
latter group can be written as a difference $\k_{\ca}([p]) - \k_{\ca}([q])$, with
$[p], [q] \in V(\ca)$.
 
While for a generic $\ca$, the semigroup $V(\ca)$ has no
cancellations, for AF algebras this happens to be the case. By defining
\be
K_{0+}(\ca) =: \k_{\ca}(V(\ca))~, \label{keya+}
\ee
the couple $(K_0(\ca), K_{0+}(\ca))$ becomes, for an AF algebra $\ca$,
an {\it ordered group} with $K_{0+}(\ca)$ the {\it positive cone}, namely
one has that 
\bea
&& K_{0+}(\ca) \ni 0~, \nonumber \\
&& K_{0+}(\ca) - K_{0+}(\ca) = K_0(\ca)~, \nonumber \\
&& K_{0+}(\ca) \cap (- K_{0+}(\ca)) = {0}~.
\eea
For a generic algebra the last property is not true and the couple
$(K_0(\ca), K_{0+}(\ca))$ is not an ordered group.

\bexam\label{ex:simple}
The group $K_0(\ca)$ for $\ca = \IC$, $\ca = \IM_k(\IC), k \in \IN$ and 
$\ca = \IM_k(\IC) \oplus \IM_{k'}(\IC), k, k' \in \IN$.\\ \\
If $\ca = \IC$, any element in $V(\ca)$ is a class of equivalent projectors in some
$\IM_n(\IC)$. Now, projectors in $\IM_n(\IC)$ are equivalent precisely when they
ranges, which are subspaces of $\IC^n$, have the same dimension. Therefore we can
make the identification
\be
V(\IC) \simeq \IN~, 
\ee
with $\IN = \{0, 1, 2, \cdots \}$ the semigroup of natural numbers. \\ \\ 
As $\IM_n(\IM_k(\IC)) \simeq \IM_{nk}(\IC)$, the same argument gives
\be
V(\IM_k(\IC)) \simeq \IN~. 
\ee
Now, the canonical group associated with the semigroup $\IN$ is just the group
$\IZ$ of integers, and we have
\be
\begin{array}{ll}
K_{0}(\IC) = \IZ~, & K_{0+}(\IC) = \IN~,\\
K_{0}(\IM_k(\IC)) = \IZ~, & K_{0+}(\IM_k(\IC)) = \IN~, ~~~ \forall ~k\in\IN~.
\end{array}
\ee  
\\
For $\ca = \IM_k(\IC) \oplus \IM_{k'}(\IC)$, the same argument for each of the two
terms in the direct sum will give
\bea
&& K_{0}(\IM_k(\IC) \oplus \IM_{k'}(\IC)) = \IZ \oplus \IZ~, \\ 
&& K_{0+}(\IM_k(\IC)
\oplus
\IM_{k'}(\IC)) = \IN \oplus \IN~, ~~~ \forall ~k, k' \in\IN~.
\eea  
\eexam

In general, the group $K_0$ has few interesting properties, notably universality.
\bprop\label{pr:kuniv}
Let $G$ be an abelian group and $\F : V(\ca) \ra G$ be a homomorphism of semigroups
such that $\F(V(\ca))$ is invertible in $G$. \\
Then, $\F$ extends uniquely to a homomorphism $\J : K_0(\ca) \ra G$, 
\be
\begin{array}{rccc}
\F : & V(\ca) & \lra & G \\
~ & ~ & ~ & ~ \\
~ & \k_{\ca} \dn & ~ & \up \J \\
~ & ~ & ~ & ~ \\
id : & K_0(\ca) & \longleftrightarrow & K_0(\ca)
\end{array}~~, ~~~~~ \J \circ \k_{\ca} = \F~.
\label{kuniv}
\ee  

\proof
First uniqueness. If $\J_1, \J_2 : K_0(\ca) \ra G$ both extend $\F$, then
$\J_1([([p], [q])]) = \J_1([p] - [q]) = \J_1(\k_{\ca}([p])) - \J_1(\k_{\ca}([q])) =
\F([p]) - \F([q]) = \J_2([([p], [q])])$, which proves uniqueness.\\
Then existence. Define $\J : K_0(\ca) \ra G$~ by ~$\J([([p], [q])]) = \F([p]) -
\F([q])$. This map is well defined because $\F([q])$ has inverse in $G$ and
because $([p], [q]) \sim ([p'], [q'])~ \iff~~ \exists ~[r] \in V(\ca)$
such that $[p] + [q'] + [r] = [p'] + [q] + [r]$, and this implies $\J([([p], [q])])
= \J([([p'], [q'])])$. Finally, $\J$ is a homomorphism and $\J(\k_{\ca}([p]) =
\J([([p], [0])]) = \F([p])$, namely $\J \circ \k_{\ca} = \F$.
\eprop

The group $K_0$ is well behaved with respect to homomorphisms
\fn{In a more sophisticated parlance, $K_0$ is a covariant functor from the category
of
$C^*$-algebras to the category of abelian groups.}. 
\bprop\label{pr:ind} 
If $\a : \ca \ra \cb$ is a homomorphism of
$C^*$-algebras, then the induced map
\be
\a_* : V(\ca) \ra V(\cb)~, ~~\a_*([a_{ij}]) =: [\a(a_{ij})]~,
\label{induc}
\ee
is a well defined homomorphism of semigroups. 
Moreover, from universality, $\a_*$ extends to a group homomorphism (denoted with
the same symbol)
\be
\a_* = K_0(\ca) \ra K_0(\cb)~. \label{keyhom}
\ee
\proof
If the matrix $(a_{ij}) \in \IM_\infty(\ca)$ is a projector, the matrix
$\a(a_{ij})$ will clearly be a projector in $\IM_\infty(\cb)$. Furthermore, if 
$(a_{ij})$ is equivalent to $(b_{ij})$, then, since $\a$ is multiplicative and
$^*$-preserving, $\a(a_{ij})$ will be equivalent to $\a(b_{ij})$. Thus $\a_* :
V(\ca) \ra V(\cb)$ is well defined and clearly a homomorphism. The last statement
follows from Proposition~\ref{pr:kuniv} with the identifications
$\F \equiv \k_{\cb} \circ \a_* : V(\ca) \ra K_0(\cb)$ so as to get for $\J$ the map
$\J \equiv \a_* : K_0(\ca) \ra K_0(\cb)$.    
\eprop

The group $K_0$ is also well behaved with respect to the process of taking
inductive limits of $C^*$-algebras, as stated by the following proposition which is
proved in \cite{W-O} and which is crucial for the calculation of the $K_0$ of AF
algebras.
\bprop\label{pr:key}
If the $C^*$-algebra $\ca$ is the inductive limit of a directed system $\{\ca_i,
\F_{ij}\}_{i,j\in\IN}$ of
$C^*$-algebras
\fn{In fact, one could substitute $\IN$ with any directed set $\L$.}, then
$\{K_0(\ca_{i}),
\F_{ij*}\}_{i,j\in\IN}$ is a directed system of groups and one can exchange the
limits,
\be
K_0(\ca) = K_0(\lim_{\ra} \ca_i) = \lim_{\ra} K_0(\ca_i)~.
\ee
Moreover, if $\ca$ is an AF algebra, then $K_0(\ca)$ is an ordered group with
positive cone given by the limit of a directed system of semigroups
\be
K_{0+}(\ca) = K_{0+}(\lim_{\ra} \ca_i) = \lim_{\ra} K_{0+}(\ca_i)~.
\ee
\eprop

\noindent
One has that as sets, 
\bea
&& K_0(\ca) = \{ (k_n)_{n \in \IN}~, k_n \in K_0(\ca_n) ~|~ \exists N_0~:~
k_{n+1} = T_n(k_n)~,~ n>N_0 \}~, \label{key} \\
&& K_{0+}(\ca) = \{ (k_n)_{n \in \IN}~, k_n \in K_{0+}(\ca_n) ~|~
\exists N_0~:~ k_{n+1} = T_n(k_n)~,~ n>N_0 \}~, \label{key+}
\eea
while the structure of (abelian) group/semigroup is inherited pointwise from the
addition in the groups/semigroups in the sequences
(\ref{key}), (\ref{key+}) respectively.

\subsubsect{The $K$-theory of the Penrose Tiling}\label{se:kpt}
The algebra $\ca_{PT}$ of the Penrose Tiling is an AF algebra which is quite far from
being postliminal, since there is an infinite number of not equivalent irreducible
representations which are faithful and then have the same kernel, namely zero which
is the only primitive ideal (the algebra $\ca_{PT}$ is indeed simple). The
construction of its
$K$-theory is rather straightforward and quite illuminating. The corresponding
Bratteli diagram is shown in Fig.~\ref{fi:penbra} \cite{Co1}. 
From Props.~(\ref{bratteli}) and~(\ref{primitive}) it is clear that $\{0\}$ is the
only primitive ideal.
\begin{figure}[htb]
\begin{center}
\begin{picture}(120,180)(20,50)
\put(60,210){\circle*{4}}
\put(30,180){\circle*{4}}
\put(30,150){\circle*{4}}
\put(30,120){\circle*{4}}
\put(30,90){\circle*{4}}
\put(30,60){\circle*{4}}
\put(90,180){\circle*{4}}
\put(90,150){\circle*{4}}
\put(90,120){\circle*{4}}
\put(90,90){\circle*{4}}
\put(90,60){\circle*{4}}
\put(60,210){\line(-1,-1){30}}
\put(60,210){\line(1,-1){30}}
\put(30,180){\line(0,-1){30}}
\put(30,150){\line(0,-1){30}}
\put(30,120){\line(0,-1){30}}
\put(30,90){\line(0,-1){30}}
\put(30,60){\line(0,-1){10}}
\put(30,180){\line(2,-1){60}}
\put(30,150){\line(2,-1){60}}
\put(30,120){\line(2,-1){60}}
\put(30,90){\line(2,-1){60}}
\put(30,150){\line(2,1){60}}
\put(30,120){\line(2,1){60}}
\put(30,90){\line(2,1){60}}
\put(30,60){\line(2,1){60}}
\put(58,214){1}
\put(21,178){1}
\put(21,148){2}
\put(21,118){3}
\put(21,88){5}
\put(21,58){8}
\put(94,178){1}
\put(94,148){1}
\put(94,118){2}
\put(94,88){3}
\put(94,58){5}
\put(60,40){$\vdots$}
\end{picture}
\end{center}
\caption{\label{fi:penbra} 
\protect{\footnotesize The Bratteli diagram for the algebra $\ca_{PT}$ of
the Penrose tiling.   }}
\end{figure}
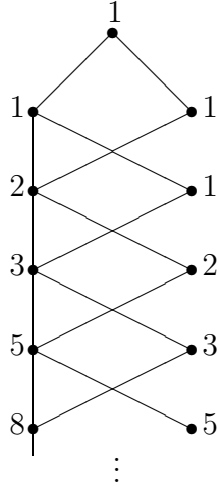\bigskip

\noindent
At each level, the algebra is given by 
\be
\ca_n = \IM_{d_n}(\IC) \oplus \IM_{d'_n}(\IC)~, ~~ n \geq 1~, \label{pt}
\ee
with inclusion 
\be\label{ptinc}
I_n : \ca_n \hookrightarrow \ca_{n+1}~,~~~
\left\{
\begin{array}{cc}
A & 0 \\
0 & B
\end{array}
\right\} \mapsto
\left\{
\begin{array}{ccc}
A & 0 & 0 \\
0 & B & 0 \\
0 & 0 & A
\end{array}
\right\}~,
~~A \in \IM_{d_n}(\IC)~, ~~B \in \IM_{d'_n}(\IC)~. 
\ee
This gives for the dimensions the recursive relations
\be
\begin{array}{l}
d_{n+1} = d_n + d'_n~, \\
d'_{n+1} = d_n~, 
\end{array}~~~ n\geq1~, ~~d_1=d'_1=1~. \label{ptdim}
\ee

From what said in the Example~\ref{ex:simple}, after the second level, the
$K$-groups are given by
\be
K_0(\ca_n) = \IZ \oplus \IZ~, ~~~K_{0+}(\ca_n) = \IN \oplus \IN~, ~~n\geq1.
\ee

The group $(K_0(\ca), K_{0+}(\ca))$ is obtained by Proposition \ref{pr:key} as the
inductive limit of the sequence of groups/semigroups
\bea
&&
K_0(\ca_1) \hookrightarrow K_0(\ca_2) \hookrightarrow K_0(\ca_3) \hookrightarrow
\cdots
\label{kappa}\\
&&
K_{0+}(\ca_1) \hookrightarrow K_{0+}(\ca_2) \hookrightarrow K_{0+}(\ca_3)
\hookrightarrow
\cdots
\label{kappa+}
\eea
The inclusions
\be
T_n : K_0(\ca_n) \hookrightarrow K_0(\ca_{n+1})~,~~
T_n : K_{0+}(\ca_n) \hookrightarrow K_{0+}(\ca_{n+1})~, \label{emb}
\ee
are easily obtained from the inclusions $I_n$ in (\ref{ptinc}), being indeed the
corresponding induced maps as in (\ref{keyhom}) $T_n = I_{n*}$. To construct the
maps $T_n$ we need the following proposition, the first part of which is just
Proposition~\ref{pr:afemb} which we repeat for clarity.
\bprop\label{pr:keyaf}
Let $\ca$ and $\cb$ be the direct sum of two matrix algebras, 
\be
\ca = \IM_{p_1}(\IC) \oplus \IM_{p_2}(\IC)~, 
~~~\cb = \IM_{q_1}(\IC) \oplus \IM_{q_2}(\IC)~. 
\ee
Then, any homomorphism $\a : \ca \ra \cb$ can be written as the direct sum
of the representations $\a_j : \ca \ra \IM_{q_j}(\IC) \simeq \cb(\IC^{q_j}),
j=1,2$. If $\p_{ji}$ is the unique irreducible representation of 
$\IM_{p_i}(\IC)$ in $\cb(\IC^{q_j})$, then $\a_j$ breaks into a direct sum of
the $\p_{ji}$. Furthermore, let $N_{ji}$ be the non-negative integers denoting
the multiplicity of $\p_{ji}$ in this sum. Then the induced homomorphism,
$\a_* = K_0(\ca) \ra K_0(\cb)$, is given by the $2 \times 2$
matrix $(N_{ij})$. 

\proof
For the first part just refer to Proposition~\ref{pr:afemb}. \\
Furthermore, given a rank $k$ projector in $\IM_{p_i}(\IC)$, the representation
$\a_j$ send it to a rank $N_{ji}k$ projector in $\IM_{q_j}(\IC)$. This proves the
final statement of the proposition. 
\eprop

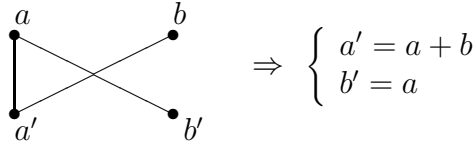
\begin{figure}[htb]
\begin{center}
\begin{picture}(200,40)(0,60)
\put(30,90){\circle*{4}}
\put(30,60){\circle*{4}}
\put(90,90){\circle*{4}}
\put(90,60){\circle*{4}}
\put(30,90){\line(0,-1){30}}
\put(30,90){\line(2,-1){60}}
\put(30,60){\line(2,1){60}}
\put(30,94){$a$}
\put(30,50){$a'$}
\put(90,94){$b$}
\put(94,50){$b'$}
\put(120,75){$\Rightarrow$}
\put(140,75)
{$
\left\{
\begin{array}{l}
a'=a+b \\
b'=a
\end{array}
\right.
$}
\end{picture}
\end{center}
\caption{\label{fi:ptmatpic} 
\protect{\footnotesize The action of the inclusion $T$. }}
\end{figure}

For the inclusion (\ref{pr:keyaf}), Proposition~\ref{pr:keyaf} gives immediately that
the maps (\ref{emb}) are both represented by the integer valued matrix
\be
T =
\left\{
\begin{array}{cc}
1 & 1 \\
1 & 0
\end{array}
\right\}~. \label{ptmat}
\ee
for any level $n$. The action of the matrix (\ref{ptmat}) can be represented
pictorially as in Fig.~\ref{fi:ptmatpic} where the couples $(a, b)$, $(a', b')$ are
both in 
$\IZ \oplus \IZ$ or $\IN \oplus \IN$.
\bigskip

Finally, we can construct the $K_0$ group.
\bprop 
The group $(K_0(\ca_{PT}), K_{0+}(\ca_{PT}))$ for the $C^*$-algebra $\ca_{PT}$ of the
Penrose tiling is given by
\bea
&& K_0(\ca_{PT}) = \IZ \oplus \IZ~, \label{ptkey} \\
&& K_{0+}(\ca_{PT}) = \{ (a, b) \in \IZ \oplus \IZ~ : 
{1+\sqrt{5} \over 2} a + b \geq 0 \}~.
\label{ptkey+}
\eea
\proof 
The result (\ref{ptkey}) follows immediately from the fact that the matrix $T$ in
(\ref{ptmat}) is invertible over the integer, its inverse being
\be
T^{-1} =
\left\{
\begin{array}{cc}
0 & 1 \\
1 & -1
\end{array}
\right\}~. \label{ptmatinv}
\ee
Now, from the definition of inductive limit we have that,
\be
K_0(\ca_{PT}) = \{ (k_n)_{n \in \IN}~, k_n \in K_0(\ca_n) ~|~ \exists N_0 ~:~
k_{n+1} = T (k_n)~, ~n>N_0 \}.
\ee
And, $T$ being a bijection, for any $k_{n+1} \in K_0(\ca_{n+1})$, there
exist a unique $k_n \in K_0(\ca_n)$ such that $k_{n+1} = T k_n$.  Thus,
$K_0(\ca_{PT}) = K_0(\ca_n) = \IZ \oplus \IZ$.\\
\\ 
As for (\ref{ptkey+}), since $T$ is {\it not} invertible over $\IN$,
$K_{0+}(\ca_{PT})
\not=
\IN
\oplus \IN$. To construct $K_{0+}(\ca_{PT})$, we study the image $T(K_{0+}(\ca_n))$
in $K_{0+}(\ca_{n+1})$. It is easily found to be 
\bea
T(K_{0+}(\ca_n)) & = & \{(a_{n+1}, b_{n+1}) \in \IN \oplus \IN~ : 
a_{n+1} \geq b_{n+1} \} \nonumber \\
                 &\not=& K_{0+}(\ca_{n+1}) ~.
\eea
Now, $T$ being injective, $T(K_{0+}(\ca_n)) = T(\IN \oplus \IN)
\simeq \IN \oplus \IN$. The inclusion of $T(K_{0+}(\ca_n))$ into
$K_{0+}(\ca_{n+1})$ is shown in  Fig.~\ref{fi:penrosekeyplus}. 
By identifying the subset $T(K_{0+}(\ca_n)) \subset K_{0+}(\ca_{n+1})$ with
$K_{0+}(\ca_n)$,  we can think of $T^{-1}(K_{0+}(\ca_{n+1}))$ as a subset
of $\IZ \oplus \IZ$ and of $T^{-1}(K_{0+}(\ca_n))$ as the standard positive cone
$\IN \oplus \IN$. The result is shown in Fig.~\ref{fi:penrosekeyminus}. 
Next iteration, namely $T^{-2}(K_{0+}(\ca_n))$ is shown in
Fig.~\ref{fi:penrosekeyminus2}. \\
{}From definition (\ref{key+}), by going to the
limit we shall have $K_{0+}(\ca_{PT}) = \lim_{m \ra \infty} T^{-m}(\IN \oplus
\IN)$ and the limit will be a subset of $ \IZ \oplus \IZ$ since $T$ is
invertible only over $\IZ$. The limit can be easily found.  
{}From the defining relation $F_{m+1} = F_m + F_{m-1}, m\geq1, $ for the 
Fibonacci numbers  (with $F_0=0, F_1=1$), it follows that
\be
T^{-m} =
(-1)^{m} \left\{
\begin{array}{cc}
F_{m-1} & -F_m \\
-F_m & F_{m+1}
\end{array}
\right\}~. \label{ptmatinvm}
\ee
Therefore, $T^{-m}$ takes the positive axis $\{(a,0) : a \geq 0\}$ to a
half-line of slope $-F_m / F_{m-1}$, and the positive axis $\{(0,b) : b \geq
0\}$  to a half-line of slope $-F_{m+1} / F_m$. Thus the positive cone
$\IN \oplus \IN$ opens into a fan-shaped wedge which is bordered by these two
half-lines. Any integer coordinate point within the wedge comes from an
integer coordinate point in the original positive cone. 
Since $\lim_{m \ra \infty} F_{m+1} / F_m = {1+\sqrt{5} \over 2}$, the limit
cone is just the half-space $\{ (a, b) \in \IZ \oplus \IZ~ :
{1+\sqrt{5} \over 2} a + b \geq 0 \}$~. Every integer coordinate point in it
belongs to some intermediate wedge and so lies in $K_{0+}(\ca_{PT})$. The latter
is shown in  Fig.~\ref{fi:penkeyzerplu}.
\eprop
\begin{figure}[htb]
\begin{center}
\begin{picture}(600,130)(-225,40)
\put(-180,50){\vector(1,0){130}}
\put(-170,40){\vector(0,1){130}}
\put(30,50){\vector(1,0){130}}
\put(40,40){\vector(0,1){130}}
\put(40,50){\vector(1,1){110}}
\put(-70,40){$a_n$}
\put(-183,150){$b_n$}
\put(140,40){$a_{n+1} = b_{n}$}
\put(18,150){$b_{n+1}$}
\put(152,151){$a_n$}
\put(-175,47){$\times$}
\put(-175,67){$\times$}
\put(-175,87){$\times$}
\put(-175,107){$\times$}
\put(-175,127){$\times$}
\put(-155,47){$\times$}
\put(-155,67){$\times$}
\put(-155,87){$\times$}
\put(-155,107){$\times$}
\put(-155,127){$\times$}
\put(-135,47){$\times$}
\put(-135,67){$\times$}
\put(-135,87){$\times$}
\put(-135,107){$\times$}
\put(-135,127){$\times$}
\put(-115,47){$\times$}
\put(-115,67){$\times$}
\put(-115,87){$\times$}
\put(-115,107){$\times$}
\put(-115,127){$\times$}
\put(-95,47){$\times$}
\put(-95,67){$\times$}
\put(-95,87){$\times$}
\put(-95,107){$\times$}
\put(-95,127){$\times$}
\put(-20,100){$\stackrel{T}{\longrightarrow}$}
\put(40,50){\circle*{3}}
\put(40,70){\circle*{3}}
\put(40,90){\circle*{3}}
\put(40,110){\circle*{3}}
\put(40,130){\circle*{3}}
\put(60,50){\circle*{3}}
\put(60,70){\circle*{3}}
\put(60,90){\circle*{3}}
\put(60,110){\circle*{3}}
\put(60,130){\circle*{3}}
\put(80,50){\circle*{3}}
\put(80,70){\circle*{3}}
\put(80,90){\circle*{3}}
\put(80,110){\circle*{3}}
\put(80,130){\circle*{3}}
\put(100,50){\circle*{3}}
\put(100,70){\circle*{3}}
\put(100,90){\circle*{3}}
\put(100,110){\circle*{3}}
\put(100,130){\circle*{3}}
\put(120,50){\circle*{3}}
\put(120,70){\circle*{3}}
\put(120,90){\circle*{3}}
\put(120,110){\circle*{3}}
\put(120,130){\circle*{3}}
\put(35,47){$\times$}
\put(55,47){$\times$}
\put(75,47){$\times$}
\put(95,47){$\times$}
\put(115,47){$\times$}
\put(55,67){$\times$}
\put(75,67){$\times$}
\put(95,67){$\times$}
\put(115,67){$\times$}
\put(75,87){$\times$}
\put(95,87){$\times$}
\put(115,87){$\times$}
\put(95,107){$\times$}
\put(115,107){$\times$}
\put(115,127){$\times$}
\end{picture}
\end{center}
\caption{\label{fi:penrosekeyplus} 
\protect{\footnotesize The image of $\IN \oplus \IN$ under $T$.   }}
\end{figure}
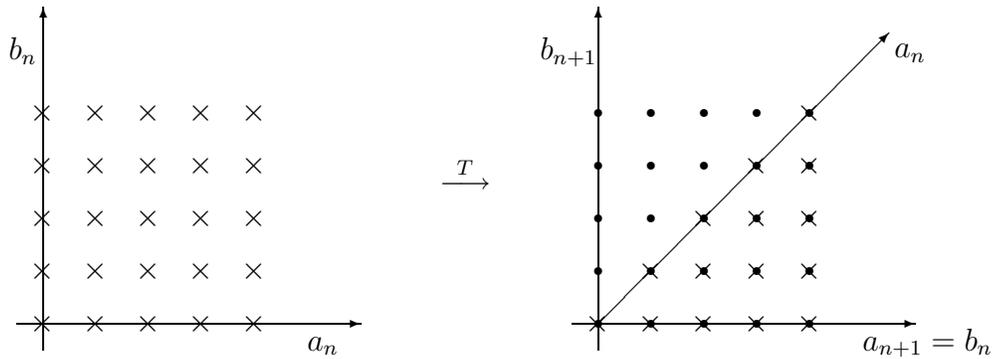
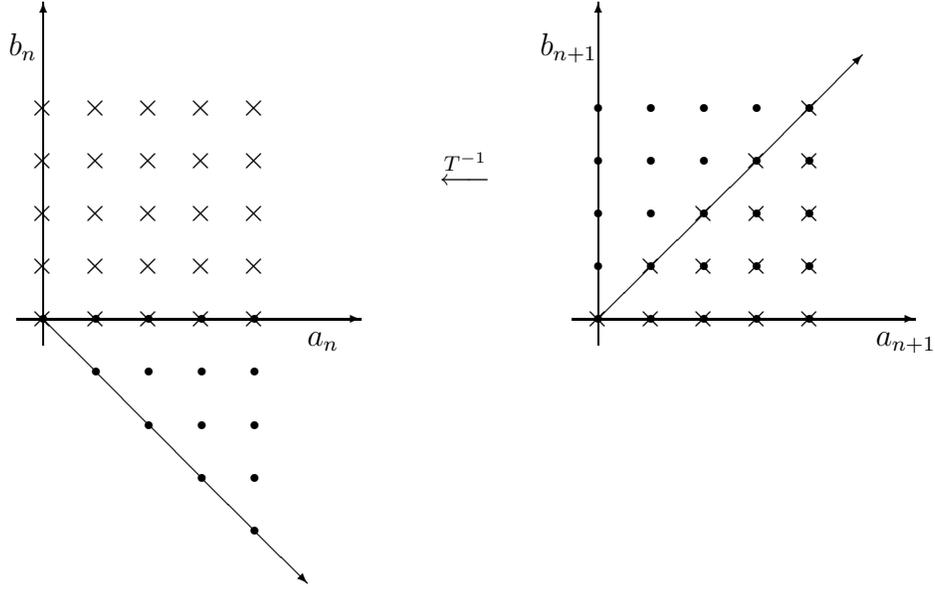
\begin{figure}[htb]
\begin{center}
\begin{picture}(600,220)(-225,-40)
\put(-180,50){\vector(1,0){130}}
\put(-170,40){\vector(0,1){130}}
\put(-170,50){\vector(1,-1){100}}
\put(30,50){\vector(1,0){130}}
\put(40,40){\vector(0,1){130}}
\put(40,50){\vector(1,1){100}}
\put(-70,40){$a_{n}$}
\put(-183,150){$b_n$}
\put(145,40){$a_{n+1}$}
\put(18,150){$b_{n+1}$}
\put(-175,47){$\times$}
\put(-175,67){$\times$}
\put(-175,87){$\times$}
\put(-175,107){$\times$}
\put(-175,127){$\times$}
\put(-155,47){$\times$}
\put(-155,67){$\times$}
\put(-155,87){$\times$}
\put(-155,107){$\times$}
\put(-155,127){$\times$}
\put(-135,47){$\times$}
\put(-135,67){$\times$}
\put(-135,87){$\times$}
\put(-135,107){$\times$}
\put(-135,127){$\times$}
\put(-115,47){$\times$}
\put(-115,67){$\times$}
\put(-115,87){$\times$}
\put(-115,107){$\times$}
\put(-115,127){$\times$}
\put(-95,47){$\times$}
\put(-95,67){$\times$}
\put(-95,87){$\times$}
\put(-95,107){$\times$}
\put(-95,127){$\times$}
\put(-170,50){\circle*{3}}
\put(-150,50){\circle*{3}}
\put(-130,50){\circle*{3}}
\put(-110,50){\circle*{3}}
\put(-90,50){\circle*{3}}
\put(-150,30){\circle*{3}}
\put(-130,30){\circle*{3}}
\put(-110,30){\circle*{3}}
\put(-90,30){\circle*{3}}
\put(-130,10){\circle*{3}}
\put(-110,10){\circle*{3}}
\put(-90,10){\circle*{3}}
\put(-90,-10){\circle*{3}}
\put(-110,-10){\circle*{3}}
\put(-90,-30){\circle*{3}}
\put(-20,100){$\stackrel{T^{-1}}{ \longleftarrow}$}
\put(40,50){\circle*{3}}
\put(40,70){\circle*{3}}
\put(40,90){\circle*{3}}
\put(40,110){\circle*{3}}
\put(40,130){\circle*{3}}
\put(60,50){\circle*{3}}
\put(60,70){\circle*{3}}
\put(60,90){\circle*{3}}
\put(60,110){\circle*{3}}
\put(60,130){\circle*{3}}
\put(80,50){\circle*{3}}
\put(80,70){\circle*{3}}
\put(80,90){\circle*{3}}
\put(80,110){\circle*{3}}
\put(80,130){\circle*{3}}
\put(100,50){\circle*{3}}
\put(100,70){\circle*{3}}
\put(100,90){\circle*{3}}
\put(100,110){\circle*{3}}
\put(100,130){\circle*{3}}
\put(120,50){\circle*{3}}
\put(120,70){\circle*{3}}
\put(120,90){\circle*{3}}
\put(120,110){\circle*{3}}
\put(120,130){\circle*{3}}
\put(35,47){$\times$}
\put(55,47){$\times$}
\put(75,47){$\times$}
\put(95,47){$\times$}
\put(115,47){$\times$}
\put(55,67){$\times$}
\put(75,67){$\times$}
\put(95,67){$\times$}
\put(115,67){$\times$}
\put(75,87){$\times$}
\put(95,87){$\times$}
\put(115,87){$\times$}
\put(95,107){$\times$}
\put(115,107){$\times$}
\put(115,127){$\times$}
\end{picture}
\end{center}
\caption{\label{fi:penrosekeyminus} 
\protect{\footnotesize The image of $\IN \oplus \IN$ under $T^{-1}$.   }}
\end{figure}\bigskip
\begin{figure}
\begin{center}
\begin{picture}(600,220)(-350,-40)
\put(-180,50){\vector(1,0){130}}
\put(-170,40){\vector(0,1){130}}
\put(-170,50){\vector(1,-1){100}}
\put(-170,50){\vector(-1,2){55}}
\put(-175,47){$\times$}
\put(-175,67){$\times$}
\put(-175,87){$\times$}
\put(-175,107){$\times$}
\put(-175,127){$\times$}
\put(-155,47){$\times$}
\put(-155,67){$\times$}
\put(-155,87){$\times$}
\put(-155,107){$\times$}
\put(-155,127){$\times$}
\put(-135,47){$\times$}
\put(-135,67){$\times$}
\put(-135,87){$\times$}
\put(-135,107){$\times$}
\put(-135,127){$\times$}
\put(-115,47){$\times$}
\put(-115,67){$\times$}
\put(-115,87){$\times$}
\put(-115,107){$\times$}
\put(-115,127){$\times$}
\put(-95,47){$\times$}
\put(-95,67){$\times$}
\put(-95,87){$\times$}
\put(-95,107){$\times$}
\put(-95,127){$\times$}
\put(-170,50){\circle*{3}}
\put(-170,70){\circle*{3}}
\put(-170,90){\circle*{3}}
\put(-170,110){\circle*{3}}
\put(-170,130){\circle*{3}}
\put(-155,27){$\times$}
\put(-135,27){$\times$}
\put(-115,27){$\times$}
\put(-95,27){$\times$}
\put(-135,7){$\times$}
\put(-115,7){$\times$}
\put(-95,7){$\times$}
\put(-95,-13){$\times$}
\put(-115,-13){$\times$}
\put(-95,-33){$\times$}
\put(-190,90){\circle*{3}}
\put(-190,110){\circle*{3}}
\put(-190,130){\circle*{3}}
\put(-210,130){\circle*{3}}
\end{picture}
\end{center}
\caption{\label{fi:penrosekeyminus2} 
\protect{\footnotesize The image of $\IN \oplus \IN$ under $T^{-2}$.   }}
\end{figure}
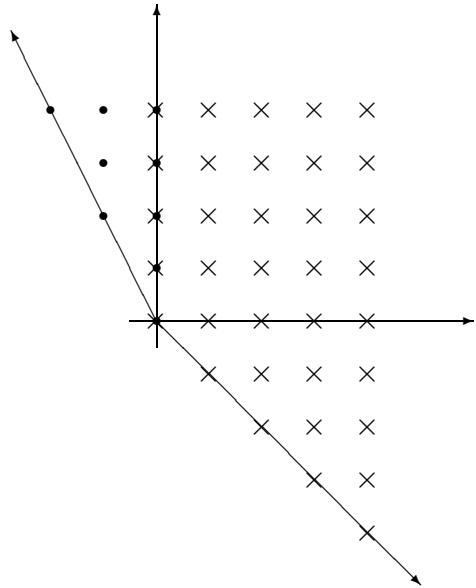\bigskip
\begin{figure}
\begin{center}
\begin{picture}(600,270)(-350,-90)
\put(-180,50){\vector(1,0){130}}
\put(-170,40){\vector(0,1){130}}
\put(-170,50){\line(3,-5){90}}
\put(-170,50){\line(-3,5){60}}
\put(-175,47){$\times$}
\put(-175,67){$\times$}
\put(-175,87){$\times$}
\put(-175,107){$\times$}
\put(-175,127){$\times$}
\put(-155,47){$\times$}
\put(-155,67){$\times$}
\put(-155,87){$\times$}
\put(-155,107){$\times$}
\put(-155,127){$\times$}
\put(-135,47){$\times$}
\put(-135,67){$\times$}
\put(-135,87){$\times$}
\put(-135,107){$\times$}
\put(-135,127){$\times$}
\put(-115,47){$\times$}
\put(-115,67){$\times$}
\put(-115,87){$\times$}
\put(-115,107){$\times$}
\put(-115,127){$\times$}
\put(-95,47){$\times$}
\put(-95,67){$\times$}
\put(-95,87){$\times$}
\put(-95,107){$\times$}
\put(-95,127){$\times$}
\put(-155,27){$\times$}
\put(-135,27){$\times$}
\put(-115,27){$\times$}
\put(-135,7){$\times$}
\put(-115,7){$\times$}
\put(-115,-13){$\times$}
\put(-135,-13){$\times$}
\put(-115,-33){$\times$}
\put(-115,-53){$\times$}
\put(-95,27){$\times$}
\put(-95,7){$\times$}
\put(-95,-13){$\times$}
\put(-95,-33){$\times$}
\put(-95,-53){$\times$}
\put(-95,-73){$\times$}
\put(-75,-100){${1+\sqrt{5} \over 2} a + b = 0$}
\put(-195,87){$\times$}
\put(-195,107){$\times$}
\put(-195,127){$\times$}
\put(-215,127){$\times$}
\end{picture}
\end{center}
\caption{\label{fi:penkeyzerplu} 
\protect{\footnotesize 
$K_{0+}(\ca_{PT})$ for the algebra of the Penrose tiling. }} 
\end{figure}
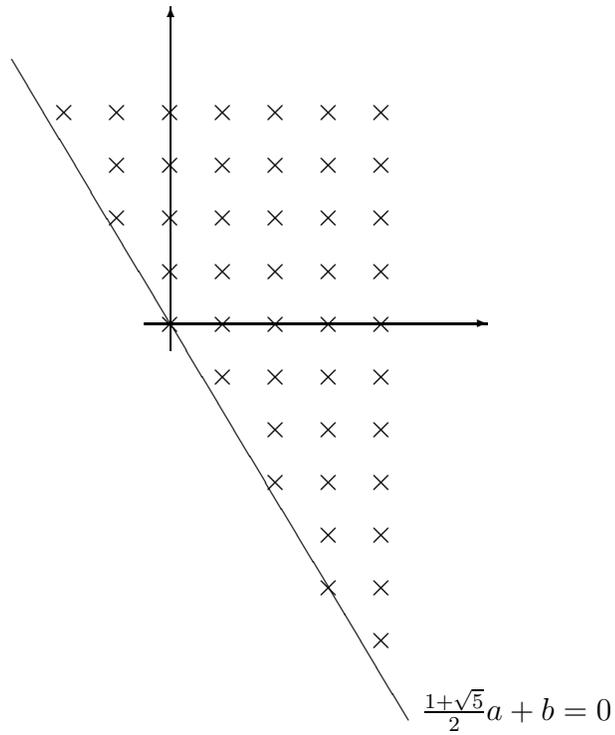\bigskip

We refer to \cite{ELTkappa} for an extensive study of the $K$-theory of
noncommutative lattices and for several examples of $K$-groups.

\vfill\eject
\subsubsect{Higher Order $K$-groups}\label{se:hok}
In order to define higher order groups, one needs to introduce the notion of
{\it suspension} of a $C^*$-algebra $\ca$: it is the $C^*$-algebra
\be
S\ca =: \ca ~\otimes~ C_0(\IR) \simeq C_0(\IR \ra \ca)~, 
\ee
where $C_0$ indicates continuous functions vanishing at infinity. Also, in the
second object, sum and product are pointwise, adjoint is the adjoint in
$\ca$ and the norm is the supremum norm $\norm{f}_{S\ca} = \sup_{x\in\IR}
\norm{f(x)}_\ca$. 

The $K$-group of order $n$ of $\ca$ is defined to be 
\be
K_n(\ca) =: K_0(S^n\ca)~, ~~~n \in \IN~.
\ee
However, the Bott periodicity theorem asserts that all $K$-groups are isomorphic to
either $K_0$ or $K_1$, so that there are really only two such groups. There are
indeed the following isomorphisms \cite{W-O}
\bea
&& K_{2n}(\ca) \simeq K_0(\ca)~, \nonumber \\
&& K_{2n+1}(\ca) \simeq K_1(\ca)~, ~~~ \forall ~n \in \IN~.
\eea

Again, AF algebras show characteristic features. Indeed, for them $K_1$
vanishes identically. 

While $K$-theory provides analogues of topological invariants for algebras, 
cyclic cohomology provides analogues of differential geometric invariants.
$K$-theory and cohomology are connected by the noncommutative Chern character in
a beautiful generalization of the usual (commutative) situation \cite{Co1}.
We regret that all this goes beyond the scope of the present notes.

As mentioned in Section~\ref{se:afa}, $K$-theory has been proved
\cite{El} to be a complete invariant which distinguishes among AF algebras if one
add to the ordered group $(K_0(\ca), K_{0+}(\ca))$ the notion of {\it scale},
the latter being defined for any $C^*$-algebra $\ca$ as
\be
\S\ca =: \{[p]~, p ~~{\rm a~ projector~ in~ \ca } \}~.
\ee
Algebras AF are completely determined, up to isomorphism, by their {\it scaled
ordered} groups, namely by triple $(K_0, K_{0+}, \S)$. The key is the fact that
scale preserving isomorphisms between the ordered groups $(K_0, K_{0+}, \S)$ of
two AF algebras are nothing but $K$-theoretically induced maps (\ref{keyhom}) of
isomorphisms between the AF algebras themselves.

\vfill\eject
\sect{The Spectral Calculus}\label{se:spc}
In this section we shall introduce the machineries of spectral calculus which
is the noncommutative generalization of the usual calculus on a manifold. As we
shall see, a crucial role is played by the Dixmier trace. 

\subsect{Infinitesimals}\label{se:mcp}

Before we proceed to illustrate Connes' theory of infinitesimals, we need few
additional facts about compact operators which we take from \cite{RS,Si} and 
state as propositions. The algebra of compact operators on the Hilbert space $\ch$
will be denoted by $\ck(\ch)$ while $\cb(\ch)$ will be the algebra of bounded
operators.

\bprop 
Let $T$ be a compact operator on $\ch$. Then, its spectrum $\s(T)$ is a
discrete set having no limit points except perhaps $\l=0$. Furthermore, any
nonzero $\l \in \s(T)$ is an eigenvalue of finite multiplicity.
\eprop 

\noindent
Notice that a generic compact operators needs not admit any eigenvalue.
\bprop 
Let $T$ be a self-adjoint compact operator on $\ch$. Then, there is a
complete orthonormal basis, $\{\f_n\}_{n \in \IN}$, for $\ch$ so that $T\f_n=\l_n\f_n$
and
$\l_n \ra 0$ as $n \ra \infty$.
\eprop
\bprop 
Let $T$ be a compact operator on $\ch$. Then, it has an uniformly
convergent (convergent in norm) expansion
\be 
T = \sum_{n \geq 0} \m_n(T) \ket{\j_n} \bra{\f_n}~, 
\ee 
where, $0 \leq \m_{j+1} \leq \m_j$, and $\{\j_n\}_{n \in \IN}, \{\f_n\}_{n \in \IN}$
are (not necessarily complete) orthonormal sets.
\eprop 

\noindent 
In this proposition one writes the polar decomposition $T=U|T|,  ~|T| = \sqrt{T^*T}$.
Then, $\{\m_n(T), \m_n \ra 0~ {\rm as} ~n\ra \infty\}$ are the non vanishing
eigenvalues of the (compact self-adjoint) operator $|T|$ arranged with repeated
multiplicity, $\{\f_n\}$ are the corresponding eigenvectors and $\j_n = U \f_n$. The
eigenvalues $\{\m_n(T)\}$ are called the {\it characteristic values} of $T$. One has
that
$\m_0(T) = \norm{T}$,  the norm of $T$.

Due to condition (\ref{compact}), compact operators are in a sense `small';
they play the r\^ole of {\it infinitesimal}. The size of the infinitesimal $T
\in \comp$ is governed by the rate of decay of the sequence $\{\m_n(T)\}$ as
$n \ra \infty$. 
\bdefi\label{de:infin} 
For any $\a \in \IR^+$, the infinitesimals of order $\a$ are all $T \in
\comp$ such that
\bea 
&&\m_n(T) = O(n^{-\a})~, ~{\rm as}~ n\ra\infty~, \nonumber \\ 
&& i.e.~~\exists ~C < \infty ~:~ \m_n(T) \leq C n^{-\a}~, 
~\forall ~n\geq 1~.
\eea
\edefi

\noindent
Given any two compact operators $T_1$ and $T_2$, there is a submultiplicative
property \cite{Si}
\be
\m_{n+m}(T_1T_2) \leq \m_n(T_1)\m_n(T_2)~,
\ee 
which, in turns, implies that the orders of infinitesimals behave well,
\be 
T_j ~~{\rm ~of ~order}~~ \a_j ~~\Rightarrow~~ T_1T_2 ~~{\rm ~of ~order}~~ 
\leq \a_1+\a_2~.
\ee 
Also, infinitesimals of order $\a$ form a  (not closed) two-sided ideal in
$\cb(\ch)$, since for any $T\in\comp$ and $B\in\cb(\ch)$, one has that \cite{Si},
\bea
\m_n(TB) \leq \norm{B} \m_n(T)~, \nonumber \\
\m_n(BT) \leq \norm{B} \m_n(T)~. \label{idpro}
\eea

\subsect{The Dixmier Trace}\label{se:dix}  
As in ordinary differential calculus one seeks for an `integral' which neglects
all infinitesimals of order $>1$. This is done with the Dixmier trace which is
constructed in such a way that
\begin{enumerate}
\item[1.] Infinitesimals of order $1$ are in the domain of the trace.
\item[2.] Infinitesimals of order higher than $1$ have vanishing trace.
\end{enumerate} The usual trace is not appropriate. Its domain is the two-sided
ideal $\cl^1$ of trace class operators. For any $T\in\cl^1$, the trace, defined
as 
\be tr ~T =: \sum_n \hs{T\x_n}{\x_n}~,
\ee 
is independent of the orthonormal basis $\{\x_n\}_{n \in \IN}$ of $\ch$ and is,
indeed, the sum of eigenvalues of $T$. When the latter is positive and compact, one
has that
\be tr ~T =: \sum_0^\infty \m_n(T)~. \label{ustr}
\ee 
In general, an infinitesimal of order $1$ is not in $\cl^1$, since the only
control on its characteristic values is that $\m_n(T) \leq C {1\over n}$~, for  some
positive constant $C$. Moreover, $\cl^1$ contains infinitesimals of order higher than
$1$. However, for (positive) infinitesimals of order $1$, the usual trace
(\ref{ustr}) is at most logarithmically divergent since
\be
\sum_0^{N-1} \m_n(T) \leq C \ln N~.
\ee 
The Dixmier trace is just a way to extract the coefficient of the logarithmic
divergence. It is somewhat surprising that this coefficient behaves as a trace
\cite{Di2}.

We shall indicate with $\inf1$ the ideal of compact operators which are infinitesimal
of order $1$. If $T\in\inf1$ is positive, one tries to define a positive functional by
taking the limit of the cut-off sums,
\be
\lim_{N\ra\infty} {1 \over \ln N} \sum_0^{N-1} \m_n(T)~. \label{limcutoff}
\ee 
There are two problems with the previous formula: its linearity and its
convergence. For any compact operator $T$, consider the sums,
\be
\s_N(T)=\sum_0^{N-1} \m_n(T)~, ~~\g_N(T)={\s_N(T) \over \ln N}~.
\ee 
They satisfy \cite{Co1},
\bea 
&& \s_N(T_1 + T_2) \leq \s_N(T_1) + \s_N(T_2)~, ~~ \forall~~ T_1, T_2~, 
\nonumber \\ 
&& \s_{2N}(T_1 + T_2) \geq \s_N(T_1) + \s_N(T_2)~, ~~ \forall~~
T_1, T_2 > 0~.
\eea 
In turn, for any two positive operators $T_1$ and $T_2$,
\bea\label{glinear}
\g_N(T_1 + T_2) \leq \g_N(T_1) + \g_N(T_2) \leq 
\g_{2N}(T_1 + T_2)(1+{\ln2 \over \ln N})~.
\eea 
From this, we see that linearity would follow from convergence. In general,
however, the sequence $\{\g_N\}$, although bounded, is not convergent. Notice
that, the eigenvalues $\m_n(T)$ being unitary invariant, so is the sequence
$\{\g_N\}$. Therefore, one gets a unitary invariant positive trace on the
positive part of
$\inf1$ for each linear form $\lim_{\o}$ on the space $\ell^\infty(\IN)$ of
bounded sequences, satisfying
\begin{itemize}
\item[1.] $\lim_{\o}\{\g_N\} \geq 0$, ~~~if $\g_N \geq 0$~.
\item[2.] $\lim_{\o}\{\g_N\} = \lim\{\g_N\}$, ~~~if $\{\g_N\}$ is convergent, with
$\lim$ the usual limit.
\item[3.] $\lim_{\o}\{\g_1,\g_1,\g_2,\g_2,\g_3,\g_3, \cdots \} =
\lim_{\o}\{\g_N\}$.
\item[3'.] $\lim_{\o}\{\g_{2N}\} = \lim_{\o}\{\g_N\}$. ~~~Scale invariance.
\end{itemize} 
Dixmier proved that there exists an infinity of such scale invariant forms
\cite{Di2,Co1}. Associated with any of it there is a trace  
\be 
tr_\o(T) = \lim{_\o} {1 \over \ln N} \sum_0^{N-1} \m_n(T)~, 
~~~ \forall ~T\geq 0~, ~T \in \inf1~.
\ee 
From (\ref{glinear}), it also follows that $tr_\o$ is additive on positive operators,
\be
tr_\o(T_1 + T_2) = tr_\o(T_1) + tr_\o(T_2)~, ~~~ \forall ~T_1, T_2 ~\geq 0~, 
~T_1, T_2 \in \inf1~.
\ee 
This, together with the fact that $\inf1$ is generated by its positive part
(see below), implies that $tr_\o$ extends by linearity to the entire $\inf1$ with 
properties,
\begin{enumerate}
\item[1.] $tr_\o(T) \geq 0$ ~~if $T \geq 0$.
\item[2.] $tr_\o(\l_1 T_1 + \l_2 T_2) = \l_1 tr_\o(T_1) + \l_2 tr_\o(T_2)$.
\item[3.] $tr_\o(BT) = tr_\o(TB)~, ~~\forall~ B \in \cb(\ch)$.
\item[4.] $tr_\o(T) = 0$~, ~if $T$ is of order higher than $1$. 
\end{enumerate} 
Property 3. follows from (\ref{idpro}). The last property follows from the fact that
the space of all infinitesimals of order higher than $1$ form a two-sided ideal whose
elements satisfy
\be
\m_n(T)=o({1 \over n})~, ~~i.e. ~~~n\m_n(T) \ra 0~, ~~{\rm as}~ n\ra\infty~.
\ee 
As a consequence, the corresponding sequence $\{\g_N\}$ is convergent and
converges to zero. Therefore, for such operators the Dixmier trace vanishes.

To prove that $\inf1$ is generated by its positive part one can use polar
decomposition and the fact that $\inf1$ is an ideal. If $T\in \inf1$, by
considering self-adjoint and anti self-adjoint part separately one can suppose
that $T$ is self-adjoint. Then, $T = U |T|$ with $|T| = \sqrt{T^2}$ and $U$ is a
sign operator, $U^2 = U$; from this $|T| = UT$ and $|T| \in \inf1$. 
Furthermore, one has the decomposition $U = U_+ - U_-$ with $U_\pm = 
{1 \over 2}(\II \pm U)$ its spectral projectors (projectors on the eigenspaces with
eigenvalue $+1$ and $-1$ respectively). Therefore, $T = U |T| = U_+|T| - U_-|T| =
U_+|T|U_+ - U_-|T|U_-$ is a difference of two positive elements in $\inf1$.

\bigskip

In many examples of interest in physics, like Yang-Mills and gravity 
theories, the sequence $\{\g_N\}$ itself converges. In these cases, the limit is
given by (\ref{limcutoff}) and does not depends on $\o$. \\
The following examples have been clarified in \cite{VG}.
\bexam\label{dixtor}
Powers of the Laplacian on the $n$-dimensional flat torus $T^n$. \\
The operator   
\be
\D f = - ({\pa ^2 \over {\pa x_1^2 }} + \cdots + {\pa ^2 \over {\pa x_n^2 }})~,
\ee
has eigenvalues $\norm{l_j}^2$ where the $l_j$'s are all points of the lattice $\IZ^n$
taken with multiplicity one. Thus, $|\D|^s$ will have eigenvalues $\norm{l_j}^{2s}$.
For the corresponding Dixmier trace, one needs to estimate $(log N)^{-1}\sum_1^N
\norm{l_j}^{2s}$ as $N \ra \infty$. Let $N_R$ be the number of lattice points in the
ball of radius $R$ centered at the origin of $\IR^n$. Then $N_R \sim vol 
\{x ~|~ \norm{x} \leq R\}$ and $N_{r-dr} - N_r \sim \O_n r^{n-1} dr$. Here 
$\O_n = 2 \p^{n/2} / \G(n/2)$ is the area of the unit sphere $S^{n-1}$. Thus,
\bea
\sum_{\norm{l} \leq R } \norm{l}^{2s} &\sim& \int_1^\infty r^{2s}(N_{r-dr} - N_r)
\nonumber \\
&=& \O_n \int_1^\infty r^{2s + n -1} d r~.
\eea
On the other side, $log N_R \sim n log R$. As $R \ra \infty$, we have to distinguish
three cases.
\begin{enumerate}
\item[~] For $s > -n/2$, 
\be
(log N_R)^{-1}\sum_{\norm{l} \leq R } \norm{l}^{2s} ~\ra~ \infty~.
\ee
\item[~] For $s < -n/2$, 
\be
(log N_R)^{-1}\sum_{\norm{l} \leq R } \norm{l}^{2s} ~\ra~ 0~.
\ee\item[~] For $s = -n/2$, 
\be
(log N_R)^{-1}\sum_{\norm{l} \leq R } \norm{l}^{-n} ~\sim~ 
{\O_n log R \over n log R} = {\O_n \over n}~.
\ee
\end{enumerate}
Therefore, the sequence $\{\g_N(|\D|^{s}) \}$ diverges for $s>-n/2$, vanishes for
$s<-n/2$ and converges for $s=-n/2$. Thus $\D^{-n/2}$ is an infinitesimal of order 
$1$, its trace being given by 
\be
tr_\o(\D^{-n/2}) = {\O_n \over n} = {2 \p^{n/2} \over n \G(n/2)}~.
\ee
\eexam
\bexam\label{dixsph}
Powers of the Laplacian on the $n$-dimensional sphere $S^n$. \\
The Laplacian operator $\D$ on $S^n$ has eigenvalues $l(l + n - 1)$ with multiplicity
\be
m_l = 
\left(
\begin{array}{c}
l+n \nonumber \\ n
\end{array}
\right)
 - 
\left(
\begin{array}{c}
l+n-2 \nonumber \\ n
\end{array}
\right)
= {(l + n - 1)!\over (n-1)! l! }{(2l + n -1) \over (l + n - 1)}~, 
\ee
where $l \in \IN$; in particular $m_0 = 1, m_1 = n +1$. One needs to estimate, as $N
\ra \infty$, the following sums
\be
log \sum_{l=0}^N m_l~, ~~~\sum_{l=0}^N m_l [l(l+n-1)]^{-n/2}~. 
\ee
Well, one finds that
\bea
\sum_{l=0}^N m_l &=& \left(
\begin{array}{c}
N+n \nonumber \\ n
\end{array}
\right)
 + 
\left(
\begin{array}{c}
M+n-1 \nonumber \\ n
\end{array}
\right) \nonumber \\
&=& {1 \over n!} (N+n-1)(N+n-2) \cdots (N+1)(2N+n) \nonumber \\
&\sim& {2 N^n \over n!}~,
\eea
from which,
\be
log \sum_{l=0}^N m_l \sim log N^n + log 2 - log n! \sim n log N~.
\ee
Furthermore,
\bea
\sum_{l=0}^N m_l [l(l+n-1)]^{-n/2} &=& 
{1 \over (n-1)!} \sum_{l=0}^N {(l + n - 1)!\over l! [l(l+n-1)]^{n/2} }
{(2l + n -1) \over (l + n - 1)} \nonumber \\
 &\sim& {2 \over (n-1)!} \sum_{l=0}^N {l^{n-1} \over [l(l+n-1)]^{n/2} }
 \nonumber \\
&\sim& {2 \over (n-1)!} \sum_{l=0}^N {l^{n-1} \over (l+ {n-1 \over 2})^{n} }
 \nonumber \\
&\sim& {2 \over (n-1)!} \sum_{l=0}^N (l+ {n-1 \over 2})^{-1}
 \nonumber \\
&\sim& {2 \over (n-1)! } log N \label{deno}.
\eea
By putting the numerator and the denominator together we finally get,
\bea
tr_\o(\D^{-n/2}) &=& lim_{N\ra\infty} ( \sum_{l=0}^N m_l [l(l+n-1)]^{-n/2} 
/ log\sum_{l=0}^N m_l) \nonumber \\
&=& lim_{N\ra\infty} {2 log N / (n-1)! \over n log N } = {2
\over n!}~.
\eea
If one replaces the exponent $-n/2$ by a smaller $s$, the series in (\ref{deno})
becomes convergent and the Dixmier trace vanishes. On the other end, if $s>\n/2$,
this series diverges faster that the one in the denominator and the corresponding
quotient diverges.
\eexam
\bexam\label{dixho}
The inverse of the harmonic oscillator.\\
The Hamiltonian of the one dimensional harmonic oscillator is given (in `momentum
space') by  $H = {1 \over 2}(\x^2 +x^2)$. It is well known that on the Hilbert
space $L^2(\IR)$ its eigenvalues are  $\m_n(H) = n + {1 \over 2}~, ~n = 0, 1, \dots $,
while its inverse  $H^{-1} = 2 (\x^2 + x^2)^{-1}$  has eigenvalues $\m_n(H^{-1}) = {2
\over 2 n + 1}$~. The sequence $\{\g_N(H^{-1}) \}$ converges and the corresponding
Dixmier trace is given by (\ref{limcutoff}),  
\be
tr_\o(H^{-1}) = \lim_{N\ra\infty} {1 \over \ln N} \sum_0^{N-1} \m_n(H^{-1}) =
\lim_{N\ra\infty} {1 \over \ln N} \sum_0^{N-1} {2 \over 2 n + 1} = 1~.
\ee 
\eexam

\subsect{Wodzicki Residue and Connes' Trace Theorem}\label{se:wrc}

The Wodzicki-Adler-Manin-Guillemin residue is the unique trace on the algebra of
pseudodifferential operators of any order which, on operators of order at most $-n$
coincides with the corresponding Dixmier trace. Pseudodifferential operators are
briefly described in Appendix~\ref{se:pdo}. In this section we shall introduce the
residue and the theorem by Connes \cite{Cot} which establishes its connection with the
Dixmier trace.
\bdefi\label{wodres}
Let $M$ be an $n$-dimensional compact Riemannian manifold. Let $T$ be a
pseudodifferential operator of order $-n$ acting on sections of a complex vector
bundles $E \ra M$. Its residue is defined by
\be\label{wodzicki1}
 Res_W T =: {1 \over n(2\p)^n } \int_{S^*M} tr_{E} ~\s_{-n}(T) d\m~.
\ee
\edefi 

\noindent
Here, $\s_{-n}(T)$ is the principal symbol: a matrix-valued function on
$T^*M$ which is homogeneous of degree $-n$ in the fibre coordinates (see
Appendix~\ref{se:pdo}). The integral is taken over the unit co-sphere 
$S^*M = \{(x, \x) \in T^*M ~:~ \norm{\x}=1 \} \subset T^*M$ with measure $d\m
=d x d \x$. The trace $tr_E$ is a matrix trace over `internal indices'
\fn{It may be worth mentioning that most authors do not include the factor ${1
\over n}$ in the definition of the residue (\ref{wodzicki1}).}. 

\bexam\label{wodtor}
Powers of the Laplacian on the $n$-dimensional flat torus $T^n$. \\
The Laplacian $\D$ is a second order operator. Then, the operator $\D^{-n/2}$ is
of order $-n$ with principal symbol $\s_{-n}(\D^{-n/2}) = \norm{\x}^{-n}$ (see
Appendix~\ref{se:pdo}), which is the constant function $1$ on ${S^*T^n}$. As a
consequence, 
\be
Res_W \D^{-n/2} = {1 \over n(2\p)^n } \int_{S^*T^n} dx d\x = {1 \over n(2\p)^n }
\O_n \int_{S^*T^n} dx = {2 \p^{n/2} \over n \G(n/2)}~.
\ee
The result coincides with the one given by the Dixmier trace in
Example~\ref{dixtor}.
\eexam
\bexam\label{wodsph}
Powers of the Laplacian on the $n$-dimensional sphere $S^n$. \\
Again the operator $\D^{-n/2}$ is of order $-n$ with principal symbol the
constant function $1$ on ${S^*T^n}$. Thus, 
\bea
Res_W \D^{-n/2} &=& {1 \over n(2\p)^n } \int_{S^*S^n} dx d\x = {1 \over n(2\p)^n }
\O_n \int_{S^*T^n} dx = {1 \over n(2\p)^n } \O_n \O_{n+1} \nonumber \\
&=& {2 \p^{n/2} \over n \G(n/2)} = {2 \over n!}~,
\eea
where we have used the formula $\G({n \over 2}) \G({n+1 \over 2}) = 2^{-n+1}
\p^{1/2} (n-1)!$. Again we see that the result coincides with the one in
Example~\ref{dixsph} obtained by taking the Dixmier trace. 
\eexam
\bexam\label{wodho}
The inverse of the one dimensional harmonic oscillator.\\
The Hamiltonian is given by  $H = {1 \over 2}(\x^2 +x^2)$. Let us
forget for the moment the fact that the manifold we are considering, $M = \IR$, is not
compact. We would like to still make sense of the (Wodzicki) residue of a suitable
negative power of $H$. Since $H$ is of order $2$, the first candidate would be
$H^{-1/2}$. From (\ref{pdrussoho}) its principal symbol is the function $\x^{-1}$.
Formula (\ref{wodzicki1}) would give $Res_W H^{-1/2} = \infty$, a manifestation of
the fact that $\IR$ is not compact. On the other side, Example~\ref{dixho} would
suggest to try $H^{-1}$. But from (\ref{pdrussoho}) we see that the symbol of
$H^{-1}$ has no term of order $-1 $ ! It is somewhat surprising that the integral of
the {\it full} symbol of $H^{-1}$ gives an answer which coincides (up to a factor
$2$) with
$tr_\o(H^{-1})$ evaluated in Example~\ref{dixho} \cite{VG},
\be
{\rm Residue}(H^{-1}) = {1 \over 2 \p} \int_{S^* \IR} \s(H^{-1}) = 
{1 \over \p} \int_{\IR} {2 \over { 1 + x^2}} = 2~.
\ee
For an explanation of the previous fact we refer to \cite{EGV}. 
\eexam

As we have already mentioned, Wodzicki \cite{Wod} has extended the formula
(\ref{wodzicki1}) to a unique trace on the algebra of pseudodifferential operator of
any order. The trace of any operator $T$ is given by the right hand side of formula
(\ref{wodzicki1}), with $\s_{-n}(T)$ the symbol of order $-n$ of $T$. In particular,
one puts  $Res_W T = 0$ if the order of $T$ is less than $-n$. As we shall see in
Section~\ref{se:gra} such general residue has been used to construct gravity models
in noncommutative geometry.

In the examples worked before we have seen explicitly that the Dixmier trace of an 
operator of a suitable type coincides with its Wodzicki residue. That the residue
coincides with the Dixmier trace for any pseudodifferential operators of order less
or equal that $-n$ have been shown by Connes \cite{Cot,Co1} (see also \cite{VG}).
\bprop\label{wodzicki} 
Let $M$ be an $n$-dimensional compact Riemannian
manifold. Let $T$ be a pseudodifferential operator of order $-n$ acting on
sections of a complex vector bundles $E \ra M$. \\ 
Then,
\begin{enumerate}
\item[1.] The corresponding operator $T$ on the Hilbert space $\ch=L^2(M,E)$ of
square integrable sections, belongs to $\inf1$.
\item[2.] The trace $tr_\o T$ does not depends on $\o$ and is (proportional to) the
residue,
\be 
tr_\o T = Res_W T =: {1 \over n (2\p)^n } \int_{S^*M} tr_{E} 
~\s_{-n}(T) d\m~.
\ee
\item[3.] The trace depends only on the conformal class of the metric on $M$.
\end{enumerate}
\proof
The Hilbert space on which $T$ acts is just $\ch = L^2(M,E)$, the space of
square-integrable sections obtained as the completion of $\G(M,E)$ with respect to the
scalar product $(u_1,u_2) = \int_M u_1^* u_2 d\m(g)~, ~d\m(g)$ being the measure
associated with the Riemannian metric on $M$. If $\ch_1, \ch_2$ are obtained from two
conformally related metric, the identity operator on $\G(M,E)$ extends to a linear
map $U : \ch_1 \ra \ch_2$ which is bounded with bounded inverse and which
transforms $T$ into $U T U^{-1}$. Since $tr_\o(U T U ^{-1}) = tr_\o(T)$, we get
$\inf1(\ch_1) \simeq \inf1(\ch_2)$ and the Dixmier trace does not changes. On the
other side, the cosphere bundle $S^*M$ is constructed by using a metric. But
since $\s_{-n}(T)$ is homogeneous of degree $-n$ in the fibre variable $\x$, the
multiplicative term obtained by changing variables just compensate the Jacobian of
the transformation and the integral in the definition of the Wodzicki residue
remains the same in each conformal class. \\  
Now, from Appendix~\ref{se:pdo}, we know that $T$ can be written as a finite sum of
operator of the form $u \mapsto \f T \j$, with $\f, \j$ belonging to a partition of
unity of $M$. Since multiplication operators are bounded on the Hilbert space $\ch$,
the operator $T$ will be in $\inf1$ if and only if all operators $\f T \j$ are.
Thus one can assume that $E$ is the trivial bundle and $M$ can be taken to be a given
$n$-dimensional compact manifold, $M = S^n$ for simplicity.
Now, it turn out that the operator $T$ can be written as $T=S(1+\D)^{-n/2}$, with $\D$
the Laplacian and $S$ a bounded operator. 
%
%
From Example~\ref{dixsph}, we know that $(1+\D)^{-n/2} \in \inf1 $, (the presence of
the identity is irrelevant since it produces only terms of lower degree), 
and this implies that $T\in \inf1$. From that example, we also have  that for 
$s < -n/2 $, the Dixmier trace of $(1+\D)^{s}$ vanishes and this implies that any
pseudodifferential operator on $M$ of order $s<-n/2$ has vanishing Dixmier trace. 
In particular, the operator of order $(-n-1)$ whose symbol is $\s(x,\x)
-\s_{-n}(x,\x) $ has vanishing Dixmier trace; as a consequence, the Dixmier trace of
$T$ depends only on the principal symbol of $T$. \\
%
%
%
%
Now, the space of all $tr_E s_{-n}(T)$ can be
identified with $C^\infty(S^*M)$.
Furthermore, the map $tr_E s_{-n}(T) \mapsto
tr_\o(T)$ is a continuous linear form, namely a distribution, on the compact
manifold $S^*M$. This distribution is positive due to the fact that the Dixmier trace
is a positive linear functional and nonnegative principal symbols correspond to
positive operators. Since a positive distribution is a measure $d m$, we can write 
$tr_\o(T) = \int_{S^*M} \s_{-n}(T) dm(x,\x)$. \\
Now, an isometry $\f : S^n \ra S^n$ will transform $\s_{-n}(T)(x,\xi)$ to
$\s_{-n}(T)(\f(x),\f^*\xi)$, $\f^*$ being the transpose of the Jacobian of $\f$, and
determines a unitary operator $U_\f$ on $\ch$ which transform $T$ to $U_\f T
U_\f^{-1}$. Since $tr_\o T = tr_\o(U_\f T U_\f^{-1})$, the measure $dm$ determined
by $tr_\o$ is invariant under all isometries of $S^n$. In particular one can take
$\f \in SO(n+1)$. But $S^* S^n$ is a homogeneous space for the action of
$SO(n+1)$ and any $SO(n+1)$-invariant measure is proportional to the volume form
on $S^* S^n$. Thus
\be
tr_\o T ~\sim~ {1 \over n(2\p)^n } \int_{S^*M} tr_{E}~\s_{-n}(T) dx d\x~ 
= Res_W T ~. 
\ee 
From Examples~\ref{dixsph} and \ref{wodsph} we sees that the proportionally
constant is just $1$. This ends the proof of the proposition.
\eprop 

\noindent 
Finally, we mention that in general there is a class $\cm$ of elements of $\inf1$ for
which the Dixmier trace does not depend on the functional $\o$. Such operators are
called {\it measurable} and in all relevant case in noncommutative geometry one deals
with measurable operators. We refer to \cite{Co1} for a characterization of $\cm$.  We
only mention that in such situations, the Dixmier trace can again be written as a
residue. If $T$ is a positive element in $\inf1$, its complex power $T^s, s \in \IC,
~\IR e ~s >1$, makes sense and is a trace class operator. Its trace
$\z(s)=tr ~T^s =\sum_{n=0}^\infty \m_n(T)^s$~,
is a holomorphic function on the half plane $\IR e ~s >1$. Connes has proved
that for $T$ a positive element in $\inf1$, $\lim_{s\ra 1^+} (s-1) \z(s) = L~$ if
and only if $tr_\o T =: \lim_{N\ra\infty} {1 \over \ln N} \sum_0^{N-1} \m_n(T) = L$.
We see that if $\z(s)$ has a simple pole at $s=1$ then, the corresponding
residue coincides with the Dixmier trace. This equality gives back
Proposition~\ref{wodzicki} for pseudodifferential operators or order at most $-n$ on
a compact manifold of dimension $n$.

\subsect{Spectral Triples}\label{se:spt} 
We shall now illustrate the basic ingredient introduced by
Connes to develop the analogue of differential calculus for
noncommutative algebras.

\bdefi\label{se:triple}
A {\rm spectral triple} $(\ca, \ch, D)$ 
\fn{The couple $(\ch, D)$ is also called a {\it $K$-cycle} over $\ca$.}
is given by an involutive algebra $\ca$ of bounded operators on
the Hilbert space $\ch$, together with a self-adjoint operator $D=D^*$ on $\ch$
with the following properties.
\begin{itemize}
\item[1.] The resolvent $(D-\l)^{-1}, ~~\l \not\in \IR$, is a compact operator
on $\ch$;
\item[2.] $[D, a]=:D a - a D \in \cb(\ch)$, for any $a\in\ca$ 
\end{itemize} 

The triple is said to be {\rm even} if there is a $\IZ_2$ grading
of $\ch$, namely an operator $\G$ on $\ch$, $\G=\G^*, \G^2=1$, such that
\bea 
&& \G D + D \G = 0~, \nonumber \\ 
&& \G a - a \G = 0~, ~~~\forall ~a \in
\ca~.
\eea 
If such a grading does not exist, the triple is said to be {\rm odd}.
\edefi

\noindent
In general, one could ask that condition 2. be satisfied only for a dense
subalgebra of $\ca$.\\ 
By the assumptions in Definition~\ref{se:triple}, the self-adjoint operator $D$ has a
real discrete spectrum made of eigenvalues, i.e. the collection $\{\l_n\}$ form a
discrete subset of $\IR$ and each eigenvalue has finite multiplicity. Furthermore,
$|\l_n| \ra
\infty$ as
$n\ra\infty$. Indeed, $(D-\l)^{-1}$ being compact, its characteristic values
$\m_n((D-\l)^{-1})\ra 0$, from which $|\l_n|=\m_n(|D|)\ra\infty$.

Various degree of regularity of elements of $\ca$ are defined using $D$ and
$|D|$. The reason for the corresponding names will be evident in the next
subsection where we shall consider the canonical triple associated with an
ordinary manifold.
To start with, $a\in\ca$ will be said to be {\it Lipschitz} if and only if the
commutator $[D, a]$ is bounded. As mentioned before in the definition of a
spectral triple, in general this condition selects a dense subalgebra of $\ca$.  
Furthermore, consider the densely defined derivation $\d$ on $\cb(\ch)$ defined
by
\be
\d(T) = [ |D|, T ]~, ~~~T\in\cb(\ch)~.
\ee 
It is the generator of the $1$-parameter group $\a_s$ of automorphism of
$\cb(\ch)$ given by
\be
\a_s(T) = e^{is|D|} T e^{-is|D|}~.
\ee
An element $a\in\ca$ is said to be 
\begin{enumerate}
\item[1.] of class $C^\infty$ if and only if the map $s\ra\a_s(a)$ is $C^\infty$.
\item[2.] of class $C^\o$ if and only if the map $s\ra\a_s(a)$ is $C^\o$.
\end{enumerate} 
Thus, $a\in\ca$ is $C^\infty$ if and only if it belongs to
$\bigcap_{n\in\IN} Dom \d^n$.

As will be evident from next Section, the spectral triples we are considering are
really `Euclidean' ones. There are some attempts to construct
spectral triples with `Minkowskian signature' \cite{Ka,Ha,Ko}. We shall not use
them in these notes.

\subsect{The Canonical Triple over a Manifold}\label{se:ctm} 

The basic example of spectral triple is constructed by means of the Dirac
operator on a closed $n$-dimensional Riemannian spin manifold $(M,g)$.   
As spectral triple $(\ca, \ch, D)$ one takes
\fn{For much of what follows one could consider spin$^c$ manifolds. The obstruction
for a manifold to have a spin$^c$ structure is rather mild and much weaker than the
obstruction to have a spin structure. For instance, any orientable four dimensional
manifold admits such structure \cite{ABS}. Then, one should accordingly modify the
Dirac operator in (\ref{dirac}) by adding a $U(1)$ gauge connection $A=dx^\m A_\m$.
The corresponding Hilbert space $\ch$ has a beautiful interpretation as the space
of square integrable Pauli-Dirac spinors \cite{FGR}.} 
\begin{enumerate}
\item[1.] $\ca=\cf(M)$ is the algebra of complex valued smooth functions on $M$.
\item[2.] $\ch=L^2(M,S)$ is the Hilbert space of square integrable sections of
the irreducible spinor bundle over $M$, its rank being equal to $2^{[n/2]}$ 
\fn{The symbol $[k]$ indicates the integer part in $k$.}.  
The scalar product in $L^2(M,S)$ is the usual
one of the measure associated with the metric $g$,
\be
(\j, \f) = \int d\m(g) \bar{\j(x)} \f(x), 
\ee
with bar indicating complex conjugation and scalar product in the spinor space being
the natural one in $\IC^{2^{[n/2]}}$.
\item[3.] $D$ is the Dirac operator associated with the Levi-Civita connection
$\o=dx^\m \o_\m$ of the metric $g$.
\end{enumerate}

First of all, the elements of the algebra $\ca$ acts as multiplicative operators on
$\ch$, 
\be 
(f \j)(x)=:f(x)\j(x)~, ~~\forall~ f\in\ca~, \j\in\ch~. \label{mul}
\ee
 
Next, let $(e_a, a = 1, \dots, n)$ be an orthonormal basis of vector fields which is
related to the natural basis $(\partial_\m, \m = 1, \dots, n)$ via the 
$n$-beins components $e_a^\m$, so that the components $\{g^{\m\n}\}$ and
$\{\h^{ab}\}$ of the curved and the flat metrics respectively, are related by,
\be
g^{\m\n} = e_a^\m e_b^\n \h^{ab}~, ~~~\h_{ab} = e_a^\m e_b^\n g_{\m\n}~.
\ee 
From now on, the curved indices $\{\m\}$ and the flat ones $\{a\}$ will run from $1$
to $n$ and as usual we sum over repeated indices. Curved indices will be lowered and
raised by the curved metric $g$, while flat indices will be lowered and raised by the
flat metric $\h$. \\
The coefficients $(\o_{\m a}^{~~b})$ of the Levi-Civita (namely metric and
torsion-free) connection of the metric $g$, defined by $\nabla_\m e_a = 
\o_{\m a}^{~~b} e_b$, are the solutions of the equations 
\be
\partial_\m e^a_\n - \partial_\n e^a_\m - \o_{\m b}^{~~a} e^b_\n + 
\o_{\n b}^{~~a}e^b_\m = 0~.
\ee

Also, let $C(M)$ be the Clifford bundle over $M$ whose fiber at $x\in M$ is just
the complexified Clifford algebra $Cliff_{\IC}(T^*_x M)$ and $\G(M, C(M))$ be
the module of corresponding sections. We get an algebra morphism 
\be
\g : \G(M, C(M)) \ra \cb(\ch)~ \label{gamma},
\ee
defined by
\be\label{gamma1}
\g(dx^\m) =: \g^\m(x) = \g^a e_a^\m~, ~~~\m = 1, \dots, n~, 
\ee 
and extended as an algebra map and by $\ca$-linearity. \\ 
The curved and flat gamma matrices $\{\g^\m(x)\}$ and $\{\g^a\}$, which we take to be
Hermitian, obey the relations
\bea\label{gamrel} 
&& \g^\m(x)\g^\n(x) + \g^\n(x)\g^\m(x) = -2g(dx^\m, dx^n) = -2g^{\m\n}~,
~~~\m, \n = 1, \dots, n~;
\nonumber \\ 
&& \g^a\g^b + \g^b\g^a = -2\h^{ab}~, ~~~a, b = 1, \dots, n~.
\eea 

The lift $\nabla^S$ of the Levi-Civita connection to the bundle of spinors is then
\be
\nabla^S_\m = \pa_\m + \o_\m^S = \pa_\m + {1 \over 2} \o_{\m a b}\g^a \g^b~.
\ee
The Dirac operator, defined by  
\be
D = \g \circ \nabla~,
\ee
can be written locally as 
\be 
D = \g(dx^\m)\nabla^S_\m = \g^\m(x)(\pa_\m+\o_\m^S) =  
\g^a e_a^\m(\pa_\m + \o_\m^S)~.
\label{dirac}
\ee 
Finally, we mention the Lichn\'erowicz formula for the square of the Dirac operator
\cite{BGV},
\be\label{lich}
D^2 = \nabla^S + {1 \over 4} R~.
\ee
Here $R$ is the scalar curvature of the metric and $\nabla^S$ is the Laplacian
operator lifted to the bundle of spinors,
\be
\nabla^S = -g^{\m\n}(\nabla^S_\m \nabla^S_\n - \G^\r_{\m\n}\nabla^S_\r)~,
\ee
with $\G^\r_{\m\n}$ the Christoffel symbols of the connection.

If the dimension $n$ of $M$ is even, the previous spectral triple is even by
taking for grading operator just the product of all flat gamma matrices,
\be
\G = \g^{n+1} = i^{n/2}\g^1 \cdots \g^n~,
\ee
which, $n$ being even, anticommutes with the Dirac operator,
\be
\G D + D \G = 0~.
\ee
Furthermore, the factor $i^{n/2}$ ensures that
\be
\G^2 = \II~, ~~~\G^* = \G~.
\ee

\bprop 
Let $(\ca, \ch, D)$ be the canonical triple over the manifold $M$ as
defined above. Then
\begin{enumerate}
\item[1.]  The space $M$ is the structure space of the algebra $\bca$ of
continuous functions on $M$, which is the norm closure of $\ca$.
\item[2.]  The geodesic distance between any two points on $M$ is given by
\be 
d(p,q) = \sup_{f\in \ca} \{ |f(p)-f(q)| ~:~ \norm{[D, f]} \leq 1 \}~,
~~\forall~ p,q \in M~. \label{dis} 
\ee
\item[3.] The Riemannian measure on $M$ is given by
\bea
&& \int_M f = c(n) ~tr_\o (f |D|^{-n})~, ~~\forall~ f \in \ca~, \nonumber \\ ~
&& ~~~~~~~~~c(n) = 2^{(n-[n/2]-1)} \p^{n/2} n\G({n \over 2})~. \label{int}
\eea
\end{enumerate}

\proof Statement 1. is just the Gel'fand-Naimark theorem illustrated in
Section~\ref{se:gnt}.\\
As for Statement 2., from the action (\ref{mul}) of $\ca$ as multiplicative
operators on $\ch$, one finds that
\be 
[D, f]\j = (\g^\m \pa_\m f)\j~, ~~~\forall ~f\in\ca~,
\ee 
and the commutator $[D, f]$ is a multiplicative operator as well,
\be 
[D, f] = (\g^\m \pa_\m f) = \g(df)~, ~~~\forall ~f \in \ca~.
\ee 
As a consequence, its norm is 
\be
\norm{[D, f]} = sup |(\g^\m \pa_\m f)(\g^\n \pa_\n f)^*|^{1/2} =  
 sup |\g^{\m\n} \pa_\m f \pa_\n f^*|^{1/2}~. \label{norcom} 
\ee 
Now, the right-hand side of (\ref{norcom}) coincides with the Lipschitz norm of 
$f$ \cite{Co1}, which is given by
\be
\norm{f}_{Lip} =: \sup_{x\not=y} {|f(x)-f(y)| \over d_\g(x,y)}~,
\ee 
with $d_\g$ the usual geodesic distance on $M$, given by the usual formula,
\be 
d_\g(x,y) = inf_{\g} \{~ {\rm length ~of ~paths} ~\g ~{\rm from} ~x ~{\rm to} ~y~\},
\ee 
Therefore, we have that
\be
\norm{[D, f]} = \sup_{x\not=y} {|f(x)-f(y)| \over d_\g(x,y)}~.
\ee 
Now, the condition $\norm{[D, f]} \leq 1$ in (\ref{dis}), automatically  gives
\be 
d(p,q) \leq d_\g(p,q)~. \label{disdis}
\ee 
To invert the inequality sign, fix the point $q$ and consider the  function
$f_{\g,q}(x) = d_\g(x,q)$. Then  $\norm{[D, f_{\g,q}]} \leq 1$, and in (\ref{dis})
this gives
\be 
d(p,q) \geq |f_{\g,q}(p) - f_{\g,q}(q)| = d_\g(p,q)~,
\ee 
which, together with (\ref{disdis}) proves Statement 2.
As a very simple example, consider $M=\IR$ and $D={d \over dx}$. Then, the 
condition $\norm{[D, f]} \leq 1$ is just $\sup |{df \over dx}| 
\leq 1$ and the sup is saturated by the function $f(x) = x + cost$  which
gives the usual distance.

The proof of Statement 3. starts with the observation that the principal symbol
of the Dirac operator is $\g(\x)$, left multiplication by $\x$, and so $D$ is a
first-order elliptic operator (see Appendix~\ref{se:pdo}).  Since any $f\in\ca$
acts as a bounded multiplicative operator, the  operator $f |D|^{-n}$ is
pseudodifferential of order $-n$. Its  principal symbol is
$\s_{-n}(x,\x)=f(x)\norm{\x}^{-n}$ which on the  co-sphere bundle $\norm{\x} =
1$ reduces to the matrix $f(x) 
\II_{2^{[n/2]}}$, $2^{[n/2]}=dim S_x$, $S_x$ being the fibre of $S$. From the
trace theorem, Prop~\ref{wodzicki}, we get 
\bea 
tr_\o (f |D|^{-n}) &=& {1\over n(2\p)^n} \int_{S^*M} tr(f(x) 
\II_{2^{[n/2]}}) dxd\x = { 2^{[n/2]} \over n(2\p)^n} (\int_{S^{n-1}} d \x)  
\int_{M} f(x) dx \nonumber \\ &=& {1 \over c(n)} \int_M f~.
\eea 
Here, $\int_{S^{n-1}} d \x = 2 \pi^{n/2} / \G(n/2)$ is the area of the unit
sphere $S^{n-1}$. This gives $c(n) = 2^{(n-[n/2]-1)} \p^{n/2} n \G(n/2)$ and 
Statement 3. is proved.
\eprop

It is worth mentioning that the geodesic distance (\ref{dis}) can also be recovered
from the Laplace operator $\nabla_g$ associated with the Riemannian metric $g$ on
$M$ \cite{FG,FGR}. One has that 
\be
d(p,q) = \sup_{f} \{ |f(p)-f(q)| ~:~ 
\norm{f \nabla f - {1 \over 2}(\nabla f^2 + f^2 \nabla)}_{L^2(M)} \leq 1 \}~,
\ee 
$L^2(M)$ being just the Hilbert space of square integrable {\it functions} on $M$.
Indeed, the operator $f \nabla f - {1 \over 2}(\nabla f^2 + f^2 \nabla)$ is just the
multiplicative operator by $g^{\m\n} \pa_\m f \pa_\n f$. Thus, much of the usual
differential geometry can be recovered from the triple
$(C^\infty(M), L^2(M), \nabla_g)$, although it is technically much more involved.

\subsect{Distance and Integral for a Spectral Triple}\label{se:dis}  

Given a general spectral triple $(\ca, \ch, D)$, there is an analogue of formula
(\ref{dis}) which gives a natural distance function on the space $\cs(\bca)$ of
states on the $C^*$-algebra $\bca$, norm closure of $\ca$. A state on $\bca$ is
any linear maps $\f:\ca\ra\IC$ which is positive, i.e. $\f(a^*a)>0$, and
normalized, i.e. $\f(\II)=1$ (see also Appendix~\ref{se:gns}). The distance
function on $\cs(\bca)$ is defined by
\be 
d(\g,\c)=:\sup_{a\in\ca}\{|\f(a)-\c(a)| ~:~ \norm{[D, a]}\leq 1\}~, 
~~~\forall~ \f, \c \in \cs(\bca)~. 
\label{algdis}
\ee

\bigskip
\noindent
In order to define the analogue of the measure integral, one needs the 
additional notion of dimension of a spectral triple.
\bdefi 
A spectral triple $(\ca, \ch, D)$ is said to be of {\rm dimension} $n>0$ (or 
$n$ summable) if $|D|^{-1}$ is an infinitesimal (in the sense of
Definition~\ref{de:infin}) of order
${1
\over n}$ or,  equivalently, $|D|^{-n}$ is an infinitesimal or order $1$.
\edefi 

\noindent
Having such a $n$-dimensional spectral triple, the {\it integral} of  any
$a\in\ca$ is defined by
\be
\int a =: {1 \over V} tr_\o a|D|^{-n}~, \label{algint}
\ee 
where the constant $V$ is determined by the behavior of the  characteristic
values of $|D|^{-n}$, namely, $\m_j\leq V j^{-1}$ for $j \ra \infty$. We see that
the r\^ole of the operator $|D|^{-n}$ is just to bring  the bounded operator $a$
into $\inf1$ so that the Dixmier trace makes  sense.
By construction, the integral in (\ref{algint}) is normalized,
\be
\int \II = {1 \over V} tr_\o |D|^{-n} = {1 \over V}
\lim_{N\ra\infty}\sum_{j=1}^{N-1} \m_j(|D|^{-n}) = 
\lim_{N\ra\infty}\sum_{j=1}^{N-1} {1 \over j} = 1~.
\ee
The operator $|D|^{-n}$ is the analogue of the volume of the
space.\\
In Section~\ref{se:spf} we shall introduce the notion of  {\it tameness}
which will make the integral (\ref{algint}) a non-negative (normalized) trace on
$\ca$, satisfying then the following relations,
\bea
&& \int a b = \int b a~, ~~~\forall ~a,b \in \ca~, \nonumber \\
&& \int a^* a \geq 0~, ~~~\forall ~a \in \ca~.  \label{proint}
\eea 

For the canonical spectral triple over a manifold $M$, its dimension coincides
with the dimension of $M$. Indeed, the Weyl formula for the eigenvalues gives for
large $j$
\cite{Gi},
\be
\m_j(|D|) ~\sim~ 2 \p ({n \over \O_n vol M}) ^{1/n} j^{1/n}~,
\ee 
$n$ being the dimension of $M$. 

\subsect{Real Spectral Triples}
In fact, one needs to introduce an additional notion, the one of {\it real structure}.
The latter is essential to introduce Poincar\'e duality and play a crucial r\^ole in
the derivation of the Lagrangian of the Standard Model \cite{Co2,Co3}. This real
structure it may be thought of as a generalized CPT operator (in fact only CP, since
we are taking Euclidean signature).
\bdefi\label{de:rest} 
Let $(\ca, \ch, D)$ be a spectral triple of dimension $n$. A {\rm real
structure} is an antilinear isometry $J : \ch \ra \ch$, with the properties
\begin{itemize}
\item[1a.] ~$J^2 = \ve(n) \II$~, 
\item[1b.] ~$J D = \ve'(n) D J$~, 
\item[1c.] ~$J\G = (i)^n \G J$~; ~if $n$ is even with $\G$ the $\IZ_2$-grading.
\item[2a.] ~$[a, b^0] = 0$~,  
\item[2b.] ~$[[D,a], b^0] = 0~, ~~b^0 = Jb^*J^*$~, ~~~for any $a,b \in \ca$~.
\end{itemize}
\edefi

\noindent
The mod $8$ periodic functions $\ve(n)$ and $\ve'(n)$ are given by
\cite{Co2} 
\bea\label{spiche}
&& \ve(n) = (1,1,-1,-1,-1,-1,1,1)~, \nonumber \\
&& \ve'(n) = (1,-1,1,1,1,-1,1,1)~,
\eea
$n$  being the dimension of the triple. The previous periodicity is a manifestation
of the so called `spinorial chessboard' \cite{BT}.

A full analysis of the previous conditions goes beyond the scope of these notes. We
only mention that $2a.$ is used by Connes to formulate Poincar\'e duality and to
define noncommutative manifolds. The map $J$ is related to Tomita(-Takesaki)
involution. Tomita theorem states that for any weakly closed $^*$-algebra of operator
$\cm$ on an Hilbert space $\ch$ which admits a cyclic and separating vector
\fn{If $\cm$ is an involutive subalgebra of $\cb(\ch)$, a vector $\x \in \ch$ is
called {\it cyclic} for $\cm$ if $\cm \x$ is dense in $\ch$. It is called
{\it separating} for $\cm$ if for any $T \in \cm$, the fact $T \x = 0$ implies $T=0$.
One finds that a cyclic vector for $\cm$ is separating for the commutant $\cm'$. If
$\cm$ is a von Neumann algebra, the converse is also true, namely a cyclic vector for
$\cm '$ is separating for $\cm$ \cite{Di1}.}, 
there exists a canonical antilinear isometric involution $J:\ch \ra\ch$ which
conjugates $\cm$ to its commutant $\cm' =: \{T \in \cb(\ch) ~|~ T a = a T ~, ~\forall
~a\in\cm\}$, namely $J \cm J^* = \cm'$. As a consequence, $\cm$ is anti-isomorphic
to $\cm'$, the anti-isomorphism being given by the map $\cm \ni a \mapsto J a^*
J^* \in \cm'$. 
The existence of the map $J$ satisfying condition
$2a.$ also turns the Hilbert space $\ch$ into a bimodules over $\ca$, the bimodules
structure being given by 
\be\label{bist}
a ~\x~ b =: a J b^* J^*~\x ~, ~~~\forall ~a,b \in \ca~.
\ee   
As for condition $2b.$, for the time being, it may be thought of to state that $D$ is
a `generalized differential operator' of order $1$. As we shall see, it will play a
crucial role in the spectral geometry described in Section~\ref{se:bsa}. It is worth
stressing that, since $a$ and $b^0$ commutes by condition $2a.$, condition $2b.$ is
symmetric, namely it is equivalent to the condition $[[D, b^0], a] = 0$, for any $a,b
\in \ca$.\\ 
If $a \in \ca$ acts on $\ch$ as a {\it left} multiplication operator, then 
$J a^* J^*$ is the corresponding {\it right} multiplication operator. 
For commutative algebras, these actions can be identified and one simply writes 
$a = J a^* J^*$. Then, condition $2b.$ reads $[[D,a], b] = 0$, for any $a,b \in \ca$,
which is just the statement that $D$ is a differential operator of order $1$.

The canonical triple associated with any (Riemannian spin) manifold has a canonical
real structure in the sense of Definition~\ref{de:rest}, the antilinear isometry $J$
being given by 
\be
J \j=: C \bar{\j}~, ~~~\forall ~\j \in \ch~,
\ee
where $C$ is the charge conjugation operator and bar indicates complex
conjugation \cite{BT}. One verifies that all defining properties of $J$ hold true.

\subsect{A Two Points Space}\label{se:tps} 
Consider a space made of two points $Y =\{1, 2\}$. The algebra $\ca$ of
continuous functions is the direct sum $\ca = \IC \oplus \IC$ and any element
$f\in\ca$ is a couple of complex numbers $(f_1, f_2)$, with $f_i = f(i)$ the
value of $f$ at the point $i$. A $0$-dimensional even spectral triple 
$(\ca, \ch, D, \G)$ is constructed as follows. The finite dimensional Hilbert
space
$\ch$ is a direct sum $\ch = \ch_1 \oplus \ch_2$ and elements of $\ca$ act as
diagonal matrices 
\be
\ca \ni f \mapsto 
\left[
\begin{array}{cc}
f_1 \II_{dim H_1} & 0 \\
0 & f_2 \II_{dim H_2}
\end{array}
\right] \in \cb(\ch)~.
\ee
We shall identify any element of $\ca$ with its matrix representation.\\  
The operator $D$ can be taken as a $2\times 2$ off-diagonal matrix, since any
diagonal element would drop from commutators with elements of $\ca$,
\be\label{tpkoper}
D =
\left[
\begin{array}{cc}
0 & M^* \\
M & 0
\end{array}
\right]~, ~~ M \in Lin(\ch_1,\ch_2)~.  
\ee 
Finally, the grading operator $\G$ is given by
\be\label{tpgrad}
\G = 
\left[
\begin{array}{cc}
\II_{dim H_1} & 0 \\
0 & - \II_{dim H_2}
\end{array}
\right]~.
\ee
With $f\in\ca$, one finds for the
commutator
\be
[D, f] = (f_2-f_1)
\left[
\begin{array}{cc}
0 & M^* \\
-M & 0
\end{array}
\right]~,
\ee
and, in turn, for its norm, $\norm{[D, f]} = |f_2-f_1| \l$ with $\l$ the largest
eigenvalue of the matrix $|M| = \sqrt{MM^*}$. Therefore, the noncommutative
distance between the two points of the space is found to be
\be
d(1,2) = sup\{|f_2-f_1| ~:~ \norm{[D, f]} ~\leq 1 \} = {1\over\l}~. 
\ee 
For the previous triple the Dixmier trace is just (a multiple of the) usual
matrix trace. \\
A real structure $J$ can be given as
\be
J \left(
\begin{array}{c}
\x \\
\bar{\h}
\end{array}
\right) = 
\left(
\begin{array}{c}
\h \\
\bar{\x}
\end{array}
\right)~, ~~~\forall ~~(\xi, \h) \in \ch_1 \oplus \ch_2~. 
\ee
One checks that $J^2  = \II, ~\G J + J \G = 0, ~D  J  - J  D  = 0$ and that all 
other requirements in the Definition~\ref{de:rest} are satisfied.

\subsect{Products and Equivalence of Spectral Triples}\label{se:pet}
We shall briefly mention two additional concepts which are useful in general and
in the description of the Standard Model, namely product and equivalence of triples.

Suppose we are given two spectral triples $(\ca_1, \ch_1, D_1, \G_1)$ and 
$(\ca_2, \ch_2, D_2)$ the first one taken to be even with $\IZ_2$-grading $\G_1$ on
$\ch_1$. The product triple is the triple $(\ca, \ch, D)$ given by
\bea
&& \ca = \ca_1 \otc \ca_2~, \nonumber \\
&& \ch = \ch_1 \otc \ch_2~, \nonumber \\
&& D = D_1 \otc \II + \G_1 \otc D_2~. 
\eea
From the definition of $D$ and the fact that $D_1$ anticommutes with $\G_1$ it
follows that
\bea
D^2 &=& {1 \over 2} \{D, D\} \nonumber \\
&=&  (D_1)^2 \otc \II + (\G_1)^2 \otc (D_2)^2 + 
          {1 \over 2} \{D_1, \G_1\} \otc D_2 \nonumber \\
&=&  (D_1)^2 \otc \II + \II \otc (D_2)^2 ~. 
\eea 
Thus, the dimensions sum up, namely, if $D_j$ is of dimension $n_j$, that is
$|D_j|^{-1}$ is an infinitesimal of order $1 / n_j, ~j = 1, 2$, then $D$ is of
dimension $n_1 + n_2$, that is $|D|^{-1}$ is an infinitesimal of order 
$1 / (n_1 + n_2)$. Furthermore, once the limiting procedure $Lim_\o$ is fixed, one
has also that \cite{Co1}, 
\be
{\G(n/2 + 1) \over {\G(n_1/2 + 1) \G(n_2/2 + 1)}} tr_\o(T_1 \otimes T_2 |D|^n) =
tr_\o(T_1 |D|^{n_1}) tr_\o(T_2 |D|^{n_2})~, 
\ee
for any $T_j \in \cb(\ch_j)$. For the particular case in which one of the triple, say
the second one, is zero dimensional so that the Dixmier trace is ordinary trace, the
corresponding formula reads
\be
tr_\o(T_1 \otimes T_2 |D|^n) = tr_\o(T_1 |D|^{n_1}) tr (T_2)~. 
\ee

The notion of equivalence of triples is the expected one. Suppose we are given two
spectral triples $(\ca_1, \ch_1, D_1)$ and  $(\ca_2, \ch_2, D_2)$, with the
associated representations $\p_j : \ca_j \ra \cb(\ch_j)~, ~j = 1,2. $ Then, the
triples are said to be equivalent if there exists a unitary operator $U : \ch_1 \ra
\ch_2$ such that $U \p_1(a) U^* = \p_2(a)$ for any $a \in \ca_1$, and $U D_1 U^* =
D_2$. If the two triples are even with grading operators $\G_1$ and $\G_2$
respectively, one requires also that $U \G_1 U^* = \G_2$. And if the two triples are
real  with real structure  $J_1$ and $J_2$ respectively, one requires also that $U
J_1 U^* = J_2$. 

\vfill\eject
\sect{Noncommutative Differential Forms}\label{se:ndf}
We shall now describe how to construct a differential algebra of forms out of a
spectral triple $(\ca, \ch, D)$. It turns out it is useful  to first introduce a
universal graded differential algebra which is associated with any  algebra $\ca$.

\subsect{Universal Differential Forms}\label{se:udf}
Let $\ca$ be an associative algebra with unit (for simplicity) over the field of
numbers $\IC$ (say).  The {\it universal differential algebra of forms} 
$\omca=\bigoplus_p\oca{p}$ is a graded algebra defined as follows. In  degree $0$ it
is equal to $\ca$, $\oca{0} = \ca$.  The space $\oca{1}$ of {\it one-forms} is
generated, as a left $\ca$-module, by symbols of degree $\d a$, $a\in\ca$, with
relations
\bea
&&\d(ab) = (\d a)b + a \d b~,~~~ \forall ~a,b \in \ca~. \label{leib} \\
&&\d(\a a + \b b) = \a \d a + \b \d b~,~~~ \forall ~a,b \in \ca~, ~~\a , \b \in
\IC~. \label{leibbis} 
\eea
Notice that relation (\ref{leib}) automatically gives $\d 1 = 0$, 
which in turn implies that $\d \IC = 0$. \\
A generic element $\o \in \oca{1}$ is a finite sum of the form
\be
\o = \sum_i a_i \d b_i~, ~~~a_i, b_i \in \ca~. 
\ee
The left $\ca$-module $\oca{1}$ can be endowed also with a structure of right
$\ca$-module by 
\be
(\sum_i a_i \d b_i) c =: \sum_i a_i (\d b_i) c = \sum_i a_i \d (b_i c) -
\sum_i a_i b_i \d c~, 
\ee
where, in the second equality we have used (\ref{leib}). The relation (\ref{leib})
is just the Leibniz rule for the map 
\be
\d : \ca \ra \oca{1}~,
\ee
which can therefore be considered as a derivation of $\ca$ with values into the
bimodule 
$\oca{1}$. The pair $(\d, \oca{1})$ is characterized by the following universal
property \cite{Bo, CE},
\bprop
Let $\cm$ be any $\ca$-bimodule and $\D : \ca \ra \cm$ any derivation, namely any map
which satisfies the rule (\ref{leib}). Then, there exists a unique bimodule
morphism $\r_\D : \oca{1} \ra \cm$ such that $\D = \r_\D \circ \d$,
\be
\begin{array}{rccc}
id : & \oca{1} & \longleftrightarrow & \oca{1} \\
~ & ~ & ~ & ~ \\
~ & \d \up & ~ & \dn \r_\D \\
~ & ~ & ~ & ~ \\
\D : & \ca & \lra & \cm \\
\end{array}~~, ~~~~~ \r_\D \circ \d = \D~.
\ee 

\proof
Notice, first of all, that for any bimodule morphism $\r : \oca{1} \ra \cm$ the
composition $\r \circ \d$ is a derivation with values in $\cm$.  
Conversely, let $\D : \ca \ra \cm$ be a derivation; then, if there exists a bimodule
morphism $\r_\D : \oca{1} \ra \cm$ such that $\D = \r_\D \circ \d$, it is unique.
Indeed, the definition of $\d$ gives
\be
\r_\D(\d a) = \D(a)~, ~~\forall ~a \in \ca~,
\ee
and the uniqueness follows from the fact that the image of $\d$ generates $\oca{1}$
as a left $\ca$-module, if one extends the previous map by
\be
\r_\D(\sum_i a_i \d b_i) = \sum_i a_i \D b_i~, ~~\forall ~a_i, b_i \in \ca~.
\label{biom}
\ee
It remains to prove that $\r_\D$ as defined in (\ref{biom}) is a bimodule morphism.
Now, with $a_i, b_i, f, g \in \ca$, by using the fact that both $\d$ and $\D$ are
derivations, one has that
\bea
\r_\D( f (\sum_i a_i \d b_i) g ) &=& \r_\D(\sum_i f a_i (\d b_i) g ) \nonumber \\
&=& \r_\D(\sum_i f a_i \d (b_i g) - \sum_i f a_i b_i \d g) \nonumber \\
&=& \sum_i f a_i \D (b_i g) - \sum_i f a_i b_i \D g \nonumber \\
&=& \sum_i f a_i (\D b_i) g \nonumber \\
&=& f (\sum_i f a_i \D b_i) g \nonumber \\
&=& f (\sum_i a_i \D b_i) g~;
\eea 
this ends the proof of the proposition.
\eprop

Let us go back to universal forms. The space $\oca{p}$ is defined as 
\be
\oca{p} = \underbrace{\oca{1} \oca{1} \cdots \oca{1} \oca{1}}_{p-times}~,
\ee
with the product of any two one-forms defined by `justapposition',
\bea
(a_0 \d a_1) (b_0 \d b_1) &=:& a_0 (\d a_1) b_0 \d b_1 \nonumber \\
&=& a_0 \d (a_1 b_0) \d b_1 -  a_0 a_1 \d b_0 \d b_1~.
\eea
Again we have used the rule (\ref{leib}). Therefore, elements of 
$\oca{p}$ are finite linear combinations of monomials of the form
\be
\o = a_0 \d a_1  \d a_2 \cdots  \d a_p~, ~~a_k\in\ca~.
\ee
The product $: \oca{p} \times \oca{q} \ra \oca{p+q}$ of any $p$-form with any $q$-form
produces a $p+q$ form and is again defined by `justapposition'  and rearranging the
result by using the relation (\ref{leib}),
\bea
(a_0 \d a_1 \cdots \d a_p) (a_{p+1} \d a_{p+2} \cdots \d a_{p+q}) &=:&
a_0 \d a_1 \cdots (\d a_p)a_{p+1}\d a_{p+2} \cdots \d a_{p+q}\nonumber \\
&=& (-1)^p a_0 a_1 \d a_2 \cdots \d a_{p+q} \nonumber \\
&+& \sum_{i=1}^p (-1)^{p-i} a_0 \d a_1 \cdots \d a_{i-1} 
\d(a_i a_{i+1}) \d a_{i+2} \cdots \d a_{p+q}~. \nonumber \\
&& ~ \label{uprod}
\eea
The algebra $\omca$ is clearly a left $\ca$-module. It is also a right $\ca$-module,
the right structures being given by
\bea
(a_0 \d a_1 \cdots \d a_p) b &=:& a_0 \d a_1 \cdots (\d a_p) b \nonumber \\
&=& (-1)^p a_0 a_1 \d a_2 \cdots \d a_p \d b \nonumber \\
&& + \sum_{i=1}^{p-1} (-1)^{p-i} a_0 \d a_1 \cdots \d a_{i-1} \d (a_i a_{i+1}) 
\d a_{i+2} \cdots \d a_p \d b~ \nonumber \\
&& + a_0 \d a_1 \cdots \d a_{p-1} \d (a_p b)~, ~~~ \forall ~ a_i, b \in \ca~.
\eea
Next, one makes the algebra $\omca$ a differential one by `extending' the {\it
differential} $\d$ to an operator 
$:\oca{p} \ra\oca{p+1}$ as a linear operator, unambiguously by 
\be
\d (a_0 \d a_1 \cdots \d a_p) =: \d a_0 \d a_1 \cdots \d a_p~. \label{uniexdi}
\ee
It is then easily seen to satisfy the basic relations
\bea
&& \d^2 = 0~, \\
&& \d (\o_1 \o_2) = \d (\o_1) \o_2 + (-1)^{p} \o_1 \d \o_2~, 
~~\o_1 \in \oca{p}~, ~\o_2 \in \omca~.
\eea

Notice that there is nothing like graded commutativity of forms, namely nothing of the
form $\o_{(p)} \o_{(q)} = (-1)^{pq} \o_{(q)} \o_{(p)}$, with $\o_{(i)} \in \oca{i}$.
  
The graded differential algebra $(\omca, \d) $ is characterized by the following
universal property \cite{Co0,Kar1},
\bprop
Let $(\G, \D)$ be a graded differential algebra, $\G = \oplus_p \G^p$, and 
let $\r : \ca \ra \G^0$ be a morphism of unital algebras. Then, there exists a unique
extension of $\r$ to a morphism of graded differential algebras $\wt{\r} : \omca \ra
\G$,
\be\label{unpr1}
\begin{array}{rccc}
\wt{\r} : & \oca{p} & \lra & \G^p \\
~ & ~ & ~ & ~ \\
~ & \d \dn & ~ & \dn \D \\
~ & ~ & ~ & ~ \\
\wt{\r} : & \oca{p+1} & \lra & \G^{p+1} \\
\end{array}~~, ~~~~~ \wt{\r} \circ \d = \D \circ \wt{\r}~.
\ee 

\proof  Given the morphism $\r : \ca \ra \G^0$, one defines $\wt{\r} : \oca{p} \ra
\G^p$ by
\be 
\wt{\r}((a_0 \d a_1 \cdots \d a_p)) =: \r(a_0) \D (\r(a_1)) \cdots \D (\r(a_p))~.
\ee
This map is uniquely defined by $\r$ since $\oca{p}$ is spanned as a left
$\ca$-module by the monomials $a_0 \d a_1 \cdots \d a_p$. Next, identity
(\ref{uprod}) and its counterpart for the elements $\r(a_i)$ and the derivation
$\D$ ensures that products are send into products. Finally, by using (\ref{uniexdi})
and the fact that $\D$ is a derivation, one has
\bea
(\wt{\r} \circ \d) (a_0 \d a_1 \cdots \d a_p) 
&=& \wt{\r} (\d a_0 \d a_1 \cdots \d a_p) \nonumber \\ 
&=& \D \r(a_0) \D (\r(a_1))\cdots \D (\r(a_p)) \nonumber \\
&=& \D ( (\r(a_0)) \D (\r(a_1)) \cdots \D (\r(a_p)) \nonumber \\ 
&=& (\D \circ \wt{\r}) (a_0 \d a_1 \cdots \d a_p)~,
\eea
which proves the commutativity of diagram (\ref{unpr1}), $\wt{\r} \circ \d = \D
\circ \wt{\r}$.
\eprop


The universal algebra $\omca$ is not very interesting from the cohomological
point of view. From the very definition of $\d$ in (\ref{uniexdi}), it follows
that all cohomology spaces $H^p(\omca) =: Ker ( \d : \oca{p} \ra \oca{p+1} ) / Im (\d
: \oca{p-1} \ra \oca{p})$ vanish, but in degree zero where $H^0(\omca)=\IC$. \\

We shall now construct explicitly the algebra $\omca$ in terms of tensor products.
Firstly, consider the submodule of $\ca \otc \ca$ given by 
\be
ker(m:\ca \otc \ca \ra \ca)~, ~~ m(a\otc b)=ab~.
\ee
This submodule is generated by elements of the form $1 \otc a - a \otc 1$ with $a
\in
\ca$. Indeed, if $\sum a_i b_i = m(\sum a_i \otc b_i) = 0$, then 
$\sum a_i \otc b_i = \sum a_i (1 \otc b_i - b_i \otc 1)$. Furthermore,
the map $\D : \ca \ra   ker(m:\ca \otc \ca \ra \ca)$ defined by 
$\D a =: 1 \otc a - a \otc 1$, satisfies the analogue
of (\ref{leib}), $\D (ab) = (\D a) b + a \D b$. There is an isomorphism of
bimodules
\bea 
\oca{1} \simeq ker(m:\ca\otc\ca\ra\ca)~,
&& \d a \leftrightarrow 1 \otc a - a \otc 1~, \nonumber \\ ~{\rm or}~ 
&& \sum a_i \d b_i \leftrightarrow \sum a_i(1 \otc b_i - b_i \otc 1)~. 
\eea
By identifying $\oca{1}$ with the space $ker(m:\ca \otc \ca \ra \ca)$ the
differential is given by
\be 
\d : \ca \ra \oca 1 ~, ~~\d a = 1 \otc a - a \otc 1~.
\ee
As for forms of higher degree, one has then,
\bea
&& \oca{p} \simeq \underbrace{\oca{1} \ota \cdots \ota \oca{1}}_{p-times}
~\subset~ \underbrace{\oca{1} \otc \cdots \otc \oca{1}}_{(p+1)-times}~, \nonumber \\
&& ~~~~~a_0 \d a_1  \d a_2 \cdots  \d a_p \mapsto a_0 (1 \otc a_1 - a_1 \otc 1)
\ota \cdots \ota (1 \otc a_p - a_p \otc 1)~, \nonumber \\
&& ~~~~~a_k\in\ca~. 
\eea
The multiplication and the bimodule structures are given by,
\bea
&& (\o_1 \ota \cdots \ota \o_p ) \cdot (\o_{p+1} \ota \cdots
\ota \o_{p+q}) =: \o_1 \ota \cdots \ota \o_{p+q}~,
\nonumber \\  
&& a \cdot (\o_1 \ota \cdots \ota \o_p ) =: 
(a \o_1) \ota \cdots \ota \o_p~, \nonumber \\ 
&& (\o_1 \ota \cdots \ota  \o_p) \cdot a =: 
\o_1 \ota \cdots \ota (\o_p a)~, ~~~~~\forall ~\o_j \in \oca{1}~, a \in \ca~.
\label{invol}
\eea
The realization of the differential $\d$ is also easily found. Firstly, consider any
one-form $\o=\sum a_i \otc b_i = \sum a_i (1 \otc b_i - b_i \otc 1)$ (since 
$\sum a_i b_i = 0)$. Its differential $\d \o \in \oca{1} \ota \oca{1}$ is given by
\bea
\d \o &=:& \sum (1 \otc a_i - a_i \otc 1) \ota (1 \otc b_i - b_i \otc 1)
\nonumber \\
&=& \sum 1 \otc a_i \otc b_i - a_i \otc1 \otc b_i - 
a_i \otc b_i \otc 1~.
\eea
Then $\d$ is extended by using Leibniz rule with respect to the product $\ota$,
\be
\d (\o_1 \ota \cdots \ota \o_p) =: \sum_{i=1}^p (-1)^{i+1} \o_1 \ota \cdots \ota 
\d \o_i \ota \cdots \ota \o_p ~, ~~~\forall ~\o_j \in \oca{1}~.
\ee
Notice that even if the algebra is commutative $f h$ and $h f$ are different with no
relations among them (there is nothing like graded commutativity).

Finally, we mention that if $\ca$ has an involution $^*$, the algebra $\omca$ is also
made an involutive algebra by defining 
\bea
(\d a)^* &=:& - \d a^* ~, ~~\forall~ a\in \ca \label{invol0} \\ 
(a_0 \d a_1 \cdots \d a_p)^* &=:& (\d a_p)^* \cdots (\d a_1)^* a_0^* \nonumber \\
&=& a^*_p \d a^*_{p-1} \cdots \d a^*_0 + \sum_{i=0}^{p-1} (-1)^{p+i}\d a^*_p
\cdots \d (a^*_{i+1}a^*_i) \cdots 
\d a^*_0~.
\eea

\subsubsect{The Universal Algebra of Ordinary Functions}\label{se:uof}

Take $\ca = \cf(M)$, with $\cf(M)$ the algebra of complex valued, continuous
functions on a  topological space $M$, or of smooth functions on a manifold $M$
(or some other algebra of functions). Then, identify (a suitable completion of) 
$\ca \otc \cdots \otc \ca$ with
$\cf (M \times \cdots \times M)$. If $f \in \ca$, then
\be\label{uoext0}
\d f (x_1, x_2) =: (1 \otc f - f \otc 1)(x_1, x_2) = f(x_2) - f(x_1)~. 
\ee
Therefore, $\oca{1}$ can be identified with the space of functions of two
variables vanishing on the diagonal. In turn, $\oca{p}$ is identified with the
set of functions $f$ of $p+1$ variables vanishing on contiguous diagonals: 
$f(x_1, \cdots, x_{k-1}, x, x, x_{k+2}, \cdots, x_{p+1}) = 0$. 
The differential is given by,
\be\label{uoext}
\d f (x_1, \cdots  x_{p+1}) =: \sum_{k=1}^{p+1} (-1)^{k-1}
f(x_1, \cdots, x_{k-1}, x_{k+1}, \cdots, x_{p+1})~. 
\ee
The $\ca$-bimodule structure is given by
\bea\label{uomod}
&& (g f)(x_1, \cdots  x_{p+1}) =: g(x_1) f (x_1, \cdots  x_{p+1})~, 
\nonumber \\  
&& (f g)(x_1, \cdots  x_{p+1}) =: f (x_1, \cdots  x_{p+1}) g(x_{p+1}) ~,
\eea 
and extends to the product of a $p$-form with a $q$-form as follows,
\be\label{uopro}
(f h)(x_1, \cdots  x_{p+q}) =: f (x_1, \cdots  x_{p+1})  
h(x_{p+1}, \cdots x_{p+q} ) ~,
\ee
Finally, the involution is simply given by
\be\label{uoinv}
f^* (x_1, \cdots  x_{p+1}) = (f (x_1, \cdots  x_{p+1}))^*~. 
\ee

\subsect{Connes' Differential Forms}\label{se:cdf}

Given a spectral triple $(\ca, \ch, D)$, one constructs an exterior algebras of
forms by means of a suitable representation of the universal algebra $\omca$ in
the algebra of bounded operators on $\ch$. The map
\bea
&&\p : \omca \lra \cb(\ch)~, \nonumber \\
&&\p(a_0 \d a_1 \cdots \d a_p) =: a_0 [D, a_1] \cdots [D, a_p]~, ~~~a_j \in \ca~,
\label{pi}
\eea
is clearly a homomorphism since both $\d$ and $[D, \cdot]$ are derivations on
$\ca$. Furthermore, since $[D, a]^* = -[D, a^*]$, one gets $\p(\o)^* =
\p(\o^*)$ for any form $\o \in \omca$ and $\p$ is a $^*$-homomorphism.  

One could think of defining forms as the image $\p(\omca)$. This is not
possible, since in general, $\p({\o}) = 0$ does not imply that$\p(\d\o)=0$. 
Such unpleasant forms $\o$ for which $\p(\o)=0$ while $\p(\d \o) \not= 0$ are called
{\it junk forms}. They have to be disposed of in order to construct a true
differential algebra and make $\p$ into a homomorphism of differential algebras.  

\bprop
Let $J_0 =: \oplus_p J^p_0$ be the graded two-sided ideal of $\omca$ given by
\be
J_0^p =: \{\o \in \oca{p}, ~\p(\o)=0~\}~.
\ee
Then, $J=J_0+\d J_0$ is a graded differential two-sided ideal of $\omca$.

\proof 
It is enough to show that $J$ is a two-sided ideal, the property $\d^2=0$
implying that it is differential. Take $\o=\o_1+\d \o_2 \in J^p$, with 
$\o_1 \in J^p~, ~\o_2 \in J^{p-1}$. If $\h \in \oca{q}$, then 
$\o \h = \o_1 \h + (\d \o_2) \h = \o_1 \h + \d (\o_2 \h) - (-1)^{p-1}
\o_2 \d \h = (\o_1 \h - (-1)^{p-1} \o_2 \d \h) + \d (\o_2 \h) \in J^{p+q}$.
Analogously, one finds that $\h \o \in J^{p+q}$.
\eprop

\bdefi
The graded differential algebra of Connes' forms over the algebra $\ca$ is
defined by 
\be
\omcad =: \omca / J \simeq \p(\omca) / \p(\d J_0)~. \label{forms}
\ee
\edefi

\noindent
It is naturally graded by the degrees of $\omca$ and $J$, the space of
$p$-forms being given by
\be
\ocad{p}=\oca{p} / J^p~. 
\label{pforms}     
\ee
Being $J$ a differential ideal, the exterior differential $\d$ defines a
differential on $\omcad$, 
\bea
&& d : \ocad{p} \lra \ocad{p+1}~, \nonumber \\
&& d [\o] =: [\d \o]~, \label{extdif}
\eea
with $\o \in \oca{p}$ and $[\o]$ the corresponding class in $\ocad{p}$.

\noindent
Let us see more explicitly the structure of the forms.
\begin{itemize}
\item[$\bullet$] $0$-forms. \\
Since we take $\ca$ to be a subalgebra of $\cb(\ch)$, we have that $J \cap \oca{0}
= J_0 \cap \ca = \{0\}$. Thus $\ocad{0}\simeq \ca$.
\item[$\bullet$] $1$-forms. \\
We have $J \cap \oca{1} = J_0 \cap \oca{1} + J_0 \cap \oca{0} = J_0 \cap
\oca{1}$. Thus, $\ocad{1}\simeq \p(\oca{1})$ and this space coincides with the
$\ca$-bimodule of bounded operators on $\ch$ of the form
\be
\o_1 = \sum_j a_0^j [D, a_1^j]~, ~~a^j_i \in \ca~.
\ee  
\item[$\bullet$] $2$-forms. \\
We have $J \cap \oca{2} = J_0 \cap \oca{2} + J_0 \cap\oca{1}$. Thus, $\ocad{2}
\simeq \p(\oca{2}) / \p(\d (j_0 \cap \oca{1}))$. Therefore, the $\ca$-bimodule
$\ocad{2}$ of $2$-forms is made of classes of elements of the kind
\be
\o_2 = \sum_j a_0^j [D, a_1^j] [D, a_2^j]~, ~~a^j_i \in \ca~,
\ee  
modulo the sub-bimodule of operators
\be
\{~~ \sum_j [D, b_0^j] [D, b_1^j] ~:~b^j_i \in \ca~, ~\sum_j b_0^j [D, b_1^j] = 0
~~\}~.
\ee
\item[$\bullet$] $p$-forms. \\
In general, the $\ca$-bimodule of $\ocad{p}$ of $p$-forms is given by
\be
\ocad{p} \simeq \p(\oca{p}) / \p(\d (j_0 \cap \oca{p-1}))~,
\ee
and is made of classes of operators of the form 
\be
\o_p = \sum_j a_0^j [D, a_1^j] [D, a_2^j] \cdots [D, a_p^j]~, ~~a^j_i \in \ca~,
\ee  
modulo the sub-bimodule of operators
\be
\{~~ \sum_j [D, b_0^j] [D, b_1^j] \cdots [D, b_{p-1}^j] ~:~b^j_i \in \ca~, 
~\sum_j b_0^j [D, b_1^j] \cdots [D, b_{p-1}^j] = 0 ~~\}~.
\ee
\end{itemize}
As for the exterior differential (\ref{extdif}) it is given by
\be
d \left[ \sum_j a_0^j [D, a_1^j] [D, a_p^j] \cdots [D, a_p^j]\right] = 
\left[ \sum_j [D, a_0^j [D, a_1^j] [D, a_2^j] \cdots [D, a_p^j] \right]~.
\ee

\subsubsect{The Usual Exterior Algebra}\label{se:uea}
The methods of previous Section, when applied to the canonical triple over an ordinary
manifold, reproduce the usual exterior algebra over the manifold. Consider the
canonical triple $(\ca, \ch, D)$ on a closed $n$-dimensional Riemannian spin$^c$
manifold
$M$ as described in Section~\ref{se:ctm}. We recall that $\ca = \cf(M)$ is the
algebra of smooth functions on $M$; $\ch=L^2(M,S)$ is the Hilbert space of square
integrable spinor fields over $M$; $D$ is the usual Dirac operator as given by
(\ref{dirac}). We see immediately that, for any $f\in\ca$,
\be
\p(\d f) =: [D, f] = \g^\m(x) \pa_\m f = \g(d_M f)~, \label{diff1form}
\ee
where $\g : \G(M, C(M)) \lra \cb(\ch)$ is the algebra morphism defined in
(\ref{gamma1}) and $d_M$ denotes the usual exterior derivative on $M$. In
general, with $f_j \in \ca$,
\be
\p(f_0 \d f_1 \dots \d f_p) =: f_0 [D, f_1] \dots [D, f_p] = 
\g(f_0 d_M f_1 \cdot \dots \cdot d_M f_p)~,
\ee
where now the differentials $d_M f_j$ are regarded as sections of the Clifford bundle
$C_1(M)$ (while $f_j$ can be thought of as sections of $C_0(M)$) and the dot
$\cdot$ denotes Clifford product in the fibers of $C(M) = \oplus_k C_k(M)$. 

Since a generic differential $1$-form on $M$ can be written as $\sum_j f_0^j d_M
f_1^j$ with $f_0^j, f_1^j \in \ca$, using (\ref{diff1form}) we can identify
Connes' $1$-forms $\ocad{1}$ with the usual differential $1$-forms $\L^1(M)$,
\be
\ocad{1} \simeq \L^1(M)~.
\ee
To be more precise, we are really identifying the space $\ocad{1}$ with the
image in $\cb(\ch)$, through the morphism $\g$, of the space $\L^1(M)$. 

Next, we analyze the  junk $2$-forms. For $f\in \ca$, consider the universal
$1$-form 
\be \a = {1 \over 2} (f \d f - (\d f) f) \not=0~, \label{junk2}
\ee
whose universal
differential is $\d \a = \d f \d f$. One easily finds that  
\bea
&& \p({\a}) = {1 \over 2} \g^\m (f \pa_\m f - (\pa_\m f) f) = 0~, \nonumber \\
&& \p({\d \a}) = \g^\m \g^\n \pa_\m f \pa_\n f = {1 \over 2} (\g^\m \g^\n + 
\g^\n \g^\m) \pa_\m f \pa_\n f = - g^{\m\n} \pa_\m f \pa_\n f
\II_{2^{[n/2]}} \not= 0~; \label{pijunk2}
\eea
here we have used (\ref{gamrel}), $g^{\m\n}$ being the components of the metric.
We conclude that the $2$-form $\d \a$ is a junk one. A generic junk
$2$-form is a combination (with coefficients in $\ca$) of forms like the one in
(\ref{junk2}). As a consequence, we infer from expression (\ref{pijunk2}) that
$\p(\d (J_0 \cap \oca{1}))$ is generated as an $\ca$-module by the matrix
$\II_{2^{[n/2]}}$. On the other side, if $f_1, f_2 \in \ca$, we have that
\bea
\g(d_M f_1 \cdot d_M f_1) &=& \g^\m \g^\n \pa_\m f_1 \pa_\n f_2 \nonumber \\
 &=& {1 \over 2}(\g^\m \g^\n - \g^\n \g^\m) \pa_\m f_1 \pa_\n f_2 + 
{1 \over 2} (\g^\m \g^\n + \g^\n \g^\m) \pa_\m f \pa_\n f \nonumber \\
&=& \g(d_M f_1 \wedge d_M f_2) - g (d_M f_1, d_M f_2) \II_{2^{[n/2]}}~.
\label{gajunk2}     
\eea
Therefore, since a generic differential $2$-form on $M$ can be written as a sum
$\sum_j f_0^j d_M f_1^j \wedge d_M f_2^j$,  with $f_0^j, f_1^j, f_2^j \in \ca$, by
using (\ref{pijunk2}) and (\ref{gajunk2}), we can identify Connes' $2$-forms
$\ocad{2}$ with the image through $\g$ of the usual differential $2$-forms
$\L^2(M)$,
\be
\ocad{1} \simeq \L^2(M)~.
\ee
The previous identifications can be made a general fact and one can identify
(through the map $\g$) 
\be
\ocad{p} \simeq \L^p(M)~.
\ee
In particular, $\ocad{p}=0$ if $p > dim M$. To establish such an identification,
we need some additional facts from Clifford bundle theory which we take from
\cite{BGV}. 

For each $m \in M$, the Clifford algebra $C_m(M)$ has a natural filtration,
$C_m(M) = \bigcup C_m^{(p)}$, where $C_m^{(p)}$ is spanned by products 
~$\x_1 \cdot \x_2 \cdot \dots \cdot \x_k,~ k\leq p, ~\x_j \in T^*_m M$. 
There is a natural graded algebra 
\be
gr C_m =: \sum_p gr_p C_m ~, ~~ gr_p C_m =  C_m^{(p)} / C_m^{(p-1)}~,
\label{graded}
\ee
with a natural projection, the {\it symbol map}, 
\be
\s_p :  C_m^{(p)} \lra gr_p  C_m~.
\ee
The graded algebra (\ref{graded}) is canonical isomorphic to the complexified
exterior algebra $\L_{\IC}(T^*_m M)$, the isomorphism being given by 
\be
\L_{\IC}(T^*_m M) \ni \x_1 \wedge \x_2 \wedge \dots \wedge \x_p \lra \s_p(\x_1
\cdot \x_2 \cdot \dots \cdot \x_p) \in gr_p C_m~.
\ee

\bprop\label{lemma}
Let  $(\ca, \ch, D)$ be the canonical triple over the manifold $M$. Then, a pair
$T_1, T_2$ of operators on $\ch$ is of the form $T_1 = \p(\o)~, T_2 = \p(\d \o)$
for some universal form $\o \in \oca{p}$, if and only if there are sections
$\r_1$ of $C^{(p)}$ and $\r_2$ of $C^{(p+1)}$, such that
\bea
T_j &=& \g(\r_j)~, ~~~j=1,2~, \nonumber \\
d_M \s_p(\r_1) &=& \s_{p+1}(\r_2)~.
\eea
\proof
If $\o = f_0 \d f_1 \dots \d f_p$, the identities $T_1 = \p(\o) = \g(f_0
\d f_1 \dots \d f_p)$ and $T_2 = \p(\o) = \g(\d f_0 \d f_1 \dots \d f_p)$ will
implies that $\r_1 = f_0 d_M f_1 \cdot  \dots \cdot d_M f_p~, ~\r_2 = d_M f_0
\cdot d_M f_1 \cdot \dots \cdot d_M f_p$, and in turn ~$\s_p(\r_1) = f_0 d_M f_1
\wedge \dots \wedge d_M f_p~, ~\s_{p+1}(\r_2) = d_M f_0 \wedge d_M f_1 \wedge
\dots \wedge d_M f_p$, and finally $d_M \s_p(\r_1)=\s_{p+1}(\r_2)$.

Conversely, if $\r_1 \in \G(C^{(p)})$ and $\r_2 \in \G(C^{(p+1)})$ are such
that $d_M \s_p(\r_1) = \s_{p+1}(\r_2)$, then $\r_2$ is determined by $\r_1$ up to an
ambiguity in $\G(C^{(p)})$. One can therefore suppose that $\r_1 = 0,  ~\r_2 \in
\G(C^{p})$. So one needs an universal form $\o \in \oca{p-1}$ such that
$\p(\o)=0, \p(\d \o) = \g(\r_2)$. Consider $\o' = {1 \over 2} (f_0 \d f_0 - 
\d f_0 f_0) \d f_1 \dots \d f_p$. Then $\p(\o')=0$ and $\p(\d \o') = 
\g (-\norm{d_M f_0}^2 d_M f_1 \cdot \dots \cdot d_M f_p$. Since terms of the type
$\norm{d_M f_0}^2 d_M f_1 \cdot \dots \cdot d_M f_p$ generate $\G(C^{(p)})$ as an
$\ca$-module, one can find  a universal form $\o \in \oca{p-1}$ with  $\p(\o)=0$
and $\p(\d \o) = \g(\r_2)$ where $\r_2$ is any given element of $\G(C^{(p)})$.
\eprop
\bprop
The symbol map $\s_p$ gives an isomorphism  
\be
\s_p : \ocad{p} \lra \G(\L_{\IC}^p T^* M)~,
\ee
which commutes with the differential.

\proof
Firstly, one identifies $\p(\oca{p})$ with $\G(C^{(p)})$ through $\g$. Then,
the previous Proposition~\ref{lemma} shows that $\p(\d(J_0 \cap \oca{p-1})) 
= ker \s_p$. If $\r \in \G(C^{(p)})$ with $\s_p(\r)=0$, then $\r_1=0$ and
$\r_2=\r$ fulfill the condition of Proposition~\ref{lemma} and there exists an
$\o\in\oca{p-1}$ such that $\r=\p(\d\o)$ and $\p(\o)=0$. 
Finally, one observe that from the definition of the symbol map, if $\r_j \in
\G(C^{p_j}), ~j=1,2$, then
\be
\s_{p_1+p_2}(\r_1 \r_2) = \s_{p_1}(\r_1) \wedge \s_{p_2}(\r_2) \in
\G(\L_{\IC}^{p_1+p_2} T^* M)~.
\ee
As a consequence, the symbol maps $\s_p$ combine to yield an isomorphism of graded
algebras
\be\label{fiso}
\s_p : \O_D(C^\infty(M)) \lra \G(\L_{\IC} T^* M)~,
\ee
which is also an isomorphism of $C^\infty(M)$-modules. 
\eprop
   
\subsubsect{Again the Two Points Space}\label{se:atp} 
As a very simple example, we shall now construct Connes' exterior algebra on the
two points space 
$Y =\{1, 2\}$ with the $0$-dimensional even spectral triple $(\ca, \ch, D)$
construct in Section~\ref{se:tps}. We already know that the associated algebra
$\ca$ of continuous function is the direct sum $\ca = \IC \oplus \IC$ and any
element
$f\in\ca$ is a couple of complex numbers $(f_1, f_2)$, with $f_i = f(i)$ the
value of $f$ at the point $i$. 

As we saw in Section~\ref{se:uof}, the space $\oca{1}$ of universal $1$-forms can
be identified with the space of functions on $Y \times Y$ which vanish on the
diagonal. Since the complement of the diagonal in $Y \times Y$ is made of
two points, namely the couples $(1,2)$ and $(2,1)$, the space $\oca{1}$ is
$2$-dimensional and a basis is constructed as follows. Consider the function
$e$ defined by $e(1) = 1, e(2) = 0$; clearly, $(1-e)(1) = 0, (1-e)(2) = 1$. A
possible basis for the $1$-forms is then given by 
\be
e\d e ~, ~~~(1-e) \d (1-e)~. 
\ee  
Their values being given by
\bea
&& (e\d e) (1,2) = -1~, ~~~((1-e) \d (1-e))(1,2) = 0 \nonumber \\
&& (e\d e) (2,1) = 0~, ~~~~~ ((1-e) \d (1-e))(2,1) = -1~. 
\eea
Any universal $1$-form $\a\in\oca{1}$ will be written as $\a = \l e\d e + \mu (1-e)
\d (1-e)$, with $\l, \mu \in \IC$. As for the differential, $\d : \ca \ra \oca{1}$, 
it is essentially a finite difference operator. For any $f\in \ca$ one finds that 
\be
\d f = (f_1 - f_2) e\d e -(f_1 - f_2)(1-e) \d (1-e) = (f_1 - f_2) \d e~.
\ee
As for the space $\oca{p}$ of universal $p$-forms, it can be identified with the
space of functions of $p+1$ variables which vanish on contiguous diagonals.
Since there are only two possible strings giving nonvanishing results, namely 
$(1,2,1,2, \cdots )$ and $(2,1,2,1, \cdots )$ the space $\oca{p}$ is two
dimensional as well and a possible basis is given by 
\be
e(\d e)^p ~, ~~~(1-e) (\d (1-e))^p~. 
\ee  
The values taken by the first basis element are 
\bea
&& (e(\d e)^p) (1,2,1,2, \cdots ) = \pm 1~, \label{orco} \\
&& (e(\d e)^p) (2,1,2,1, \cdots ) = 0~; 
\eea
in (\ref{orco}) the plus (minus) sign occurs if the number of contiguous
couples $(1,2)$ is even (odd). As for the second basis element we have
\bea
&& ((1-e) (\d (1-e))^p)(1,2,1,2, \cdots ) = 0~, \\
&& ((1-e) (\d (1-e))^p)(2,1,2,1, \cdots ) = \pm 1~, \label{orco1}  
\eea
in (\ref{orco1}) the plus (minus) sign occurs if the number of contiguous
couples $(2,1)$ is even (odd).

We pass now to Connes' forms. We recall that the finite dimensional Hilbert space
$\ch$ is a direct sum $\ch = \ch_1 \oplus \ch_2$; elements of $\ca$ act as
diagonal matrices $\ca \ni f \mapsto {\rm diag} (f_1 \II_{dim H_1}, 
f_2 \II_{dim H_2})$; $D$ is an off diagonal operator 
${\scriptstyle            
 \addtolength{\arraycolsep}{-.5\arraycolsep}
 \renewcommand{\arraystretch}{0.5}
 \left[ \begin{array}{cc}
 \scriptstyle 0 & \scriptstyle M^* \\
 \scriptstyle M  & \scriptstyle 0 \end{array} \scriptstyle\right]}, 
M \in Lin(\ch_1,\ch_2)$~. \\
It is immediate to find 
\bea
&& \pi(e \d e) =: e [D, e] = 
\left[
\begin{array}{cc}
0 & - M^* \\
0 & 0
\end{array}
\right]~, \nonumber \\
&& \pi((1-e) \d (1-e)) =: (1-e)[D, 1-e] = 
\left[
\begin{array}{cc}
0 & 0\\
-M & 0
\end{array}
\right]~, 
\eea 
and the representation of a generic $1$-form $\a = \l e\d e + \mu (1-e) \d
(1-e)$ is given by
\be
\pi(\a) = 
- \left[
\begin{array}{cc}
0 & \l M^* \\
\mu M & 0
\end{array}
\right]~.
\ee
As for the representation of $2$-forms one gets,
\bea
&&\pi(e \d e \d e) =: e [D, e][D, e] = 
\left[
\begin{array}{cc}
-M^*M & 0 \\
0 & 0
\end{array}
\right]~, \nonumber \\
&&\pi((1-e) \d (1-e) \d (1-e)) =: (1-e)[D, 1-e][D, 1-e] = 
\left[
\begin{array}{cc}
0 & 0\\
0 & -M M^*
\end{array}
\right]~,~~~~~~ 
\eea 
In particular the operator $\pi(\d \a)$ is readily found to be
\be
\pi(\d \a) = 
- (\l + \mu) \left[
\begin{array}{cc}
M^* M & 0 \\
0 & M M^*
\end{array}
\right]~,
\ee
from which we infer that there are no junk $1$-forms. In fact, there are no junk
forms whatsoever. Even forms are represented by diagonal operators while odd forms
are represented by off diagonal ones.

\subsect{Scalar Product of Forms}\label{se:spf}

In order to define a scalar product for forms, we need another definition
which was introduced in \cite{VG}.
\bdefi An $n$-dimensional spectral triple $(\ca, \ch, D)$ is defined to be {\rm
tame} if, for any $T\in \p(\omca)$ and $S \in \cb(\ch)$, one has that
\be
tr_\o(ST |D|^{-n}) = tr_\o(S |D|^{-n} T)~,
\ee
with $tr_\o$ denoting the Dixmier trace. 
\edefi

\noindent
From tameness and the cyclic property of $tr_\o$, the following three traces
coincides and can be taken as a definition of an inner product on $\p(\oca{p})$,
\be
\hs{T_1}{T_2}_p~ =: tr_\o(T_1^* T_2 |D|^n) = tr_\o(T_1^* |D|^n T_2) = 
tr_\o(T_2 |D|^n T_1^*)~,~~~\forall ~T_1, T_2 \in \p(\oca{p})~. \label{innfor}   
\ee
Forms of different degree are defined to be orthogonal. In particular, for $p=0$
one gets a positive trace on $\ca$ as it was alluded to at the end of
section~\ref{se:dis}.\\ 
Let now $\wt{\ch}_p$ be the corresponding completion of $\p(\oca{p})$. 
With $a \in \ca$ and $T_1, T_2 \in \p(\oca{p})~$, we shall get 
\bea
&&\hs{a T_1}{a T_2}_p = tr_\o(T_1^* a^* |D|^n a T_2) = tr_\o(T_2 |D|^n T_1^* a^* a)~,
\\ 
&&\hs{T_1 a}{T_2 a}_p = tr_\o(a^* T_1^* |D|^n T_2 a) = tr_\o(a^* a T_1^* |D|^n
T_2)~. 
\eea 
As a consequence, the unitary group $\cu(\ca)$ of $\ca$, 
\be
\cu(\ca) =: \{ u \in \ca ~|~ u^* u = u u^* = 1 \}~,
\ee
has two commuting unitary representations $L$ and $R$ on $\wt{\ch}_p$ given by left
and right multiplications. Now, being $\p(\d(J_0\cap\oca{p-1})$ a submodule of
$\p(\oca{p})$, its closure in $\wt{\ch}_p$ is left invariant by these two
representations. Let $P_p$ be the orthogonal projection of $\wt{\ch}_p$, with
respect to the inner product (\ref{innfor}), which projects onto the orthogonal
complement of $\p(\d(J_0\cap\oca{p-1}))$. Then $P_p$ commutes with $L(a)$ and $R(a)$,
if $a \in \cu(\ca)$ and so for any $a\in\ca$. 
Define $\ch_p=P_p\wt{\ch}_p$; this space also coincides with the completion of the
Connes' forms $\ocad{p}$. The left and right representations of $\ca$ on
$\wt{\ch}_p$ reduce to algebra representation on $\ch_p$ which extend the left
and right module action of $\ca$ on $\ocad{p}$.

\bigskip

As an example, consider again the algebra $\ca = C^\infty(M)$ and the associated
canonical triple $(\ca, \ch, D)$ over a manifold $M$ of dimension $n=dim M$.
Then, one one can prove that this triple is tame \cite{VG}. Furthermore,
\bprop
With the canonical isomorphism between $\omcad$ and $\G(\L_{\IC} T^* M)$
described in Sec.~\ref{se:uea}, the inner product on $\ocad{p}$ is proportional
to the Riemannian inner product on $p$-forms,
\be
\hs{\o_1}{\o_2}_p~ = (-1)^{p}{2^{[n/2] + 1 - n} \p^{-n/2} \over n\G(n/2)}
\int_M \o_1
\wedge ^*\o_2~, ~~~\forall ~\o_1,
\o_2
\in
\ocad{p}\simeq\G(\L_{\IC} T^* M)~. \label{rip}  
\ee 

\proof If $T\in\oca{p}$ and $\r\in \G(C^p)$, with $\p(T)=\g(\r)$, we have that
$P_p \p(T)=\g(\o) \in \ch_p$, with $\o$ the component of $\r$ in $\G(C^p \ominus
C^{p-1})$. Using the trace theorem \ref{wodzicki}, we get 
\begin{eqnarray*}
\hs{\g(\o_1)}{\g(\o_2)}_p &=& tr_\o (\o_1^* \o_2 |D|^{-n}) \\
&=& {1 \over n(2\p)^n} \int_{S^* M} tr \s_{-n} (\g(\o_1)^* \g(\o_2) |D|^{-n}) \\
&=& {1 \over n(2\p)^n} (\int_{S^{n-1}}d\x) 
\int_{M} tr (\g(\o_1)^* \g(\o_2) dx \\ 
&=& {2^{1-n}\p^{-n/2} \over n \G(n/2)} \int_{M} tr (\g(\o_1)^* \g(\o_2)) dx \\ 
&=& (-1)^{p}{2^{[n/2]+1-n}\p^{-n/2} \over n \G(n/2)} \int_{M} \o_1 \wedge ^* \o_2
.
\end{eqnarray*}
The last equality follows from the explicit (partially normalized) trace in the
spin representation. Indeed, 
\be
\begin{array}{l}
\o_j = {1 \over p!} \o^{(j)}_{\m_1 \cdots \m_p} dx^{\m_1} \wedge
\cdots \wedge dx^{\m_p}, ~~~ j = 1,2~, ~~~ \Rightarrow \\
~ \\
\g(\o_j) = {1 \over p!} \o^{(j)}_{\m_1 \cdots \m_p} \g^{\m_1} \wedge \cdots
\wedge \g^{\m_p} = {1 \over p!} \o^{(j)}_{\m_1 \cdots \m_p} e^{\m_1}_{a_1}
\cdots e^{\m_p}_{a_p} \g^{a_1} \wedge \cdots \wedge \g^{a_p}, ~~~ \Rightarrow \\
~ \\
tr (\g(\o_1)^* \g(\o_2)) = (-1)^{p} 2^{[n/2]} \o^{(1)*}_{\m_1 \cdots \m_p} 
\o^{(2)}_{\n_1 \cdots \n_p} e^{\m_1}_{a_1} \cdots e^{\m_p}_{a_p} 
e^{\n_1}_{b_1} \cdots e^{\n_p}_{b_p} \h^{a_1b_1} \cdots \h^{a_pb_p} \\ 
~ \\
~~~~~~~~~~~~~~~~~~~~~ = (-1)^{p} 2^{[n/2]} \o^{(1)*}_{\m_1 \cdots \m_p}
\o^{(2)}_{\n_1 \cdots \n_p} g^{\m_1\n_1}\cdots g^{\m_p\n_p}~,           
\end{array}
\ee
from which one gets $tr (\g(\o_1)^* \g(\o_2)) dx = 
(-1)^{p} 2^{[n/2]} \o_1\wedge ^* \o_2$~.

\eprop

\vfill\eject
\sect{Connections on Modules}\label{se:conn}
As an example of the general situation, we shall start by
describing the analogue of `electromagnetism', namely the algebraic theory of
connections (vector potentials) on a rank one trivial bundle (with fixed
trivialization).

\subsect{Abelian Gauge Connections}\label{se:electro}
Suppose we are given a spectral triple $(\ca, \ch, D)$ out of which we construct
the algebra $\omcad = \oplus_p \ocad{p}$ of forms. We shall also take it to be
tame and of dimension $n$. 
\bdefi\label{pot}
A vector potential $V$ is a self-adjoint element of $\ocad{1}$. The corresponding
field strength is the two-form $\q \in \ocad{2}$ defined by
\be
\q = d V + V^2~. 
\label{cur}
\ee
\edefi

\noindent
Thus, $V$ is of the form $V = \sum_j a_j [D, b_j], ~a_j, b_j \in \ca$ with
$V$ self-adjoint, $V^*=V$. Notice that, although $V$ can be written in several ways
as a sum, its exterior derivative $d V \in \ocad{2}$ is defined unambiguously, though,
again it can be written in several ways as a sum, $d V = \sum_j [D, a_j] [D, b_j]$,
modulo junk. The curvature $\q$ is self-adjoint as well. It is evident that $V^2$ is
self-adjoint if $V$ is. As for $d V$, we have,
\be\label{djunk}
d V - (d V)^* = \sum_j [D, a_j] [D, b_j] - \sum_j [D, b^*_j] [D, a^*_j]~. 
\ee 
On the other side, from $V^* = - \sum_j[D, b^*_j] a^*_j = - \sum_j[D, b^*_j a^*_j]
+ \sum_j b^*_j[D, a^*_j]$ and $ V- V^* = 0$, we get that the following is a junk
2-form,
\be
j_2 = d V - d V^* = \sum_j [D, a_j] [D, b_j] - \sum_j [D, b^*_j] [D, a^*_j]~. 
\ee
But $j_2$ is just the right-hand side of (\ref{djunk}), and we infer that,
modulo junk forms, $d V = (d V)^*$.
     
\bdefi
The unitary group $\cu(\ca)$ acts on the vector potential $V$ with the usual
affine action
\be\label{trapot}
(V, u) \lra V^u =: u V u^* + u [D, u^*]~, ~~u \in \cu(\ca)~. 
\ee
\edefi

\noindent
The curvature $\q$ will then transform with the adjoint action,
\bea
\q^u 
&=& d V^u + (V^u)^2 \nonumber \\
&=& du V u^* + u dV u^* - u V du^* + du [D, u^*] +
u V^2 u^* + \nonumber \\
&~& ~~~~~ + u V [D, u^*] + u [D, u^*] u V u^* + u [D, u^*] u [D, u^*] 
\nonumber \\
&=& \dots \nonumber \\
&=& u ( d V + V^2) u^*~,
\eea
namely
\be\label{tracur}
(\q, u) \lra \q^u = u \q u^*~, ~~u \in \cu(\ca)~.  
\ee

We can now introduce the analogue of the Yang-Mills functional.
\bprop\label{abeym}
1.~~~ The functional
\be
YM(V) =: \hs{d V + V^2}{d V + V^2}_2 ~,
\ee
is positive, quartic and invariant under gauge transformations
\be
V \lra V^u =: u V u^* + u [D, u^*]~, ~~u \in \cu(\ca)~. 
\ee
2.~~~ The functional
\be
I(\a) =: tr_\o (\p (\d \a + \a^2))^2 |D|^{-n})~, 
\ee
is positive, quartic and invariant on the space $\{\a \in \oca{1} ~|~ 
\a = \a^* \}$, under gauge transformations
\be
\a \lra \a^u =: u \a u^* + u \d u^*~, ~~u \in \cu(\ca)~. 
\ee
3.~~~
\be
YM(V) = inf ~\{ I(\a) ~|~ \p(\a) = V \}~.
\ee

\proof 
Statements 1. and 2. are consequences of properties of the Dixmier trace
for a tame triple and of the fact that both $d V + V^2$ and $\d \a + \a^2$
transform `covariantly' under gauge transformation. As for statement 3., it
follows from the nearest-point property of an orthogonal projector: as an
element of $\ch_2$, $d V + V^2$ is equal to $P(\p (\d \a + \a^2))$ for any $\a
\in \oca{1}$ such that $\p(\a) = V$. Since the ambiguity in $\p(\d \a)$ is
exactly $\p(\d(J_0 \cap \oca{1})$, one gets 3.
\eprop

\noindent
Point 3. of Prop.~\ref{abeym} just states that the ambiguity in the
definition of the curvature $\q = d V + V^2$ can be ignored by taking the
infimum $YM(V) = Inf~ \{tr_\o \q^2 |D|^{-n} \}$ over all possibilities for $\q = d V
+ V^2$, the exterior derivative $d V = \sum_j [D, a_0^j] [D, a_1^j]$ being
ambiguous.

As already mentioned, to consider the module $\ce = \ca$ is just the analogue of of
considering a rank one trivial bundle with fixed trivialization so that one can
identify the section of the bundle with the complex-valued functions on the base.

\subsubsect{The Usual Electromagnetism}\label{se:uel}
For the canonical triple $(\ca, \ch, D)$ over the manifold $M$, consider a
$1$-form $V \in \L^1(M)$ and a universal $1$-form $\a \in \oca{1}$ such that
$\s_1(\p(\a)) = V$. Then $\s_2(\p(\d \a)) = d_M V$. From proposition
\ref{lemma}, for any two such $\a$'s, the corresponding operators $\p(\d \a)$
differ by an element of $\p(\d(J_0 \cap \oca{1}) = ker \s_2$. Then, by using
(\ref{rip})
\bea
YM(V) &=& inf ~\{ I(\a) ~|~ \p(\a) = V \} = \hs{ d_M V}{d_M V }_2~ \nonumber \\
&=& {2^{[n/2] + 1 - n} \p^{-n} \over n\G(n/2)} \int \norm{d_M V}^2 dx~,
\eea
which is (proportional to) the usual abelian gauge action.

\subsect{Universal Connections}\label{se:ucmo}
We shall now introduce the notion of connection on a (finite projective) module.
We shall do it with respect to the universal calculus $\omca$ introduced in
Section~\ref{se:udf} as this is the prototype for any calculus. So, to be
precise, by connection we really mean universal connection although we drop the
adjective universal whenever there is no risk of confusion.

\bdefi\label{de:unco}
A (universal) connection on the right $\ca$-module $\ce$ is a $\IC$-linear map
\be
\nabla : \ce \ota \oca{p} \lra \ce \ota \oca{p+1}~, \label{unab}
\ee
defined for any $p \geq 0$, and satisfying the Leibniz rule
\be
\nabla(\o \r) = (\nabla \o) \r + (-1)^{p} \o \d \r ~, 
~~\forall ~\o \in \ce \ota \oca{p}~, ~\r \in \omca~.  \label{ulei}
\ee
\edefi

\noindent
In this definition, the adjective universal refers to the use of the
universal forms and to the fact that a connection constructed for any calculus can be
obtained by a universal one via a projection much in the same way as any calculus can
be obtained from the universal one. In Proposition~\ref{0surj} we shall
explicitly construct the projection for the Connes' calculus.

\noindent 
A connection is completely determined by its restriction 
$\nabla : \ce \ra \ce \ota \oca{1}$, which satisfy
\be
\nabla (\h a) = (\nabla \h) a + \h \ota \d a ~, 
~~\forall ~\h \in \ce~, ~a \in \ca~.  \label{ulei0}
\ee
This is then extended by using Leibniz rule (\ref{ulei}).
\bprop\label{ucurva}
The composition $\nabla^2 = \nabla \circ \nabla : \ce \ota \oca{p} 
\lra \ce \ota \oca{p+2}$ is $\omca$-linear.

\proof By condition (\ref{ulei}) one has
\bea
\nabla^2 ({\o \r}) &=& \nabla \left( (\nabla \o) \r + (-1)^{p} \o \d \r \right)
\nonumber \\
&=& (\nabla ^2 \o) \r + (-1)^{p+1} (\nabla \o) \d \r + (-1)^{p} (\nabla \o) \d \r 
+ \o \d^2 \r \nonumber \\
&=& (\nabla ^2 \o) \r ~. 
\eea
\eprop

\noindent
The restriction of $\nabla^2$ to $\ce$ is the {\it curvature} 
\be
\q : \ce \ra \ce \ota \oca{2}
\ee 
of the connection. By (\ref{ulei}) it is $\ca$-linear, $\q(\h a) = \q(\h)a$ for any
$\h\in\ce, a \in \ca$, and satisfies
\be
\nabla^2(\h \ota \r) = \q(\h) \r~, ~~\forall ~\h \in \ce~, ~\r \in \omca~. 
\ee

Since $\ce$ is projective, any $\ca$-linear map $: \ce \ra \ce \ota \omca$ can be
thought of as a matrix with entries in $\omca$ or as an element in $End_{\ca}\ce
\ota \omca$. In particular, the curvature $\q$ can be thought of as an element of 
$End_{\ca}\ce\ota \oca{2}$. Furthermore, by viewing any element of $End_{\ca}\ce
\ota \omca$ as a map $: \ce \ra \ce \ota \omca$, the connection $\nabla$ on
$\ce$ determine a connection $[\nabla, ~\cdot~ ]$ on $End_{\ca}\ce$ by
\bea
&&[\nabla, ~\cdot~ ] : End_{\ca}\ce \ota \oca{p} \lra End_{\ca}\ce \ota \oca{p+1}~,
\nonumber \\
&&[\nabla, \a ] =: \nabla \circ \a - \a \circ \nabla ~, 
~~~\forall ~\a \in End_{\ca}\ce \ota \oca{p}.
\eea

\bprop\label{ubianchi}
The curvature $\q$ satisfies the following Bianchi identity,
\be
[\nabla, \q ] = 0~.  \label{ubia}
\ee
\proof Since $\q : \ce \ra \oca{2}$, the map $[\nabla, \q ]$ makes sense. And,
$[\nabla, \q ] = \nabla \circ \nabla^2 - \nabla^2 \circ \nabla = \nabla^3 -
\nabla^3 = 0$~.  
\eprop

Connections always exists on a projective module. Consider, to start with, the case
of a free module $\ce = \IC^N \otc \ca \simeq \ca^N$. Forms with values in $\IC^N
\otc \ca$ can be identified canonically with
\be
\IC^N \otc \omca = (\IC^N \otc \ca) \ota \omca ~\simeq~ (\omca)^N~.
\ee
Then, a connection is given by the operator 
\be
\nabla_0 = \II \otimes \d : \IC^N \otc \oca{p} \lra \IC^N \otc \oca{p+1}~.
\ee
If we think of $\nabla_0$ as acting on $(\omca)^N$ we can represent it as the
operator $\nabla_0 = (\d, \d, \cdots, \d)$  ($N$-times). \\
Consider then a generic projective module $\ce$, and let $p : \IC^N \otc \ca
\ra \ce$ and $\l : \ce \ra \IC^N \otc \ca$ be the corresponding projection and
inclusion maps as in Section~\ref{se:mod}. On $\ce$ there is a connection
$\nabla_0$ given by the composition
\be  \label{ugras0}
\ce \ota \oca{p} ~\stackrel{\l}{\lra}~ \IC^N \otc \oca{p} 
~\stackrel{\II\otimes \d}{\lra}~ \IC^N \otc \oca{p+1} 
~\stackrel{p}{\lra}~ \ce \ota \oca{p+1}~, 
\ee
where we use the same symbol to denote the natural extension of the maps $\l$ and
$p$ to $\ce$-valued forms. The connection defined in (\ref{ugras0}) is called the
{\it Grassmann connection} and is explicitly given by
\be\label{ugras1}
\nabla_0 = p \circ (\II \otimes \d ) \circ \l~. 
\ee
In the following, we shall simply indicate it by 
\be\label{ugras}
\nabla_0 = p \d.    
\ee
In fact, it turns out that the existence of a connection on the module $\ce$ is
completely equivalent to its being projective \cite{CQ}.
\bprop\label{unicon}
A right module has a connection if and only if it is projective.

\proof 
Consider the exact sequence of right $\ca$-modules
\be  \label{1mod}
0 \lra \ce \ota \oca{1} ~\stackrel{j}{\lra}~ \ce \otc \ca  
~\stackrel{m}{\lra}~ \ce \lra 0~,  
\ee
where $j( \h \d a) = \h \otimes a - \h a \otimes 1$ and $m(\h \otimes a) = \h a$;
both these maps are (right) $\ca$-linear. Now, as a sequence of vector spaces,
(\ref{1mod}) admits a splitting given by the section $s_0(\h) = \h \otimes 1$ of
$m$, $ m \circ s_0 = id_{\ce}$. Furthermore, all such splittings form an affine
space which is modeled over the space of linear maps from the base space $\ce$ to
the subspace $j(\ce \ota \oca{1})$. This means that there is a one to one
correspondence between linear sections $s : \ce \ra \ce \otc \ca$ of $m$  
( $m \circ s = id_{\ce}$ ) and linear maps $\nabla : \ce \ra \ce \ota \oca{1}$ 
given by 
\be   
s = s_0 + j \circ \nabla~, ~~~
s(\h) = \h \otimes 1 + j(\nabla \h)~, ~~~ \forall ~\h \in \ce~.
\ee
Since 
\be
s(\h a) - s(\h) a = \h a \otimes 1 - \h \otimes a + j(\nabla (\h a)) - 
j(\nabla (\h))a = j(\nabla (\h a) - \nabla (\h) a - \h \d a)~,
\ee
and $j$ being
injective, we see that $\nabla$ is a connection if and only if $s$ is a right
$\ca$-module map,
\be
\nabla (\h a) - \nabla (\h) a - \h \d a = 0 ~~~\iff ~~~s(\h a) - s(\h) a = 0~.
\ee
 But such module maps exists if and only if $\ce$ is projective : any right
module map $s: \ce \ra \ce \otc \ca$ such that $m \circ s = \II_{\ce}$
identifies $\ce$ with a direct summand of the free module $\ce \otc \ca$, the
corresponding idempotent being $p = s \circ m$.  
\eprop 

\noindent
The previous proposition also says that the space $CC(\ce)$ of all universal
connections on $\ce$ is an affine space modeled on $End_{\ca} \ota \oca{1}$.
Indeed, if $\nabla_1, \nabla_2$ are two connections on $\ce$, they difference is
$\ca$-linear,
\be
(\nabla_1 - \nabla_2)(\h a) = ((\nabla_1 - \nabla_2)(\h )) a ~, 
~~ \h \in \ce~, ~a \in \ca~,
\ee
so that $\nabla_1 - \nabla_2 \in End_{\ca} \ota \oca{1}$. By using (\ref{ugras})
and (\ref{end1}) any connection can be written as 
\be
\nabla = p \d + \a ~, \label{uconn}
\ee
where $\a$ is any element in $\IM_{\ca}(\ca) \ota \oca{1}$ such that $\a = \a p
= p \a = p \a p$.  The matrix of $1$-forms $\a$ as in (\ref{uconn}) is called the
{\it gauge potential} of the connection $\nabla$.
For the corresponding curvature $\q$ of $\nabla$ we get
\be
\q = p \d \a + \a^2 + p \d p \d p~. \label{ucurv} 
\ee
Indeed, 
\bea
\q(\h) = \nabla^2(\h) &=& (p \d + \a)(p \d \h + \a \h) \nonumber \\
&=& p \d (p \d \h) + p \d(\a \h) + \a p \d \h + \a^2 \h \nonumber \\ 
&=& p \d (p \d \h) + p \d \a \h + \a^2 \h \nonumber \\  
&=& (p \d p \d p + p \d \a + \a^2)(\h)~, 
\eea
since, by using $p \h = p$ and $p^2 =p$, one has that  
\bea
p \d (p \d \h) &=& p \d (p \d (p \h)) \nonumber \\ 
&=& p \d (p \d p \h + p \d \h ) \nonumber \\
&=& p \d p \d p \h - p \d p \d \h + p \d p \d \h \nonumber \\
&=& p \d p \d p \h~.
\eea 

\bigskip

With any connection $\nabla$ on the module $\ce$ there is associated a
{\it dual connection} $\nabla '$ on the dual module $\ce '$. Notice first, that
there is a pairing 
\be
(\cdot, \cdot) : \ce ' \times \ce \lra \ca~, ~~~~(\f, \h) =:\f(\h)~, 
\ee
which, due to (\ref{dualmodu}), with respect to the right-module structures, has
the following property
\be\label{pali}
(\f \cdot a, \h \cdot b) = a^* (\f, \h) b ~, 
~~~\forall ~\f \in \ce ', \h \in \ce, a,b \in \ca~. 
\ee
Therefore, it can be extended to maps
\bea\label{expa}
&&(\cdot, \cdot) : \ce ' \ota \omca \times \ce \lra \ca~,
~~~(\f \cdot \a, \h ) = \a^* (\f, \h)~, \nonumber \\
&&(\cdot, \cdot) : \ce ' \times \ce \ota \omca \lra \ca~,
~~~(\f , \h \cdot \b) = (\f, \h) \b~, 
\eea 
for any $\f \in \ce '; \h \in \ce; \a, \b \in \omca$.

Let us suppose now that we have a connection $\nabla$ on $\ce$. The dual
connection 
\be
\nabla ': \ce ' \ra \ce ' \ota \oca{1}~,
\ee
is defined by
\be\label{duco}
\d (\f, \h) = - (\nabla ' \f, \h) + (\f, \nabla ' \h)~, ~~~~
\forall ~\f \in \ce ', \h \in \ce~.
\ee
It is easy to check right-Leibniz rule
\be
\nabla '(\f \cdot a) = (\nabla '\f) a + \f \ota \d a ~, 
~~\forall ~\f \in \ce '~, a \in \ca~.  \label{uleid}
\ee
Indeed, for any $\f \in \ce ', a \in \ca, \h \in \ce$,
\be
\begin{array}{rclccc}
\d (\f \cdot a, \h) & = & - (\nabla ' (\f \cdot a), \h) + 
(\f \cdot a, \nabla '\h) & \Longrightarrow & {\rm ~by} & \ref{pali} \\
\d a^* (\f, \h) + a^* \d (\f, \h) & = & - (\nabla ' (\f \cdot a), \h) + 
a^* (\f , \nabla '\h) & \Longrightarrow & {\rm ~by} & \ref{duco} \\
\d a^* (\f, \h) - a^* \d (\nabla ' \f, \h) & = & - (\nabla ' (\f \cdot a), \h)
& \Longrightarrow & {\rm ~by} & \ref{invol0} \\
-(\d a)^* (\f, \h) - a^* \d (\nabla ' \f, \h) & = & - (\nabla ' (\f \cdot a), \h)
& \Longrightarrow & {\rm ~by} & \ref{expa} \\
(\f \ota \d a, \h) + ((\nabla ' \f )\cdot a , \h) & = & (\nabla ' 
(\f \cdot a), \h)~, & ~ & ~ & ~ \\
\end{array}
\ee
from which (\ref{uleid}) follows. 

\subsect{Connections Compatible with Hermitian Structures}\label{se:cch}
Suppose now, we have a Hermitian structure $\hs{ \cdot}{\cdot}$ on the
module $\ce$ as defined in Section~\ref{se:hsp}. A connection $\nabla$ on
$\ce$ is said to be {\it compatible with the Hermitian structure} if the
following condition is satisfied \cite{Co1},
\be\label{ucompat}
-\hs{\nabla \h}{x} + \hs{\h}{\nabla \x} = \d \hs{\h}{\x}~, ~~~\forall ~\h, \x \in
\ce~.
\ee
Here the Hermitian structure is extended to linear maps (denoted with the same symbol) 
$:~\ce \ota \oca{1} \times \ce \ra \oca{1}$ and $:~\oca{1} \ota \ce \times \ce \ra
\oca{1}$ by
\bea\label{uhext}
&& \hs{ \h \ota \o}{\x }~ = \o^* \hs{\h}{\x} ~, \nonumber \\
&& \hs{\h}{\x \ota \o }~ = ~\hs{\h}{\x} \o ~, ~~~\forall ~\h, \x \in \ce~, ~\o \in
\oca{1}~.
\eea
Also, the minus sign in the left hand side of eq.~(\ref{ucompat}) is due to the
choice  $ (\d a)^* = - \d a^*$ which we have made in (\ref{invol0}).

Compatible connections always exist. As explained in Section~\ref{se:hsp}, any
Hermitian structure on $\ce = p \ca^N$ can be written as $\hs{\h}{\x} =
\sum_{j=1}^N \h_j^* \x_j$ with $\h = p \h = (\h_1, \cdots, \h_N)$ and the same
for $\x$. Then the Grassman connection (\ref{ugras}) is compatible, since 
\bea 
\d \hs{\h}{\x} &=& \d (\sum_{j=1}^N \h_j^* \x_j) \nonumber \\
&=& \sum_{j=1}^N \d \h_j^* ~\x_j + \sum_{j=1}^N \h_j^* ~\d \x_j 
   =  - \sum_{j=1}^N (\d \h_j)^* ~\x_j + \sum_{j=1}^N \h_j^* ~\d \x_j 
\nonumber \\  &=& -\hs{\d \h}{p \x} + \hs{p \h}{\d \x} \nonumber \\
&=& -\hs{p \d \h}{\x} + \hs{\h}{p \d \x} \nonumber \\
&=& -\hs{\nabla_0 \h}{\x} + \hs{\h}{\nabla_0 \x}~.
\eea
For a general connection (\ref{uconn}), the compatibility with the Hermitian
structure reduces to
\be
\hs{ \a \h}{\x } - \hs{\h}{\a \x }= 0~, ~~~ \forall ~\h, \x \in \ce~, 
\ee
which just says that the gauge potential is Hermitian,
\be
\a ^* = \a~. 
\ee
We still use the symbol $CC(\ce)$ to denote compatible universal connections on 
$\ce$.

\subsect{The Action of the Gauge Group}\label{se:uagg}
Suppose we are given a Hermitian finite projective $\ca$-module $\ce = p \ca^N$. 
Then, the {\it algebra of endomorphisms} of $\ce$ is defined as
\be
End_\ca(\ce) = \{\f : \ce \ra \ce ~|~ \f(\h a) = \f(\h)a ~, ~\h \in \ce~, ~a \in
\ca \}~.
\ee
It is clearly an algebra under composition. It also admits a natural involution
$^* : \ce \ra \ce$ determined by
\be
\hs{T^*\h}{\x} =: \hs{\h}{T \x}~, ~~~ \forall ~T \in End_\ca(\ce)~, 
~~\h, \x \in \ce~. 
\ee
With this involution, there is an isomorphism
\be
End_\ca(\ce) \simeq p \IM_N(\ca) p~, \label{end1} 
\ee 
namely, elements of $End_\ca(\ce)$ are matrices $m \in \IM_N(\ca)$ which commutes
with the idempotent $p$, $p m = m p$.

The group $\cu(\ce)$ of {\it unitary automorphisms} of $\ce$ is the subgroup of
$End_\ca(\ce)$ given by
\be
\cu(\ce) = \{ u \in End_\ca(\ce) ~|~ u u^* = u^* u = 1 \}~. \label{unit}
\ee
In particular, we have that $\cu_N(\ca) =: \cu(\ca^N) = \{ u \in \IM_N(\ce) ~|~ 
u u^* = u^* u = 1 \}$~. Also, there is an isomorphism
$\cu_N(C^\infty(M)) \simeq C^\infty(M, U(N))$, with $M$ a smooth manifold and
$U(N)$ the usual $N$-dimensional unitary group. In general, if $\ce = p\ca^N$
with $p^*=p$, one gets that $\cu(\ce) = p \cu(\ca^N) p$.

The group $\cu(\ce)$ of unitary automorphisms of 
the module $\ce$, defined in (\ref{unit}) plays
the r\^ole of the infinite dimensional group of gauge transformations.
Indeed, there is a natural action of such group on the space $CC(\ce)$
of universal compatible connections on $\ce$. It is given by
\be
(u, \nabla) \lra \nabla^u =: u \nabla u^*~, ~~~ \forall ~u\in \cu(\ce), 
~\nabla \in CC(\ce)~.\label{ugcon}
\ee
It is then straightforward to check that the curvature transforms in a 
covariant way
\be
(u, \q) \lra \q^u =: u \q u^*~, \label{ugcur}
\ee
since, evidently, $\q^u = (\nabla^u)^2 = u \nabla u^* u \nabla u^* = u \nabla^2
u^* = u \q u^*$. \\
As for the gauge potential, one has the
usual affine transformation
\be
(u, \a) \lra \a^u =: u p \d u^* + u \a u^*~. \label{ugpot}
\ee
Indeed, for any $\h\in\ce$,  
\bea 
\nabla^u (\h) &=& u( p \d + \a) u^* \h = u p \d (u^* \h) + u \a u^* \h \nonumber \\
&=& p u (u^* \d \h ) + u p (\d u^* ~\h ) + u \a u^* ~~~~{\rm using}~~~~ up = pu
\nonumber \\  
&=& p \d \h + (u p \d u^* + u \a u^*) \h  \nonumber \\
&=& ( p \d + \a^u ) \h~, 
\eea
which gives (\ref{ugpot}) for the transformed potential.

\subsect{Connections on Bimodules}\label{se:bico}

In constructing gravity theories  one needs to introduce the analogues of linear
connections. These are connections defined on the bimodule of $1$-forms which plays
the role of the cotangent bundle. Since this module is in fact a bimodule, it seems
natural to exploit both left and right module structures
\fn{As we shall see in Section~\ref{se:gra}, gravity theories have been constructed
which use only one structure (the right one, although it would be completely
equivalent to use the left one). In this context, the usual Einstein gravity has been
obtained as a particular case.}.

One of the ideas which have been proposed \cite{Mo} is that of a `braiding' which, 
generalizing the permutation of forms, flips two element of a tensor product so as to
make possible a {\it left} Leibniz rule once a {\it right} Leibniz rule is satisfied.

Then, let $\ce$ be an $\ca$-bimodule which is left and right projective, endowed with
a right connection, namely a linear map $\nabla : \ce \ra \ce \ota \oca{1}$ which
obeys the right Leibniz rule (\ref{ulei0}). 
\bdefi
Given a bimodule isomorphism, 
\be
\s : \oca{1} \ota \ce \longrightarrow \ce \ota \oca{1}~, \label{biiso} 
\ee
the couple $(\nabla, \s)$ is said to be {\rm compatible} if and only if a left
Leibniz rule of the form
\be
\nabla (a\h) =  (\nabla \h)a + \s (\d a \ota \h)~, ~~~ 
\forall ~ a\in \ca~, ~\h \in \ce~. \label{sileib}
\ee
is satisfied.
\edefi

\noindent
We see that the role of the map $\s$ is to bring the one form $\d a$ to the
`right place'. Notice that in general $\s$ needs not square to the identity, namely,
$\s
\circ \s \not= \II$. Several examples of such connections have been constructed for
the case of linear connection, namely $\ce = \oca{1}$ (see \cite{Mas} and
references therein). 
 
To get a bigger space of connections  a weaker condition has been proposed in
\cite{DHLS} where the compatibility has been required to be satisfied only on the
center of the bimodule. Recall first of all that the center $\cz(\ce)$ of a bimodule
$\ce$ is the bimodule defined as
\be
\cz(\ce) =: \{ \h \in \ce ~|~ a \h = \h a ~, ~~ \forall a \in \ca \}~.
\ee
Now, let $\nabla^L$ be a {\it left connection}, namely a linear map 
$: \ce \ra \oca{1} \ota \ce$ satisfying the left Leibniz rule
\be
\nabla^L (a \h ) = \d a \ota \h + a \nabla^L \h~, ~~~
\forall ~a \in A, ~\h \in E, 
\ee
and let $\nabla^R$ be a right connection,  namely a linear map $: \ce \ra \ce \ota
\oca{1}$  satisfying the right Leibniz rule
\be
\nabla^R (\h a) =  (\nabla^R \h) a + \h \ota \d a~, ~~~
\forall ~a \in A, ~\h \in E. 
\ee
\bdefi
With $\s$ a bimodule isomorphism as in (\ref{biiso}), a pair $(\nabla^L,\nabla^R)$ is
said to be {\rm $\s$-compatible} if and only if 
\be\label{compatible} 
\nabla^R \h = (\sigma \circ \nabla^L) \h ~, \forall ~ \h \in \cz(\ce)~.
\ee
\edefi

\noindent
By requiring that the condition $\nabla^R = \s \circ \nabla^L$ be satisfied
on the whole bimodule~$\ce$, one can equivalently think of a pair 
$(\nabla^L, \nabla^R)$ as a right connection $\nabla^R$ fulfilling the additional left
Leibniz rule (\ref{sileib}) so reproducing the previously described situation
\fn{ In \cite{CQ} a connection on a bimodule is also defined as a pair consisting of
a left and right connection. There, however, there is no $\s$-compatibility condition
while the additional conditions of $\nabla\sp L$ being a right $\ca$-homomorphism and
$\nabla\sp R$ being a  left $\ca$-homomorphism is imposed. These latter conditions,
are not satisfied in the classical commutative case $\cz(\ce) = \ce = \O^1(M)$.}.

We finish by mentioning that other categories of relevant bimodules have been
introduced, notably the one of {\it central bimodules}. We refer to \cite{DM} for
details and for a theory of connections on these bimodules.

\vfill\eject
\sect{Field Theories on Modules}\label{se:ftm}

In this section we shall describe how to construct field theoretical  models in
the algebraic noncommutative framework developed by  Connes. Throughout the
section, the basic ingredient will be a  spectral triple $(\ca, \ch, D)$ which
we take to be tame and of dimension $n$. Associated with it there is the algebra
$\omcad = \oplus_p \ocad{p}$  of forms as constructed in Section~\ref{se:cdf}
with exterior differential $d$. 

\subsect{Yang-Mills Models}\label{se:ymm}  
The theory of connections on any (finite projective) $\ca$-module $\ce$, 
with respect to the differential calculus $(\omcad , d)$ is, {\it mutatis 
mutandis}, formally the same as the theory of universal connections 
developed in Section~\ref{se:ucmo}. 

\bdefi\label{ycon} 
A connection on the $\ca$-module $\ce$ is a $\IC$-linear map
\be\label{nab}
\nabla : \ce \ota \ocad{p} \lra \ce \ota \ocad{p+1}~, 
\ee 
satisfying Leibniz rule
\be
\nabla(\o \r) = (\nabla \o) \r + (-1)^{p} \o d \r ~,  ~~\forall ~\o \in 
\ce \ota \ocad{p}~, ~\r \in \omcad~.  \label{lei}
\ee
\edefi

\noindent
Then, the composition $\nabla^2 = \nabla \circ \nabla : \ce \ota \ocad{p} 
\ra \ce \ota \ocad{p+2}$ is $\omcad$-linear and its restriction to $\ce$ is 
the {\it curvature} $F : \ce \ra \ce \ota \ocad{2}$ of the connection. The
curvature is $\ca$ linear, $F(\h a) = F(\h) a$, for any $\h\in\ce, a\in \ca$,  and
satisfies, 
\be
\nabla^2(\h \ota \r) = F(\h) \r~, ~~\forall ~\h \in \ce~, ~\r \in \omcad~. 
\ee
As before, thinking of the curvature $F$ as an element of 
$End_{\ca}\ce\ota \ocad{2}$, it satisfies Bianchi identity,
\be 
[\nabla, F ] = 0~.  \label{bia}
\ee
As already said previously, connections always exists on a projective module. 
If $\ce = p \ca^N$, it is possible to write any connection as
\be
\nabla = p d + A ~, \label{conn}
\ee 
where $A$ is any element in $\IM_{\ca}(\ca) \ota \ocad{1}$ such that $A = A p
= p A = p A p$.  The matrix of $1$-forms $A$ is called the {\it gauge potential}
of  the connection $\nabla$. For the corresponding curvature $F$ we get
\be 
F = p d A + A^2 + p d p d p~. \label{curv} 
\ee
The space $C(\ce)$ of all connections on $\ce$ is an affine space modeled on
$End_{\ca} \ota \ocad{1}$. 

The compatibility of the connection $\nabla$ with respect to an Hermitian
structure on $\ce$ is expressed exactly as in Section~(\ref{se:cch}),
\be\label{compat} 
-\hs{\nabla \h}{\x} + \hs{\h}{\nabla \x} = d \hs{\h}{\x}~, ~~\forall~
\h, \x \in \ce~,
\ee 
with the Hermitian structure extended as before to linear maps  
$:~\ce \ota \ocad{1} \times \ce \ra \ocad{1}$ and $:~\ocad{1} \ota \ce \times
\ce \ra \ocad{1}$, by
\bea\label{hext} 
&& \hs{ \h \ota \o}{\x} = \o^* \hs{\h}{\x} ~, \nonumber \\ 
&& \hs{ \h}{\x \ota \o } = \hs{\h}{x} \o ~, ~~~~\forall ~\h, \x \in \ce, 
~\o \in \ocad{1}~.
\eea
The connection (\ref{conn}) is compatible with the Hermitian structure 
$\hs{\h}{\x} = \sum_{j=1}^N \h_j^* \x_j$ on $\ce = p\ca^N$ 
($\h = (\h_1, \cdots, \h_N) =  p \h$ and the same for $\x$), 
provided the gauge potential is Hermitian, \be A ^* = A~. 
\ee

The action of the group $\cu(\ce)$ of unitary automorphisms of the module $\ce$
on the space $C(\ce)$ of compatible connections on $\ce$ it is given by
\be 
(u, \nabla) \lra \nabla^u =: u \nabla u^* ~, ~~~ \forall ~u \in \cu(\ce), 
~\nabla \in C(\ce)~. 
\label{gcon}
\ee 
Then, the gauge potential and the curvature transform in the usual way
\bea &&(u, A) \lra A^u = u [ p d  + A ] u^*~, \label{gpot} 
\\ &&(u, F) \lra F^u =: u F u^*~, ~~~\forall ~u \in \cu(\ce). \label{gcur}
\eea

The following proposition clarifies in which sense the connections defined in
\ref{de:unco} are universal.

\bprop\label{0surj}  The representation $\p$ in (\ref{pi}) can be extended to a
surjective map 
\be
\II \otimes \p : CC(\ce) \lra C(\ce)~, \label{surj}
\ee 
namely, any compatible connection is the composition of $\p$ with a universal
compatible connection.

\proof 
By construction, $\p$ is a surjection from $\oca{1}$ to $\p(\oca{1}) \simeq
\ocad{1}$. Then, we get a surjection $\II \otimes \p : End_{\ca}\ota\oca{1} 
\ra End_{\ca}\ota\ocad{1}$. Finally, define $\II \otimes \p (p \circ \d) = 
p \circ d$ to get the desired surjection $: CC(\ce) \lra C(\ce)$.
\eprop

\bigskip

By using the Hermitian structure on $\ce$ together with an ordinary
matrix trace over `internal indexes', one can construct an inner  product on
$End_{\ca}$. By combining this product with the inner product  on $\ocad{2}$
given in (\ref{innfor}), one has than a natural inner  product on the space
$End_{\ca}\ce\ota \ocad{2}$. Since the curvature $F$  is an element of such a
space, the following definition makes sense. 
\bdefi 
The Yang-Mills action for the connection $\nabla$ with curvature $F$ is
given 
\be 
YM(\nabla) = \hs{F}{F}_2 ~. \label{ym}
\ee
\edefi 

\noindent
By its very construction it is invariant under gauge transformations 
(\ref{gpot}) and (\ref{gcur}).

\bigskip

Consider now the tensor product $\ce \ota \ch$. This space can be made a Hilbert
space by combining the Hermitian structure on $\ce$ with the scalar product on
$\ch$,
\be\label{modhil} 
( \h_1 \ota \j_1 , \h_2 \ota \j_2 ) =: (\j_1 , \hs{\h_1}{\h_2}
\j_2 )~, ~~~
\forall ~\h_1, \h_2 \in \ce, ~~\j_1, \j_2 \ch~.  
\ee
Then, by using the projection (\ref{pi}) we get a projection 
\be\label{modhilpro}
\II_{\ce} \otimes \p : \ce \ota \omcad \lra \cb(\ce \ota \ch)~,
\ee 
and an inner product on $(\II_{\ce} \otimes \p) (\ce \ota \oca{p})$ by
\be 
\hs{T_1}{T_2}_p = tr_{\o} T_1^* T_2 | \II_{\ce} \otimes D | ^{-n}~,
\ee 
which is the analogue of the inner product (\ref{innfor}). The corresponding
orthogonal projector $P$ has a range which can be identified with $\ce \ota
\ocad{p}$. 

If $\nabla_{un} \in CC(\ce)$ is any universal connection with curvature
$\q_{un}$,  one defines a pre-Yang-Mills action $I(\nabla_{un})$ by,
\be\label{pym} 
I(\nabla_{un}) = tr_{\o} \p(\q_{un})^2 | \II \otimes D | ^{-n}~.
\ee
Then, one has the analogue of proposition (\ref{abeym})
\bprop\label{ymmin} 
For any compatible connection $\nabla \in C(\ce)$, one has that
\be 
YM(\nabla) = inf \{ I(\nabla_{un}) ~|~ \p(\nabla_{un}) = \nabla \}~.
\ee
\proof It really goes as the analogous proof of Proposition~\ref{abeym}.
\eprop

It is also possible to define a {\it topological action} and extend the usual
inequality between Chern classes of vector bundles and the value of the
Yang-Mill action on an arbitrary connection on the bundle. First observe that
from definition (\ref{ym}) of the Yang-Mills action functional, if $D$ is
replaced by $\l D$, then $YM(\nabla)$ is replaced by $|\l| ^{4-n}YM(\nabla)$.
Therefore, it will have chances to be related to `topological invariants' of
finite projective modules only if $n=4$. Let us then assume that our spectral
triple is four dimensional. We also need it to be even with a $\IZ_2$ grading
$\G$. With these ingredients, one defines two traces on the algebra
$\oca{4}$, 
\bea
&&\t(a_0 \d a_1 \cdots \d a_4) = tr_w (a_0 [D, a_1] \cdots [D, a_1] |D|^{-4})~,
~~~ a_j \in \ca \nonumber \\
&&\F(a_0 \d a_1 \cdots \d a_4) = tr_w (\G a_0 [D, a_1] \cdots [D, a_1]
|D|^{-4})~,~~~ a_j \in \ca~.
\eea
By using the projection (\ref{modhilpro}) and an `ordinary trace over internal
indices' and by substituting $\G$ with $\II_{\ce} \otimes \G$ and
$|D|^{-4}$ with $\II_{\ce} \otimes |D|^{-4}$, the previous traces can be
extended to traces $\wt{\t}$ and $\wt{\F}$ on $End_\ca \ce \ota \oca{4}$.
Then, by the very construction (\ref{pym}), one has that
\be\label{ymtr}
I(\nabla_{un}) = \wt{\t}(\q_{un}^2)~, ~~~ \forall ~\nabla_{un} \in CC(\ce)~, 
\ee
with $\q_{un}$ the curvature of $\nabla_{un}$. Furthermore, since the operator
$\II_\ce \pm \G$ is positive and anticommutes with $\pi(\oca{4})$
\fn{We recall that $\G$ commutes with elements of $\ca$ and anticommutes with
$D$.}, one can prove an inequality \cite{Co1}
\be
\wt{\t}(\q_{un}^2) \geq |\wt{\F}(\q_{un}^2)| ~, ~~\forall ~\nabla_{un} \in CC(\ce)~.
\ee
In turn, by using \ref{ymtr}) and (\ref{ymmin}), one gets the inequality
\be
YM(\nabla) \geq |\wt{\F}(\q_{un}^2)| ~, ~~~ \pi(\nabla_{un}) = \nabla~. 
\ee  

It turn out that $\wt{\F}(\q_{un}^2)$ is a closed {\it cyclic cocycle} and its
topological interpretation in terms of topological invariants of finite
projective modules follows from the pairing between $K$-theory and cyclic
cohomology. Indeed, the value of $\wt{\F}$ does not depend on the particular
connection and one could evaluate it on the curvature $\q_0 = p d p d p$ of
the Grassmannian connection. More, it depends only on the stable
isomorphic class $[p] \in K_0(\ca)$. We refer to \cite{Co1} for details. In
the next section, we shall show that for the canonical triple over an ordinary
four dimensional manifold, the term $\wt{\F}(\q_{un}^2)$ reduces to the usual
topological action.

\subsubsect{The Usual Gauge Theory}\label{se:uym} 
For simplicity we shall consider the case when $n=4$. For the canonical triple
$(\ca, \ch, D, \G)$ over the (four dimensional) manifold $M$ as described in
Section~\ref{se:ctm}, consider  a matrix $A$ of usual $1$-forms and a universal
connection  $\nabla = p \d + \a $ such that $\s_1(\p(\a)) = \g(A)$.
Then $P(\p(\q)) = P(\p(\d \a + \a^2)) = \g(F)$ with $F = d_M A + A \wedge A$. By
using eq.~(\ref{rip}), with an additional  matrix trace over the `internal
indices',  we get  
\bea 
YM(A) &=& inf \{ I(\a) ~|~ \p(\a) = A \}  \nonumber \\ 
&=& {1\over 8 \p^2} \int_M \norm{F}^2 dx~,
\eea
namely, the usual Yang-Mills action of the gauge potential $A$.\\
More explicitly, let $\a = \sum_j f_j \d g_j $. Then, we have 
\bea
&& \pi(\a) = \g^\m A_\m~,~~~ A_\mu = \sum_j f_j \pa_\mu g_j~, \nonumber \\
&& P(\p(\d \a + \a^2)) = \g^{\m\n} F_{\m\n}~,~~~ \g^{\m\n} = {1 \over 2}
(\g^\m \g^\n - \g^\n \g^\m)~. 
\eea
By using the trace theorem \ref{wodzicki} again, (with an additional matrix trace
$Tr$ over the `internal indices') one gets
\bea
YM(A) &=:& {1\over 8 \p^2} 
\int_M tr(\g^{\m\n} \g^{\r\s}) Tr(F_{\m\n}F_{\r\s}) dx \nonumber \\
&=:& {1\over 8 \p^2} 
\int_M g^{\m\s} g^{\n\r} Tr(F_{\m\n}F_{\r\s}) dx \nonumber \\
&=:& {1\over 8 \p^2}  \int_M Tr(F \wedge * F)~.
\eea
With the same token, we get for the topological action
\bea
Top(A) &=:& {1\over 8 \p^2} 
\int_M tr(\G \g^{\m\n} \g^{\r\s}) Tr(F_{\m\n}F_{\r\s}) dx \nonumber \\
&=:& - {1\over 8 \p^2}  
\int_M \ve^{\m\n\r\s} Tr(F_{\m\n}F_{\r\s}) dx \nonumber \\
&=:& - {1\over 8 \p^2} \int_M Tr(F \wedge F)~,
\eea
namely the usual topological action. \\
Here we have used the following (normalized) traces of gamma matrices
\bea
&& tr(\g^\m\g^\n\g^\r\g^\s) = ( g^{\m\n} g^{\r\s} - g^{\m\r} g^{\n\s} + 
g^{\m\s} g^{\n\r} )\\ 
&&
tr(\G \g^\m\g^\n\g^\r\g^\s) = - \ve^{\m\n\r\s}~.  
\eea

\subsubsect{Yang-Mills on a Two Points Space}\label{se:ytp}
We shall first study all modules on the two points space $Y=\{1,2\}$
described in Section~\ref{se:tps}. The associated algebra is $\ca =
\IC\oplus\IC$. The generic module $\ce$ will be of the form $\ce = p
\ca^{n_1}$, with $n_1$ a positive integer, and $p$ a $n_1 \times n_1$
idempotent matrix with entries in $\ca$. The most general such an
idempotent can be written as a diagonal matrix of the form
\be
p = {\rm diag}[ \underbrace{(1,1), \cdots, (1,1)}_{n_1}, 
\underbrace{(1,0), \cdots, (1,0)}_{n_1-n_2},
 ]~, 
\ee 
with $n_2 \leq n_1$. Therefore, the module $\ce$ can be thought of as $n_1$
copies of $\IC$ on the point $1$ and $n_2$ copies of $\IC$ on the point $2$,
\be\label{tpmodu}
\ce = \IC^{n_1} \oplus \IC^{n_2}~.
\ee
The module is trivial if and only if $n_1 = n_2$. There is a {\it topological
number} which measures the triviality of the module and that, in this case, turns
out to be proportional to $n_1-n_2$. From eq. (\ref{curv}), the curvature of
the Grassmannian connection on $\ce$ is just $F_0 = p d p d p$. The mentioned
topological number is then
\be\label{topo}
c(\ce) =: tr \G F_0^2 = tr \G (p d p d p)^2 = tr \G p (d p)^4~.
\ee
Here $\G$ is the grading matrix given by (\ref{tpgrad}) and, the spectral triple
being $0$-dimensional, the Dixmier trace reduces to ordinary trace
\fn{In fact, in (\ref{topo}), $\G$ should really be $\II \otimes \G$.}. 
This is really the same as the topological action $\F(\q_{un}^2)$ encountered
Section~\ref{se:ymm}. It takes some little algebra to find that, for a module of the 
form (\ref{tpmodu}), one has
\be
c(\ce) = tr(M^* M)^4 (n_1 - n_2)~,
\ee
where $M$ is the matrix appearing in the corresponding operator $D$ as in
(\ref{tpkoper}).

\bigskip

Let us now turn to gauge theories. First recall that from the analysis of
Section~\ref{se:atp} there are no junk forms and that Connes' forms are the
image of universal forms through $\pi$, $\omcad = \pi(\omca)$ with $\pi$
injective. We shall consider the simple case of `trivial $1$-dimensional bundle
over' $Y$, namely we shall take as module of sections just $\ce=\ca$.  A vector
potential is then a self-adjoint element $A \in \ocad{1}$ and is determined by a
complex number $\F\in\IC$,
\be\label{fptp}
A=\left[
\begin{array}{cc}
0 & \bar{\F} M^* \\
\F M & 0
\end{array}
\right]~. 
\ee     
If $\a$ is the universal form such that $\pi(\a)=A$, then
\be
\a = -\bar{\F} e\d e - \F (1-e)\d (1-e)~,
\ee
and its curvature is
\be
\d \a + \a^2 = -(\bar{\F} + \F + |\F|^2 ) \d e \d e~.
\ee
Finally, the Yang-Mills curvature turns out to be
\be\label{2pym}
YM(A) =: tr \pi(\d \a + \a^2) ^2 = 2 tr (M^*M)^2 ~( |\F + 1|^2 -1)^2~.
\ee
The gauge group $\cu(\ce)$ is the group of unitary elements of $\ca$, namely 
$\cu(\ce) = U(1) \times U(1)$. Any of its elements $u$ can be represented
as a diagonal matrix
\be
u=\left[
\begin{array}{cc}
u_1 & 0 \\
0 & u_2
\end{array}
\right]~, ~~~ |u_1|^2 = 1~, ~~|u_2|^2 = 1~. 
\ee     
Its action, $A^u = u A u^* + u d u^*$, on the gauge potential results in
multiplication by $u_1^*u_2$ on the variable $\F + 1$,
\be
(\F + 1)^u = (\F + 1)u_1^*u_2~,
\ee 
and the action (\ref{2pym}) is gauge invariant.\\
We see that in this example, the action $YM(A)$ reproduces the usual situation
of broken symmetry for the `Higgs field' $\F + 1$~: there is a $S^1$-worth of
minima which are acted upon nontrivially by the gauge group. This fact has been
used in \cite{CL} in a reconstruction of the Standard Model. The Higgs field has
a geometrical interpretation: it is the component of a gauge connection along
an `internal' discrete direction made of two points.

\subsect{The Bosonic Part of the Standard Model}\label{se:osm}

There are excellent review papers on the derivation of the Standard Model using 
noncommutative geometry, notably \cite{VG, MGV} and \cite{kastler} and we do not
feel the need to add more to those. Rather we shall only overview the main features.
Here we limit ourself to the bosonic content of the model while postponing to
following sections the description of the fermionic part.

In \cite{CL}, Connes and Lott computed the Yang-Mills action $YM(\nabla)$ for a
space which is the product of a Riemannian spin manifold $M$ by an `discrete'
internal space $Y$ consisting of two points. One constructs the product, as
described in Section~\ref{se:pet}, of the the canonical triple $(C^\infty(M),
L^2(M,S), D_S, \G_5)$ on a Riemannian four dimensional spin manifold by the finite
triple 
$(\IC \oplus \IC, \ch_1 \oplus \ch_2, D_F)$ described in Sections~\ref{se:tps}
and~\ref{se:ytp}. The product triple is then given by
\bea
&& \ca =: C^\infty(M) \otimes (\IC \oplus \IC) \simeq C^\infty(M) \oplus
C^\infty(M)~,
\nonumber \\ && \ch =: L^2(M,S) \otimes (\ch_1 \oplus \ch_2) \simeq L^2(M,S) \otimes
\ch_1 \oplus L^2(M,S) \otimes \ch_2~, \nonumber \\
&& D =: D_S \otimes \II + \G_5 \otimes D_F
\eea 
A nice feature of the model is a geometric interpretation of the Higgs field which
appears as the component of the gauge field in the internal direction. Geometrically
one has a space $M \times Y$ with two sheets which are at a distance of the order of
the inverse of the mass scale of the theory (which appears in the operator $D_F$ for
the finite part). Differentiation in the space $M \times Y$ consists of
differentiation on each copy of $M$ together with a finite difference operation in the
$Y$ direction. A gauge potential $A$ decomposes as a sum of an ordinary differential
part $A^{(1,0)}$ and a finite difference part $A^{(0,1)}$ which gives the Higgs field.
To get the full bosonic standard model one has to take for the finite part the algebra
\cite{Co2} 
\be\label{finalgsta}
\ca_F = \IC \oplus \IH \oplus \IM_3(\IC)~,
\ee
$\IH$ being  the algebra of quaternions. The unitary elements of this algebra form
the group $U(1) \times SU(2) \times U(3)$. The finite Hilbert space $\ch_F$ is the
fermion space of leptons, quarks and their antiparticles $\ch_F = \ch_F^+ \oplus 
\ch_F^- = \ch_\ell^+ \oplus \ch_q^+ \oplus \ch_{\bar{\ell}}^- \oplus
\ch_{\bar{q}}^-$. As for the finite Dirac operator $D_F$ is by
\be\label{finalgdir}
D_F = 
\left[
\begin{array}{cc}
Y & 0 \\
0 & \bar{Y}
\end{array}
\right]~,
\ee
with $Y$ the Yukawa coupling matrix. The real structure $J_F$ defined by 
\be
J_F \left(
\begin{array}{c}
\x \\
\bar{\h}
\end{array}
\right) = 
\left(
\begin{array}{c}
\h \\
\bar{\x}
\end{array}
\right)~, ~~~\forall ~~(\xi, \h) \in \ch_F^+ \oplus \ch_F^-~, 
\ee
exchanges fermions with antifermions and it is such that $J^2_F = \II, ~\G_F J_F +
J_F \G_F = 0, ~D_F J_F - J_F D_F = 0$. Next, one defines an action of the algebra
(\ref{finalgsta}) so as to meet the other requirements in the
Definition~\ref{de:rest} of a real structure. For details on this we refer to
\cite{Co2,MGV} as well as for details on the construction of the full bosonic
Standard Model action starting from the Yang-Mills action $YM(\nabla)$ on a `the
rank one trivial' module associated with the product geometry
\bea\label{oldsm}
&& \ca =: C^\infty(M) \otimes \ca_F ~, \nonumber \\ 
&& \ch =: L^2(M,S) \otimes \ch_F~, \nonumber \\
&& D =: D_S \otimes \II + \G_5 \otimes D_F~.
\eea
The product triple has a real structure given by 
\be\label{rsoldsm}
J = C \otimes J_F~, 
\ee
with $C$ the charge-conjugation operation on $L^2(M,S)$ and $J_F$ the real structure
of the finite geometry.\\ 
The final model has problems, notably unrealistic mass relations \cite{MGV} and a
disturbing fermion doubling, the removal of which causes the loss of degrees of
freedom \cite{LMMS2}. It is worth mentioning that while the standard model can be
obtained from noncommutative geometry, most model of the Yang-Mills-Higgs type cannot
\cite{SZ,IS,LMMS1}.

\subsect{The Bosonic Spectral Action}\label{se:bsa}
Recently, in \cite{Co3} Connes  has proposed a new interpretation of gauge degrees of
freedom as the `inner fluctuations' of a noncommutative geometry. This fluctuations
replace the operator $D$, which gives the `external geometry', by $D + A + JAJ^*$,
where $A$ is the gauge potential and $J$ is the real structure. In fact, there is
also a purely geometrical (spectral) action, depending only on the spectrum of the
operator $D$, which, for a suitable algebra (noncommutative geometry of the Standard
Model) gives the Standard Model Lagrangian coupled to gravity. 

Observe first that if $M$ is a smooth (paracompact) manifold, than the group
$Diff(M)$ of diffeomorphisms of $M$, is isomorphic to the group 
$Aut(C^\infty(M))$ of ($^*$-preserving) automorphisms of the algebra
$C^\infty(M)$ \cite{AMR}. Here $Aut(C^\infty(M))$ is the collection of all
invertible, linear maps $\a$ from $C^\infty(M)$ into itself such that
$\a(fg) = \a(f)\a(g)$ and $\a(f^*) = (\a(f))^*$, for any $f,g\in
C^\infty(M)$; $Aut(C^\infty(M))$ is a group under map composition. The
relation between a diffeomorphism $\vf \in Diff(M)$ and the corresponding
automorphism $\a_\vf \in Aut(C^\infty(M))$ is via pull-back
\be
\a_\vf(f)(x) =: f(\vf^{-1}(x))~, ~~~\forall ~f \in C^\infty(M)~, ~x\in M~.
\ee    

If $\ca$ is any noncommutative algebra (with unit) one defines the group $Aut(\ca)$
exactly as before: and $\vf(\II) = \II$, for any $\vf \in Aut(\ca)$. This group will
be the analogue of the group of diffeomorphism of the (virtual) noncommutative space
associated with $\ca$.  Now, with any element $u$ of the unitary group $\cu(\ca)$ of
$\ca$, $\cu(\ca) = \{u \in \ca~, ~u u^* = u^* u = \II \}$, there is an {\it inner
automorphism} $\a_u \in Aut(\ca)$ defined by
\be
\a_u(a) = u a u^*~, ~~~\forall ~a \in \ca~.
\ee 
One can easily convince oneself that $\a_{u^*} \circ \a_u = \a_u \circ \a_{u^*} =
\II_{Aut(\ca)}$, for any $u\in \cu(\ca)$. 
The subgroup $Inn(\ca) \subset Aut(\ca)$ of all inner automorphisms is a normal
subgroup. First of all, any automorphism will preserve the groups of unitaries in
$\ca$. If $u\in \cu(\ca)$ and $\vf \in Aut(\ca)$, then $\vf(u) (\vf(u))^* =
\vf(u) \vf(u^*) = \vf(u u^*) = \vf(\II) = \II$; analogously $(\vf(u))^* \vf(u) =
\II$ and $\vf(u) \in \cu(\ca)$. Furthermore, 
\be\label{normal}
\a_{\vf(u)} = \vf \circ \a_u \circ \vf^{-1} \in Inn(\ca)~, ~~~ \forall ~\vf \in
Aut(\ca)~, ~\a_u \in Inn(\ca)~. 
\ee
Indeed, with $a \in \ca$, for any  $\vf \in Aut(\ca)$ and $\a_u \in Inn(\ca)$ 
one finds
\bea
\a_{\vf(u)}(a) &=& \vf(u) a \vf(u^*) \nonumber \\
&=& \vf(u) \vf(\vf^{-1}(a) \vf(u^*) \nonumber \\
&=& \vf(u \vf^{-1}(a) u^*) \nonumber \\
&=& (\vf \circ \a_u \circ \vf^{-1})(a)~,
\eea
from which one gets (\ref{normal}). \\
By indicating with $Out(\ca) =: Aut(\ca) /
Inn(\ca)$ the outer automorphisms, we have a short exact sequence of groups
\be\label{autseq}
\II_{Aut(\ca)} ~\lra~ Inn(\ca) ~\lra~ Aut(\ca) ~\lra~ Out(\ca) ~\lra~
\II_{Aut(\ca)}~.
\ee
For any commutative $\ca$ (in particular for $\ca = C^\infty(M)$) there are no
nontrivial inner automorphisms and $Aut(\ca) \equiv Out(\ca)$  (in particular
$Aut(\ca) \equiv Out(\ca) \simeq Diff(M)$).

The interpretation that emerges is that the group $Inn(\ca)$ will give `internal'
gauge transformations while the group $Out(\ca)$ will give `external' diffeomorphisms.
In fact, gauge degrees of freedom are the `inner fluctuations' of the  noncommutative
geometry. This is due to the following beautiful fact. Consider the real triple 
$(\ca, \ch, \p, D)$, where we have explicitly indicated the representation $\p$
of the algebra $\ca$ on the Hilbert space $\ch$. The real structures is provided by
the antilinear isometry $J$ with properties as in Definition~\ref{de:rest}. Any
inner automorphism $\a_u \in Inn(\ca)$ will produce a new representation $
\p_u =: \p \circ \a_{u}$ of $\ca$ in $\ch$. It turns out that the
replacement of the representation is equivalent to the replacement of the operator
$D$ by 
\be
D_u = D + A + \ve' J A J^{*}~,
\ee
where $A = u [D, u^* ]$ and $\ve' = \pm 1$ from (\ref{spiche}) according to the
dimension of the triple. If the dimension is four, then $\ve' = 1$; in the
following we shall limit to this case, the generalization being straightforward. \\
This result is to important and beautiful that we shall restate it as a Proposition.
\bprop\label{inneract}
For any inner automorphism $\a_u \in Inn(\ca)$, with $u$ unitary, the triples $(\ca,
\ch, \p, D, J)$ and $(\ca, \ch, \p \circ \a_{u}, D + u [D, u^* ] + J u [D, u^* ]
J^{*}, J)$ are equivalent, the intertwiner unitary operator being given by
\be\label{uniope}
U = u J u J^*~.
\ee
\proof
Note first that
\be\label{unreal}
 U J U^*  = J~.
\ee 
Indeed, by using properties from the Definition~\ref{de:rest} of a real structure,
we have, 
\bea
U J U^* &=& u J u J^* J J u^* J^* u^* \nonumber \\
&=& \pm u J u J^* u^* J^* u^* \nonumber \\ 
&=& \pm J u J^* u u^* J^* u^* \nonumber \\ 
&=& J~.  
\eea
Furthermore, by dropping again the symbol $\pi$, we have to check that
\bea
&& U a U^*  = \a_{u}(a)~, ~~~\forall ~a\in \ca~, \label{uneq1} \\
&& U D U^*  = D_u~. \label{uneq2}
\eea
As for (\ref{uneq1}), for any $a \in \ca$ we have,
\bea
U a U^* &=& u J u J^* a J u^* J^* u^* \nonumber \\
&=& u J u J^* J u^* J^* a u^* ~~~~~{\rm by ~2a. ~in ~Definition~\ref{de:rest}}
\nonumber \\
&=& u a u^* \nonumber \\
&=& \a_{u }(a)~,
\eea
which proves (\ref{uneq1}). Next, by using properties $1b.$ and $2a., 2b.$ of
Definition~\ref{de:rest} (and their analogues with $J$ and $J^*$ exchanged) the left
hand side of (\ref{uneq2}) is given by
\bea
U D U^* &=&  u J u J^*  D J u^* J^* u^* \nonumber \\
 &=&  u J u D u^* J^* u^* \nonumber \\ 
 &=&  u J u (u^* D + [D, u^*]) J^* u^* \nonumber \\ 
 &=&  u J D J^* u^* + u J u [D, u^*] J^* u^* \nonumber \\ 
 &=&  u D u^* + J J^* u J u [D, u^*] J^* u^* \nonumber \\ 
 &=&  u (u^* D + [D, u^*]) + J u J^* u J [D, u^*] J^* u^* \nonumber \\ 
 &=&  D + u [D, u^*] + J u [D, u^*] J^* u J J^* u^* \nonumber \\ 
 &=&  D + u [D, u^*] + J u [D, u^*] J^* ~, \nonumber \\ 
\eea
and (\ref{uneq2}) is proven.
\eprop

\noindent
The operator $D_u$ is interpreted as the product of the perturbation of the
`geometry' given by the operator $D$, by `internal gauge degrees of freedom' given
by the gauge potential $A = u^* [D, u]$. A general {\it internal perturbation of
the geometry} is provided by
\be\label{gpert}
D ~\mapsto~ D_A = D + A + J A J^*~, 
\ee  
where $A$ is an arbitrary {\it gauge potential}, namely an arbitrary Hermitian
operator, $A^* = A$, of the form
\be
A = \sum_j a_j [D, b_j]~, ~~~a_j, b_j ~\in \ca~.
\ee

The dynamics of the coupled gravitational and gauge degrees of freedom is
governed by a {\it  spectral action principle}. The action is a `purely geometric'
one depending only on the spectrum of the self-adjoint operator $D_A$ \cite{Co3,CC}, 
\be\label{spac}
S_B(D,A) = tr_\ch(\c({D_A^2 \over \L^2}))~.
\ee 
Here $tr_\ch$ is the {\it usual} trace in the Hilbert space $\ch$, $\L$ is a `cut
off parameter' and $\c$ is a suitable  function which cut off all eigenvalues of
$D_A^2$ larger than $\L^2$. 

The computation of the action (\ref{spac}) is conceptually simple although technically
it may be involved. One has just to compute the square of the Dirac operator with
Lichn\'erowicz' formula \cite{BGV} and the trace with a suitable heat kernel
expansions \cite{Gi}, to get an expansion in terms of powers of the parameter
$\L$. The action (\ref{spac}) is interpreted in the framework of Wilson's
renormalization group approach to field theory: it gives the {\it bare} action with
{\it bare coupling constants}. There exists a cut off scale $\L_P$ which regularizes
the action and where the theory is geometric. The renormalized action will have the
same form as the bare one with bare parameters replaced by physical parameters
\cite{CC}. \\
In fact, a full analysis is rather complicated and there are several caveats
\cite{EGV}. 

In Section~\ref{se:gra} we shall work out in detail the action  for the
usual gravitational sector while here we shall indicate how to work out it for a 
generic gauge fields and in particular for the bosonic sector of the standard model.
\\ 
We first proceed with the `mathematical aspects'.

\bprop\label{spacgainv}
The spectral action (\ref{spac}) is invariant under the gauge action of the inner
automorphisms given by
\be\label{spaction}
A \mapsto A^u =: u A u^* + u [D, u^*]~, ~~~\forall ~u \in Inn(\ca)~.
\ee
\proof
The proof amount to show that 
\be\label{trgadi}
D_{A^u} = U D_A U^*~,
\ee
with $U$ the unitary operator in (\ref{uniope}), $U = u J u J^*$. Now, given
(\ref{spaction}), it turns out that
\bea
D_A &=:& D + A^u + J A^u J^* \nonumber \\
&=& D + u [ D,u^*] + J [ D,u^*] J^* + u A u^* + J u A u^* J^* \nonumber \\
&=& D_u + u A u^* + J u A u^* J^*.
\eea
In Proposition~\ref{inneract} we have already proved that $D_u = U D U^* $, 
eq.(~\ref{uneq2}). To prove the rest, remember that $A$ is of the form $A =\sum_j a_j
[D, b_j]$ with $a_j, b_j \in \ca$. But, from properties $2a.$ and $2b.$ of
Definition~\ref{de:rest}, it follows that $[A, J c^* J^*] = 0$, for any $c \in \ca$.
By using this fact and properties $2a.$ and $2b.$ of Definition~\ref{de:rest} (and
their analogues with $J$ and $J^*$ exchanged) we have that,
\bea
U A U^* &=&  u J u J^* A J u^* J^* u^* \nonumber \\
&=&  u J u J^*  J u^* J^* A u^* \nonumber \\
&=& u A u^*~. 
\eea
\bea
U (J A J^*) U^*  &=& u J u J^* J A J^* J u^* J^* u^* \nonumber \\
&=& u J u A u^* J^* u^* J J^* \nonumber \\
&=& u J u A J^* u^* J u^* J^* \nonumber \\
&=& u J u  J^* u^* J A u^* J^* \nonumber \\
&=& J u  J^* u u^* J A u^* J^* \nonumber \\
&=& J u A u^* J^* ~. 
\eea
The previous two results together with (~\ref{uneq2}) prove eq.~(\ref{trgadi}) and,
in turn, the proposition.
\eprop

Before proceedings, let us observe that for commutative algebras, the internal
perturbation $A + J A J^*$ of the metric in (\ref{gpert}) vanish. From what we said
after Definition~\ref{de:rest}, for commutative algebras one can write $a = J a^*
J^*$ for any $a\in \ca$, which amount to identify the left multiplicative action by
$a$ with the right multiplicative action by $J a^*J^*$ (always possible if $\ca$ is
commutative). Furthermore,
$D$ is a differential operator of order $1$, namely $[[D,a],b]] = 0$ for any $a,b \in
\ca$. Then, with 
$A=\sum_j a_j [D, b_j]$, $A^* = A$, we get 
\bea
J A J^* &=& \sum_j J a_j [D, b_j] J^*  
 = \sum_j J a_j J J^* [D, b_j] J^*  \nonumber \\ 
&=& \sum_j a_j^* J [D, b_j] J^* 
 = \sum_j a_j^* [D, J b_j J^*]  \nonumber \\
&=& \sum_j a_j^* [D, b_j^* ]   
 = \sum_j [D, b_j^* ] a_j^*  \nonumber \\
&=& - ( a_j \sum_j [D, b_j] )^* 
 = - A^*~,
\eea
and, in turn, $A + J A J^* = A - A^* = 0$.

\bigskip

In the usual approach to gauge theories, one constructs connections on a
principal bundle $P \ra M$ with structure group a finite dimensional Lie group $G$. 
Associated with this bundle there is a sequence of infinite dimensional
(Hilbert-Lie) groups which looks remarkably similar to the sequence (\ref{autseq})
\cite{gauge,Tr},
\be
\II ~\lra~ \cg ~\lra~ Aut(P) ~\lra~ Diff(M) ~\lra~ \II~.
\ee
Here $Aut(P)$ is the group of automorphism of the total space $P$, namely
diffeomorphisms of $P$ which commutes with the action of $G$, and $\cg$ is  the
subgroup of vertical automorphisms, identifiable with the group of gauge 
transformations $\cg \simeq C^\infty(M,G)$. \\
Thus, here is the recipe to construct a spectral gauge theory corresponding to the
structure group $G$ or equivalently to the gauge group $\cg$ \cite{CC}: 
\begin{enumerate}
\item look for an algebra $\ca$ such that $Inn(\ca) \simeq \cg$;
\item construct a suitable spectral triple `over' $\ca$;
\item compute the spectral action (\ref{spac}).
\end{enumerate}
The result would be a gauge theory of the group $G$ coupled with gravity of the
diffeomorphism group $Out(\ca)$ (with additional extra terms). 

For the standard model we have $G= U(1) \times SU(2) \times SU(3)$. It turns out 
that the relevant spectral triple is the one in (\ref{oldsm}), (\ref{rsoldsm}). 
In fact, as already mentioned in Section~\ref{se:osm}, for this triple the structure
group would be $U(1) \times SU(2) \times U(3)$; however the computation of $A + J A
J^*$ removes the extra $U(1)$ part from the gauge fields. The associated spectral
action has been computed in \cite{CC} and in full details in \cite{IKS}. The result
is the Yang-Mill-Higgs part of the standard model coupled with Einstein gravity plus a
cosmological term, a term of Weyl gravity and a topological term. 
Unfortunately the model still suffers from the problems alluded at the end of
Section~\ref{se:osm}: namely unrealistic mass relations and an unphysical fermion
doubling.

\subsect{Fermionic Models}\label{se:fem}  
It is also possible to construct the analogue of a gauged Dirac operator by a
`minimal coupling' recipe and an associated action.

If we have a gauge theory on the trivial module $\ce = \ca$ as in
Sec.~\ref{se:electro}, then a gauge potential is just a self-adjoint element 
$A \in \ocad{1}$ which transforms under the unitary group $\cu(\ca)$ by 
(\ref{trapot}),
\be\label{trapot1} 
(A, u) \lra A^u = u A u^* + u [D, u^*]~, ~~~\forall ~u \in \cu(\ca)~.
\ee 
Then, the following expression in gauge invariant,
\be\label{ferm0} 
I_{Dir}(A, \j) =: \hs{ \j}{(D + A) \j} ~, ~~~\forall ~\j \in Dom(D) \subset 
\ch~, ~~A \in \ocad{1}~,
\ee 
where the action of the group $\cu(\ca)$ on $\ch$ is by restriction of the
action of $\ca$. Indeed, for any $\j \in \ch$, one has that 
\bea\label{giferm} (D + A^u) u \j &=& (D + u [D, u^*] + u A u^* ) u \j \nonumber
\\ &=& D(u \j) + u (D u^* - u^* D)(u \j) + u A \j \nonumber \\ &=& u D u^* (u
\j) + u A\j \nonumber \\ &=& u (D + A) \j~, 
\eea   from which the invariance of (\ref{ferm0}) follows.

The generalization to any finite projective module $\ce$ over $\ca$ 
endowed with a Hermitian structure, needs extra care but is 
straightforward. In this case
one considers the Hilbert space $\ce \ota \ch$ of `gauged spinors' introduced in
the previous section and with scalar product given in (\ref{modhil}). The action
of the group
$End_{\ca}(\ce)$ of endomorphisms of $\ce$ extends to an action on $\ce \ota \ch$
by
\be\label{auteh}
\f (\h \otimes \j) =: \f(\h) \otimes \j~, ~~~\forall ~\f \in End_{\ca}(\ce)~,
~~\h \otimes \j \in \ce \ota \ch~.
\ee 
In particular, the unitary group $\cu(\ce)$ yields a unitary action on $\ce
\ota
\ch$,
\be\label{unieh} 
u(\h \otimes \j) =: u(\h) \otimes \j~, ~~ u \in \cu(\ce)~, 
~~\h \otimes \j \in \ce \ota \ch~,
\ee 
since
\bea 
(u(\h_1 \otimes \j_1), u(\h_2 \otimes \j_2) )  &=&  ( \j_1 , \hs{u(\h_1)}{
u(\h_2)} \j_2 ) \nonumber \\  &=&  ( \j_1 , \hs{\h_1}{\h_2} \j_2 ) \nonumber \\ 
&=& (\h_1 \otimes \j_1, \h_2 \otimes \j_2 )~, \nonumber \\  
& & ~~~\forall ~u \in \cu(\ce)~, ~~\h_i \otimes \j_i \in \ce  \ota \ch~, ~i = 1,2~.
\eea

If $\nabla : \ce \ra \ce \ota \ocad{1}$ is a compatible connection on $\ce$, the
associated `gauged Dirac operator' $D_{\nabla}$ on the Hilbert space $\ce \ota
\ch$ is defined by
\be\label{gadi} 
D_{\nabla}(\h \otimes \j) = \h \otimes D \j + (( \II \otimes \p)
\nabla_{un} \h) \j~,  ~~~ \h \in \ce~, ~~ \j \in \ch~,
\ee 
where $\nabla_{un}$ is any universal connection on $\ce$ which projects onto
$\nabla$.

If $\ce=p\ca^N$, and $\nabla_{un}=p \d + \a$, then the operator in (\ref{gadi})
can be written as  
\be\label{gadi1} 
D_{\nabla} = p D + \p (\a)~,
\ee 
with $D$ acting component-wise on $\ca^N \otimes \ch$. Since $\p(\a)$ is a
self-adjoint operator, from (\ref{gadi1}), we see that $D_{\nabla}$ is a
self-adjoint operator on $\ce \ota \ch$ with domain $\ce \ota Dom D$.
Furthermore, since any two universal connections projecting on $\nabla$ differ
by $\a_1 -  \a_2 \in ker \p$, the right-hand side of (\ref{gadi}) depends only
on $\nabla$. Notice that one cannot write directly $(\nabla \h) \j$ since 
$\nabla \h$ is not an operator on $\ce \ota \ch$.

\bprop The gauged Dirac action
\be\label{diac} 
I_{Dir}(\nabla, \J) =: \hs{ \J}{D_{\nabla} \J}~, ~~~ \forall 
~\J \in \ce \ota Dom D~, ~~~ \nabla \in C(\ce)~.
\ee 
is invariant under the action (\ref{unieh}) of the unitary group
$\cu(\ce)$.  

\proof The proof goes along the same line of (\ref{giferm}). For any $\J \in \ce
\ota\ch$, one has that 
\bea (p D + \p( \a ^u) ) u \J &=& ( p D + \p( u \d u^* + u \a u^* )) u \J
\nonumber \\ &=& p D(u \J) + u (D u^* - u^* D)(u \J) + u \p (\a) \J \nonumber \\
&=& p D(u \J) + p u (D u^* - u^* D)(u \J) + u \p (\a) \J \nonumber \\ &=& p u D
u^* (u \J) + u \p(\a) \J \nonumber \\ &=& u p D u^* (u \J) + u \p(\a) \J
\nonumber \\ &=& u (p D + \p(\a)) \J~, 
\eea   which implies the invariance of (\ref{diac}).
\eprop

\subsubsect{Fermionic Models on a Two Points Space}\label{se:ftp}
As a very simple example, we shall construct the fermionic Lagrangian (\ref{ferm0})
on the two point space $Y$ studied in Sections~\ref{se:tps} and \ref{se:ytp}, 
\be
I_{Dir}(A, \j) =: \hs{ \j}{(D + A) \j } ~, ~~~\forall ~\j \in Dom(D) \subset 
\ch~, ~~A \in \ocad{1}~,
\ee 
As seen in Section~\ref{se:tps}, the finite dimensional Hilbert
space $\ch$ is a direct sum $\ch = \ch_1 \oplus \ch_2$ and the operator $D$ is an
off-diagonal matrix
\be
D =
\left[
\begin{array}{cc}
0 & M^* \\
M & 0
\end{array}
\right]~, ~~ M \in Lin(\ch_1,\ch_2)~.  
\ee 
In this simple example $Dom(D) = \ch$. 
On the other-side, the generic gauge potential on the trivial module
$\ce=\ca$ is given by (\ref{fptp}),
\be
A=\left[
\begin{array}{cc}
0 & \bar{\F} M^* \\
\F M & 0
\end{array}
\right]~, ~~~ \F \in \IC~.
\ee     
Summing up, the gauged Dirac operator is the matrix
\be
D + A = \left[
\begin{array}{cc}
0 & (1+\bar{\F}) M^* \\
(1+\F) M & 0
\end{array}
\right]~, 
\ee     
which gives for the action $I_{Dir}(A, \j)$ a Yukawa-type term coupling the fields
$(1+\F)$ and $\j$ and invariant under the gauge group $\cu(\ce) = U(1) \times U(1)$. 

\subsubsect{The Standard Model}\label{se:oldfstm}

Let us now put together the Yang-Mill action (\ref{ym}) with the fermionic one in
(\ref{diac}),
\be\label{stan}
\begin{array}{llll}
I(\nabla, \J) & = & YM(\nabla) + I_{Dir}(\nabla, \J) & ~ \\
~ & = & \hs{F_\nabla}{F_\nabla}_2 + \hs{ \J}{D_{\nabla} \J }~, ~~~
\forall & ~\nabla \in C(\ce)~, \\ 
& ~ & & ~\J \in \ce \ota Dom D~.
\end{array}
\ee

Consider then the canonical triple $(\ca, \ch, D)$ on a Riemannian spin 
manifold. By taking $\ce = \ca$, the action (\ref{stan}) is just the Euclidean
action of massless quantum electrodynamics. If $\ce = \ca^N$, the action
(\ref{stan})   is the Yang-Mills action for $U(N)$ coupled with a massless fermion
in the fundamental representation of the gauge group $U(N)$ \cite{Co2}.  

In \cite{CL}, the action (\ref{stan}) for a product space of a Riemannian spin
manifold $M$ by an `discrete' internal space $Y$ consisting of two points. They
obtained the full Lagrangian of the standard model. An improved version which uses
a real spectral triple and done by means of a spectral action along the lines of
Section~\ref{se:bsa} will be briefly described in next Section.

\subsect{The Fermionic Spectral Action}\label{se:fsa}

Consider a real spectral triple $(\ca, \ch, D, J)$. And recall from
Section~\ref{se:bsa} the interpretation of gauge degrees of freedom as `inner
fluctuations' of a noncommutative geometry, fluctuations which replace the operator
$D$ by $D + A + JAJ^*$, where $A$ is the gauge potential. 

Well, the fermionic spectral action is just given by 
\be\label{spacfer}
S_F(\j, A, J) =: \hs{\j}{D_A \j} = \hs{\j}{D + A + J A J^*) \j}~, 
\ee
with $\j \in \ch$. The previous action again depends only on spectral properties of
the triple. \\
By using the $\ca$-bimodule structure on $\ch$ in (\ref{bist}), we get an `adjoint
representation' of the unitary group $\cu(\ca)$ by unitary operators on $\ch$,
\be\label{unacspi}
\ch \times \cu(\ca) \ni (\j, u) \ra \j^u =: u ~\x~ u^* =  u J u J^*~ \j \in \ch~. 
\ee   
That this action preserves the scalar product, namely $\hs{\j^u}{\j^u} =
\hs{\j}{\j}$, follows from the fact that both $u$ and $J$ act as isometries.
\bprop\label{spacfergainv}
The spectral action (\ref{spacfer}) is invariant under the gauge action of the inner
automorphisms given by (\ref{unacspi}) and (\ref{spaction}),
\be
S_F(\j^u, A^u, J) = S_F(\j, A, J)~, ~~~\forall ~u \in \cu(\ca)~.
\ee
\proof
By using the result (\ref{trgadi}) $D_{A^u} = U D_A U^*$, with $U = u J u J^*$, we
get
\bea
S_F(\j^u, A^u, J) &=& \hs{\j^u}{D_{A^u} \j^u} \nonumber \\
&=& \hs{\j J u^* J^* u}{ U D_A U^*) u J u J^*~ \j} \nonumber \\
&=& \hs{\j J u^* J^* u}{ u J u J^*~ D_A J u^* J^* u  u J u J^*~ \j} \nonumber \\
&=& \hs{\j}{D_{A} \j} \nonumber \\
&=& S_F(\j, A, J)
\eea
\eprop

For the spectral triple of the standard model in (\ref{oldsm}), (\ref{rsoldsm}),
(\ref{finalgsta}),  (\ref{finalgdir}), the action (\ref{spacfer}) gives the
fermionic sector of the standard model \cite{Co2,MGV}. It is worth stressing that
although the noncommutative fermionic multiplet $\j$ transforms by the adjoint
representation (\ref{unacspi}) of the gauge group, the physical fermion fields will
transform in the fundamental representation while the antifermions will transform
in the conjugate one.

\vfill\eject
\sect{Gravity Models}\label{se:gra}

We shall describe three possible approaches (two, in fact, since as we
shall see the first two are really the same) to the construction of gravity
models in noncommutative geometry which, while agreeing for the canonical
triple associated with an ordinary manifold (and reproducing the usual
Einstein theory), seem to give different answers for more general examples. \\
As a general remark, we should like to mention that a noncommutative
recipe to construct  gravity theories (at least the usual Einstein
one)  has to consider the metric as a dynamical variable not given a priori.
In particular, one should not start with the Hilbert space $\ch = L^2(M,S)$ of
spinor fields whose scalar product uses a  metric on $M$ which, therefore,
would plays the role of a background metric. The beautiful result by Connes
\cite{Co2} which we recall in the following Section goes exactly in this
direction. A possible alternative way has been devised in \cite{LR2}.        

\subsect{Gravity \`a la Connes-Dixmier-Wodzicki}
The first scheme to construct gravity models in noncommutative geometry, and
in fact to reconstruct the full geometry out of the algebra $C^\infty(M)$, is
based on the use of the Dixmier trace and the Wodzicki residue
\cite{Co3}, which we have studied at length in
Sections~\ref{se:dix} and~\ref{se:wrc}.
\bprop
Suppose we have a smooth compact manifold $M$ without boundary and of dimension
$n$. Let $\ca = C^\infty(M)$ and $D$ just a `symbol' for the time being. Let
$(\ca_\p , D_\p)$ be a unitary representation of the couple $(\ca, D)$ as
operators on an Hilbert space $\ch_\p$ endowed with an operator $J_\p$, such
that $(\ca_\p , D_\p, \ch_\p, J_\p)$ satisfy all axioms of a real spectral
triple given in Section~\ref{se:spt}. \\
Then,
\begin{itemize}
\item[a)]
There exists a unique Riemannian metric $g_\p$ on $M$ such that the
geodesic distance between any two points on $M$ is given by
\be 
d(p,q) = \sup_{a\in \ca} \{ |a(p) - a(q)| ~:~ \norm{[D_\p,
\p(a)]}_{\cb(\ch_\p)} \leq 1 \}~, ~~\forall~ p,q \in M~.  
\ee 
\item[b)]
The metric $g_\p$ depends only on the unitary equivalence class of the
representation $\p$. The fibers of the map $\p \mapsto g_\p$ from unitary
equivalence classes of representations to metrics form a finite collection of
affine spaces $\ca_\s$ parameterized by the spin structures $\s$ on $M$.
\item[c)]
The action functional given by the Dixmier trace  
\be
G(D) = tr_\o (D^{n-2})~,
\ee
is a positive quadratic form with a unique minimum $\p_\s$ on on each $\ca_\s$.
\item[d)]
The minimum $\p_\s$ is the representation of $(\ca, D)$ on the Hilbert space
of square integrable spinors $L^2(M,S_\s$); $\ca_\s$ acts by multiplicative
operators and $D_\s$ is the Dirac operator of the Levi-Civita connection.
\item[e)]
At the minimum $\p_\s$, the values of $G(D)$ coincides with the Wodzicki
residue of $D_\s^{n-2}$ and is proportional to the Hilbert-Einstein action of
general relativity
\bea\label{wgrav} 
G(D_\s) = Res_W(D_\s^{n-2}) &=:& {1 \over n (2\pi)^n} \int_{S^* M} tr
(\s_{-n}(x,\x)) dx d\x \nonumber \\ 
&=& c_n \int_M R dx ~, \nonumber \\
& ~ & \nonumber \\
c_n &=& {(n-2) \over 12}{2^{ [n/2] - n/2} \over (2\pi)^{n/2} } \G({n \over 2}
+1) ^{-1}~.
\eea 
Here,
\be
\s_{-n}(x,\x) = {\rm part ~of ~order} ~-n ~{\rm of ~the ~total ~symbol ~of}
~D_\s^{n-2}~,
\ee
$R$ is the scalar curvature of the metric of $M$ and $tr$ is a normalized
Clifford trace. 
\item[f)]
If there is no real structure $J$, one has to replace spin above by
spin$^c$. Uniqueness of point c) is lost and the minimum of the functional
$G(D)$ is reached on a linear subspace of $\ca_\s$ with $\s$ a fixed spin$^c$
structure. This subspace is parameterized by the $U(1)$ gauge potentials
entering in the spin$^c$ Dirac operator. Point d) and c) still hold. In
particular the extra terms coming from the $U(1)$ gauge potential drop out
from the gravitational action $G(D_\s)$.
\end{itemize}

\proof 
At the moment, a complete proof of this theorem goes beyond our means (and the
scope of these notes). We only mention that for $n=4$ equality (\ref{wgrav})
was proved by `brute force' in \cite{Ka} by means of symbol calculus of
pseudodifferential operators. There it was also proved that the results
does not depends upon the extra contributions coming from the $U(1)$ gauge
potential. In \cite{KW}, equality (\ref{wgrav}) was proved in any dimension
by realizing that $Res_W(D_\s^{n-2})$ is (proportional) to the integral of the
second coefficient of the heat kernel expansion of $D_\s^2$. It is this fact
that relates the previous theorem to the spectral action for gravity as we
shall see in the next section.
\eprop

\noindent
Finally, we mention, with Connes, that the fact that $\ca$ is the
algebra of smooth functions on a manifold can be recovered a posteriori as
well. Connes axioms allow to recover the spectrum of $\ca$ as a smooth manifold
(a smooth submanifold of $\IR^N$ for a suitable $N$) \cite{Co2}.

\subsect{Spectral Gravity}\label{spgra}
In this section we shall compute the spectral action (\ref{spac}) described in
Section~\ref{se:bsa} for the  purely gravitational sector. Consider then the
canonical triple $(\ca, \ch, D)$  on a closed $n$-dimensional Riemannian spin 
manifold  $(M,g)$ which we have described in Section~\ref{se:ctm}.  We recall
that $\ca=C^\infty(M)$ is the algebra of complex valued smooth  functions on
$M$; $\ch=L^2(M,S)$ is the Hilbert space of square integrable  sections of the
irreducible, rank $2^{[n/2]}$ spinor bundle over $M$; finally, $D$ is the Dirac
operator of the Levi-Civita spin connection.

The action we need to compute is 
\be\label{spac1}
S_G(D,\L) = tr_\ch(\c({D^2 \over \L^2}))~.
\ee 
Here $tr_\ch$ is the usual trace in the Hilbert space $\ch=L^2(M,S)$,
$\L$ is the cutoff parameter and $\c$ is a suitable cutoff function which
cut off all eigenvalues of $D^2$ larger than $\L^2$. As already mentioned this
action depends only on the spectrum of $D$. 

Before we proceed let us spend few words on the problem of {\it spectral
invariance versus diffeomorphism invariance}. Let us indicate by $spec(M,D)$
the spectrum of the Dirac operator with each eigenvalue repeated according to
its multiplicity. Two manifolds $M$ and $M'$ are called {\it isospectral} if
$spec(M,D) = spec(M,D)$ 
\fn{In fact, one usually take the Laplacian instead of the Dirac operator.}.
From what said, the action (\ref{spac1}) is a {\it spectral invariant}.
Now, it is well know that one  cannot {\it hear the shape of a drum}
\cite{Kac,Mil} (see also \cite{Gi,drum} and references therein), namely there
are manifold which are isospectral without being isometric (the converse
is obviously true). Thus, spectral invariance is {stronger} that
usual diffeomorphism invariance.

The Lichn\'erowicz formula (\ref{lich}) gives the square of the Dirac
operator
\be\label{lich1}
D^2 = \nabla^S + {1 \over 4} R~.
\ee
with $R$ the scalar curvature of the metric and $\nabla^S$ the Laplacian
operator lifted to the bundle of spinors,
\be
\nabla^S = -g^{\m\n}(\nabla^S_\m \nabla^S_\n - \G^\r_{\m\n}\nabla^S_\r)~;
\ee
and $\G^\r_{\m\n}$ are the Christoffel symbols of the connection.\\
The heat kernel expansion \cite{Gi,CC}, allows to write the action
(\ref{spac1}) as an expansion
\be
S_G(D,\L) = \sum_{k \geq 0} f_k a_k(D^2 / \L^2)~, 
\ee 
where the coefficients $f_k$ are given by
\bea
&& f_0 = \int_0^\infty \c(u) u d u~, \nonumber \\
&& f_2 = \int_0^\infty \c(u) d u~, \nonumber \\
&& f_{2(n+2)} = (-1)^n \c^{(n)}(0)~, ~~ n \geq 0~, 
\eea
and $\c^{(n)}$ denotes the $n$-th derivative of the function $\c$ with respect
to its argument. \\
The Seeley-de Witt coefficients $a_k(D^2 / \L^2)$ vanishes for odd
values of $k$. The even ones are given as integrals
\be\label{sdwc}
a_k(D^2 / \L^2) = \int_M a_k(x; D^2 / \L^2) \sqrt{g} d x~.
\ee
The first three coefficients, for even $k$, are given by
\bea
&& a_0(x; D^2 / \L^2) = (\L^2)^2 ~(4\p)^{-n/2} ~tr \II_{2^{[n/2]}}~, 
\nonumber \\ 
&& a_2(x; D^2 / \L^2) = (\L^2)^1 ~(4\p)^{-n/2} ~(-{R \over 6} + E) 
~tr \II_{2^{[n/2]}}  \nonumber \\ 
&& a_4(x; D^2 / \L^2) = (\L^2)^0 ~(4\p)^{-n/2} {1 \over 360} 
~(- 12 R_{;\m}^{~\m} + 5 R^2 - 2 R_{\m\n} R^{\m\n} \nonumber \\ 
&& ~~~~~~~~~~~~~~~~~~~~~~~~~ 
- {7 \over 4} R_{\m\n\r\s} R^{\m\n\r\s} - 60 R
E + 180 E^2  + 60 E_{;\m}^{~\m}) ~tr \II_{2^{[n/2]}} ~.
\eea
Here $ R_{\m\n\r\s}$ are the component of the Riemann tensor, $R_{\m\n}$
the component of the Ricci tensor and $R$ is the scalar curvature. As for $E$,
it is given by $E =: D^2 - \nabla^S = {1 \over 4} R$. By substituting back in
(\ref{sdwc}) and by taking the integrals we get
\bea
&& a_0(D^2 / \L^2) = (\L^2)^2 ~{ 2^{[n/2]} \over (4\p)^{n/2}} 
~\int_M \sqrt{g} d x ~, \nonumber \\ 
&& a_2(D^2 / \L^2) = (\L^2)^1 ~{ 2^{[n/2]} \over (4\p)^{n/2}} 
{1 \over 12} ~\int_M \sqrt{g} d x ~R ~, \nonumber \\ 
&& a_4(D^2 / \L^2) = (\L^2)^0 ~{ 2^{[n/2]} \over (4\p)^{n/2}} {1 \over 360} 
~\int_M \sqrt{g} d x ~( 3 R_{;\m}^{~\m} + {5 \over 4} R^2 \nonumber \\
&& ~~~~~~~~~~~~~~~~~~~~~~~~~~~~~~~~~~~ 
- 2 R_{\m\n} R^{\m\n}  - {7 \over 4} R_{\m\n\r\s} R^{\m\n\r\s} ) ~.
\eea
Summing up, the action (\ref{spac1}) turns out to be
\bea\label{spac2}
S_G(D,\L) &=& tr_\ch(\c({D^2 \over \L^2})) \nonumber \\
&=& (\L^2)^2 f_0 ~{ 2^{[n/2]} \over (4\p)^{n/2}} 
~\int_M \sqrt{g} d x   \nonumber \\ 
&+& (\L^2)^1 f_2 ~{ 2^{[n/2]} \over (4\p)^{n/2}} 
{1 \over 12} ~\int_M \sqrt{g} d x ~R   \nonumber \\
&+& (\L^2)^0 f_4 ~{ 2^{[n/2]} \over (4\p)^{n/2}} {1 \over 360} 
~\int_M \sqrt{g} d x ~( 3 R_{;\m}^{~\m} + {5 \over 4} R^2 \nonumber \\
&& ~~~~~~~~~~~~~~~~~~~~~~~~~~~~~~~~~~~ 
- 2 R_{\m\n} R^{\m\n}  - {7 \over 4} R_{\m\n\r\s} R^{\m\n\r\s} ) \nonumber \\
&+& O((\L^2)^{-1})~. 
\eea
The action is dominated by the first term, a huge cosmological constant. \\ 
By using for $\c$ the characteristic value of the interval $[0,1]$, namely
$\c(u) = 1, ~u\leq 1, ~\c(u) = 0, ~u\geq 1$, possibly `smoothed out' at $u=1$,
we get 
\bea
&& f_0 = 1 / 2 ~, \nonumber \\
&& f_2 = 1~, \nonumber \\
&& f_{2(n+2)} = 0~, ~~ n \geq 0~,
\eea
and the action (\ref{spac2}) becomes
\be
S_G(D,\L) = (\L^2)^2 {1 \over 2} ~{ 2^{[n/2]} \over (4\p)^{n/2}} 
~\int_M \sqrt{g} d x + (\L^2)^1  ~{ 2^{[n/2]} \over (4\p)^{n/2}} 
{1 \over 12} ~\int_M \sqrt{g} d x ~R ~.
\ee
In \cite{LR1} the following trick was suggested to eliminate the cosmological
term: replace the function $\c$ by $\wt{\c}$ defined as
\be
\wt{\c}(u) = \c(u) - a \c(b u)~,
\ee
with $a,b$ any two numbers such that $a = b^2$ and $b \geq 0, b \not= 1$.
Indeed, one easily finds out that,
\bea
&& \wt{f}_0 =: \int_0^\infty \wt{\c}(u) u d u = (1 - {a \over b^2}) f_0 = 0~,
\nonumber \\ 
&& \wt{f}_2 =: \int_0^\infty \wt{\c}(u) d u = (1 - {a \over b}) f_2 ~, 
\nonumber \\ 
&& \wt{f}_{2(n+2)} =: (-1)^n \wt{\c}^{(n)}(0) = (-1)^n (1 - a b^n)
\c^{(n)}(0)~, ~~ n \geq 0~. 
\eea
The action (\ref{spac1}) becomes
\be\label{spacmod}
\wt{S}_G(D,\L) = (1-{a \over b}) f_2 (\L^2)^1  ~{ 2^{[n/2]} \over
(4\p)^{n/2}}  {1 \over 12} ~\int_M \sqrt{g} d x ~R ~ + O((\L^2)^{0}).
\ee

We finish by mentioning that in \cite{LR1}, in the
spirit of spectral gravity, the eigenvalues of the Dirac operator, which are
diffeomorphic invariant functions of the geometry and therefore true observable
in general relativity, have been taken as a set of variables for an invariant
description of the dynamics of the gravitational field. The Poisson brackets of
the eigenvalues was computed and found in terms of the energy-momentum of the
eigenspinors and of the propagator of the linearized Einstein equations. The
eigenspinors energy-momenta form the Jacobian of the transformation of
coordinates from metric to eigenvalues, while the propagator appears as the
integral kernel giving the Poisson structure. The equations of motion of
the modified action (\ref{spacmod}) are satisfied if the trans Planckian
eigenspinors scale linearly with the eigenvalues: this requirement approximate
Einstein equations.

As already mentioned, there exist isospectral manifolds which fail
to be isometric. Thus, the eigenvalues of the Dirac operator cannot be used
to distinguish among such manifolds (should one really do that from a physical
point of view?).  A complete analysis of this problem and of its consequences
should await another time.

\subsect{Linear Connections}\label{se:lc}

A different approach to gravity theory, developed in \cite{CFF,CFG}, is based
on a theory of {\it linear connections} on an analogue of the cotangent bundle
in the noncommutative setting. It turns out the the analogue of the cotangent
bundle is more appropriate that the one of tangent bundle. One could define the
(analogue) of `the space of sections of the tangent bundle' as the space of
derivations
$Der(\ca)$ of the algebra $\ca$. However, in many cases this is not a very
useful notion since there are algebras with too few derivations. Moreover,
$Der(\ca)$ is not an $\ca$-module but a module only over the center of $\ca$.
For models constructed along these lines we refer to \cite{Mad}.

We shall now briefly describe the notion of linear connection. There are
several tricky technical points mainly related to Hilbert spaces closure of
space of forms. We ignore them here while referring to \cite{CFF,CFG}.

Suppose then, we have a spectral triple $(\ca, \ch, D)$ with associated
differential calculus $(\omcad , d)$. The space $\ocad{1}$ is the analogue of
the `space of sections of the cotangent bundle'. It is naturally a right
$\ca$-module and we furthermore assume that it is also projective of finite
type. 

In order to develop `Riemannian geometry', one need the `analogue' of a 
metric on $\ocad{1}$. Now, there is a canonical Hermitian structure 
$\hs{ \cdot}{\cdot}_D~ : \ocad{1} \times \ocad{1} \ra \ca$ which is uniquely
determined by the triple  $(\ca, \ch, D)$. It is given by,
\be\label{rmet}
\hs{ \a}{\b}_D ~=: P_0 (\a^* \b) \in \ca~,~~~ \a, \b \in \ocad{1}~, 
\ee
where $P_0$ is the orthogonal projector onto $\ca$ determined by the scalar
product (\ref{innfor}) as in Section~\ref{se:spf} 
\fn{In fact the left hand side of (\ref{rmet}) is in the completion of $\ca$.}.
The map (\ref{rmet}) satisfies properties (\ref{her1}-\ref{her2}) which
characterizes an hermitian structure. It is also weakly nondegenerate, namely 
$\hs{ \a}{\b}_D ~= 0$ for any $\a \in \ocad{1}$ implies that $\b = 0$. It does
not, in general, satisfy the strong nondegeneracy expressed in terms of the
dual module $(\ocad{1})'$ as in Section~\ref{se:hsp}. Such a property it is
assumed to hold. Therefore, if $(\ocad{1})'$ is the dual module, we assume
that the Riemannian structure in (\ref{rmet}) determines an isomorphism of
right $\ca$-modules,
\be\label{isom}
\ocad{1} \lra (\ocad{1})'~, ~~~ \a \mapsto \hs{ \a}{\cdot}_D~.   
\ee 
 
We are now ready to define a linear connection. It is formally the same as in
the definition \ref{ycon} by taking $\ce=\ocad{1}$. 
\bdefi\label{lico} 
A linear connection on $\ocad{1}$ is a $\IC$-linear map
\be
\nabla : \ocad{1} \lra \ocad{1} \ota \ocad{1}~, 
\ee satisfying Leibniz rule
\be
\nabla(\a a) = (\nabla \a) a + \a d a ~,  ~~\forall ~\a \in \ocad{1}~, 
~a \in \ca~.  \label{lile}
\ee
\edefi

\noindent
Again, one can extend it to a map $\nabla : \ocad{1} \ota \ocad{p} 
\ra \ocad{1} \ota \ocad{p+1}$ and the {\it Riemannian curvature} of $\nabla$
is then the $\ca$-linear map given by
\be\label{ricu}
R_\nabla =: \nabla^2 ~:~ \ocad{1} \ra \ocad{1} \ota \ocad{1}~.
\ee
The connection $\nabla$ is said to be {\it metric} if it is compatible with the
Riemannian structure $\hs{\cdot}{\cdot}_D$ on $\ocad{1}$, namely if it
satisfies the relation,
\be\label{ricomp} 
-\hs{\nabla \a}{\b}_D + \hs{\a}{\nabla \b}_D~ = d \hs{\a}{\b}_D~, ~~\forall~
\a, \b \in \ocad{1}~.
\ee 

Next, one defines the {\it torsion} of the connection $\nabla$ as the map
$T_\nabla : \ocad{1} \ra \ocad{2}$ given by
\be\label{tors}
T_\nabla = d - m \circ \nabla~, 
\ee
where $m : \ocad{1} \ota \ocad{1} \ra \ocad{2} $ is just multiplication, 
$m(\a \ota \b) = \a \b$. One easily checks (right) $\ca$-linearity so that
$T_\nabla$ is a `tensor'. For an ordinary manifold with linear connection,
definition(\ref{tors}) yields the dual (i.e. the cotangent space version) of
the usual definition of torsion. 

A connection $\nabla$ on $\ocad{1}$ is a {\it Levi-Civita connection} if it is
compatible with the Riemannian structure $\hs{\cdot}{\cdot}_D$ on $\ocad{1}$
and its torsion vanishes. Contrary to what happens in the ordinary differential
geometry, a Levi-Civita connection needs not exist for a generic spectral
triple or there may exist more than one such connection.

Next, we derive {\it Cartan structure equations}. For simplicity, we shall
suppose that $\ocad{1}$ is a free module with a basis 
$\{E^A, A = 1, \cdots N \}$ so that any element $\a \in \ocad{1}$ 
can be written as $\a = E^A \a_A$. The basis is taken to be 
orthonormal with respect to the Riemannian structure $\hs{\cdot}{\cdot}_D$,
\be\label{orth}   
\hs{E^A}{E^B}_D~ = \h^{AB}~, ~~~\h^{AB} = diag(1, \cdots, 1)~, 
~~~ A, B = 1, \cdots, N ~.
\ee

A connection $\nabla$ on $\ocad{1}$ is completely determined by the
{\it connection $1$-forms} $\O_A^{~~B} \in \ocad{1}$ which are defined by,
\be\label{cofo}
\nabla E^A = E^B \otimes \O_B^{~~A}~, ~~~ A=1, \dots, N.
\end{equation}
The components of torsion $T^A \in \ocad{2}$ and curvature $R_A^{~~B} \in
\ocad{2}$ are defined by 
\bea\label{tori}
&&T_\nabla(E^A) = T^A ~, \nonumber \\
&&R_\nabla(E^A) = E^B \otimes R_B^{~~A}~, ~~~ A=1, \dots, N. 
\eea
By using definitions (\ref{tors}) and (\ref{ricu}) one gets the structure
equations,
\bea
T^A &=& d E^A - E^B \O_B^{~~A}~, ~~~ A=1, \dots, N~, \label{cart1} \\
R_A^{~~B} &=& d \O_A^{~~B} + \O_A^{~~C} \O_C^{~~B}~, 
~~~ A,B = 1, \dots, N. \label{cart2}
\eea
The metricity conditions (\ref{ricomp}), for the connection 
$1$-forms now reads,
\be
-\O_C^{~~A}{^*} \h^{CB} + \h^{AC} \O_C^{~~B} = 0~.
\ee
As mentioned before, metricity and vanishing of torsion do not fix uniquely
the connection. Sometimes, one imposes additional constrains by requiring
that the connection $1$-forms are Hermitian,
\be
\O_A^{~~B}{^*} = \O_A^{~~B}~. 
\ee
  
The components of a connection, its torsion and its Riemannian curvature
transform in the `usual' way under a change of orthonormal basis for $\ocad{1}$. 
Consider then a new basis $\{\wt{E}^A, A = 1, \cdots N \}$ of $\ocad{1}$. 
The relation between the two basis is given by
\be\label{cb}
\wt{E}^A = E^B (M^{-1})_B^{~~A}~, ~~~ E^A = \wt{E}^B M_B^{~~A}~,
\ee 
with the obvious identities,
\be\label{inve}
M_A^{~~C} (M^{-1})_C^{~~B} = (M^{-1})_A^{~~C}M_C^{~~B} = \d_A^B~,
\ee
which just says that the matrix $M = (M_B^{~~A}) \in \IM_N(\ca)$ is invertible
with inverse given by $M^{-1} = ((M^{-1})_B^{~~A})$ . By requiring that the new
basis be orthonormal with respect to
$\hs{\cdot}{\cdot}_D$ we get, 
\bea   
\h^{AB} &=& \hs{E^A}{E^B}_D ~=~ \hs{\wt{E}^P M_P^{~~A}}{\wt{E}^Q M_Q^{~~B}}_D
~= (M_P^{~~A})^* \hs{\wt{E}^P}{\wt{E}^Q }_D M_Q^{~~B}~, \nonumber \\
&=& (M_P^{~~A})^* \h^{PQ} M_Q^{~~B}~. 
\eea
From this and (\ref{inve}) we obtain the identity
\be
(M^{-1})_A^{~~B} = \h_{AQ} (M_P^{~~Q})^* \h^{PB}~,
\ee
or $M^* = M^{-1}$. By using again (\ref{inve}), we infer that $M$ is a unitary
matrix, $M M^* = M^* M =~\II$, namely an element in $\cu_N(\ca)$. \\
It is now straightforward to find the transformed components of the connection,
its curvature and its torsion
\bea
&& \wt{\O}_A^{~~B} = M_A^{~~P} \O_P^{~~Q} (M^{-1})_Q^{~~B} +
M_A^{~~P} d (M^{-1})_P^{~~B}~,  \label{trco} \\ 
&& \wt{R}_A^{~~B} = M_A^{~~P} R_P^{~~Q}  (M^{-1})_Q^{~~B}~,  \label{trrc} \\ 
&& \wt{T}^A =T^B (M^{-1})_B^{~~A}~.  \label{trto} 
\eea

Let us consider now the basis $\{\ve_A, A = 1, \cdots, N\}$ of $(\ocad{1})'$,
dual to the basis $\{E^A\}$,
\be
\ve_A(E^B) = \d^B_A~.
\ee  
By using the isomorphism (\ref{isom}) for the element $\ve_A$, there is an 
$\wh{\ve}_A \in \ocad{1}$ determined by
\be
\ve_A(\a) = ~\hs{\wh{\ve}_A}{\a}_D~, ~~~\forall ~\a \in \ocad{1}~, 
~~~ A=1, \dots, N. 
\ee
One finds that 
\be
\wh{\ve}_A = E^B \h_{BA}~, ~~~ A=1, \dots, N,
\ee
and under a change of basis as in (\ref{cb}), they transform as 
\be
\wt{\wh{\ve}}_A = \wh{\ve}_B (M_A^{~~B})^*~, ~~~ A=1, \dots, N.
\ee

The {\it Ricci $1$-forms} of the connection $\nabla$ are defined by
\be\label{ricc}
R_A^\nabla = P_1 (R_A^{~~B}(\wh{\ve}_B)^*) ~\in \ocad{1}~, 
~~~ A = 1, \cdots, N~. 
\ee 
As for the {\it scalar curvature}, it is defined by 
\be\label{scal}
r_\nabla = P_0( E ^A R_A^\nabla) = P_0(E ^A P_1 (R_A^{~~B}\wh{\ve}_B)^*) 
\in \ca~. 
\ee
The projectors $P_0$ and $P_1$ are again the orthogonal projectors
on the space of zero and one forms determined by the scalar
product (\ref{innfor}). It is straightforward to check that the scalar
curvature does not  depend on the particular orthonormal basis of
$\ocad{1}$. Finally, the {\it Hilbert-Einstein} action is given by
\be\label{he}
I_{HE}(\nabla) = tr_\o r |D|^{-n} =  tr_\o 
E ^A R_A^{~~B}\wh{\ve}_B^* |D|^{-n}~. 
\ee

\subsubsect{The Usual Einstein Gravity}

Let us consider the canonical triple  $(\ca, \ch, D)$ 
on a closed $n$-dimensional Riemannian spin manifold 
$(M,g)$ which we have described in Section~\ref{se:ctm}. 
We recall that $\ca=C^\infty(M)$ is the algebra of complex valued smooth 
functions on $M$; $\ch=L^2(M,S)$ is the Hilbert space of square integrable 
sections of the irreducible spinor bundle over $M$; finally, $D$ is the 
Dirac operator of the Levi-Civita spin connection, which locally can be written
as
\bea
D &=& \g^\m(x)\pa_\m + {\rm ~lower ~order ~terms} \nonumber \\
&=& \g^a e_a^\m \pa_\m + {\rm ~lower ~order ~terms}~.
\label{ridirac}
\eea
The `curved' and `flat' Dirac matrices are related by
\be
\g^\m(x) = \g^a e_a^\m~, ~~~ \m=1, \dots, n, 
\ee
and obey the relations
\bea
&& \g^\m(x)\g^\n(x) + \g^\n(x)\g^\m(x) = -2g^{\m\n}~, ~~~ \m,\n = 1, \dots, n,
\nonumber \\
&& \g^a\g^b + \g^b\g^a = -2\h^{ab}~, ~~~ a,b = 1, \dots, n.
\eea
We shall take the matrices $\g^a$ to be hermitian. \\
The $n$-beins $e_a^\m$ relate the components of the curved and flat 
metric, as usual by,
\be
e_a^\m g_{\m\n} e_b^\n = \h_{ab}~, ~~~ e_a^\m \h^{ab} e_b^\n = g^{\m\n}~.
\ee
Finally, we recall that, from the analysis of Section~\ref{se:uea}, generic
elements $\a \in \ocad{1}$ and $\b \in \ocad{2}$ can be written as 
\bea
&& \a = \g^a \a_a = \g^\m \a_\m~, ~~~\a_a = e_a^\m \a_\m~, \nonumber \\
&& \b = {1 \over 2} \g^{ab} \b_{ab} = {1 \over 2}\g^{\m\n} \b_{\m\n}~, 
~~~\b_{ab} = e_a^\m e_a^\n \b_{\m\n}~,   
\eea
with $\g^{ab} = {1 \over 2}(\g^a \g^b - \g^b \g^a)$ and 
$\g^{\m\n} = {1 \over 2}(\g^\m \g^\n - \g^\n \g^\m)$. The module $\ocad{1}$
is projective of finite type and we can take as orthonormal basis
\be
E^a = \g^a~, ~~~\hs{E^a}{E^b}~ = tr \g^a \g^b = \h^{ab}~, ~~~ a,b = 1, \dots,
n,
\ee
with $tr$ a normalized Clifford trace. Then, the dual basis $\{\ve_a \}$ of
$(\ocad{1})'$ is given by,
\be
\ve_a(\a) = \a_a = e_a^\m \a_\m~,
\ee
and the associated $1$-forms $\wh{\ve}_a$ are found to be 
\be
\wh{\ve}_a = \g^a \h_{ab}~.
\ee 
Hermitian connection $1$-forms are of the form
\be
\O_a^{~b} = \g^c \o_{c a}^{~~b} = \g^\m \o_{\m a}^{~~b}~.
\ee
Then, metricity and vanishing of torsion read respectively
\bea
&& \g^\m ( \o_{\m c}^{~~a} \h^{cb} + \h^{ac} \o_{\m c}^{~~b}) = 0~,
\label{lc1} 
\\  && \g^{\m\n} ( \pa_\m e^a_\n  - e^b_\m \o_{\n b}^{~~a}) = 0~. 
\label{lc2} 
\eea
The sets of matrices $\{ \g^\m \}$ and $\{ \g^{\m\n} \}$ being independent,
conditions (\ref{lc1}) and (\ref{lc2}) require the vanishing of the terms in
parenthesis and, in turn, these just say that the coefficients $\o_{\m a}^{~~b}$
(or equivalently $\o_{c a}^{~~b}$) determine the Levi-Civita connection of the
metric $g^{\m\n}$ \cite{Th}.\\
The $2$-forms of curvature can then be written as 
\be
R_a^{~~b} = {1 \over 2} \g^{cd} R_{cd a}^{~~~b}~, 
\ee  
with $R_{cd a}^{~~~b}$ the components of the Riemannian tensor of the connection 
$\o_{c a}^{~~b}$. \\
As for the Ricci $1$-forms, they are given by
\be
R_a = P_1 ( R_a^{~~b} \wh{\ve}_a^* ) = {1 \over 2} \g^{cd}\g^f
R_{cd a}^{~~~b} \h_{fb}~.   
\ee
It takes some little algebra to find
\be
R_a = - {1 \over 2} \g^c R_{cb a}^{~~~b}~.  
\ee
The scalar curvature is found to be 
\be
r =: P_0(\g^a R_a) = - {1 \over 2} P_0(\g^a \g^c) R_{cb a}^{~~~b} 
= \h^{ac} R_{cb a}^{~~~b}~,  
\ee
which is just the usual scalar curvature \cite{Th}. 

\subsubsect{Other Gravity Models}

In \cite{CFF,CFG}, the action (\ref{he}) was computed for a Connes-Lott space
$M \times Y$, product of a Riemannian, four-dimensional, spin manifold $M$
by an discrete internal space $Y$ consisting of two points. The Levi-Civita
connection on the module of $1$-forms depends on a Riemannian metric on $M$
and a real scalar field which determines the distance between the two-sheets.
The action (\ref{he}) contains the usual integral of the scalar curvature of
the metric   on $M$, a minimal coupling for the scalar field to such a metric,
and a kinetic term for the scalar field. \\
The Wodzicki residue methods applied to the same space yields a Hilbert-Space
action which is the sum of the usual term for the metric of $M$ together with
a term proportional to the square of the scalar field. There is no kinetic term
for the latter \cite{KW}.\\
A somewhat different model of geometry on the Connes-Lott space $M \times Y$
was presented in \cite{LNW}. The final action is just the Kaluza-Klein
action of unified gravity-electromagnetism and consists of the usual gravity
term, a kinetic term for a minimally coupled scalar field and an electromagnetic
term.  

\vfill\eject
\sect{Quantum Mechanical Models on Noncommutative Lattices}\label{se:qmm}

As a very simple example of a quantum mechanical system treated with
techniques of noncommutative geometry on noncommutative lattices, we shall
construct the $\theta$-quantization of a particle on a lattice for the circle.
We shall do so by constructing an appropriate `line bundle' with connection.
We refer to \cite{ncl} and  \cite{bbllt} for more details and additional field
theoretical examples. In particular, in \cite{bbllt} we derived the Wilson's
actions for gauge and fermionic fields and analogues of topological and
Chern-Simons actions.

The real line $\IR^1$ is the universal covering space of the circle $S^1$, 
the fundamental group $\pi_1(S^1) = \IZ$ acting on $\IR^1$ by translation
\be
\IR^1 \ni x \rightarrow x + N ~ , ~ N \in \IZ ~ . 
\ee
The quotient space of this action is $S^1$ and the the projection 
$ : \IR^1 \ra S^1$ is given by $ \IR^1 \ni x \ra e^{i2\p x} \in S^1$. 

Now, the domain of a typical Hamiltonian for a particle on $S^1$ needs
not consist of functions on $S^1$. Rather it can be obtained from
functions $\psi_{\theta}$ on $\IR^1$ transforming under an irreducible
representation of $\pi(S^1) = \IZ$,
\be
\r_{\q} : N \rightarrow e^{iN\q} \label{7.3}
\ee
according to
\be
\psi_{\q}(x+N) = e^{iN\q} \psi_{\q}(x) ~ . 
\ee
The domain $D_{\q}(H)$ for a typical Hamiltonian $H$ then consists of these
$\psi_{\q}$ restricted to a fundamental domain $0 \leq x \leq 1$ for the
action of $\IZ$, and subjected to a differentiability requirement:
\be
D_{\q}(H) = \{ \psi_{\q} : \psi_{\q}(1) = e^{i\q} \psi_{\q}(0)~;~~
\frac{d\psi_{\q}(1)}{dx} = e^{i\q} \frac{d\psi_{\q}(0)}{dx} \} ~. \label{dom}
\ee
In addition, $H\psi_{\q}$ must be square integrable for the measure on $S^1$
used to define the scalar product of wave functions. \\  
One obtains a distinct quantization, called $\q$-quantization, for each choice
of $e^{i\q}$.

Equivalently, wave functions could be taken to be  single-valued functions on
$S^1$ while adding a `gauge potential' term to the Hamiltonian. To be more
precise, one constructs a line bundle over $S^1$ with a connection one-form
given by $i\q d x$. If the Hamiltonian with the domain (\ref{dom}) is
$-d^2 / d x^2$, then the Hamiltonian with the domain $D_0(h)$
consisting of single valued wave functions is $-(d / d x + i \q)^2$.

There are similar quantization possibilities for a noncommutative lattice for
the circle as well \cite{ncl}. One constructs the algebraic analogue of the
trivial bundle on the lattice endowed with a gauge connection which is such
that the corresponding Laplacian has an approximate spectrum reproducing the
`continuum' one in the limit.

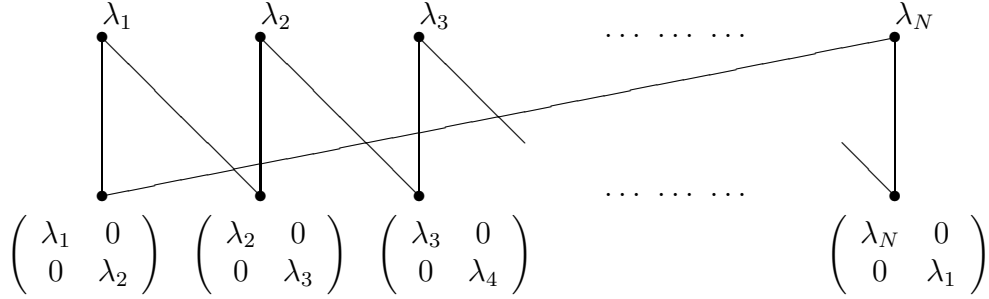
\begin{figure}[t]
\begin{center}
\begin{picture}(340,105)(-50,-70)
\put(-30,30){\circle*{4}}
\put(30,30){\circle*{4}}
\put(90,30){\circle*{4}}
\put(270,30){\circle*{4}}
\put(-30,-30){\circle*{4}}
\put(30,-30){\circle*{4}}
\put(90,-30){\circle*{4}}
\put(270,-30){\circle*{4}}
\put(160,-30){$\dots$}
\put(180,-30){$\dots$}
\put(200,-30){$\dots$}
\put(160,30){$\dots$}
\put(180,30){$\dots$}
\put(200,30){$\dots$}
\put(-30,-30){\line(5,1){300}}
\put(-30,30){\line(0,-1){60}}
\put(30,30){\line(0,-1){60}}
\put(90,30){\line(0,-1){60}}
\put(270,30){\line(0,-1){60}}
\put(-30,30){\line(1,-1){60}}
\put(30,30){\line(1,-1){60}}
\put(90,30){\line(1,-1){40}}
\put(270,-30){\line(-1,1){20}}
\put(-30,35){$\l_1$}
\put(30,35){$\l_2$}
\put(90,35){$\l_3$}
\put(270,35){$\l_N$}
\put(-70,-55){
$\left(\begin{array}{cc}
 \l_1 & 0 \\
0 & \l_2
\end{array}
\right)$}
\put(0,-55){
$\left(\begin{array}{cc}
 \l_2 & 0 \\
0 & \l_3
\end{array}
\right)$}
\put(70,-55){
$\left(\begin{array}{cc}
 \l_3 & 0 \\
0 & \l_4
\end{array}
\right)$}
\put(240,-55){
$\left(\begin{array}{cc}
 \l_N & 0 \\
0 & \l_1
\end{array}
\right)$}
\end{picture}
\caption{\label{fi:appcir2n}
\protect{\footnotesize $P_{2N}(S^1)$ for the approximate algebra $\cc(\ca)$.}}
\end{center}
\vskip.5cm
\end{figure}

As we have seen in Section~\ref{se:ncl}, the algebra $\ca$ associated with any
noncommutative lattice of the circle is rather complicate and involves
infinite dimensional operators on direct sums of infinite dimensional
Hilbert space. In turn, this algebra $\ca$, being AF (approximately finite
dimensional), can indeed be approximated by algebras of matrices. The simplest
approximation is just a commutative algebra $\cc(\ca)$ of the form 
\be
\cc(\ca) \simeq \IC^N = \{ c=(\l_1 ,\l_2, \cdots, \l_N ) ~:~ \l_i \in \IC \}~.
\label{calg}
\ee
The algebra (\ref{calg}) can produce a noncommutative lattice
with $2N$ points by considering a particular class of not necessarily
irreducible representations as in Fig.~\ref{fi:appcir2n}.
In that Figure, the top points correspond to the irreducible one
dimensional representations
\be
\p_i : \cc(\ca) \ra \IC~, ~~~ c \mapsto \p_i(c) = \l_i~, ~~i = 1, \cdots, N~.
\ee  
As for the bottom points, they correspond to the reducible two dimensional
representations
\be
\pi_{i+N} : \cc(\ca) \ra \IM_{2}(\IC)~, ~~~ c \mapsto \p_{i+N}(c) =
\left(
\begin{array}{cc}
\l_i & 0 \\
0 & \l_{i+1}
\end{array}
\right)~, ~~i = 1, \cdots, N~,
\ee
with the additional condition that $N+1 = 1$. The partial order, or
equivalently the topology, is determined by inclusion of the corresponding
kernels as in Section~\ref{se:ncl}. 

By comparing Fig.~\ref{fi:appcir2n} with the corresponding
Fig.~\ref{fi:cirfun}, we see that by trading $\ca$ with $\cc(\ca)$, all
compact operators have been put to zero. A better approximation would be
obtained by approximating compact operators with finite dimensional matrices
of increasing rank.

The finite projective module of sections $\ce$ associated with the
trivial bundle is just $\cc(\ca)$ itself:
\be
\ce = \IC^N = \{ \h = (\m_1 ,\m_2, \cdots, \m_N ) ~:~ \m_i \in \IC \}~.
\label{cmod} 
\ee
The action of $\cc(\ca)$ on $\ce$ is simply given by
\be
\ce \times \cc(\ca) \ra \ce~, ~~~(\h, c) 
\mapsto \h c = (\h_1 \l_1, \h_2 \l_2 \cdots \h_N \l_N)~. \label{ceact}
\ee
On $\ce$ there is a $\cc(\ca)$-valued Hermitian structure $\langle
\cdot,\cdot \rangle$,
\be
\langle \h' , \h \rangle :=
(\h_1'^{*} \h_1, \h_2'^{*} \h_2, \cdots, \h_N'^{*} \h_N)~ \in ~\cc(\ca)~.
\label{cherm}
\ee

Next, we need a $K$-cycle $(\ch, D)$ over $\cc(\ca)$. We take for $\ch$ just
$\IC^N$ on which we represents elements of $\cc(\ca)$ as diagonal matrices
\be
\cc(\ca) \ni c \mapsto \mbox{diag}(\l_1, \l_2, \dots \l_N) \in \cb(\IC^N)
\simeq \IM_{N}(\IC)~. 
\ee
Elements of $\ce$ will be realized in the same manner,
\be
\ce \ni \h \mapsto \mbox{diag}(\h_1, \h_2, \dots \h_N) \in \cb(\IC^N)
\simeq \IM_{N}(\IC)~. 
\ee
Since our triple $(\cc(\ca), \ch, D)$ will be zero dimensional, the
($\IC$-valued) scalar product associated with the Hermitian structure
(\ref{cherm}) will be taken to be the following one
\be
(\h', \h) = \sum_{j=1}^N \h'^*_j \h_j = tr \langle \h' , \h \rangle~,
~~~\forall ~\h', \h \in \ce~.
\label{csca}
\ee

By identifying $N + j$ with $j$, we take for the operator $D$ the
$N \times N$ self-adjoint matrix with elements
\be
D_{ij} = \frac{1}{\sqrt{2}\epsilon} (m^* \d_{i+1,j} + m \d_{i,j+1})~,
~ i,j = 1, \cdots, N~, \label{7.14.3}
\ee
where $m$ is any complex number of modulus one, $m m^* =1$.

As for the connection one form $\r$ on the bundle $\ce$, we take it to be the
hermitian matrix with elements
\bea
&&\r_{ij} =  \frac{1}{\sqrt{2}\epsilon}
(\s^* m^* \d_{i+1,j} + \s m \d_{i,j+1})~, \nonumber \\
&&~~~ \s = e^{- i \q / N} - 1~,
~~i,j = 1, \cdots, N~.
\label{7.14.4}
\eea
One checks that, modulo junk forms, the curvature of $\r$ vanishes, namely
\be
d \r + \r^2 = 0~. \label{7.14.5}
\ee
It is also possible to prove that $\r$ is a `pure gauge', that is
that there exists a $c \in \cc(\ca)$ such that $\r = c^{-1} d c$, only for
$\theta = 2 \p k$, with $k$~ any integer. If $c = \mbox{diag}(\l_1 ,\l_2,
\ldots ,\l_N )$, any such $c$ will be given by $\l_1 = \l~, ~\l_2 = 
e^{i 2\p k / N } \l~, ... , ~\l_j = e^{i 2\p k (j-1) / N } \l~, ...,
~\l_N = e^{i 2\p k (N - 1) / N } \l~, \l$ not equal to $0$
(these properties are the analogues of the properties of the
connection $i \q d x$ in the `continuum' limit). 

The covariant derivative $\nabla_\q$ on $\ce$, $\nabla_\q : \ce \ra \ce
\otimes_{\cc(\ca)} \O^1(\cc(\ca))$  is then given by
\be
\nabla_\q \h = [D, \h] + \r \h~, ~~~ \forall ~\h \in \ce~.
\label{ccovder}
\ee
In order to define the Laplacian $\D_\q$ one first introduces a `dual' operator
$\nabla_q^*$ via 
\be
(\nabla_q \h', \nabla_q \h) = (\h', \nabla_q^{*} \nabla_q \h)~, 
~~~\forall ~\h', \h \in \ce. \label{cdcovder}
\ee
The Laplacian $\D_\q$ on $\ce$, $\D_\q : \ce \ra \ce$, can then be defined by
\be
\D_\q \eta = - q (\nabla_q)^{*} \nabla_q \eta ~,~~~ \forall ~\h \in \ce~,
\label{clap}
\ee
where $q$ is the orthogonal projector on $\ce$ for the scalar product
$( \cdot, \cdot )$ in (\ref{csca}). This projection operator is readily seen
to be given by
\be
(q M)_{ij} = M_{ii}\delta_{ij}~,~~~\mbox{no summation on}~i~,
\label{7.14.7}
\ee
with $M$ any element in $\IM_{N}(\IC)$. Hence, the action of $\D_\q$ on the
element $\h = (\h_1, \cdots, \h_N)$~, $\h_{N+1} = \h_1$, is explicitly given by
\bea
 (\D_\q \h)_{ij} &=& - (\nabla_q^{*} \nabla_q \h)_{ii}
\delta_{ij}~, \nonumber \\
- (\nabla_q^{*} \nabla_q\h)_{ii} &=&
\left\{ - \left[ D, [D, \h ] \right] - 2 \r [D,
\h] - \r^2 \h \right\}_{ii} \nonumber \\
&=& \frac{1}{\epsilon^2}
\left[ e^{-i\q/N} \h_{i-1} - 2 \h_i +  e^{i\q/N} \h_{i+1} \right]~;
~~~i = 1, 2, \cdots , N~.
\label{7.15}
\eea
The associated eigenvalue problem
\be
\D_\q \h = \l \h 
\ee
has solutions
\bea
\l &=& \l_k = \frac{2}{\epsilon ^2}
\left[\cos (k + \frac{\theta}{N}) -1 \right]~, \label{ceigen} \\
\h &=& \h^{(k)}
= \mbox{diag}(\h^{(k)}_1, \h^{(k)}_2, \cdots, \h^{(k)}_N)~,
~~~k = m \frac{2\p}{N}~,~ m = 1, 2, \cdots, N~, 
\eea
with each component $\h^{(k)}_j$ having an expression of the form 
\be
\h^{(k)}_j = A^{(k)} e^{i kj} + B^{(k)}e^{-i kj}~,~~~
A^{(k)}, B^{(k)}\in \IC~. 
\ee
We see that the eigenvalues (\ref{ceigen}) are an approximation to the
continuum answers $-4k^2~, k \in \IR$.

\vfill\eject

\section*{Appendices}

\appendix

\sect{Basic Notions of Topology}\label{se:bnt}
In this appendix we gather few fundamental notions regarding the notions of
topology and topological spaces while referring to \cite{HY,Ke}.

A {\it topological space} is a set $S$ together with a collection 
$\t = \{O_\a \}$ of subsets of $S$, called {\it open sets}, which satisfy the
following axioms
\begin{enumerate}
\item[~] $O_1$. ~~~The union of any number of open sets is an open set. 
\item[~] $O_2$. ~~~The intersection of a finite number of open sets is an open set.
\item[~] $O_3$. ~~~Both $S$ and the empty set $\emptyset$ are open. 
\end{enumerate}
Topology allows one to define the notion of continuous map. 
A map $f : (S_1, \t_1) \ra (S_2, \t_2)$ between two topological spaces is
defined to be {\it continuous} if the inverse image $f^{-1}(O)$ is open in
$S_1$ for any open $O$ in $S_2$. A continuous map $f$ which is a bijection and such
that $f^{(-1)}$ is continuous as well is called a {\it homeomorphism}. 

Having a topology on a space, one can define the notion of limit point of a subset. 
A point $p$ is a {\it limit point} of a subset $X$ of $S$ if every open set
containing $p$ contains at least another point of $X$ distinct from $p$.\\ 
A subset $X$ of a topological space $S$ is called {\it closed} if the
complement $S \setminus X$ is open. It turns out that the subset $X$ is closed if
and only if it contains all its limit points.

The collection $\{C_\a\}$ of all closed subsets of a topological space $S$, satisfy
properties which  are dual to the corresponding ones for the open sets.
\begin{enumerate}
\item[~] $C_1$. ~~~The intersection of any number of closed sets is a closed set. 
\item[~] $C_2$. ~~~The union of a finite number of closed sets is a closed set.
\item[~] $C_3$. ~~~Both $S$ and the empty set $\emptyset$ are closed. 
\end{enumerate}   
One could then give a topology on a space by giving the collection of closed
sets.

The {\it closure} $\bar{X}$ of a subset $X$ of a topological space $(S, \t)$ is
the intersection of all closed set containing $X$. It is evident that
$\bar{X}$ is the smallest closed set containing $X$ and that $X$ is closed if
and only if $\bar{X} = X$. It turns out that a topology on a set $S$ can be
given by means of a {\it closure operation}. Such an operation is an assignment
of a subset $\bar{X}$ of $S$ to any subset $X$ of $S$, in such a manner that the
following {\it Kuratowski closure axioms} are true
\begin{enumerate}
\item[~] $K_1$. ~~~$\bar{\emptyset} = \emptyset$~.
\item[~] $K_2$. ~~~$X \subseteq \bar{X}$~. 
\item[~] $K_3$. ~~~$\bar{\bar{X}} = \bar{X}$~. 
\item[~] $K_4$. ~~~$\bar{X \bigcup Y} = \bar{X} \bigcup \bar{Y}$~. 
\end{enumerate}
If $\s$ is the family of all subset $X$ of $S$ for which $\bar{X} = X$ and $\t$
is the family of all complements of members of $\s$, then $\t$ is a topology for
$S$, and $\bar{X}$ is the $\t$-closure of $X$ for any subset of $S$. Clearly,
$\s$ is the family of closed sets. 

A topological space $S$ is said to be a {\it $T_0$-space} if: given any two
points of $S$, at least one of them is contained in an open set not containing
the other. This can also be stated by saying that for any couple of points, at
least one of the points is not a limit point of the other. In such a space,
there may be sets consisting of a single point which are not closed.
 
A topological space $S$ is said to be a {\it $T_1$-space} if: given any two
points of $S$, each of them lies in an open set not containing the other. This
requirement implies that each point (and then, by $C_2$ above, every finite
set) is closed. This is often taken as a definition of $T_1$-space. 

A topological space $S$ is said to be a {\it $T_2$-space} or a {\it Hausdorff
space} if: given any two points of $S$, there are {\it disjoint} open sets each
containing just one of the two points.\\
It is clear that the previous conditions are in an increasing order of strength
in the sense that being $T_2$ implies being $T_1$ and being $T_1$ implies being
$T_0$. 

A family $\cu$ of sets is a {\it cover} of a (topological) space if $S = \bigcup
\{X, X \in \cu \}$. The family is an {\it open cover} of $S$ if any member of $\cu$ is
an open set. The family is a {\it finite cover} is the number of members of
$\cu$ is finite. It is a {\it locally finite} cover if and only if every $x \in S$
has a neighborhood that meets only a finite number of members of the family. \\ 
A topological space $S$ is called {\it compact} if any open cover of $S$ has a finite
subcover of $S$. A topological space $S$ is called {\it locally compact} if any point
of $S$ has at least one compact neighborhood. A compact space is automatically
locally compact. If $S$ is a locally compact space which is also Hausdorff, then the
family of closed compact neighborhoods of any point is a base for its neighborhood
system.

The {\it support} of a real or complex valued function $f$ on a topological space
$S$ is the closure of the set $K_f = \{x \in S ~|~ f(x) \not= 0 \}$. The function
$f$ is said to have compact support if $K_f$ is compact. The collection of all
continuous functions on $S$ whose support is compact is denoted by $C_c(S)$. \\
A real or complex valued function $f$ on a locally compact Hausdorff space $S$ is
said to {\it vanish at infinity} if for every $\e > 0$ there exists a compact set 
$K \subset S$ such that $|f(x)| < \e$ for all $x \notin K $. The collection of all
continuous functions on $S$ which vanishes at infinity is denoted by $C_0(S)$.
Clearly $C_c(S) \subset C_0(S)$, and the two classes coincides if $S$ is compact.
Furthermore, one can prove that $C_0(S)$ is the completion of $C_c(S)$ relative to
the supremum norm (\ref{suno}) described in Example~\ref{ex:cofu} \cite{Ru}.

A continuous map between two locally compact Hausdorff spaces $f: S_1 \ra S_2$ is
called {\it proper} if and only if for any compact subset $K$ of $S_2$, the inverse
image $f^{-1}(K)$ is a compact subset of $S_1$.

A space which contains a dense subset is called {\it separable}. A topological space
which has a countable basis of open sets is called {\it second-countable} (or 
{\it completely separable}). 

A topological space $S$ is called {\it connected} if it is not the union of two
disjoint, nonempty open set. Equivalently, if the only sets in $S$ that are both
open and closed are $S$ and the empty set. A subset $C$ of the topological space $S$
is called a {\it component} of $S$, provided that $C$ is connected and maximal,
namely is not a proper subset of another connected set in $S$. One can prove that
any point of $S$ lies in a component. A topological space is a called {\it totally
disconnected} if the (connected) component of each point consists only of the point
itself. The {\it Cantor set} is a totally disconnected space. In fact, any totally
disconnected, second countable, compact hausdorff space is homeomorphic to a subset
of the Cantor set.

If $\t_1$ and $\t_2$ are two topologies on the space $S$, one says that $\t_1$
is {\it coarser} than $\t_2$ (or that $\t_2$ is {\it finer} than $\t_1$) if and
only if
$\t_1 \subset \t_2$, namely if and only if any subset of $S$ which is
open in $\t_1$ it is also open in $\t_2$. Given two topologies on the 
space $S$ it may happen that neither of them is coarser (or finer) than the
other. The set of all possible topologies on the same space is a partially
ordered set whose {\it coarsest} element is the topology in which only
$\emptyset$ and $S$ are open, while the {\it finest} element is the topology
in which all subsets of $S$ are open (this topology is called the discrete
topology).   

\vfill\eject
\sect{The Gel'fand-Naimark-Segal Construction}\label{se:gns} 

A {\it state} on the $C^*$-algebra $\ca$ is a linear functional
\be
\f : \ca \lra \IC~,
\ee 
which is positive and of norm one, namely it satisfies
\bea 
&&\f (a^* a) \geq 0~, ~~\forall ~a \in \ca~, \nonumber \\ 
&&\norm{\f} = 1~.
\eea 
Here the norm of $\f$ is defined as usual by $\norm{\f} = sup \{ |\f(a)|
~:~ \norm{a} \leq 1 \}$. If $\ca$ has a unit (we always assume this is 
the case) the positivity implies that
\be\label{posi}
\norm{\f} = \f(\II) = 1~.
\ee
The set $\cs(\ca)$ of all states of $\ca$ is clearly a convex space,  since
$\l \f_1 + (1-\l) \f_2 \in \cs(\ca)$, for any $\f_1, \f_2 \in 
\cs(\ca)$ and $0 \leq \l \leq 1$. Elements at the boundary of $\cs(\ca)$ are 
called {\it pure states}, namely, a states $\f$ is called pure if it 
cannot be written as the convex combination of (two) other states. The 
space of pure states is denoted by $\cp\cs(\ca)$. If the algebra $\ca$ is
abelian, a pure state is the same as a character and the space
$\cp\cs(\ca)$ is just the space $\ha$ of characters of $\ca$; endowed
with the Gel'fand topology is a Hausdorff (locally compact) topological space. 

With each state $\f \in \cs(\ca)$ there is associated a representation 
$(\ch_\f, \p_\f$) of $\ca$, called the Gel'fand-Naimark-Segal (GNS) 
representation. The procedure to construct such a representation is also 
called the GNS construction which we shall now briefly describe \cite{Di1,Mu}.

Suppose then that we are given a state $\f \in \cs(\ca)$ and consider the space
\be
\cn_\f = \{ a \in \ca ~|~ \f(a^* a) = 0 \}~.
\ee
By using the fact that $\f(a^* b^* b a) \leq \norm{b} ^2 \f(a^* a)$, one 
infers that $\cn_\f$ is a closed (left) ideal of $\ca$. The space 
$ \ca / \cn_\f $ of equivalence classes is made a pre-Hilbert space by 
defining a scalar product by
\be
\ca / \cn_\f \times  \ca / \cn_\f \lra \IC~, ~~~
(a + \cn_\f, b + \cn_\f) \mapsto \f(a^* b)~.
\ee
The scalar product is clearly independent of the representatives in the
equivalence classes. \\
The Hilbert space $\ch_\f$ completion of $\ca / \cn_\f$ is the space of the 
representation. Then, to any $a \in \ca$ one associates an operator
$\p(a) 
\in \cb( \ca / \cn_\f)$ by
\be
\p(a) (b + \cn_\f) =: ab + \cn_\f~.
\ee
Again, this action does not depends on the representative. 
From $\norm{\p(a) (b + \cn_\f)}^2 = \f(b^* a^* ab) \leq \norm{a}^2 
\f(b^* b) = \norm{b + \cn_\f}^2$ one gets $\norm{\p(a)} \leq \norm{a}$ and 
in turn, $\p(a) \in \cb(\ca / \cn_\f)$. There is a unique extension of 
$\p(a)$ to an operator $\p_\f(a) \in \cb(\ch_\f)$. Finally, one easily 
checks the
algebraic properties $\p_\f(a_1a_2) = \p_\f(a_1)\p_\f(a_2)$ and 
$\p_\f(a^*) = (\p_\f(a))^*$ and one gets a $^*$-morphism (a representation) 
\be\label{gnsrep}
\p_\f : \ca \lra \cb(\ch_\f)~, ~~~a \mapsto \p_\f(a)~.
\ee

It turns out that any state $\f$ is a {\it vector state}, namely there 
exists a vector $\xi_\f \in \ch_\f$ with the property,
\be\label{veid}
(\xi_\f, \pi_\f(a)\xi_\f) = \f(a)~, \forall ~a \in \ca~.
\ee
Such a vector is defined by
\be
\xi_\f =: [\II] = \II + \cn_\f~,
\ee
and is readily seen to verify (\ref{veid}). Furthermore, the set
$\{\pi_\f(a) \xi_\f ~|~ a \in \ca\}$ is just the dense set 
$\ca / \cn_\f$ of equivalence classes. This fact is stated by saying that 
the vector $\xi_\f$ is a {\it cyclic vector} for the representation
$(\ch_\f, \pi_\f)$. By construction, and by (\ref{posi}), the cyclic vector 
is of norm one, $ \norm{\xi_\f}^2 = \norm{\f} = 1$.

The cyclic representation $(\ch_\f, \pi_\f, \xi_\f)$ is unique up to unitary 
equivalence. If $(\ch_\f', \pi_\f', \xi_\f')$ is another cyclic  representation such
that $(\xi_\f', \pi'_\f(a)\xi_\f') = \f(a)$,  for all $a \in \ca$, then there exists
a unitary operator $U : \ch_\f  \ra \ch_\f'$ such that 
\bea
&& U^{-1} \pi_\f'(a) U = \pi_\f(a)~, ~~\forall ~a \in \ca~, \nonumber \\
&& U \xi_\f = \xi_\f'~.
\eea
The operator $U$ is just defined by $U \pi_\f(a) \xi_\f = \pi_\f'(a) \xi_\f'$ for any
$a \in \ca$. Then,  the properties of the state $\f$ ensure that $U$ is well defined
and  preserves the scalar product.

It is easy to see that the representation $(\ch_\f, \xi_\f)$ is irreducible 
if and only if every non zero vector $\xi \in \ch_\f$ is cyclic so that 
there are no nontrivial invariant subspaces. It is somewhat surprising 
that this happens exactly when the state $\f$ is pure \cite{Di1}.
\bprop
Let $\ca$ be a $C^*$-algebra. Then,
\begin{itemize}
\item[1.]
A state $\f$ on $\ca$ is pure if and only if the associated GNS representation
$(\ch_\f, \p_\g)$ is irreducible.
\item[2.] Given  a pure state $\f$ on $\ca$ there is a canonical bijection between
rays in the associated Hilbert $\ch_\f$ and the equivalence class of $\f$,
\[
C_\f = \{\j ~{\rm ~pure ~state ~on}~ \ca ~|~ \p_\j ~{\rm equivalent ~to}~ \p_\f \}~.
\]
\end{itemize}
\eprop 

\noindent
The bijection of point 2. of previous preposition is explicitly given by
associating with any $\x \in \ch_\f~, \norm{\x} = 1$, the state on $\ca$ given by
\be
\j(a) = (\xi, \pi_\f(a)\xi)~, ~~\forall ~a \in \ca~, 
\ee
which is seen to be pure. As said before, the representation $(\ch_\f, \pi_\f)$ being
associative, each vector of $\ch_\f$ is cyclic; this in turn implies that the
representation associated with the state $\j$ is equivalent to $(\ch_\f, \pi_\f)$. 

\bigskip

As a simple example, we consider the algebra ${\bf M}_2(\IC)$ with the two pure 
states constructed in Section~\ref{se:nca},
\be
\f_1(  
\left[
\begin{array}{cc} a_{11} & a_{12} \\ a_{21} & a_{22} 
\end{array}
\right]) = a_{11}~,~~~
\f_2(  
\left[
\begin{array}{cc} a_{11} & a_{12} \\ a_{21} & a_{22} 
\end{array}
\right]) = a_{22}~.
\ee

\noindent
As we mentioned before, the corresponding representations are equivalent. 
We shall show that they are both equivalent to the the defining two 
dimensional one.\\
The ideals of elements of `vanishing norm' of the states $\f_1, \f_2$ are 
respectively,
\be
\cn_1 = \left\{  
\left[
\begin{array}{cc} 0 & a_{12} \\ 0 & a_{22} 
\end{array}
\right] \right\} ~,~~~~~
\cn_2 = \left\{  
\left[
\begin{array}{cc} a_{11} & 0 \\ a_{21} & 0 
\end{array}
\right] \right\}~.
\ee 
The associated Hilbert spaces are then found to be
\bea
&& \ch_1 = \left\{  
\left[
\begin{array}{cc} x_1 & 0 \\ x_2 & 0 
\end{array}
\right] \right\} ~\simeq~ \IC^2 =
\left\{  
X = \left(
\begin{array}{c} x_1 \\ x_2  
\end{array}
\right) \right\}~, ~~ \hs{X}{X'} = x_1^* x_1' + x_2^* x_2'~.  
\nonumber \\
&& \ch_2 = \left\{  
\left[
\begin{array}{cc} 0 & y_1 \\ 0 & y_2  
\end{array}
\right] \right\} ~\simeq~ \IC^2 =
\left\{  
X = \left(
\begin{array}{c} y_1 \\ y_2  
\end{array}
\right) \right\}~, ~~ \hs{Y}{Y'} = y_1^* y_1' + y_2^* y_2'~.  
\eea
As for the action of any element $A \in {\bf M}_2(\IC)$ on $\ch_1$ and 
$\ch_2$, we get 
\bea
&& \p_1(A)  
\left[
\begin{array}{cc} x_1 & 0 \\ x_2 & 0 
\end{array}
\right] =
\left[
\begin{array}{cc} a_{11}x_1 + a_{12}x_2 & 0 \\ 
a_{21}x_1 + a_{22}x_2 & 0 
\end{array} \right]
\equiv A \left(
\begin{array}{c} 
x_1  \\ 
x_2  
\end{array}
\right) ~, \nonumber \\
&& \p_2(A)  
\left[
\begin{array}{cc} 0 & y_1 \\ 0 & y_2  
\end{array}
\right] =
\left[
\begin{array}{cc} 0 & a_{11}y_1 + a_{12}y_2 \\ 
0 & a_{21}y_1 + a_{22}y_2 
\end{array} \right]
\equiv A \left(
\begin{array}{c} 
y_1 \\ 
y_2  
\end{array}
\right) ~. \label{repr}
\eea
The two cyclic vectors are given by 
\be
\xi_1 = 
\left(
\begin{array}{c} 
1 \\ 
0  
\end{array}
\right)~, ~~~
\xi_2 = 
\left(
\begin{array}{c} 
0 \\ 
1  
\end{array}
\right)~. 
\ee
The equivalence of the two representations is provided by the off-diagonal
matrix 
\be
U = \left[
\begin{array}{cc}
0 & 1 \\
1 & 0 
\end{array}
\right]~,
\ee
which interchange $1$ and $2$~, $U \xi_1 = \xi_2$. 
In fact, by using the fact that for an irreducible representation any non 
vanishing vector is cyclic, from (\ref{repr}) we see that the two 
representation can indeed be identified. \\

\vfill\eject
\sect{Hilbert Modules}\label{se:hm}

The theory of Hilbert modules is a generalization of the theory of Hilbert spaces and
it is the natural framework for the study of modules over a $C^*$-algebra $\ca$
endowed with hermitian $\ca$-valued inner products. Hilbert modules have been (and
are) used in a variety of applications, notably for the notion of strong Morita
equivalence.  The subject started with the works \cite{Rimor} and \cite{Pa}. We refer
to \cite{W-O}  for a very nice introduction while we report on the fundamentals of
the theory.  Throughout this appendix, $\ca$ will be a $C^*$-algebra (almost
always unital) with and its norm will be denoted simply by $\norm{\cdot}$.
\bdefi\label{phm}
A {\rm right pre-Hilbert module} over $\ca$ is a right $\ca$-module $\ce$ endowed with
an $\ca$-valued {\it hermitian structure}, namely a sesquilinear form 
$\hs{~}{~}_\ca : \ce \times \ce \ra \ca$, which is conjugate linear in the first
variable and such that 
\bea
&& \hs{\h_1}{\h_2 a}_\ca = \hs{\h_1}{\h_2}_\ca a~, \label{hsp1}\\ 
&& \hs{\h_1}{\h_2}_\ca^* = \hs{\h_2}{\h_1}_\ca~,   \label{hsp2}\\ 
&& \hs{\h}{\h}_\ca \geq 0~, ~~\hs{\h}{\h}_\ca = 0~ \Leftrightarrow ~\h = 0~,
\label{hsp3}
\eea
for all $\h_1, \h_2, \h \in \ce$, $a \in \ca$. 
\edefi

\noindent
By the property (\ref{hsp3}) in the previous definition the element
$\hs{\h}{\h}_\ca$ is self-adjoint. As in ordinary Hilbert spaces, the property
(\ref{hsp3}) provides a {\it generalized Cauchy-Schwartz inequality}
\be
\hs{\h}{\x}^*_\ca \hs{\h}{\x}_\ca \leq \norm{\hs{\h}{\h}_\ca} \hs{\x}{\x}_\ca~,
~~~\forall ~\h,\x \in \ce~, \label{chi1}
\ee
which in turns, implies
\be
\norm{\hs{\h}{\x}_\ca}^2 \leq \norm{\hs{\h}{\h}_\ca} \norm{\hs{\x}{\x}_\ca}~,
~~~\forall ~\h,\x \in \ce~, \label{chi2}
\ee
By using these properties and the norm $\norm{\cdot}$ in $\ca$ one can
defines a norm in $\ce$.
\bdefi
The {\rm norm} of any element $\h \in \ce$ is defined by
\be
\norm{\h}_\ca =: \sqrt{\norm{\hs{\h}{\h}}}~.
\ee
\edefi

\noindent
Then, one can prove that $\norm{\cdot}_\ca$ satisfies all properties
(\ref{normprop}) of a norm.
\bdefi\label{hm}
A {\rm right Hilbert module} over $\ca$ is a right pre-Hilbert module $\ce$ which is
complete with respect to the norm $\norm{\cdot}_\ca$.
\edefi

\noindent
By completion any right pre-Hilbert module will give a right Hilbert module.
\\ It is clear that Hilbert modules over $\IC$ are ordinary Hilbert spaces. 

A structure of {\it left} (pre-)Hilbert module on a left $\ca$-module $\ce$ is
provided by an $\ca$-valued Hermitian structure $\hs{~}{~}_\ca$ on $\ce$ which is
conjugate linear in the second variable and the condition (\ref{hsp1}) is
replaced by
\be
\hs{a\h_1}{\h_2}_\ca = a\hs{\h_1}{\h_2}_\ca~, ~~~\forall~  \h_1, \h_2, \in \ce, 
~a \in \ca~. \label{hspl}
\ee
 
In the following, unless stated otherwise, by Hilbert module we shall mean a right
one. It is straightforward to pass to equivalent statements concerning left modules.

Given any Hilbert module $\ce$ over $\ca$, the closure of the linear span of
$\{\hs{\h_1}{\h_2}_\ca~, ~\h_1, \h_2 \in \ce\}$ is an ideal in $\ca$. If this ideal
is the whole of $\ca$ the module $\ce$ is called a {\it full Hilbert module}
\fn{Rieffel call it an {\it $\ca$-rigged space.}}. \\
It is worth noticing that, contrary to what happens in an ordinary Hilbert space,
Pythagoras equality is non valid in a generic Hilbert module $\ce$. 
If $\h_1,\h_2$ are any two orthogonal elements in $\ca$, namely
$\hs{\h_1}{\h_2}_\ca = 0$,  in general one has that $\norm{\h_1 + \h_2}_\ca^2 \not=
\norm{\h_1}_\ca^2 +
\norm{\h_2}_\ca^2$. Indeed, properties of the norm only assure that
$\norm{\h_1 + \h_2}_\ca^2 \leq \norm{\h_1}_\ca^2 + \norm{\h_2}_\ca^2$.

An `operator' on a Hilbert module needs not admits an adjoint.
\bdefi
Let $\ce$ be an Hilbert module over the $C^*$-algebra $\ca$. A continuous
$\ca$-linear maps $T:\ce\ra\ce$ is said to be {\rm adjointable} if there exists a
map $T^*:\ce\ra\ce$ such that 
\be
\hs{T^* \h_1}{\h_2}_\ca ~=~ \hs{\h_1}{T \h_2}_\ca~, ~~~\forall ~\h_1, \h_2 \in
\ce~. 
\ee
The map $T^*$ is called the adjoint of $T$. We shall denote by $End_\ca(\ce)$ the
collection of all continuous $\ca$-linear adjointable maps. Elements of
$End_\ca(\ce)$ will be also called {\it endomorphisms} of $\ce$.
\edefi

\noindent
One can prove that if $T\in End_\ca(\ce)$, then its adjoint  $T^*\in
End_\ca(\ce)$ with $(T^*)^* = T$. Also, if both $T$ and $S$ are in  $End_\ca(\ce)$,
then $TS \in End_\ca(\ce)$ with $(TS)^* = S^* T^*$. Finally, endowed with this
involution and with the operator norm
\be
\norm{T} =: sup\{\norm{T \h}_\ca ~:~ \norm{\h}_\ca \leq 1\}~, 
\ee
the space $End_\ca(\ce)$ becomes a $C^*$-algebra of bounded operators:
$\hs{T\h}{T\h}_\ca \leq \norm{T}^2 \hs{\h}{\h}_\ca$. Indeed, $End_\ca(\ca)$ is
complete if $\ce$ is.

There are also the analogue of {\it compact endomorphisms} which are obtained as usual
from `endomorphisms of finite rank'. For any $\h_1, \h_2 \in \ce$ an endomorphism 
$\ket{\h_1}\bra{\h_2}$ is defined by
\be\label{bk}
\ket{\h_1}\bra{\h_2} (\x) =: \h_1 \hs{\h_2}{\x}_\ca~, ~~~\forall ~\x \in \ce~. 
\ee
Its adjoint is just given by
\be
(\ket{\h_1}\bra{\h_2})^* = \ket{\h_2}\bra{\h_1}~, ~~~\forall ~\h_1, \h_2 \in \ce~.
\ee
One can check that 
\be
\norm{\ket{\h_1}\bra{\h_2}}_\ca \leq \norm{\h_1}_\ca \norm{\h_2}_\ca~, 
~~~\forall ~\x \in \ce~. \label{boco}
\ee
Furthermore, for any $T\in End_\ca(\ce)$ and any $~\h_1, \h_2,
\x_1,
\x_2
\in
\ce$, one has the expected composition rules
\bea
&& T \circ \ket{\h_1}\bra{\h_2} = \ket{T \h_1}\bra{\h_2}~, \\
&& \ket{\h_1}\bra{\h_2} \circ T = \ket{\h_1}\bra{T ^* \h_2}~, \\
&& \ket{\h_1}\bra{\h_2} \circ \ket{\x_1}\bra{\x_2} = 
\ket{\h_1 \hs{\h_2}{\x_1}_\ca} \bra{\x_2} = \ket{\h_1} \bra{ \hs{\h_2}{\x_1}_\ca
\x_2}~. 
\eea
From this rule, we get that the linear span of the endomorphisms of the form
(\ref{bk}) is a self-adjoint two-sided ideal in $End_\ca(\ce)$. The norm closure
in $End_\ca(\ce)$ of this two-sided ideal is denoted by $End^0_\ca(\ce)$; its
elements are called {\it compact endomorphisms} of $\ce$.

\bexam\label{ex:hba}
The Hilbert module $\ca$.\\
The $C^*$-algebra $\ca$ can be made into a (full) Hilbert Module by considering
it a {\it right} module over itself and with the following Hermitian structure
\be
\hs{~}{~}_\ca : \ce \times \ce \ra \ca~,~~ \hs{a}{b}_\ca =: a^* b~, ~~~\forall
~a,b \in
\ca~.
\ee 
The corresponding norm coincides with the norm of $\ca$ since from the norm property
(\ref{ss2}), $\norm{a}_\ca = \sqrt{\norm{\hs{a}{a}_\ca}} = \sqrt{\norm{a^*a}} =
\sqrt{\norm{a}^2} =
\norm{a}$. Thus, $\ca$ is complete also as a Hilbert module. 
Furthermore, the algebra $\ca$ being unital, one finds that $End_\ca(\ca) \simeq
End^0_\ca(\ca) \simeq \ca$, with the latter acting as multiplicative operators on
the {\it left} on itself. In particular, the isometric isomorphisms $End^0_\ca(\ca)
\simeq
\ca$ is given by 
\be
End^0_\ca(\ca) \ni \sum_k \l_k \ket{a_k} \bra{\b_k} ~\lra~
\sum_k \l_k a_k \b^*_k~, ~~~ \forall ~\l_k \in \IC~, ~a_k, b_k \in \ca~. 
\ee  
\eexam
\bexam\label{ex:hban}
The Hilbert module $\ca^N$.\\
Let $\ca^N = \ca \times \cdots \times \ca$ be the direct sum of $N$ copies of
$\ca$. It is made a full Hilbert module over $\ca$ with module action and
hermitian product given by
\bea
&& (a_1, \cdots, a_N) a =: (a_1 a, \cdots, a_N a) ~, \\
&& \hs{(a_1, \cdots, a_N)}{(b_1, \cdots, b_N)}_\ca =: \sum_{k=1}^n a_k^* b_k~,
\label{hbanhp}
\eea 
for all $a, a_k, b_k \in \ca$. The corresponding norm is
\be
\norm{(a_1, \cdots, a_N)}_\ca =: \norm{\sum_{k=1}^n a_k^* a_k}~.
\ee
That $\ca^N$ is complete in this norm is a consequence of the completeness of $\ca$
with respect to its norm. Indeed, if $(a^\a_1, \cdots, a^\a_N)_{\a\in\IN}$ is a
Cauchy sequence in $\ca^N$, then, for each component, $(a^\a_k)_{\a\in\IN}$
is a Cauchy sequence in $\ca$. The limit of $(a^\a_1, \cdots, a^\a_N)_{\a\in\IN}$ in
$\ca^N$ is just the collection of the limits from each component.

Since $\ca$ is taken to be unital, the unit vectors $\{e_k\}$ of $\IC^N$ form an
orthonormal basis for $\ca^N$ and each element of $\ca^N$ can be written uniquely
as $(a_1, \cdots, a_N) = \sum_{k=1}^N e_k a_k$ giving an identification $\ca^N \simeq
\IC^N \otc \ca$. As already mentioned,  in spite of the
orthogonality of the basis elements, one has that 
$\norm{(a_1, \cdots, a_N)}_{\ca} =: \norm{\sum_{k=1}^n a_k^* a_k} \not= 
\sum_{k=1}^n \norm{ a_k^* a_k}$. 
Parallel to the situation of the previous example, the algebra $\ca$ being unital,
one finds that $End_\ca(\ca^N) \simeq End^0_\ca(\ca^N) \simeq \IM_n(\ca)$. Here
$\IM_n(\ca)$ is the algebra of $n \times n$ matrices with entries in $\ca$; it acts
on the left on $\ca^N$. The isometric isomorphisms $End^0_\ca(\ca^N) \simeq
\IM_n(\ca)$ is now given by 
\be
End^0_\ca(\ca) \ni \ket{(a_1, \cdots, a_N)} \bra{(b_1, \cdots, b_N)} ~\lra~
\left(
\begin{array}{ccc}
a_1 b^*_1 & \cdots & a_1 b^*_N \\
\vdots & ~ & \vdots \\
a_N b^*_1 & \cdots & a_N b^*_N \\
\end{array}
\right)~, ~~~\forall ~a_k, b_k \in \ca~, 
\ee  
which is extended by linearity.
\eexam
\bexam
The sections of an Hermitian complex vector bundle.\\
Let $\ca = C(M)$ be the commutative $C^*$-algebra of complex-valued continuous
functions on the locally compact Hausdorff space $M$. Here the norm is the sup norm
as in (\ref{suno}). Given a complex vector bundle $E \ra M$, the collection $\G(E,M)$
of its continuous sections is a $C(M)$-module. This module is made a Hilbert module
if the bundles carries a Hermitian structure, namely a Hermitian scalar product 
$\hs{~}{~}_{E_p} : E_p \times E_p \ra \IC$ on each fibre $E_p$, which varies
continuously over $M$ (the space $M$ being compact, this is always the case, any such
a structure being constructed by standard arguments with a partition of unit). 
The $C(M)$-valued Hermitian structure on $\G(E,M)$ is then given by
\be
\hs{\h_1}{\h_2}(p) = ~\hs{\h_1(p)}{\h_2(p)}_{E_p}~, ~~\forall ~\h_1, \h_2
\in \G(E,M)~, ~p\in M~. 
\ee   
The module $\G(E,M)$ is complete for the associated norm. It is also full since the
linear span of  $\{\hs{\h_1}{\h_2}~, ~\h_1, \h_2 \in \G(E,M)\}$ is dense in $C(M)$.
Furthermore, one can prove (see later) that $End_{C(M)}(\G(E,M)) \simeq
End_{C(M)}^0(\G(E,M)) = \G(End E, M)$ is the $C^*$-algebra of continuous sections of
the endomorphism bundle $End E \ra M$ of $E$.

If $M$ is only locally compact, one has to consider the algebra $C_0(M)$
of complex-valued continuous functions vanishing at infinity and the corresponding
module $\G_0(E,M)$ of continuous sections vanishing at infinity which again can be
made a full Hilbert module as before. But now $End_{C(M)}(\G_0(E,M)) = \G_b(End E,
M)$, the algebra of bounded sections,  while $End_{C(M)}^0(\G_0(E,M)) = \G_0(End
E,M)$, the algebra of sections vanishing at infinity.
\eexam

It is worth mentioning that not every Hilbert module over $C(M)$ arises in the manner
described in the previous example. From the Serre-Swan theorem described in
Section~\ref{se:mod}, one obtains only (and all) projective modules of finite type. 
Now, there is a beautiful characterization of projective modules $\ce$ over a
$C^*$-algebra $\ca$ in terms of the compact operators $End^0(\ce)$ \cite{Rimor1,Mis},
\bprop\label{algpro}
Let $\ca$ be a unital $C^*$-algebra. 
\begin{enumerate}
\item Let $\ce$ be a Hilbert module over $\ca$ such that $\II_\ce \in End^0(\ce)$
(so that $End(\ce)=End^0(\ce)$). Then, the underlying right $\ca$-module is
projective of finite type.
\item Let $\ce$ be a projective module of finite type over $\ca$. Then, there exist 
$\ca$-valued hermitian structures on $\ce$ for which $\ce$ becomes a
Hilbert module and one has that $\II_\ce \in End^0(\ce)$. Furthermore, given any two 
$\ca$-valued hermitian structures $\hs{~}{~}_1$ and $\hs{~}{~}_2$, on $\ce$, there
exists an invertible endomorphism $T$ of $\ce$ such that
\be
\hs{\h}{\x}_2 = \hs{T\h}{T\x}_1~, ~~~ \forall ~\h,\x \in \ce~.
\ee
\end{enumerate}
\proof
To prove point 1., observe that by hypothesis  there are two finite strings
$\{\x_k\}, \{\z_k\}$ of elements of $\ce$ such that 
\be
\II_\ce = \sum_k \ket{\x_k} \bra{\z_k}~. 
\ee
Then, for any $\h \in \ce$, one has that 
\be
\h = \II_\ce \h = \sum_k \ket{\x_k} \bra{\z_k}\h = \sum_k \x_k  \hs{\z_k}{\h}_\ca~,
\ee
and $\ce$ is finitely generated by the string $\{\x_k\}$. If $N$ is the length of
the strings $\{\x_k\}, \{\z_k\}$, one can embed $\ce$ as a direct summand of
$\ca^N$, proving that $\ce$ is projective. The embedding and the surjection maps are 
defined respectively by
\bea
&& \l : \ce \ra \ca^N~, ~~~ \l(\h) = (\hs{\z_1}{\h}_\ca, \cdots, \hs{\z_N}{\h}_\ca)~,
\nonumber \\
&& \r : \ca^N \ra \ce~, ~~~ \r((a_1, \cdots, a_N)) = \sum_k \x_k a_k~. 
\eea
Then, for any $\h \in \ce$, $\r \circ \l (\h) = \r((\hs{\z_1}{\h}_\ca, \cdots,
\hs{\z_N}{\h}_\ca)) = \sum_k \x_k \hs{\z_k}{\h}_\ca = \sum_k \ket{\x_k} \bra{\z_k}(\h)
=\II_\ce(\h)$, namely $\r \circ \l = \II_\ce$ as required. The projector $p = \l
\circ \r$ identifies $\ce$ as $p\ca^N$. 

To prove point 2., observe that, the module $\ce$ being a direct summand of the  free
module $\ca^N$ for some $N$, the restriction of the Hermitian structure (\ref{hbanhp})
on the latter to the submodule $\ce$ makes it a Hilbert module. Furthermore, if $\r :
\ca^N \ra \ce$ is the surjection associated with $\ce$, the image $\e_k = \r(e_k),
k = 1, \dots N$, of the free basis $\{e_k\}$ of $\ca^N$ described in
Example~\ref{ex:hban} is a (not free) basis of $\ce$. Then the identity $\II_\ce$
can be written as
\be
\II_\ce = \sum_k \ket{\e_k} \bra{\e_k}~, 
\ee
and is an element of $End_\ca^0(\ce)$.
\eprop

\vfill\eject
\sect{Strong Morita Equivalence}\label{se:sme}

In this  Appendix, we describe the notion of strong Morita equivalence
\cite{Rimor, Rimor1} between two $C^*$-algebras. This really boils down to an
equivalence between the corresponding representation theories. We refer to the
previous Appendix~\ref{se:hm} for the fundamentals of Hilbert modules over a
$C^*$-algebra.
\bdefi\label{moriequi}
Let $\ca$ and $\cb$ be two $C^*$-algebras. We say that they are {\rm
strongly Morita equivalent} if there exists a {\rm $\cb$-$\ca$ equivalence Hilbert
bimodule $\ce$}, namely a module $\ce$ which is at the same time a right Hilbert
module over $\ca$ with $\ca$-valued Hermitian structure $\hs{~}{~}_\ca$, and a left
Hilbert module over $\cb$ with $\cb$-valued Hermitian structure $\hs{~}{~}_\cb$ such
that
\begin{enumerate}
\item 
The module $\ce$ is full both as a right and as a left Hilbert module;
\item
The Hermitian structure are compatible, namely
\be
\hs{\h}{\x}_\cb \z = \h \hs{\x}{\z}_\ca~, ~~~\forall ~\h, \x, \z \in \ce~;
\ee
\item
The left representation of $\cb$ on $\ce$ is a continuous $^*$-representation
by operators which are bounded for $\hs{~}{~}_\ca$,  namely $\hs{b\h}{b\h}_\ca
\leq \norm{b}^2 \hs{\h}{\h}_\ca$. \\
Similarly, the right representation of $\ca$ on $\ce$ is a continuous
$^*$-representation by operators which are bounded for $\hs{~}{~}_\cb$,  namely
$\hs{\h a}{\h a}_\cb \leq \norm{a}^2 \hs{\h}{\h}_\cb$. 
\end{enumerate}
\edefi
\bexam
For any full Hilbert module $\ce$ over the $C^*$-algebra $\ca$, the latter is
strongly Morita equivalent to the $C^*$-algebra $End^0_\ca(\ce)$ of compact
endomorphisms of $\ce$. If $\ce$ is projective of finite type so that by
Proposition~\ref{algpro} $End^0_\ca(\ce) = End_\ca(\ce)$, the algebra $\ca$ 
is strongly Morita equivalent to the whole $End_\ca(\ce)$.\\
Consider then a full {\it right} Hilbert module $\ce$ on the algebra $\ca$ with
$\ca$-valued Hermitian structure $\hs{~}{~}_\ca$. Now, $\ce$ is a {\it left} module
over the $C^*$-algebra $End^0_\ca(\ce)$.  A structure of left Hilbert module
is constructed by inverting definition (\ref{bk}) so as to produce an
$End^0_\ca(\ce)$-valued Hermitian structure on $\ce$,
\be
\hs{\h_1}{\h_2}_{End^0_\ca(\ce)} =: \ket{\h_1}\bra{\h_2}~, 
~~~\forall ~\h_1, \h_2 \in \ce~. \label{hscom}
\ee 
It is straightforward to check that the previous structure satisfies all
properties of a left structure including conjugate linearity in the second variable.
From the very definition of compact endomorphisms, the module $\ce$ is full also as
a module over $End^0_\ca(\ce)$ so that requirement $1.$ in the
Definition~\ref{moriequi} is satisfied. Furthermore, from definition \ref{bk} one
has that for any $\h_1, \h_2, \x \in \ce$,
\be
\hs{\h_1}{\h_2}_{End^0_\ca(\ce)} \x =: \ket{\h_1}\bra{\h_2}(\x) =
\h_1 \hs{\h_2}{\x}_\ca~,
\ee
so that also requirement $2.$ is met. Finally, the left action of $End^0_\ca(\ce)$
on  $\ce$ as $\ca$-module is by bounded operator. And, for any $a \in \ca, ~\h,
\x \in \ce$, one has that
\bea
\hs{\hs{\h a}{\h a}_{End^0_\ca(\ce)} \x}{\x}_\ca &=& 
\hs{(\h a) \hs{\h a}{\x}_\ca}  {\x}_\ca  \nonumber \\
&=& \hs{\h a a^* \hs{\h}{\x}_\ca}  {\x}_\ca  \nonumber \\
&=& \hs{\h}{\x}^*_\ca a a^* \hs{\h}{\x}_\ca  \nonumber \\
&\leq & \norm{a}^2 \hs{\h}{\x}^*_\ca \hs{\h}{\x}_\ca  \nonumber \\
&\leq & \norm{a}^2 \hs{\h \hs{\h}{\x}_\ca}{\x}_\ca \nonumber \\
&\leq & \norm{a}^2 \hs{\hs{\h}{\h}_{End^0_\ca(\ce)} \x} {\x}_\ca~,
\eea
from which we get
\be
\hs{\h a}{\h a}_{End^0_\ca(\ce)} \leq  \norm{a}^2  \hs{\h}{\h}_{End^0_\ca(\ce)}~,
\ee
which is the last requirement of Definition~\ref{moriequi}. 
\eexam
  
Given any $\cb$-$\ca$ equivalence Hilbert bimodule $\ce$ one can exchange the role
of $\ca$ and $\cb$ by constructing the associated {\it complex conjugate} 
\fn{Not to be confused with the dual module as introduced in eq.~(\ref{dual}).}
$\ca$-$\cb$ equivalence Hilbert bimodule $\wt{\ce}$ with a {\it right}
action of $\ca$ and a {\it left} action of
$\ca$. As an additive groups $\wt{\ce}$ is identified with $\ce$ and any element of
it will be denoted by $\wt{\h}$, with $\h\in\ce$. Then one gives a conjugate action
of $\ca$, $\cb$ (and complex numbers) with corresponding Hermitian structures. The
left action by $\ca$ and the right action by $\cb$ are defined by
\bea
&& a \cdot \wt{\h} =: \wt{\h a^*}~, ~~~ \forall ~a\in \ca~, \wt{\h} \in \wt{\ce}~,\\
&& \wt{\h} \cdot b =: \wt{b^* \h}~, ~~~ \forall ~b\in \cb~, \wt{\h} \in
\wt{\ce}~,
\eea 
and are readily seen to satisfy  the appropriate properties. As for the Hermitian
structures, they are given by
\bea
&& \hs{\wt{\h}_1}{\wt{\h}_2}_\ca =: \hs{\h_1}{\h_2}_\ca~, \\
&& \hs{\wt{\h}_1}{\wt{\h}_2}_\cb =: \hs{\h_1}{\h_2}_\cb~, ~~~\forall
~ \wt{\h}_1}, {\wt{\h}_2 \in \ce~.
\eea
Again one readily checks that the appropriate properties, notably conjugate
linearity in the second and first variable respectively, are satisfied as well as
all the other requirements for an $\ca$-$\cb$ equivalence Hilbert bimodule.

\bigskip

As already mentioned, two strongly Morita equivalent $C^*$-algebras have
equivalent representation theory. We sketch this fact in the following while referring  to
\cite{Rimor,Rimor1} for more details.

Suppose then that we are given two strongly Morita equivalent $C^*$-algebras $\ca$ and
$\cb$ with $\cb$-$\ca$ equivalence bimodule $\ce$. Let $(\ch, \p_\ca)$ be a
representation of $\ca$ on the Hilbert space $\ch$. The algebra $\ca$ acts with
bounded operators on the left on $\ch$ via $\pi$. This action can be used to
construct another Hilbert space
\be
\ch' =: \ce \ota \ch~, ~~~\h a \ota \j - \h \ota \p_\ca(a) \j = 0~, ~~~\forall
~a\in\ca, ~\h\in\ce, ~\j\in\ch~,
\ee
with scalar product
\be
(\h_1 \ota \j_1, \h_2 \ota \j_2) =: (\j_1, \hs{\h_1}{\h_2}_\ca \j_2)_\ch~,
~~~\forall ~\h_1, \h_2 \in\ce, ~\j_1, \j_2 \in\ch~.  
\ee
A representation $(\ch', \p_\cb)$ of the algebra $\cb$ is constructed by
\be
\p_\cb(b)(\h \ota \j) =: (b \h) \ota \j~, ~~~\forall ~b\in\ca, ~\h \ota \j\in\ch'~.
\ee
This representation is unitary equivalent to the representation $(\ch, \p_\ca)$. If
one starts with a representation of $\ca$, by using the conjugate $\ca$-$\cb$
equivalence bimodule $\wt{\ce}$ one constructs an equivalent representation of $\ca$.
Therefore, there is an equivalence between the category of representations of the
algebra $\ca$ and the category of representations of the algebra $\cb$

As a consequence, strong Morita equivalent $C^*$-algebras $\ca$ and $\cb$
have the same space of classes of (unitary equivalent) irreducible representations.
Furthermore, there exists also an isomorphism between the lattice of
two-sided ideals of $\ca$ and $\cb$ and  a homeomorphism between the spaces of
primitive ideals of $\ca$ and $\cb$.

In particular,  if a $C^*$-algebra $\ca$ is strongly Morita equivalent to some
commutative $C^*$-algebra, from the results of Section~\ref{se:gnt}, the latter is
unique and is the $C^*$-algebra of continuous functions vanishing at infinity on the
space $M$ of irreducible representations of $\ca$. \\ For any integer $n$, the algebra
$\IM_n(\IC) \otimes C_0(M) \simeq \IM_n(C_0(M))$ is strongly Morita equivalent to the
algebra $C_0(M)$.

We finish by mentioning that if $\ca$ and $\cb$ are two separable $C^*$-algebras and 
$\ck$ is the $C^*$-algebra of compact operators on an infinite dimensional separable
Hilbert space,  then one proves \cite{BGR} that the algebras $\ca$ and $\cb$ are
strongly Morita equivalent if and only if $\ca \otimes \ck$ is isomorphic to
$\cb \otimes \ck$. 

\vfill\eject
\sect{Partially Ordered Sets}\label{se:pos}

Here we gather few facts about partially ordered set taken mainly from
\cite{St}.
\bdefi
A {\rm partially ordered set} (or {\it poset} for short) $P$ is a set endowed with a
binary relation $\preceq $ which satisfies the following
axioms:
\begin{enumerate}
\item[~] $P_1$. ~~~ $x \preceq x$~, ~~ for all $x \in P$; ~~({\rm reflexivity}) 
\item[~] $P_2$. ~~~ $x \preceq y$ ~and~ $y \preceq x ~~\Rightarrow ~x=y$;
~~({\rm antisymmetry})
\item[~] $P_3$. ~~~ $x \preceq y$ ~and~ $y \preceq z ~~\Rightarrow ~x \preceq z$. ~~
({\rm transitivity})
\end{enumerate}
\edefi

\noindent
The relation $\preceq$ is called a {\it partial order} and the set $P$ will
be said to be partially ordered. The relation $x \preceq y$ is also read $x$ precedes
$y$.  The obvious notation $x\prec y$ will mean $x \preceq y$ and $x \not= y$;
$x \succeq y$ will mean $y \preceq x$ and $x \succ y$ will mean $y \prec x$. 
Two elements $x,y$ of $P$ are said to be {\it comparable} if $x \preceq y$ or $y
\preceq x$; otherwise they are {\it incomparable} (or {\it not comparable}). 
A subset $Q$ of $P$ is called a {\it subposet} of $P$ if it is endowed with the
induced order, namely for any $x,y \in Q$ one has $x \preceq_Q y$ in $Q$ if
and only if $x \preceq_P y$ in $P$.

An element $x \in P$ is called {\it maximal} if there is no other $y \in P$ such
that $x \prec y$. An element $x \in P$ is called {\it minimal} if there is no other
$y \in P$ such that $y \prec x$. Notice that $P$ may admit more that one maximal
and/or minimal point. One says that $P$ admits a $\hat{0}$ if there exists an
element $\hat{0} \in P$ such that $\hat{0} \preceq x$ for all $x \in P$. Similarly,
$P$ admits a $\hat{1}$ if there exists an element $\hat{1} \in P$ such that 
$x \preceq \hat{1}$ for all $x \in P$. 

\bexam
Any collection of sets can be partially ordered by inclusion. In particular,
throughout the paper we have considered at length the collection of all primitive
ideals of a $C^*$-algebras. 
\eexam 
\vskip-.75cm
\bexam
As mentioned in the previous Appendix, the set of all possible topologies on the same
space $S$ is a partially ordered set. If $\t_1$ and $\t_2$ are two topologies on the
space $S$, one puts $\t_1 \preceq \t_2$ if and only if $\t_1$ is coarser than $\t_2$.
The corresponding poset has a $\hat{0}$, the  coarsest topology, in which only
$\emptyset$ and $S$ are open, and a $\hat{1}$, the finest topology, in which all
subsets of $S$ are open.   
\eexam

Two posets $P$ and $Q$ are {\it isomorphic} if there exists an {\it order preserving
bijection} $\f : P \ra Q$, that is $x \preceq y$ in $P$ if and only if $\f(x) \preceq
\f(y)$ in $Q$, whose inverse is also order preserving.

For any relation $x \preceq y$ in $P$, we get a (closed) interval defined by 
$[x,y] = \{z \in P ~:~ x \preceq z\preceq y \}$. The poset $P$ is called {\it
locally finite} if every interval of $P$ is finite (it consists of a finite number
of elements). 

If $x,y \in P$, we say that $y$ covers $x$ if $x \prec y$ and no element
$z \in P$  satisfies $x \prec z \prec y$. A locally finite poset is completely
determined by its cover relations. 

The {\it Hasse diagram} of a (finite) poset $P$ is a graph whose vertices are the
elements of $P$ drawn in such a manner that if $x \prec y$ then $y$ is `above' $x$;
furthermore, the {\it links} are the cover relations, namely, if $y$ covers $x$ then 
a link is drawn between $x$ and $y$. One does not draw links which would be implied
by transitivity. In Section~\ref{se:ncl} we showed few Hasse diagrams. 

A {\it chain} is a poset in which any two elements are comparable. A subset $C$ of a
poset $P$ is called a chain (of $P$) if $C$ is a chain when regarded as a subposet
of $P$. The {\it length} $\ell(C)$ of a finite chain is defined as $\ell(C) =
|C|-1$, with $|C|$ the number of elements in $C$. The {\it length} (or {\it rank})
of a finite poset $P$ is defined as $\ell(P)$ =: max $\{ \ell(C) ~|$ is a chain of
$P \}$. If every maximal chain of $P$ has the same length $n$, one says that $P$ is
{\it graded of rank $n$}. In this case there is a unique {\it rank function} $\r : P
\ra \{0, 1, \dots, n\}$ such that $\r(x) = 0$ if $x$ is a minimal element and
$\r(y) = \r(x) + 1$, if $y$ covers $x$. The point $x\in P$ is said to be of {\it
rank $i$} if $\r(x) = i$.

If $P$ and $Q$ are posets, their {\it cartesian product} is the poset $P \times Q$
on the set $\{(x, y) ~:~ x \in P, y \in Q \}$ such that $(x,y) \preceq (x',y')$ in
$P \times Q$ if $x \preceq x'$ in $P$ and $y \preceq y'$ in $Q$. To draw the Hasse
diagram of $P \times Q$, one draws the diagram of $P$,, replace each element $x$ of
$P$ by a copy $Q_x$ of $Q$ and connects corresponding elements of $Q_x$ and $Q_y$
(by identifying $Q_x \simeq Q_y$) if
$x$ and $y$ are connected in the diagram of $P$.

Finally we mention that the {\it dual} of a poset $P$ is the poset $P^*$ on the
same set as $P$, but such that $x \preceq y$ in $P^*$ if and only if $y \preceq x$ in
$P$. If $P$ and $P^*$ are isomorphic, then $P$ is called {\it self-dual}.

If $x,y$ belong to a poset $P$, an {\it upper bound} of $x$ and $y$ is an element
$z\in P$ for which $x \preceq z$ and $x \preceq y$. A {\it least upper bound} of $x$ 
and $y$ is an upper bound $z$ of $x$ and $y$ such that any other upper bound $w$ of 
$x$ and $y$ satisfies $z \preceq w$. If a least upper bound of $x$ and $y$ exists,
then it is unique and it is denoted $x$ \join $y$, `{\it $x$ join $y$}'. Dually one
can define the greatest lower bound $x$ \meet $y$, `{\it $x$ meet $y$}', when it
exists. A {\it lattice} is a poset $L$ for which every pair of elements has a join
and a meet. In a lattice the operations \join and \meet satisfy the following
properties
\begin{itemize}
\item[1.] they are associative, commutative and idempotents (namely $x$ \join $x$ =
$x$ \meet $x$ = $x$); 
\item[2.] $x$ \meet ($x$ \join $y$) $ = x = $ $x$ \join ($x$ \meet $y$)
~~~(absorbation laws);
\item[3.] $x$ \meet $y = x ~\Leftrightarrow~ x$ \join $y ~\Leftrightarrow~ x
\preceq y$.
\end{itemize}
All finite lattices have the element $\hat{0}$ and the element $\hat{1}$.
%
%
%
\vfill\eject
\sect{Pseudodifferential Operators}\label{se:pdo}
We shall give a very sketchy overlook of some aspects of the theory of pseudo
differential operators  while referring to \cite{LM,Ta} for details.

Suppose we are given a rank $k$ vector bundle $E \ra M$ with $M$ a 
compact manifold of dimension $n$. We shall denote by $\G(E)$ the 
$C^\infty(M)$-module of corresponding smooth sections.\\
A {\it differential operator of rank $m$} is a linear operator
\be
P : \G(M) \lra \G(M)~,
\ee
which, in local coordinates $x=(x_1,\cdots,x_n)$ of $M$, is written as
\be
P = \sum_{|\a|\leq m} A_\a (x) (-i)^{|\a|} { \pa^{|\a|} \over \pa 
x^{\a}}~,~~~ {\pa^{|\a|} \over \pa x^{\a}} =  
{\pa^{\a_1} \over \pa x_1^{\a_1}} \circ \cdots \circ 
{\pa^{\a_n} \over \pa x_1^{\a_n}} ~.
 \ee
Here $\a = (\a_1, \cdots, \a_n), 0 \leq \a_j \leq n$, is a 
multi-index of cardinality $|\a| = \sum_{j=1}^n \a_j$. 
Each $A_\a$ is a $k \times k$ matrix of smooth functions on $M$ and 
$A_\a \neq 0$ for some $\a$ with $|\a|=m$. 

Consider now an element $\xi$ of the cotangent space $T^*_x M$, 
$\xi = \sum_j \xi_j d x_j$. 
The {\it complete symbol} of $P$ is defined by the 
following polynomial function in the components $\xi_j$.
\be
p^P(x,\xi) = \sum_{j=0}^m p^P_{m-j}(x,\xi)~,~~~ 
p^P_{m-j}(x,\xi) = \sum_{|\a|\leq (m-j)} A_\a (x) \xi^{\a}~, 
\ee
and the leading term is called the {\it principal symbol}
\be
\s^P(x,\xi) = p^P_m(x,\xi) = \sum_{|\a| = m} A_\a (x) \xi^{\a}~, 
\ee
here $\xi^\a = \xi_1^{\a_1} \cdots \xi_n^{\a_n}$. Hence, for each cotangent
vector $\xi\in T^*_xM$, the principal symbol  gives a map 
\be
\s^P(\xi) : E_x \lra E_x~,
\ee
where $E_x$ is the fibre of $E$ over $x$.
If $\t : T^*M \ra M$ is the cotangent bundle of $M$ and 
$\t^* E$ the pullback of the bundle $E$ to $T^*M$, then, the principal 
symbol $\s^P$ determines in an invariant manner a (fibre preserving) bundle 
homomorphism of $\t^*E$, namely an element of $\G(\t^* End E \ra T^* M)$. 

The differential operator $P$ is called {\it elliptic} if its principal symbol
$\s^P(\xi) : E_x \ra E_x$ is invertible for any non  zero cotangent vector $\xi \in
T^*M$. If $M$ is a Riemannian manifold with metric $g=(g^{\m\n})$, since $\s^P(\xi)$
is polynomial in $\xi$, being elliptic is equivalent to the fact that the linear 
transformation $\s^P(\x) : E_x \ra E_x$ is invertible on the cosphere bundle
\be
S^*M = \{ (x, \xi) \in T^*M ~:~ g^{\m\n}\xi_\m\xi_\n = 1\}~.
\ee 
\bexam
The Laplace-Beltrami operator $\D : C^\infty(M) \ra C^\infty(M)$ of a 
Riemannian metric $g=(g_{\m\n})$ on $M$, in local coordinates is written 
as 
\be
\D f = - \sum_{\m\n} g^{\m\n} {\pa ^2 f \over {\pa x^\m \pa x^\n} } + 
~{\rm ~lower ~order ~term}~.
\ee
As for its principal symbol we have,
\be
\s^\D(\xi) = \sum_{\m\n} g^{\m\n} \xi_\m \xi_\n = \norm{\xi}^2~,
\ee
which is clearly invertible for any non zero cotangent vector $\xi$.
Therefore, the Laplace-Beltrami operator is an elliptic second order 
differential operator.
\eexam
\bexam
Suppose now that $M$ is a Riemannian spin manifold as in
Section~\ref{se:ctm}. The corresponding Dirac operator can be written 
locally as,
\be
D = \g(dx^\m)\pa _\m + ~{\rm ~lower ~order ~term}~,
\ee
and $\g$ is the algebra morphism defined in (\ref{gamma1}).
Then, its principal symbol is just `Clifford multiplication' by $\xi$,
\be
\s^D(\xi) = \g({\xi})~.
\ee
By using (\ref{gamrel}) one gets $\g({\xi})^2 = - \norm{\xi}^2 Id$, and the 
symbol is certainly invertible for $\xi\not=0$.
Therefore, the Dirac operator is an elliptic first order 
differential operator.
\eexam

\bigskip
By using its symbol, the action of the operator $P$ on a local section $u$
of the bundle $E$ can be written as a Fourier integral,
\bea\label{four}
&& (P u)(x) = {1 \over (2\p)^{n/2}} \int e^{i \hs{\xi}{x}} p(x,\xi) 
\hat{u}(\xi) d \xi~, \nonumber \\
&& ~~~~~ \hat{u}(\xi) = {1 \over (2\p)^{n/2}} \int e^{- i \hs{\xi}{x}} 
u(x) d x~,
\eea
with $\hs{\xi}{x} = \sum_{j=1}^n \xi_j x_j$. \\
One uses formula (\ref{four}) to define {\it pseudodifferential operators},
taking $p(x,\xi)$ to belong to a more general class of symbols. The problems
is to control the growth of powers in $k$. 
We shall suppose, for simplicity, that we have a trivial vector bundle over 
$\IR^n$ of rank $k$.  

With $m \in \IR$, one defines the symbol class $Sym^m$ to consist of
matrix-valued smooth functions $p(x,\xi)$ on $\IR^n \times \IR^n$, with the
property that, for any $x$-compact $K \subset \IR^n$ and any multi-indices $\a, \b$,
there exists a constant $C_{K\a\b}$ such that 
\be\label{regu}
|D^\b_x D^\a_\xi p(x,\xi)| \leq C_{K\a\b} (1 + |\xi|)^{m-|\a|},
\ee
with $D^\b_x = (-i)^{|\b|} \pa^{|\b|} / \pa  x^{\b}$ and 
$D^\a_\xi = (-i)^{|\a|} \pa^{|\a|} / \pa  \xi^{\a}$. Furthermore, the function
$p(x,\xi)$ has an `asymptotic expansion'  given by
\be\label{asym}
p(x,\xi) \sim \sum_{j=0}^\infty p_{m-j}(x,\xi)~.
\ee
where $p_{m-j}$ are matrices of smooth functions on $\IR^n \times \IR^n$, 
homogeneous in $\xi$ of degree $(m-j)$,
\be
p_{m-j}(x,\l \xi) = \l^{m-j} p_{m-j}(x, \xi)~, ~~~|\xi| \geq 1, ~\l \geq 1~. 
\ee
The asymptotic condition (\ref{asym}) means that for any integer $N$, the difference
\be
p(x,\xi) - \sum_{j=0}^N p_{m-j}(x,\xi) = F^N(x, \xi)~
\ee
satisfies a regularity condition condition similar to (\ref{regu}): for any
$x$-compact $K\in \IR^n$ and any multi-indices $\a, \b$ there exists a constant
$C_{K\a\b}$ such that 
\be
|D^\b_x D^\a_\xi F^N(x,\xi)| \leq C_{K\a\b} (1 + |\xi|)^{m -(N+1) - |\a|}~. 
\ee
Thus, $F^N \in Sym^{m-N-1}$ for any integer $N$.\\
As we said before, any symbol $p(x,\xi) \in Sym^m$ defines a pseudodifferential
operator $P$ of order $m$ by formula (\ref{four}) where now $u$ is a section of the
rank $k$ trivial bundle over $\IR^n$ and can therefore be identified with a
$\IC^k$-valued smooth function on $\IR^n$. The space of all such operators is denoted
by $\J DO_m$.  
Let $P\in \J DO_m$ with symbol $p \in Sym^m$. Then, the {\it principal symbol} of
$P$ is the residue class $\s^P = [p] \in Sym^m / Sym^{m-1}$. One can prove that
the principal symbol transforms under diffeomorphisms as a matrix-valued function
on the cotangent bundle of $\IR^n$. 

The class $Sym^{-\infty}$ is defined by $\bigcap_{m}Sym^m$ and the corresponding
operators are called {\it smoothing operators}, the space of all such operators being
denoted by $\J DO_{-\infty}$. An smoothing operator $S$ has an integral representation
with smooth kernel, namely its action on a section $u$ can be written as 
\be
(Pu)(x) = \int K(x,y) u(y) d y ~,
\ee
where $K(x,y)$ is a smooth function on $\IR^n \times \IR^n$ (with compact support).
One is really interested in equivalence classes of pseudodifferential operators, two
operators $P, P'$ being declared equivalent if $P - P'$ is a smoothing operator.

Given $P \in \J DO_m$ and $Q \in \J DO_\m$ with symbols $p(x,\x)$ and $q(x,\x)$
respectively, the composition $R = P \circ Q \in \J DO_{m+\m}$ has symbol with
asymptotic expansion
\be
r(x,\x) \sim \sum_\a {i^{|\a|} \over \a! } D^\a_\x p(x,\x) D^\a_x q(x,\x)~. 
\ee
In particular, the leading term $|\a| = 0$ of previous expression shows that the
principal symbol of the composition is the product of the principal symbols of the 
factors
\be
\s^R(x,\x) = \s^P(x,\x) \s^Q(x,\x)~.
\ee
Given $P \in \J DO_m$, its formal adjoint $P^*$ is defined by 
\be
(Pu, v)_{L^2} = (u, P^*)_{L^2},
\ee
for all section $u,v$ with compact support. Then, $P^* \in \J DO_m$ and, if $P$ has
symbol $p(x,\x)$, the operator $P^*$ has symbol $p^*(x,\x)$ with asymptotic expansion
\be
p^*(x,\x) \sim \sum_\a {i^{|\a|} \over \a! } D^\a_\x D^\a_x (p(x,\x))^*~,
\ee
with $^*$ on the right-hand side denoting matrix Hermitian conjugation 
$(p(x,\x))^* = \bar{p(x,\x)}~^t$, $^t$ being matrix transposition. Again, by
taking the leading term $|\a| = 0$, we see that the principal symbol $\s^{P^*}$ of
$P^*$ is just the Hermitian conjugate $(\s^{P})^*$ of the principal symbol of $P$.
As a consequence, the principal symbol of a positive pseudodifferential operator
$R=P^*P$ is nonnegative.

An operator $P \in \J DO_m$ with symbol $p(x,\x)$ is said to be {\it elliptic} if its
principal symbol $\s^P \in Sym^m / Sym^{m-1}$ has a representative which, as a
matrix-valued function on $T^*\IR^n$ is pointwise invertible outside the zero section
$\x=0$ in $T^*\IR^n$. An elliptic (pseudo-)differential operator $P\in \J DO_m$
admits an inverse modulo smoothing operators. This means that there exist a pseudo
differential operator $Q \in \J DO_{-m}$ such that
\bea
&& P Q - \II = S_1~, \nonumber \\
&& Q P - \II = S_2~,
\eea 
with $S_1$ and $S_2$ smoothing operators. The operator $Q$ is called a {\it
parametrix} for $P$.

The general situation of pseudodifferential operators acting on sections of a
nontrivial vector bundle $E \ra M$, with $M$ compact, is worked out with suitable
partitions of unity. An operator $P$ acting on $\G(E \ra M)$ is a pseudodifferential
operator of order $m$, if and only if the operator $u \mapsto \f P (\j u)$ is a
pseudodifferential operator of order $m$ for any $\f, \j \in  C^\infty(M)$ which are
supported in trivializing charts for $E$. The operator $P$ is then recovered from its
components via a partition of unity.   Although the symbol of the operator $P$ will
depends on the charts, exactly as it happens for ordinary differential operators, its
principal symbol $\s^P$  has an invariant meaning as a mapping from $T^*M$ into
endomorphisms of $E \ra M$.  Thus, ellipticity has an invariant meaning and an
operator $P$ is called  elliptic if its principal symbol $\s^P$ is pointwise
invertible  off the zero section of $T^*M$.  Again, if $M$ is  a Riemannian manifold
with metric $g=(g^{\m\n})$, since $\s^P(\xi)$ is homogeneous in $\xi$, being elliptic
means that the linear  transformation $\s^P(\x) : E_x \ra E_x$ is invertible on the
cosphere bundle $S^*M \subset T^* M$.
%
%
%

\bexam\label{inverserusso}
Consider the one dimensional Hamiltonians given, in `momentum space' by 
\be
H(\x, x) = \x^2 + V(x)~,
\ee
with $V(x) \in C^\infty(\IR)$. It s clearly a differential operator of order $2$. The
following are associated pseudodifferential operators of order $-2, 1, -1$
respectively \cite{russo}, 
\bea\label{pdrusso}
&& (\x^2 + V)^{-1} = \x^{-2} - V \x^{-4} + 2 V^{(1)} \x^{-5} + ... ~, \nonumber \\
&& (\x^2 + V)^{1/2} = \x  + {V \over 2} \x^{-1} - {V^{(1)} \over 4} \x^{-2} + ... ~,
\nonumber \\
&& (\x^2 + V)^{-1/2} = \x^{-1} - {V \over 2} \x^{-3} + {3 V^{(1)} \over 4} \x^{-4} +
... ~,
\eea
where $V(k)$ is the $k$-th derivative of $V$ with respect to its argument. \\
In particular, for the one dimensional harmonic oscillator $V(x) = x^2$. The
pseudodifferential operators in (\ref{pdrusso}) become,  
\bea\label{pdrussoho}
&& (\x^2 + x^2)^{-1} = \x^{-2} - x^2 \x^{-4} + 4x \x^{-5} + ... ~, \nonumber \\
&& (\x^2 + x^2)^{1/2} = \x  + {x^2 \over 2} \x^{-1} - {x \over 2} \x^{-2} + ... ~.
\nonumber \\
&& (\x^2 + x^2)^{-1/2} = \x^{-1} - {x^2 \over 2} \x^{-3} + {3 x \over 2}
\x^{-4} + ... ~.
\eea
\eexam

\vfill\eject
\bibliographystyle{unsrt}

\end{document}